\documentclass[review]{elsarticle}

\pdfoutput=1
\usepackage{graphicx}
\usepackage{amsfonts, amsmath, amssymb}
\usepackage{booktabs,siunitx}
\usepackage[svgnames,table]{xcolor}
\usepackage[tableposition=above]{caption}
\usepackage{pifont}
\usepackage{bm}
\usepackage{cancel}
\usepackage{caption}
\usepackage{color}
\usepackage{enumerate}
\usepackage{float}
\usepackage{hyperref}
\usepackage[english]{babel}
\usepackage[margin=1in]{geometry}
\usepackage{mathtools}
\usepackage{setspace}
\usepackage{subfigure}
\usepackage{lineno}

\newcommand \D [2]{\frac{\partial #1}{\partial #2}}

\renewcommand{\vec}[1]{\bm{\mathrm{#1}}}
\newcommand{\V}[1]{\bm{\mathrm{#1}}}

\def \div{\nabla \cdot \mbox{}}
\def \grad{\nabla}

\def \x{\vec{x}}
\def \y{\vec{y}}

\def \n{\vec{n}}

\def \u{\vec{u}}

\def \e{\vec{e}}

\def \I{\vec{I}}
\def \F{\vec{F}}
\def \N{\vec{N}}

\def \U{\vec{U}}
\def \L{\vec{L}}

\def \Sb{S_b}
\def \Sbt{S_b(t)}

\def \Vbt{V_b(t)}
\def \Vb{V_b}

\def \C{\vec{C}}

\def \cF{\vec{\mathcal{F}}}

\def \F{\vec{F}}

\def \g{\vec{g}}

\def \I{\vec{I}}
\def \Ib{\I_{\text{b}}}

\def \Mb{\text{M}_{\text{b}}}
\def \N{\vec{N}}
\def \Nx{N_x}
\def \Ny{N_y}
\def \Nz{N_z}

\def \Omegal{\Omega_{\text{l}}}
\def \Omegag{\Omega_{\text{g}}}

\def \R{\vec{R}}
\def \U{\vec{U}}
\def \Ub{\U_{\text{b}}}

\def \Ur{\U_{\text{r}}}
\def \W{\vec{W}}

\def \Wr{\W_{\text{r}}}

\def \X{\vec{X}}
\def \Xcom{\X_{\text{com}}}

\def \cS{\vec{\mathcal{S}}}
\def \cJ{\vec{\mathcal{J}}}
\def \cH{\mathcal{H}}
\def \cT{\mathcal{T}}

\def \e{\vec{e}}

\def \f{\vec{f}}

\def \fc{\f_{\text{c}}}
\def \fs{\f_{\text{s}}}

\def \half{\frac{1}{2}}
\def \3half{\frac{3}{2}}
\def \5half{\frac{5}{2}}

\def \mul{\mu_{\text{l}}}
\def \mus{\mu_{\text{s}}}
\def \mug{\mu_{\text{g}}}
\def \n{\vec{n}}

\def \nref{n_{\text{ref}}}
\def \ncells{n_{\text{cells}}}
\def \ncycles{n_{\text{cycles}}}

\def \rhol{\rho_{\text{l}}}
\def \rhos{\rho_{\text{s}}}
\def \rhog{\rho_{\text{g}}}
\def \s{\vec{s}}

\def \sgn{\textrm{sgn}}
\def \u{\vec{u}}

\def \uw{u_{\text{w}}}
\def \vw{v_{\text{w}}}

\def \x{\vec{x}}

\def \xL{x_L}

\def \xU{x_U}

\def \div{\nabla \cdot \mbox{}}
\def \grad{\nabla}

\def \dt{\Delta t}
\def \dx{\Delta x}
\def \dy{\Delta y}
\def \dz{\Delta z}
\def \delV{\Delta V}

\def \delU{\Delta \vec{U}}

\def \Ds{{\mathrm d}\s}

\def \dS{\,\mathrm{dS}}

\def \Dx{{\mathrm d}\x}

\def \ds{\Delta s}
\def \dt{\Delta t}
\def \dx{\Delta x}

\def \ndot{\n \cdot}

\newcommand{\upperRomannumeral}[1]{\uppercase\expandafter{\romannumeral#1}}

\begin{document}

\begin{frontmatter}
	
\title{A DLM immersed boundary method based wave-structure interaction solver for high density ratio multiphase flows}
\author[Northwestern1]{Nishant Nangia}
\author[Northwestern1,Northwestern2]{Neelesh A. Patankar\corref{mycorrespondingauthor}}
\ead{n-patankar@northwestern.edu}
\author[SDSU]{Amneet Pal Singh Bhalla\corref{mycorrespondingauthor}}
\ead{asbhalla@sdsu.edu}

\address[Northwestern1]{Department of Engineering Sciences and Applied Mathematics, Northwestern University, Evanston, IL}
\address[Northwestern2]{Department of Mechanical Engineering, Northwestern University, Evanston, IL}
\address[SDSU]{Department of Mechanical Engineering, San Diego State University, San Diego, CA}
\cortext[mycorrespondingauthor]{Corresponding author}

\begin{abstract}
In this paper we present a robust immersed boundary (IB) method for high density ratio multiphase flows 
that is capable of modeling complex wave-structure interaction (WSI) problems arising in marine and coastal 
engineering applications. The IB/WSI methodology is enabled by combining the distributed Lagrange multiplier (DLM) method of 
Sharma and Patankar (2005)~\cite{Sharma2005} with a robust level set method based multiphase flow 
solver. The fluid solver integrates the conservative form of  the variable-coefficient incompressible Navier-Stokes equations 
using a hybrid preconditioner and ensures consistent transport of mass and momentum at a discrete level. 
The consistent transport scheme preserves the numerical stability of the method in the presence of 
large density ratios found in problems involving air, water, and an immersed structure. The air-water interface is captured by 
the level set method on an Eulerian grid, whereas the free-surface piercing immersed structure is represented 
on a Lagrangian mesh. The Lagrangian structure is free to move on the background Cartesian grid 
without conforming to the grid lines. The fluid-structure interaction (FSI) coupling is 
mediated via Peskin's regularized delta functions in an implicit manner, which obviates the need to integrate the 
hydrodynamic stress tensor on the complex surface of the immersed structure. The IB/WSI numerical scheme is 
implemented within an adaptive mesh refinement (AMR) framework, in which the Lagrangian structure and the air-water 
interface are embedded on the finest mesh level to capture the thin boundary layers and  the vortical structures arising 
from WSI. We use a well balanced gravitational force discretization that eliminates spurious velocity currents in 
the hydrostatic limit due to density variation in the three phases (air, water and solid).
We also show that using a non-conservative and an inconsistent fluid solver can lead to catastrophic failure of the numerical scheme 
for large density ratio variations that are prevalent in WSI applications.
An effective wave generation and absorption
technique for a numerical wave tank is presented and used to simulate a benchmark case of water wave distortion due
to a submerged structure.
The numerical scheme is tested 
on several benchmark WSI problems from numerical and experimental literature in both two and three dimensions
to demonstrate the applicability of the IB/WSI method to practical marine and coastal engineering 
problems.
\end{abstract}


\end{frontmatter}

\section{Introduction}

Wave-structure interaction phenomena are critical design considerations for marine engineers to ensure 
the safe operability of coastal and offshore structures. In marine and coastal engineering 
applications, complex floating structures, such as floating oil platforms, wave energy converter (WEC)
devices, and foundations of offshore wind turbines, are subject to wave loading, wave run-up, wave scattering, 
and wave breaking effects, which can severely damage or affect the performance of these structures. 

Recently, development of marine renewable energy has received renewed interest within the scientific community
due to fluctuating oil prices and the negative impact of fossil fuels on the environment.
It is estimated that 2.11 $\pm$ 0.05 TW of coastal 
wave energy is available globally, with equal amounts in the Northern and Southern hemispheres~\cite{Gunn12}. Simulations of WECs can 
help increase their power extraction capacity by interrogating the underlying physics.
However, many existing numerical models of WECs 
are based on simplified flow physics, i.e. by assuming inviscid potential flow equations that are mostly linear~\cite{Dean1991,Eriksson2005} 
or weakly nonlinear~\cite{Madsen98,Booij99}. Viscous drag in such models is generally accounted for by 
using Morison's equation~\cite{Morison1950}, which is valid only for slender 
offshore structures. Moreover, Morison's equation has been obtained empirically from experimental measurements 
for limited wave conditions and is not valid over all flow regimes~\cite{Chen14,Sarpkaya2010}. Therefore, these 
methods cannot handle free-surface and wave breaking effects around the structure, which are highly 
nonlinear in nature. Neglecting realistic sea or ocean conditions 
can lead to suboptimal design for WEC devices. Fully-resolved wave-structure interaction simulations of WECs 
are closer to reality as they model all three phases, but are considerably costlier than potential flow models in turnaround time. 

Traditionally, fluid-structure interaction (FSI) problems involving the full incompressible Navier-Stokes (INS) system of equations have been 
modeled by using Arbitrary Lagrangian-Eulerian (ALE) methods~\cite{Hu01} on body conforming grids. The main advantage of an ALE-like 
approach is that the boundary conditions on the fluid-structure~\cite{Hu01} or the fluid-fluid~\cite{Ramaswamy1987} interface can be satisfied exactly.  
For single-phase FSI applications, ALE methods can be used to obtain high-resolution results, albeit at the cost of 
frequent re-meshing of the entire computational domain due to the structural displacement~\cite{Kern06}.  However for WSI applications 
where the air-water interface undergoes non-smooth and non-continuous topological changes due to wave-breaking processes, 
the application of ALE methods is not practical.

To overcome these limitations, fictitious domain~\cite{Patankar2000,Sharma2005} 
or immersed boundary (IB) methods~\cite{Peskin02}, combined with 
level-set~\cite{Osher1988} or volume of fluid (VOF) approaches~\cite{Hirt1981} are gaining popularity for both single phase FSI 
applications and multiphase FSI applications~\cite{Zhang2010,Calderer2014,Pathak16,Bihs2016,Wang2017,Patel2018}. There are two major implementation categories of the IB method --- diffuse and sharp. 
In the diffuse IB approach, the fluid equations are extended inside the structure domain so that regular and fast 
Cartesian solvers can be used to solve the INS equations everywhere in the computational domain. An additional body force is applied in the structure domain, 
which is conveniently represented on a Lagrangian mesh to constrain the motion of the fluid occupying the solid region as a rigid body motion. 
The most efficient way to compute the FSI body force is through distributed Lagrange multipliers (DLM), a method pioneered 
by Patankar et al.~\cite{Patankar2000}.  A fractional time stepping approach is used to impose the DLM-based rigidity constraint, which is suitable 
for moderate to high Reynolds number flows~\cite{Sharma2005,Bhalla13}. For zero Reynolds number Stokes flow, the 
DLM or constraint force needs to be computed simultaneously by solving an extended saddle point system, along with 
fluid velocity and pressure degrees of freedom. This is because Stokes flow is a purely 
elliptic system describing a force equilibration process and any fractional time stepping scheme introduces a large numerical error in its solution. 
Kallemov et al.~\cite{Kallemov16} and Usabiaga et al.~\cite{Usabiaga17} describe an efficient preconditioner for the 
monolithic fluid-DLM solver. The FSI coupling for diffuse IB methods is mediated via Peskin's regularized delta functions, in which the 
Lagrangian DLM force is \emph{spread} onto the background Eulerian grid and the fluid velocity is \emph{interpolated} onto the Lagrangian mesh. The use of regularized delta functions smears the fluid-structure interface over a few grid cells 
(according to the delta function support), which makes the interface diffuse rather than keeping it sharp. In this work we use an 
efficient fractional time stepping, diffuse DLM approach to model WSI. Sharp IB methods, on the other hand, imposes 
the velocity of the fluid-structure interface at the nearby ``IB nodes". This is achieved by fitting a spatial polynomial 
(linear or quadratic) through the solid interface and fluid nodes. The INS equations are solved only at the fluid nodes, with the 
IB nodes 
acting as velocity boundary conditions. The velocity and pressure values for the interior solid nodes are zeroed-out 
during the solution procedure, which then creates ``punctured" domain effects.
The most notable sharp IB method implementations 
and their extensions have been carried by Borazjani et al.~\cite{Borazjani13}, Mittal et al.~\cite{RMittal08}, 
Udaykumar et al.~\cite{Uday01}, and Tseng and Ferziger~\cite{YHTseng03}.           

There are several advantages and disadvantages to both diffuse and sharp IB methods. For example, diffuse IB methods
permit a continuous solution of velocity and pressure in the entire domain, which eliminates ``spurious force 
oscillations" (SFO) in the time histories of the integrated drag and lift quantities for the moving immersed bodies. In contrast, 
spurious force oscillations are an outstanding issue for the sharp IB 
methodology because of the punctured domain effect~\cite{Lee11,Martins17,Patel2018}. Since the solution is continuous 
throughout the domain for diffuse IB methods, there are no issues with ``fresh" and ``dead"
fluid cells when the structure changes it 
location in the domain, which is an challenging issue for sharp IB methods.
Diffuse IB methods also allow for an implicit coupling of the fluid and 
structure domains without requiring hydrodynamic stress tensor computations on the (possibly complex) surface of the immersed 
structure~\cite{Nangia17}. In contrast, sharp IB methods compute pointwise hydrodynamic forces on the immersed surface 
and often require several fluid and structure solver iterations to 
converge to a stable solution within a single time step~\cite{Calderer2014,Borazjani2008}. The main disadvantage of diffuse 
IB methods is the smearing of the fluid-structure 
interface over few grid cells, which reduces the accuracy of the solution near the interface. The order of accuracy for diffuse IB methods is generally 
between one and two; the former for non-smooth and the latter for sufficiently smooth FSI problems.
In contrast, sharp IB methods retain full second-order accuracy by sharply resolving the fluid-structure interface. Diffuse IB methods are also known to produce non-smooth 
pointwise hydrodynamic stress on the immersed surface even though the net hydrodynamic force and torque are smooth and SFO-free.  
This issue can however be mitigated by interpolating the hydrodynamic stress sufficiently far away from the fluid-structure interface. 
The lack of geometric information for the immersed surface also makes the implementation of wall functions required for turbulence 
modeling difficult for diffuse IB methods. For sharp IB methods, application of Robin-type boundary conditions and 
implementing wall functions is quite natural. The SFO in sharp IB methods can be mitigated by increasing the grid resolution 
and using larger time steps. However, very refined meshes can make the simulations extremely expensive and the use of large time 
steps can make them unstable unless fully implicit time stepping schemes are used for the INS equations.  The demarcation of grid nodes 
into ``IB nodes", ``fluid nodes" and ``solid nodes" is a computationally taxing task as well and a novice procedure to reconstruct 
the IB node velocity (from interface and fluid nodes) can lead to numerical instabilities for certain geometric configurations of the interface relative to 
the background Cartesian grid~\cite{Borazjani2008,YHTseng03,RMittal08}.
We remark that in spite of the aforementioned shortcomings 
of the diffuse and sharp IB methods, both have been applied successfully to solve complicated engineering problems.  
Combined with level set or volume of fluid methods that can capture the air-water interface on 
Eulerian grids, these IB methods allow for an efficient solution of topologically complex WSI problems.

An issue that is unique to WSI or two-phase multiphase flows is the presence of highly contrasting density ratios in the computational domain.
High density ratio multiphase flows are known to develop numerical instabilities whenever convection is the dominant 
physical process~\cite{Patel2018,Nangia2018,Patel2017,Raessi2008,Raessi2012,Pathak16,Desjardins2010,Ghods2013}. 
Recently the multiphase community (including us) has proposed several stabilizing remedies for convection-dominated, 
high density ratio multiphase flows, for solvers based on both
volume of fluid and level set methods~\cite{Nangia2018,Patel2017,Pathak16}. 
The underlying cause of the instability 
is the inconsistent transport of mass and momentum at a discrete level. In this work we achieve a consistent transport of mass and momentum by solving an additional mass balance equation using a strong-stability preserving Runge-Kutta (SSP-RK3) integrator~\cite{Gottlieb2001}. The mass flux that updates the density variable is also used to construct a discrete convective operator for the momentum equation. This necessarily requires solving the conservative form of the mass balance and momentum equations. The strong coupling between (discrete) mass and momentum convective operators preserves the stability of the numerical scheme for density ratios as high as $10^6$. Our multiphase flow solver is based on the level set method, which makes the implementation of the proposed IB/WSI methodology relatively easier (than VOF methods) on locally refined meshes. We employ a hybrid preconditioner that solves the velocity and pressure degrees of freedom simultaneously, i.e., we do not use a projection-method (which is an operator-splitting approach) to solve the INS equations~\cite{Cai2014}. Only the distributed Lagrange multipliers for the FSI coupling are imposed via operator-splitting. For computational efficiency the air-water interface and the immersed structure are resolved on the finest mesh level, whereas the rest of the computational domain is resolved on progressively coarser grids. Therefore, we are able to capture important flow features at a substantially reduced computational cost, especially in 3D. Since we extend the fluid equations inside the solid domain and since the density of the structure is different than surrounding fluid (almost always heavier than air for WSI applications), the gravitational body force can produce spurious velocity currents near the fluid-structure interface for certain cases. Similarly, due to the density contrast of air and water, spurious velocity currents can also form near the air-water interface. In this work we employ a well-balanced gravity force discretization that eliminates such spurious currents near the air-water-solid interface even in the hydrostatic limit. Section~\ref{sec_ex_wellbalance} provides a numerical example that highlights this problem and shows the numerical ``fix". 

The remainder of the paper is organized as follows. We first introduce the continuous and discrete system of equations in Secs.~\ref{sec_cont_eqs} and~\ref{sec_discrete_eqs}, respectively. Next we discuss the solution methodology in Sec.~\ref{sec_sol_method}. Section~\ref{sec_solid_materials} comments on the well-balanced gravity force implementation and Sec.~\ref{sec_ex_wellbalance} presents the corresponding numerical example. Software implementation is described in Sec.~\ref{sec_sfw}. Section~\ref{sec_wsi} describes the implementation of a numerical wave tank based on the level set method, and demonstrates the interaction of a Stokes second-order wave in the presence of a submerged structure. Finally, more complicated three-phase flow examples that demonstrate the applicability of the proposed IB/WSI methodology 
to simulate free-surface piercing and floating structures are presented in Sec.~\ref{sec_examples}. We also contrast the consistent results from the conservative flow solver against 
the unstable results obtained from an inconsistent and non-conservative flow solver to highlight the importance of consistent mass and momentum 
transport for practical WSI applications. Wherever possible, simulation results from locally refined grids are presented.

\section{The continuous equations of motion} \label{sec_cont_eqs}
\subsection{Multiphase constraint immersed boundary formulation}
We begin by stating the governing equations for a multiphase fluid-structure system
occupying a fixed region of space $\Omega \subset \mathbb{R}^d$, for $d = 2$ or $3$
spatial dimensions. In the immersed boundary formulation, a fixed Eulerian coordinate
system $\x = (x_1, \ldots, x_d) \in \Omega$ is used to describe the momentum equation and divergence-free condition
for both the fluid and structure. It is convenient to employ a Lagrangian description of the immersed body configuration,
in which $\s = (s_1, \ldots s_d) \in B$ denotes the fixed material coordinate system 
attached to the structure and $B \subset \mathbb{R}^d$ is the Lagrangian curvilinear 
coordinate domain.
The position of the immersed structure occupying a volumetric 
region $\Vbt \subset \Omega$ at time $t$ is denoted by $\X (\s,t)$.
In contrast with the previous formulation of the DLM or constraint immersed boundary method~\cite{Bhalla13},
we allow for a spatially and temporally varying density $\rho(\x,t)$ and dynamic viscosity $\mu(\x,t)$,
implying that the structure can be heavier or lighter than the surrounding fluids.
Hence, the equations of motion for the coupled fluid-structure system in \emph{conservative} form are

\begin{align}
\D{\rho \u(\x,t)}{t} + \div \rho\u(\x,t)\u(\x,t) &= -\grad p(\x,t) + \div \left[\mu \left(\grad \u(\x,t) + \grad \u(\x,t)^T\right) \right]+ \rho\g + \fs(\x,t) + \fc(\x,t) , \label{eqn_momentum}\\
  \div \u(\x,t) &= 0, \label{eqn_continuity} \\
\fc(\x,t)  &= \int_{B} \F(\s,t) \, \delta(\x - \X(\s,t)) \, \Ds, \label{eqn_F_f} \\
  \U(\s,t) &= \int_{\Omega} \u(\x,t) \, \delta(\x - \X(\s,t)) \, \Dx, \label{eqn_u_interpolation} \\
   \D{\X}{t} (\s,t) &= \U(\s,t). \label{eqn_body_motion} 
\end{align}

Eqs.~\eqref{eqn_momentum} and \eqref{eqn_continuity} are the incompressible 
Navier-Stokes momentum and continuity equations written in Eulerian form, in which $\u(\x,t)$ is the velocity, 
$p(\x,t)$ is the pressure, and $\fc(\x,t)$ is the Eulerian constraint force density, which is non-zero 
only in the structure region. The gravitational acceleration is denoted by $\g = (g_1, \ldots, g_d)$,
and $\fs(\x,t)$ is the continuum surface tension force. The interactions between Eulerian and Lagrangian
quantitates are facilitated by Dirac delta function kernels, in which the $d$-dimensional delta function 
is $\delta(\x) = \Pi_{i=1}^{d}\delta(x_i)$. Eq.~\eqref{eqn_F_f} converts the Lagrangian 
force density $\F(\s,t)$ into an equivalent Eulerian density $\fc(\x,t)$, in an operation called \emph{force spreading}.
 Eq.~\eqref{eqn_u_interpolation} determines the physical velocity of each Lagrangian 
material point from the background Eulerian velocity field in an operation called \emph{velocity interpolation}.
This ensures that the immersed structure moves 
according to the local value of the velocity field $\u(\x,t)$ (Eq.~\eqref{eqn_body_motion}), and thus the
no-slip condition is satisfied at fluid-solid interfaces. Using short-hand notation, the force spreading operation
is denoted by $\fc = \cS[\X] \, \F$, in which $\cS[\X]$ is the \emph{force-spreading operator} and the velocity
interpolation operation is denoted by $\D{\X}{t} = \U = \cJ[\X] \, \u$, in which $\cJ[\X]$ is the
\emph{velocity-interpolation operator}.
It can be shown that if $\cS$ and $\cJ$ are taken to be adjoint operators, i.e. $\cS = \cJ^{*}$, then 
the Lagrangian-Eulerian coupling conserves energy~\cite{Peskin02}.

The specific rigidity constraint imposed within the structure domain,
written in Lagrangian form, is given by
\begin{equation}
\label{eq_vel_constraint}
\frac{1}{2}\left[\grad \U(\s,t) + \grad \U(\s,t)^T\right] = 0,
\end{equation}
which states that the body has zero deformation rate and must undergo a rigid body motion~\cite{Patankar2000}.
In the present work, we compute a discrete approximation
to the constraint force $\F(\s,t)$, although a numerical method that enforces this constraint exactly for a range
of Reynolds numbers (including zero Reynolds number Stokes flow) has been described by one of us 
in~\cite{Kallemov16,Usabiaga17}.

Note that the momentum equation (Eq.~\eqref{eqn_momentum}) can also be cast to an equivalent \emph{non-conservative}
form. However, it has been shown that direct discretization of the non-conservative form can lead to numerical
instabilities for high density ratio multiphase flows~\cite{Raessi2008, Raessi2012, Desjardins2010, Ghods2013, Nangia2018}.
The differences between the conservative and non-conservative flow solvers will be discussed in later sections.

\subsection{Interface tracking for material properties}
\label{sec_cont_ls}
Next, we describe the governing equations for tracking and transporting material properties.
Suppose a liquid of density $\rhol$ and viscosity $\mul$ occupies a region $\Omegal(t) \subset \Omega$, while
a gas of density $\rhog$ and viscosity $\mug$ occupies a region $\Omegag(t) \subset \Omega$. The codimension-$1$
interface between these two fluids is denoted by $\Gamma(t) = \Omegal \cap \Omegag$ can be tracked as the zero contour
of a scalar function $\phi(\x,t)$, which is the so-called level set function~\cite{Osher1988,Sethian2003,Sussman1994},
\begin{equation}
\Gamma(t) = \{\x \in \Omega \mid \phi(\x,t) = 0\}. \label{eq_interface}
\end{equation}
Level set methods are particularly well-suited for tracking liquid-gas interfaces
undergoing complex topological changes and are relatively simple to implement in both two and three
spatial dimensions, and on locally refined meshes. It is also useful to define an additional level set 
function $\psi(\x,t)$ to track the boundary of the immersed
structure $\Sb(t) = \partial \Vb(t)$. Using this auxiliary field, the density $\rhos$ and the viscosity $\mus$ in the solid region
can be readily prescribed in the Eulerian regions occupied by the solid $\Vb(t) \subset \Omega$.
Both level set functions are passively advected by the incompressible fluid velocity,
which in conservative form reads
\begin{align}
\D{\phi}{t} + \div \phi \u &= 0, \label{eq_ls_fluid_advection} \\
\D{\psi}{t} + \div \psi \u &= 0. \label{eq_ls_solid_advection}
\end{align}
The material properties including density and viscosity in the three phases are determined as a function of these two scalar fields by 
\begin{align}
\rho (\x,t) &= \rho(\phi(\x,t), \psi(\x,t)), \label{eq_rho_ls}\\
\mu (\x,t) &= \mu(\phi(\x,t), \psi(\x,t)) \label{eq_mu_ls}.
\end{align}
The discretized form of Eqs.~\eqref{eq_rho_ls} and~\eqref{eq_mu_ls} are defined in Section~\ref{sec_reinit}
using regularized Heaviside functions.

One particularly useful level set function is the \emph{signed distance function}, which can be prescribed as initial
conditions to Eqs.~\eqref{eq_ls_fluid_advection} and~\eqref{eq_ls_solid_advection}
\begin{align}
\phi\left(\x, 0\right) &= 
\begin{cases} 
       \min\limits_{\y \in \Gamma(0)} \|\x-\y\|,  & \x \in \Omegag(0), \\
        -\min\limits_{\y \in \Gamma(0)} \|\x-\y\|,  & \x \in \Omegal(0), \label{eq_signed_phi}
\end{cases} \\
\psi\left(\x, 0\right) &= 
\begin{cases} 
       \min\limits_{\y \in \Sb(0)} \|\x-\y\|,  & \x \not \in \Vb(0), \\
        -\min\limits_{\y \in \Sb(0)} \|\x-\y\|,  & \x \in \Vb(0). \label{eq_signed_psi}
\end{cases}
\end{align}
However, we note that $\phi$ and $\psi$ generally will not remain signed
distance functions under advection by Eqs.~\eqref{eq_ls_fluid_advection} 
and~\eqref{eq_ls_solid_advection}. A \emph{reinitialization}
or redistancing procedure is used to maintain the signed distance property at every time step.
When the fluid properties are determined from the level set fields,
we need initial conditions for $\phi$ and $\psi$ but not for $\rho$ or $\mu$.

\section{Spatial discretization} \label{sec_discrete_eqs}
This section describes the discrete form of the governing equations for the coupled
fluid-structure system. Eulerian quantities are discretized on a staggered Cartesian
grid, whereas Lagrangian quantities are approximated on a collection of immersed markers
that can be arbitrarily positioned on the grid. A regularized version of the Dirac delta
function is used to facilitate the velocity interpolation and force spreading operations.
Therefore, we are \emph{not} employing a body-conforming mesh to the fluid-structure
interface since the structure markers need not conform to the Eulerian grid. 

Throughout this section, we describe the discretization for $d = 3$ spatial dimensions;
the discretization in two spatial dimensions is analogous.
For simplicity, we describe the case for which there is no local grid refinement in the domain,
although this is not a limitation of the present formulation. Details on adaptive mesh
refinement are delegated to Sec.~\ref{sec_amr}. Finally, we note that evaluating the discrete operators
described in this section near boundaries of the computational domain and locally refined mesh boundaries
requires the specification of adjacent ``ghost" cells. For more details on the treatment of boundary conditions
and coarse-fine interfaces, we refer readers to~\cite{Bhalla13,Nangia2018,Griffith2009,Griffith2012}.
\subsection{Eulerian discretization}
\label{eulerian_discretization}
We employ a staggered grid discretization for quantities described in the Eulerian frame.
A $\Nx \times \Ny \times \Nz$ Cartesian grid covers the physical, rectangular domain
$\Omega$ with mesh spacing $\dx$, $\dy$, and $\dz$ in each direction.
Without loss of generality, we assume that the bottom left corner of the domain
is situated at the origin $(0,0,0)$. Therefore, each cell center of the grid has
position $\x_{i,j,k} = \left((i + \half)\dx,(j + \half)\dy,(k + \half)\dz\right)$
for $i = 0, \ldots, \Nx - 1$, $j = 0, \ldots, \Ny - 1$, and $k = 0, \ldots, \Nz - 1$.
For a given cell $(i,j,k)$, $\x_{i-\half,j,k} = \left(i\dx,(j + \half)\dy, (k + \half)\dz\right)$ is the physical location of the cell face 
that is half a grid space away from $\x_{i,j,k}$ in the $x$-direction,
$\x_{i,j-\half,k} =\left((i + \half)\dx,j\dy,(k + \half)\dz\right)$ is the physical location of the cell 
face that is half a grid cell away from  $\x_{i,j,k}$ in the $y$-direction,
and $\x_{i,j,k-\half} =\left((i+\half)\dx,(j + \half)\dy,k\dz\right)$ is the physical location of the cell 
face that is half a grid cell away from  $\x_{i,j,k}$ in the $z$-direction.
The pressure degrees of freedom are approximated at cell centers and are
denoted by $p_{i,j,k}^{n} \approx p\left(\x_{i,j,k},t^{n}\right)$, in which $t^n$ is the time at time step $n$.
Similarly, the flow and structure level set functions are also defined at cell centers and are denoted
by $\phi_{i,j,k}^{n} \approx \phi\left(\x_{i,j,k}, t^n\right)$ and
$\psi_{i,j,k}^{n} \approx \psi\left(\x_{i,j,k}, t^n\right)$, respectively.

Velocity components are staggered and defined on their respective cell faces:
$u_{i-\half,j,k}^{n} \approx u\left(\x_{i-\half,j,k}, t^{n}\right)$, 
$v_{i,j-\half,k}^{n} \approx v\left(\x_{i,j-\half,k}, t^{n}\right)$,
and $w_{i,j,k-\half}^{n} \approx w\left(\x_{i,j,k-\half}, t^{n}\right)$.
The components of various body forces on the right-hand side of the momentum equation
(Eq.~\eqref{eqn_momentum}) are similarly approximated on respective faces of the staggered grid.
The density and viscosity are approximated at cell centers and
are denoted by $\rho_{i,j,k}^{n} \approx \rho\left(\x_{i,j,k},t^{n}\right)$
and  $\mu_{i,j,k}^{n} \approx \mu\left(\x_{i,j,k},t^{n}\right)$.
These quantities are interpolated onto the required degrees of freedom as needed~\cite{Nangia2018}.

Standard second-order finite differences are used to approximate spatial derivative
operators and are denoted with $h$ subscripts; i.e. $\grad \approx \grad_h$. The full description
of these staggered grid discretizations have been recorded in various prior studies and we refer
readers to~\cite{Nangia2018,Cai2014,Griffith2009,Harlow1965,Guermond2006} for more details. 

\subsection{Lagrangian discretization}
\begin{figure}[]
  \centering
  \subfigure[Continuous domain]{
    \includegraphics[scale = 0.32]{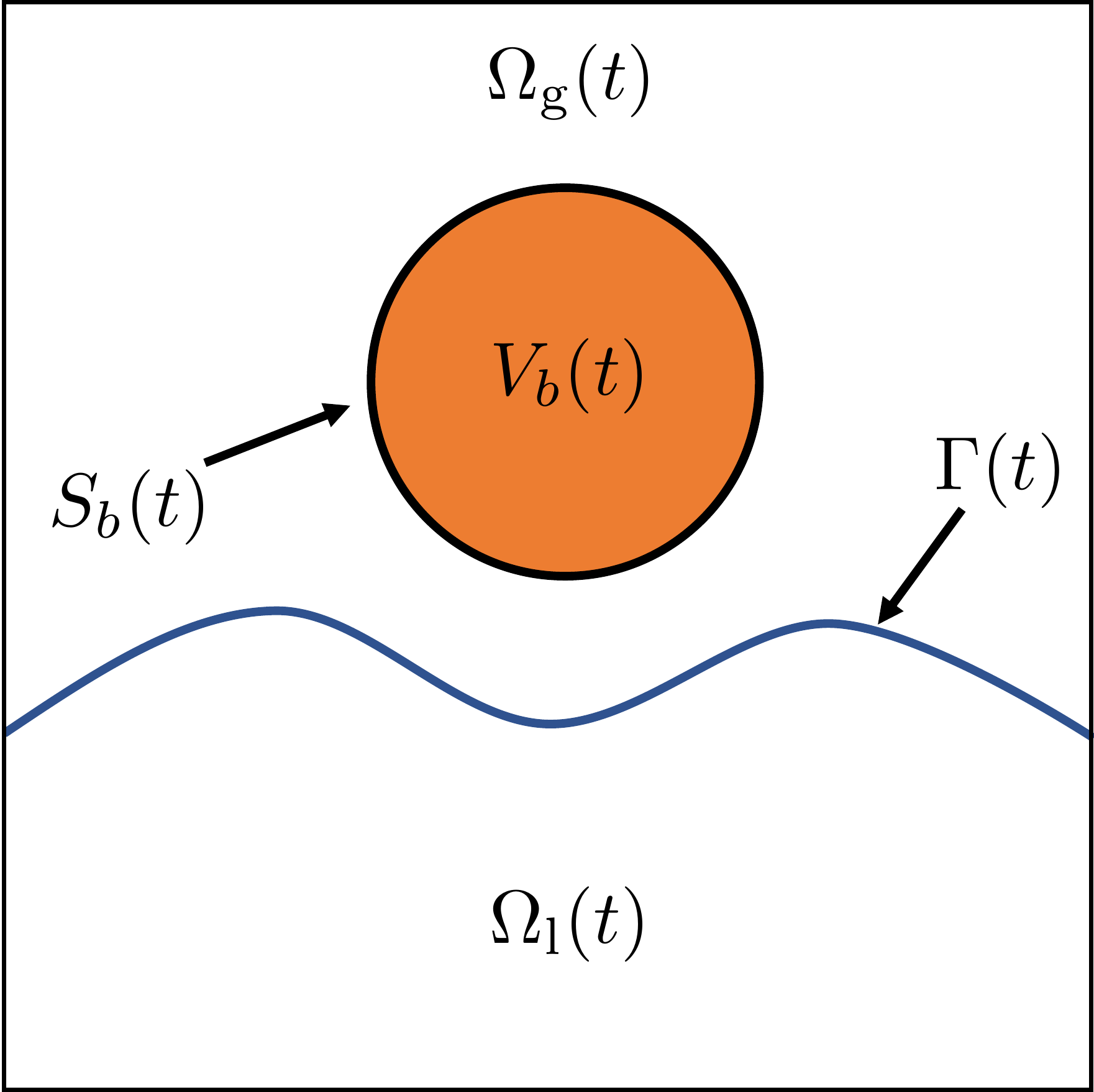}
    \label{ib_continuous}
  }
   \subfigure[Discretized domain]{
    \includegraphics[scale = 0.32]{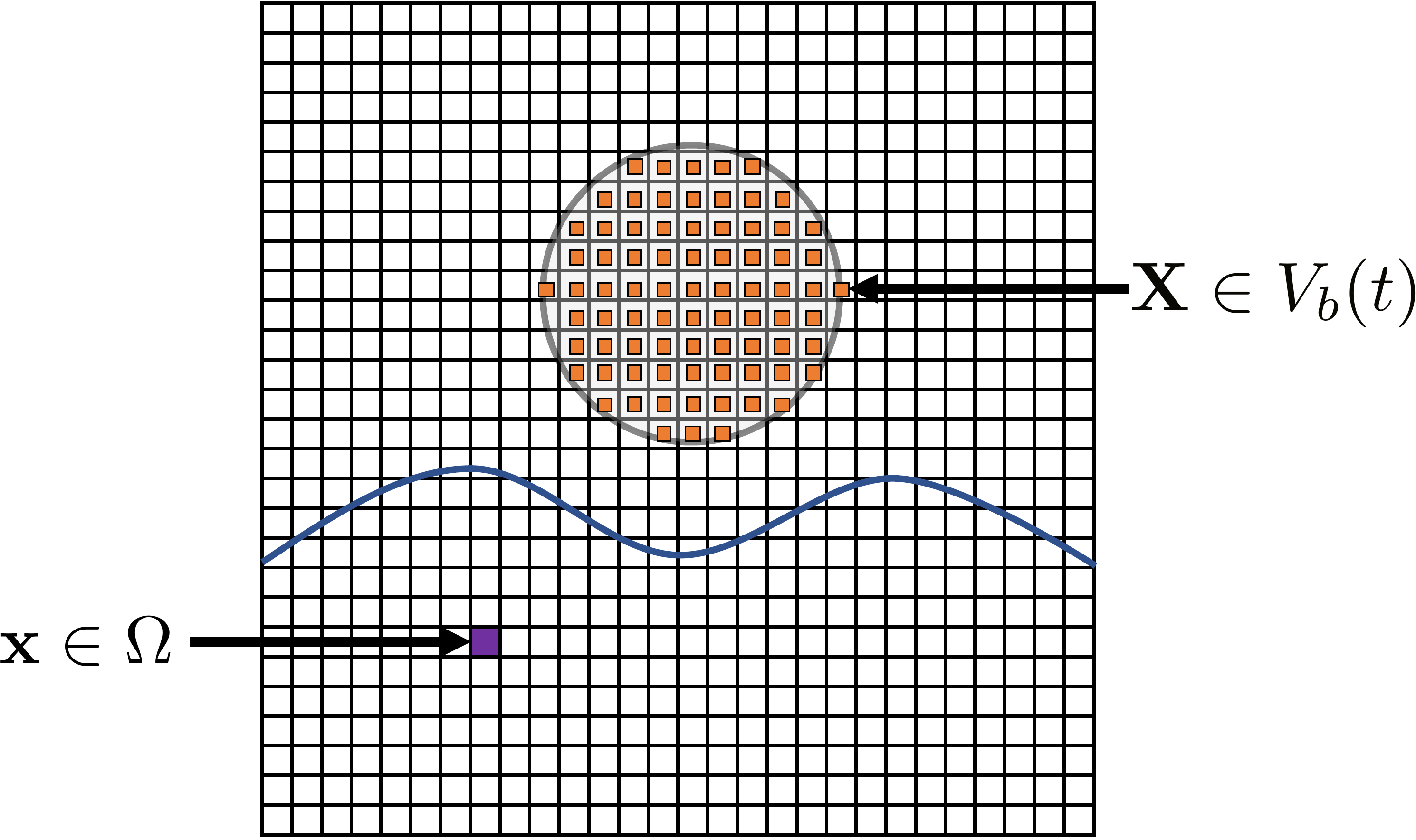}
    \label{ib_discrete}
  }
  \subfigure[Single grid cell \& Lagrangian marker]{
    \includegraphics[scale = 0.4]{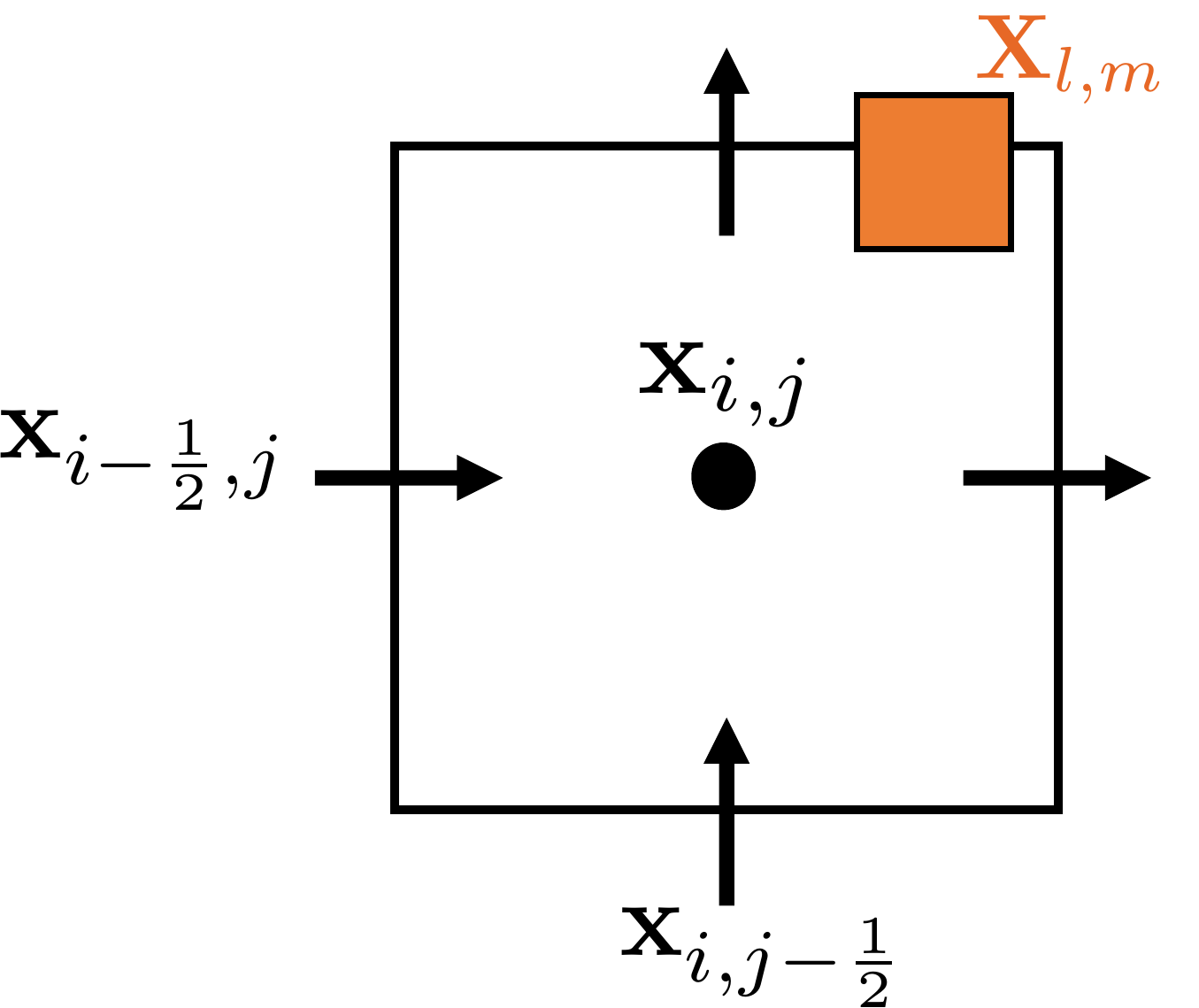}
    \label{grid_cell}
  }
  \caption{\subref{ib_continuous} Sketch of the immersed structure contained within a domain
  containing liquid and gas phases.
  \subref{ib_discrete} Numerical discretization of the domain $\Omega$ into
  Eulerian grid cells ($\blacksquare$, purple) and Lagrangian markers ($\blacksquare$, orange).
  \subref{grid_cell} A single Cartesian grid cell on which the components of the velocity field $\u$
  are approximated on the cell faces ($\rightarrow$, black); the pressure $p$ and level sets $\phi$ and $\psi$ 
  are approximated on the cell center ($\bullet$, black);
  and the Lagrangian quantities are approximated on the marker point ($\blacksquare$, orange), which can
  be arbitrarily placed on the Eulerian grid.}
  \label{fig_ib_diagram}
\end{figure}

Quantities attached to the structure are described in a Lagrangian frame
on immersed markers that are free to arbitrarily cut through the background Cartesian mesh.
These nodes are indexed by $(l,m,n)$ with curvilinear mesh spacings $(\ds_1,\ds_2,\ds_3)$.
An arbitrary quantity can be discretely approximated on a marker points as
$\Phi^n_{l,m,n} \approx \Phi(\s_{l,m,n}, t^n) = \Phi(l\ds_1, m\ds_2,n\ds_3, t^n)$ at time $t^n$.
Henceforth the position, velocity, and force of a marker point are denoted as 
$\X_{l,m,n}$, $\U_{l,m,n}$, and $\F_{l,m,n}$. In this work we only consider rigid bodies
without any constitutive model applied in the structure domain,
and therefore explicit mesh connectivity information is not needed~\cite{Bhalla13}.
See Fig.~\ref{fig_ib_diagram} for a sketch of the discretization in two spatial dimensions.

\subsection{Lagrangian-Eulerian interaction}
Finally, the transfer of quantities between the Eulerian and Lagrangian coordinate systems
requires discrete approximations to the velocity interpolation and force spreading operations to be defined.
We briefly summarize them here to complete the description of the spatial discretization.
\subsubsection{In the interior domain}
For a given fluid velocity defined on faces of the staggered grid, the discretized velocity interpolation
operation for a particular configuration of Lagrangian markers (i.e. $\U = \cJ_h[\X] \u)$ away from the 
physical boundary follows the standard treatment 
\begin{align}
U_{l,m,n} & = \sum_{\x_{i-\half,j,k} \in \Omega} u_{i-\half,j,k} \delta_h\left(\x_{i-\half,j,k} - \X_{l,m,n}\right)  \dx\dy\dz, \\
V_{l,m,n} & = \sum_{\x_{i,j-\half,k} \in \Omega} v_{i,j-\half,k} \delta_h\left(\x_{i,j-\half,k} - \X_{l,m,n}\right)  \dx\dy\dz, \\
W_{l,m,n} & = \sum_{\x_{i,j,k-\half} \in \Omega} w_{i,j,k-\half} \delta_h\left(\x_{i,j,k-\half} - \X_{l,m,n}\right)  \dx\dy\dz,
\end{align}
in which $ \delta_h(\x)$ is a regularized version of the $d$-dimensional Dirac delta function
based on a four-point kernel function~\cite{Peskin02}. For a given force density defined on 
Lagrangian markers, the discretized force spreading operation $\f = \cS_h[\X] \F$ reads
\begin{align}
(f_1)_{i-\half,j,k} & = \sum_{\X_{l,m,n} \in \Vb} (F_1)_{l,m,n} \delta_h\left(\x_{i-\half,j,k} - \X_{l,m,n}\right) \ds_1\ds_2\ds_3, \\
(f_2)_{i,j-\half,k} & = \sum_{\X_{l,m,n} \in \Vb} (F_2)_{l,m,n} \delta_h\left(\x_{i,j-\half,k} - \X_{l,m,n}\right) \ds_1\ds_2\ds_3, \\
(f_3)_{i,j,k-\half} & = \sum_{\X_{l,m,n} \in \Vb} (F_3)_{l,m,n} \delta_h\left(\x_{i,j,k-\half} - \X_{l,m,n}\right) \ds_1\ds_2\ds_3.
\end{align}
We refer readers to~\cite{Bhalla13,Peskin02} for more details on properties and implementation of the grid transfer
operations.
\subsubsection{Near the physical boundary}
When a Lagrangian marker is near the physical boundary, the support of the standard IB kernel extends beyond the 
computational domain. In this case $\cS$ and $\cJ$ operators are modified to $\cS_{\text{BC}}$ and $\cJ_\text{BC}$, 
respectively, to satisfy the discrete adjointness property $\cS_{\text{BC}}  = \cJ_\text{BC}^{*}$ near the physical boundary. 
Briefly, $\cJ_\text{BC}$ is obtained by first filling the ghost cell values abutting the physical domain to satisfy the imposed boundary 
conditions (say for velocity) and then using the standard  weights of the $\cJ$ operator to interpolate onto the 
Lagrangian marker. The adjoint spreading operator near the boundary $\cS_{\text{BC}}$ is obtained by first spreading 
to ghost (and interior) cells beyond the physical boundary and then adding back values to the interior cells by identifying 
their mirror images in the ghosted region. More details on this construction can be found in the Appendix of 
Kallemov et al.~\cite{Kallemov16}.
 
\subsection{Adaptive mesh refinement}
\label{sec_amr}
Some cases presented in this work make use of a structured adaptive mesh refinement (SAMR)
framework to discretize the multiphase fluid-structure interaction equations. 
These discretization approaches describe the computational domain as composed of
multiple grid levels, which is hereafter known as a \emph{grid hierarchy}.
Assuming uniform and isotropic mesh refinement, a grid hierarchy with $\ell$ levels
and coarsest grid spacings $\dx_0$, $\dy_0$, and $\dz_0$
has grid spacings $\dx_\textrm{min} = \dx_0/\nref^{\ell-1}$, 
$\dy_\textrm{min} = \dy_0/\nref^{\ell-1}$, and $\dz_\textrm{min} = \dz_0/\nref^{\ell-1}$ on the finest grid level, in which
$\nref$ is the integer refinement ratio between levels.
Although not considered here, both the numerical method and software implementation allow for general refinement ratios.

The locally refined meshes can be static, in that they occupy a fixed region in the domain $\Omega$,
or adaptive, in that some criteria of interest is used to ``tag'' coarse cells for refinement.
In our current implementation, cells are refined based on two criteria: 1)~if the local magnitude of vorticity 
$\|\omega\|_{i,j,k} = \|\grad \times \u\|_{i,j,k}$ exceeds a relative threshold and 2)~if the 
flow level set function $\phi_{i,j,k}$ is within some threshold of zero.
This ensures that the important dynamics (e.g., regions of high velocity gradients or the multiphase interfaces)
are always approximated on the most resolved mesh. Additionally, we find that restricting the liquid-gas
interface to the finest grid level can greatly mitigate spurious mass changes typically seen in level set methods. 
We also note that the immersed structure is always placed on the finest grid level, ensuring adequate accuracy
near the fluid-solid interface.
We refer readers to prior work by Griffith~\cite{Griffith2012} for additional details on the AMR discretization methods,
which includes a description of the refine and coarsen operations carried out during hierarchy regridding and
a treatment of the coarse-fine interface ghost cells.


\section{Solution methodology} \label{sec_sol_method}
Our strategy for solving the coupled fluid-structure interaction system of equations is
similar to that of Bhalla et al.~\cite{Bhalla13}. The numerical method relies on a time-splitting approach,
in which we first solve the incompressible Navier-Stokes equations (Eqs.~\eqref{eqn_momentum}
and~\eqref{eqn_continuity}) without accounting for the constraints associated with the motion of the
immersed body. We then correct the velocity field to comply with the constrained Lagrangian velocity field via
a projection step. This section also describes additional complexities related to the multiphase nature
of the problems considered in this work.

\subsection{Interface tracking and reinitialization}
\label{sec_reinit}
As described in Sec.~\ref{sec_cont_ls}, two level set functions are defined for the present
numerical method: 1)~the scalar field $\phi(\x,t)$ whose zero contour represents the liquid-air
interface $\Gamma(t)$ and 2)~the scalar field $\psi(\x,t)$ whose zero contour represents
the boundary of the immersed structure $\Sb(t)$. The transition between different materials
on the Eulerian grid can be completely described by these two level set functions. Indeed if $\phi$
and $\psi$ represent signed distance functions to their respective interfaces, we can define smoothed
Heaviside functions that have been regularized over $\ncells$ grid cells on either side of the interfaces
(assuming $\dx = \dy = \dz)$,

\begin{align}
\label{eq_heaviside}
\widetilde{H}^{\text{flow}}_{i,j,k} &= 
\begin{cases} 
       0,  & \phi_{i,j,k} < -\ncells \dx,\\
        \frac{1}{2}\left(1 + \frac{1}{\ncells \dx} \phi_{i,j,k} + \frac{1}{\pi} \sin\left(\frac{\pi}{ \ncells \dx} \phi_{i,j,k}\right)\right) ,  & |\phi_{i,j,k}| \le \ncells \dx,\\
        1,  & \textrm{otherwise},
\end{cases} \\
\widetilde{H}^{\text{body}}_{i,j,k} &= 
\begin{cases} 
       0,  & \psi_{i,j,k} < -\ncells \dx,\\
        \frac{1}{2}\left(1 + \frac{1}{\ncells \dx} \psi_{i,j,k} + \frac{1}{\pi} \sin\left(\frac{\pi}{ \ncells \dx} \psi_{i,j,k}\right)\right) ,  & |\psi_{i,j,k}| \le \ncells \dx,\\
        1,  & \textrm{otherwise},
\end{cases}
\end{align}
in which we have assumed that the number of transition cells is the same across $\Gamma$ and $\Sb$. This is not
an inherent limitation of the numerical method, but is true for all the cases considered in the present work.
A given material property $\zeta$ (such as $\rho$ or $\mu$) is then set in the whole domain using a two-step process.
First, the material property in the ``flowing" phase is set via the liquid-gas level set function
\begin{equation}
\label{eq_ls_flow}
\zeta^{\text{flow}}_{i,j,k} = \zeta_\text{l} + (\zeta_\text{g} - \zeta_\text{l}) \widetilde{H}^{\text{flow}}_{i,j,k}.
\end{equation}
Next, the material property is set on cell centers throughout the computational domain, taking into account the solid phase
\begin{equation}
\label{eq_ls_solid}
\zeta_{i,j,k} = \zeta_\text{s} + (\zeta^{\text{flow}}_{i,j,k} - \zeta_\text{s}) \widetilde{H}^{\text{body}}_{i,j,k}.
\end{equation}
Hence the solid level set always takes precedent over the flow phase. Note that we have assumed that
the liquid phase is represented by negative $\phi$ values and the solid phase is represented by negative
$\psi$ values, without loss of generality.

\begin{figure}[]
  \centering
  \subfigure[Material properties in the ``flowing" phases]{
    \includegraphics[scale = 0.4]{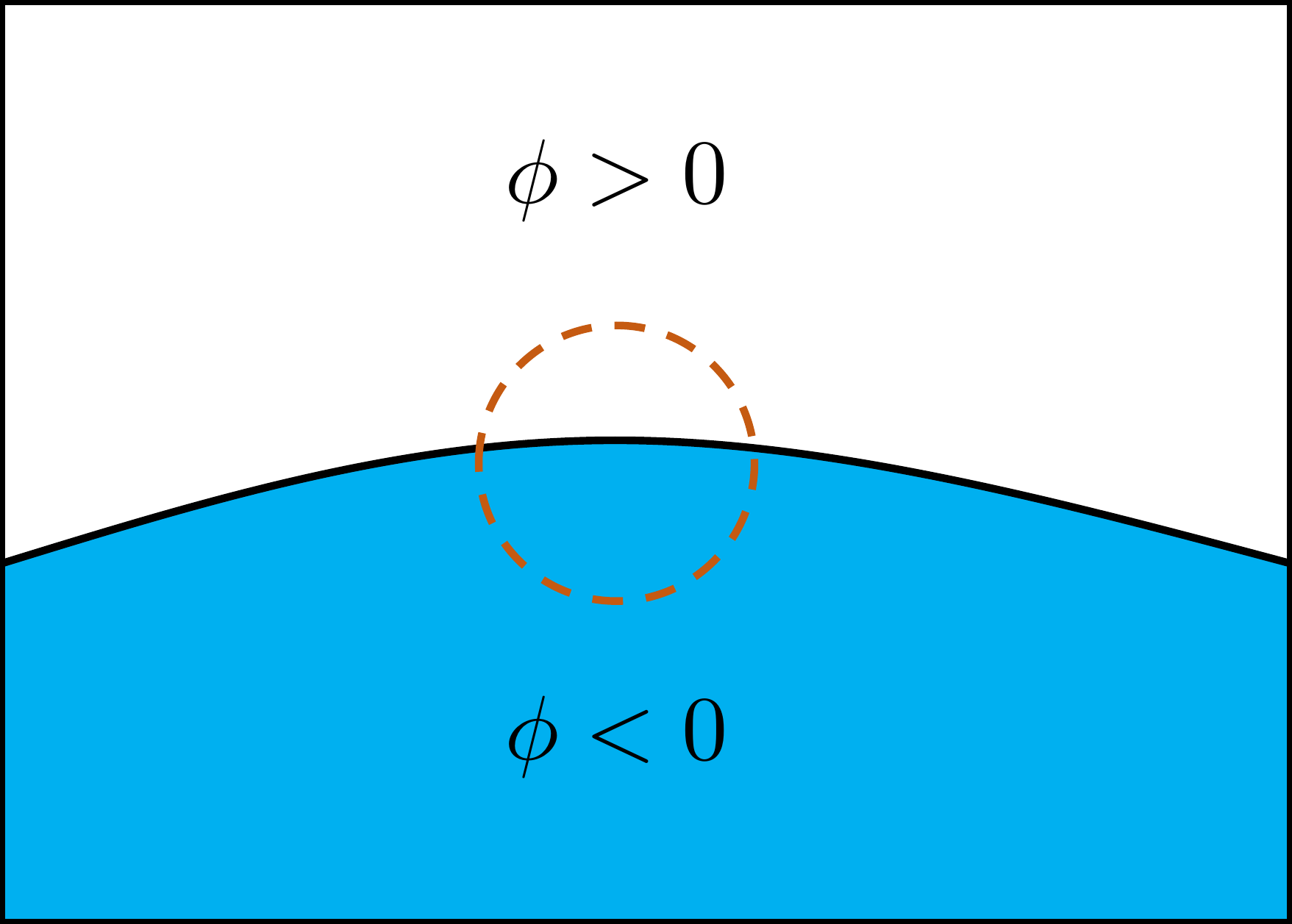}
    \label{ls_flow_diagram}
  }
   \subfigure[Material properties in the entire domain]{
    \includegraphics[scale = 0.4]{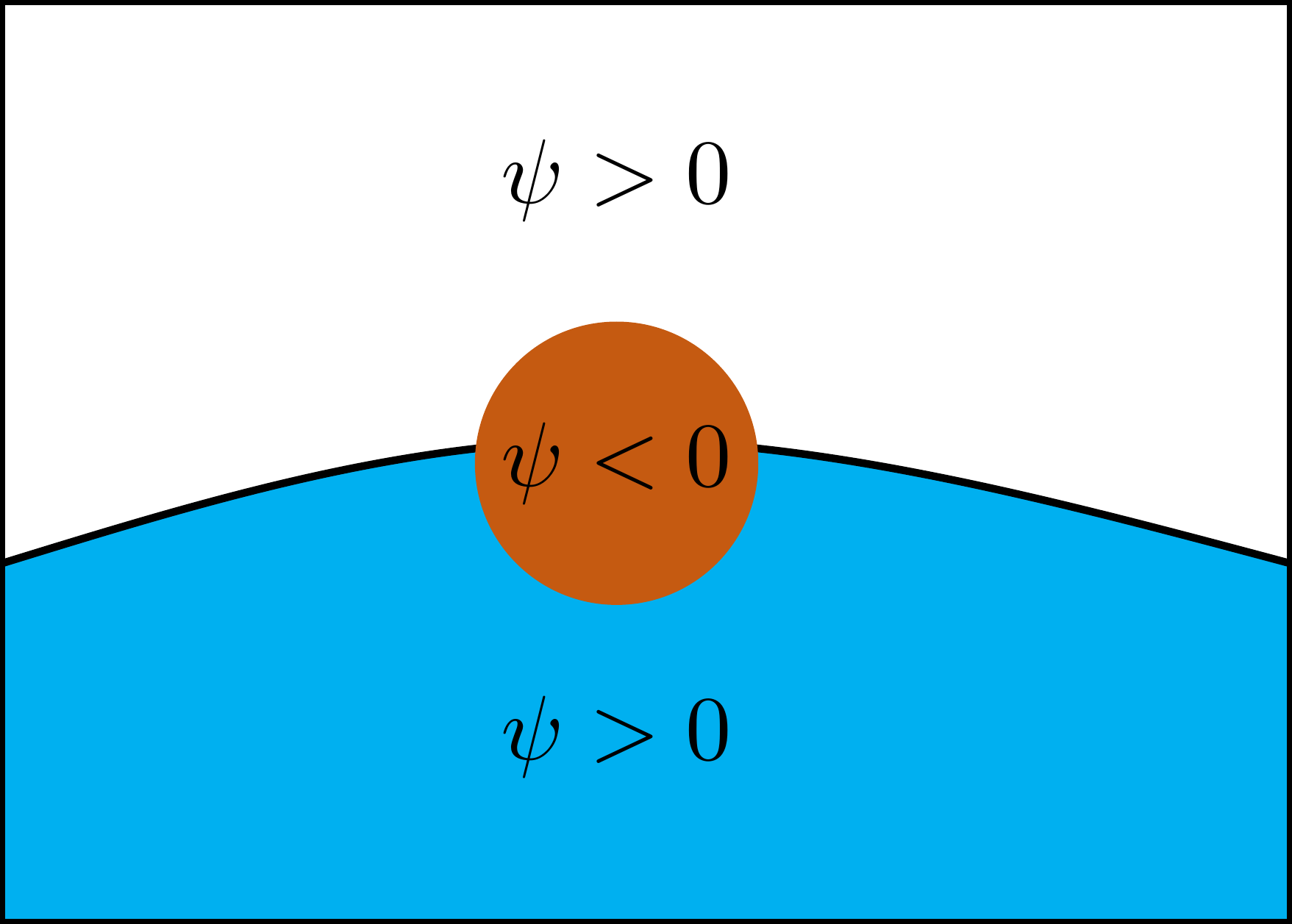}
    \label{ls_solid_diagram}
  }
  \caption{Sketch of the two-stage process for setting the density and viscosity in the computational domain.
  \subref{ls_flow_diagram} Material properties are first prescribed in the ``flowing" phase based on the liquid-gas level set
  function $\phi$ (---, black) and ignoring the structure level set function $\psi$ (\texttt{---}, orange).
  \subref{ls_solid_diagram} Material properties are then corrected in the phase occupied by the immersed body.}
  \label{fig_ls_diagram}
\end{figure}

Even if $\phi$ and $\psi$ are initially set to be the signed distance function from their respective
interfaces, they are not guaranteed to retain the signed distance property under linear advection,
Eqs.~\eqref{eq_ls_fluid_advection} and~\eqref{eq_ls_solid_advection}. Let $\widetilde{\phi}^{n+1}$
denote the flow level set function following an advective transport after
time stepping through the interval $\left[t^{n}, t^{n+1}\right]$. The flow level set is \emph{reinitialized}
to obtain a signed distance field $\phi^{n+1}$ by computing a steady-state solution to the Hamilton-Jacobi
equation

\begin{align}
&\D{\phi}{\tau} + \sgn\left(\widetilde{\phi}^{n+1}\right)\left(\|\grad \phi \| - 1\right) = 0, \label{eq_eikonal} \\
& \phi(\x, \tau = 0) = \widetilde{\phi}^{n+1}(\x), \label{eq_eikonal_init}
\end{align}
which will yield a solution to the Eikonal equation $\|\grad \phi \|  = 1$ at the end of each time step. 
We refer the readers to~\cite{Nangia2018} for more details on the specific discretization of
Eqs.~\eqref{eq_eikonal} and~\eqref{eq_eikonal_init}, which employs second-order
ENO finite differences combined with a subcell-fix method described by Min~\cite{Min2010},
and an immobile interface condition described by Son~\cite{Son2005}.

Since we only consider relatively simple body geometries in the present work, we can make use
of the positions of the Lagrangian markers to reinitialize the structure level set function.
As an example, let a volumetric sphere body with radius $R$ be made up of $N_\text{s}$ Lagrangian markers.
At time $t^{n+1}$, its center of mass can be computed as
\begin{equation}
\label{eq_COM}
\Xcom^{n+1} = \frac{1}{N_\text{s}} \sum_{\X_{l,m,n} \in \Vb^{n+1}} \X_{l,m,n}^{n+1},
\end{equation}
and the structure level set function can be directly recomputed as $\psi^{n+1}(\x_{i,j,k}) = \|\x_{i,j,k} - \Xcom^{n+1}\| - R$.
(Note that whenever $n$ appears as a superscript, it refers to a time step number, whereas $n$ as a subscript refers
to the indexing of Lagrangian particles.)
For more complicated immersed bodies, one can make use of constructive solid geometry (CGS) concepts or 
R-functions (see Shapiro~\cite{Shapiro2007}) to determine analytical expressions for various signed distance 
functions~\footnote{R-functions tend to smooth sharp corners of geometries. We prefer CGS over R-functions wherever the former is applicable.}.
In the present work, we always reinitialize both level set functions every time step.


\subsection{Full time stepping scheme}
\label{sec_temporal_scheme}
Next, we describe the temporal discretization over the interval $\left[t^n, t^{n+1}\right]$
for the coupled fluid-structure equations of motion. We employ $\ncycles$ cycles of fixed-point iteration
per time step, with $\ncycles = 2$ being used for all the cases in the present work. Note that $k$ appears
in superscript to denote the cycle number. The full time stepping scheme consists of three major operations:
\begin{enumerate}
\item Advect the signed distance functions, obtaining $\phi^{n+1,k+1}$ and $\psi^{n+1,k+1}$, and the cell-centered viscosity $\mu^{n+1,k+1}$ using the signed distance functions~\footnote{We 
first set the cell-centered viscosity and then use harmonic averaging to interpolate onto the appropriate degrees of freedom.}. 
\item Solve the incompressible Navier-Stokes equations, obtaining $\widetilde{\u}^{n+1,k+1}$ and $p^{~n+\half,k+1}$.
\item Enforce the rigidity constraint, obtaining $\X^{n+1,k+1}$ and the corrected fluid velocity $\u^{n+1,k+1}$.
\end{enumerate}
At the beginning of each time step we set $k = 0$, with $\u^{n+1,0} = \u^{n}$, $p^{n+\half,0} = p^{n-\half}$,
$\phi^{n+1,0} = \phi^{n}$, $\psi^{n+1,0} = \psi^{n}$, and $\X^{n+1,0} = \X^{n}$. At the initial time step $n = 0$,
these quantities are obtained using the prescribed initial conditions. The midpoint, time-centered approximations
to these quantities are given by $\u^{n+\half,k} = \half\left(\u^{n+1,k} + \u^{n}\right)$,
$\widetilde{\u}^{n+\half,k} = \half\left(\widetilde{\u}^{n+1,k} + \u^{n}\right)$,
$\phi^{n+\half,k} = \half\left(\phi^{n+1,k} + \phi^{n}\right)$, $\psi^{n+\half,k} = \half\left(\psi^{n+1,k} + \psi^{n}\right)$,
and $\X^{n+\half,k} = \half\left(\X^{n+1,k} + \X^{n}\right)$.
Below, we describe in detail the solution methodology for all three steps.

\subsubsection{Scalar advection}
The level set functions are updated by discretizing Eqs.~\eqref{eq_ls_fluid_advection}
and~\eqref{eq_ls_solid_advection}, which reads
\begin{align}
\frac{\phi^{n+1,k+1} - \phi^{n}}{\dt} + Q\left(\u^{n+\half,k}, \phi^{n+\half,k}\right) &= 0, \\
\frac{\psi^{n+1,k+1} - \psi^{n}}{\dt} + Q\left(\u^{n+\half,k}, \psi^{n+\half,k}\right) &= 0,
\end{align}
in which $Q(\cdot,\cdot)$ represents an explicit piecewise parabolic method (xsPPM7-limited)
approximation to the linear advection terms on cell centers.
We refer the readers to~\cite{Griffith2009,Rider2007} for more details on the numerical implementation
of this flux limiter. Homogenous Neumann boundary conditions for $\phi$ and $\psi$ are imposed on $\partial \Omega$,
using a standard ghost value treatment~\cite{Harlow1965}.

\subsubsection{Incompressible Navier-Stokes solver: Conservative and consistent transport formulation}
The incompressible Navier-Stokes equations Eqs.~\eqref{eqn_momentum}
and~\eqref{eqn_continuity} are discretized and solved for in conservative form as
\begin{align}
	&\frac{\breve{\V \rho}^{n+1,k+1} \widetilde{\u}^{n+1,k+1} - { \V \rho}^{n} \u^n}{\dt} + \C^{n+1,k} = -\grad_h p^{n+\half, k+1}
	+ \left(\L_{\mu} \widetilde{\u}\right)^{n+\half, k+1}
	+ \breve{ \V \rho}^{n+1,k+1}\g + \fs^{n+\half, k+1}, \label{eq_c_discrete_momentum}\\
	& \grad_h \cdot \widetilde{\u}^{n+1,k+1} = 0 \label{eq_c_discrete_continuity},
\end{align}
in which $\C^{n+1,k}$ is an explicit cubic upwind interpolation (CUI-limited)~\cite{Roe1982,Waterson2007,Patel2015}
approximation to the nonlinear convection term,
and $\left(\L_{\mu} \widetilde{\u}\right)^{n+\half, k+1} =  \half\left[\left(\L_{\mu} \widetilde{\u}\right)^{n+1,k+1} + \left(\L_{\mu} \u\right)^n\right]$
is a semi-implicit approximation to the viscous strain rate with
$\left(\L_{\mu}\right)^n = \grad_h \cdot \left[\mu^{n} \left(\grad_h \u + \grad_h \u^T\right)^n\right]$.
The above time-stepping scheme with $\ncycles = 2$ is similar to a combination of
explicit midpoint rule for the convective term and Crank-Nicolson for the viscous terms.
We note that the newest approximation to viscosity $\mu^{n+1,k+1}$ is obtained via the procedure
described in Eqs.~\eqref{eq_ls_flow} and~\eqref{eq_ls_solid}. The newest approximation
to density $\breve{\V \rho}^{n+1,k+1}$ in Eq.~\eqref{eq_c_discrete_momentum} is obtained by solving a 
discretized mass update equation directly on the faces of the staggered grid from the previous time step 
and level set synchronized density field $\V \rho^{n}$ (obtained after averaging $\phi^{n}$ and $\psi^{n}$ onto faces).
The discretized density update equation is solved using 
the third-order accurate strong stability preserving Runge-Kutta (SSP-RK3) time integrator~\cite{Gottlieb2001} as follows
	\begin{align}
	& \breve{\V \rho}^{(1)} = \V \rho^{n} - \dt \R\left(\u^{n}_\text{adv}, \V \rho^{n}_\text{lim}\right), \label{eq_rk1}\\
	& \breve{\V  \rho}^{(2)} = \frac{3}{4} \V \rho^{n} + \frac{1}{4} \breve{\V \rho}^{(1)} - \frac{1}{4} \dt \R\left(\u^{(1)}_\text{adv}, \breve{\V \rho}^{(1)}_\text{lim}\right), \label{eq_rk2} \\
	& \breve{\V \rho}^{n+1, k+1} = \frac{1}{3} \V \rho^n + \frac{2}{3} \breve{\V \rho}^{(2)} - \frac{2}{3} \dt \R\left(\u^{(2)}_\text{adv}, \breve{\V \rho}^{(2)}_\text{lim}\right) \label{eq_rk3}.
	\end{align} 
	Here $\R\left(\u_{\text{adv}}, \V \wp_{\text{lim}}\right) \approx \left[\left(\div \left(\u_\text{adv} \V \wp_\text{lim}\right)\right)_{i-\half, j,k}, \left(\div \left(\u_\text{adv} \V \wp_\text{lim}\right)\right)_{i, j-\half,k}, \left(\div \left(\u_\text{adv} \V \wp_\text{lim}\right)\right)_{i, j,k-\half}\right]$ 
	is an explicit CUI-limited approximation to the linear density 
	advection term; $\V \wp$ is either $\V \rho$ or $\breve{\V \rho}$. We distinguish $\breve{\V \rho}$, the density vector obtained 
	via the SSP-RK3 integrator, from $\V \rho$, the density vector that is set from the level set fields. 
	The subscript ``adv'' indicates the interpolated advective velocity on the faces of face-centered 
	control volume, and the subscript ``lim'' indicates the limited value (see Nangia et al. in~\cite{Nangia2018} for details on obtaining 
	advective and flux-limited fields). We remark that the density integration procedure is occurring \emph{within} the overall 
	fixed-point iteration scheme. We have found it to be \emph{crucial} to use appropriately interpolated and extrapolated velocities 
	to maintain the second-order accuracy of the INS scheme.  To wit, for the first cycle ($k = 0$), the velocities are
	\begin{align}
		& \u^{(1)} = 2 \u^n - \u^{n-1}, \\
		& \u^{(2)} = \3half \u^n - \half \u^{n-1}.
	\end{align}
	For all remaining cycles ($k > 0$), the velocities are
	\begin{align} 
		& \u^{(1)} = {\u}^{n+1,k}, \\
		& \u^{(2)} = \frac{3}{8} {\u}^{n+1,k} + \frac{3}{4} \u^{n} - \frac{1}{8} \u^{n-1}.
	\end{align}
	Notice that $\u^{(1)}$ is an approximation to ${\u}^{n+1}$, and $\u^{(2)}$ is an approximation to ${\u}^{n+\half}$. Similarly, $\breve{\V \rho}^{(1)}$ is an approximation to $\breve{\V \rho}^{n+1}$, and $\breve{\V \rho}^{(2)}$ is an approximation to $\breve{\V \rho}^{n+\half}$. 
	To ensure consistent transport of mass and momentum fluxes, the convective derivative in Eq.~\eqref{eq_c_discrete_momentum} 
	is given by
	\begin{equation}
	\C\left(\u^{(2)}_\text{adv}, \breve{\V \rho}^{(2)}_\text{lim}\u^{(2)}_\text{lim}\right) \approx 
	\begin{bmatrix}
		\left(\div \left(\u^{(2)}_\text{adv} \breve{\V \rho}^{(2)}_\text{lim} u^{(2)}_\text{lim}\right)\right)_{i-\half,j,k} \\ 
		\left(\div \left(\u^{(2)}_\text{adv} \breve{\V \rho}^{(2)}_\text{lim} v^{(2)}_\text{lim}\right)\right)_{i,j-\half,k} \\
		\left(\div \left(\u^{(2)}_\text{adv} \breve{\V \rho}^{(2)}_\text{lim} w^{(2)}_\text{lim}\right)\right)_{i,j,k-\half}
	\end{bmatrix}
	\end{equation}
	which uses the same velocity $\u^{(2)}_{\text{adv}}$ and density $\breve{\V \rho}^{(2)}_{\text{lim}}$ used to 
	update $\breve{\V \rho}^{n+1}$ in Eq~\eqref{eq_rk3}. This is the key step required to strongly couple the mass and momentum 
	convective operators. Results presented in Sec.~\ref{sec_examples} demonstrate that the consistent discretization is stable for practical 
	air-water density ratio of $10^3$ and produce significantly more accurate results than the inconsistent discretization for realistic 
	three phase WSI simulations.
	
	
\subsubsection{Incompressible Navier-Stokes solver: Non-conservative and inconsistent transport formulation}
One can directly use the face-centered density field $\V \rho^{n+1,k+1}$ obtained through the updated level set 
information, $\phi^{n+1,k+1}$ and $\psi^{n+1,k+1}$, and integrate the INS equations from $[t^n,t^{n+1} ]$. 
In this scenario the time stepping scheme reads

\begin{align}
	&\V \rho^{n+1,k+1} \left(\frac{\widetilde{\u}^{n+1,k+1} - \u^n}{\dt} + \N^{n+\half,k}\right)  = -\grad_h p^{n+\half, k+1}
	+ \left(\L_{\mu} \widetilde{\u}\right)^{n+\half, k+1}
	+ \V \rho^{n+1,k+1}\g + \fs^{n+\half, k+1}, \label{eq_nc_discrete_momentum}\\
	& \grad_h \cdot \widetilde{\u}^{n+1,k+1} = 0 \label{eq_nc_discrete_continuity},
\end{align}
in which $\N^{n+\half,k}$ is an explicit CUI-limited 
approximation to the nonlinear convection term in non-conservative form (i.e. 
$\N^{n+\half,k} \approx \grad \cdot (\u^{n+\half,k}\u^{n+\half,k})$)~\cite{Nangia2018}. 
Integrating INS equations in the above manner decouples the mass and momentum advection
and this results in an inconsistent transport of mass flux in the two discrete operators.  

The performance of these two solvers are compared for some of the numerical examples considered
in Sec.~\ref{sec_examples}. In particular, we will show that the non-conservative solver is numerically
unstable for highly contrasting air-water density ratios. When stable, both schemes are second-order accurate in time.
The continuum surface tension~\cite{Brackbill1992}
force $\fs^{n+\half, k+1}$ is computed as a function of the flow level set field
$\phi^{n+\half,k+1}$, and its treatment is described in~\cite{Nangia2018}.

\subsubsection{Incompressible Navier-Stokes solver}
We obtain the updated velocity $\widetilde{\u}^{n+1,k+1}$ and pressure $p^{n+\half, k+1}$ fields
by \emph{simultaneously} solving Eqs.~\eqref{eq_c_discrete_momentum} and~\eqref{eq_c_discrete_continuity} 
(or Eqs.~\eqref{eq_nc_discrete_momentum} and~\eqref{eq_nc_discrete_continuity}) using
the flexible GMRES (FGMRES) Krylov solver~\cite{Saad93} preconditioned by a variable-coefficient
projection method that is hybridized with a local-viscosity solver~\cite{Griffith2009,Cai2014}.
The solvers have been shown to be second-order accurate in space and to converge for density and viscosity
ratios of up to $10^6$~\cite{Nangia2018}.
Unless otherwise stated, a relative convergence tolerance of $10^{-10}$
is specified for the FGMRES solver, which leads to a converged solution
in between $1$ and $7$ iterations for all of the cases considered here.

\subsubsection{Rigid body projection}
In general, the velocity field computed from the conservative
(Eqs.~\eqref{eq_c_discrete_momentum} and~\eqref{eq_c_discrete_continuity})
and non-conservative (Eqs.~\eqref{eq_nc_discrete_momentum} and~\eqref{eq_nc_discrete_continuity}) 
flow solvers will not satisfy the constraints placed in the structure 
domain (Eq.~\eqref{eq_vel_constraint}). To correct the velocity in $\Vb(t)$, we carry out the following
\emph{projection} step~\cite{Bhalla13}

\begin{equation}
\label{eq_constraint_projection}
{\V \wp}^{n+1,k+1} \left(\frac{\u^{n+1,k+1} - \widetilde{\u}^{n+1,k+1}}{\dt}\right) = \fc^{n+1,k+1},
\end{equation}
in which $\fc^{n+1,k+1}$ is the Eulerian constraint force that imposes the rigidity constraint.
This force can be computed by spreading the Lagrangian constraint force
$\F_{l,m,n}^{n+1,k+1} = \frac{\rho_s}{\dt} \delU^{n+1,k+1}_{l,m,n}$,
which is constructed using the difference between the desired body velocity and the interpolated uncorrected fluid velocity:
\begin{align}
\fc^{n+1,k+1}
&=\cS_h\left[\X^{n+\half, k}\right]\F^{n+1,k+1} \nonumber \\
&= \frac{{\V \wp}^{n+1,k+1}}{\dt}\cS_h\left[\X^{n+\half, k}\right] \delU^{n+1,k+1} \nonumber \\
& = \frac{{\V \wp}^{n+1,k+1}}{\dt}\cS_h\left[\X^{n+\half, k}\right] \left(\Ub^{n+1,k+1} - \cJ_h\left[\X^{n+\half,k}\right]\widetilde{\u}^{n+1,k+1}\right) \label{eq_constraint_force}.
\end{align}
This force is nonzero only in the structure domain.
A correction of this type ensures that the fluid velocity $\u^{n+1,k+1}$ in $\Vb(t)$ 
approximately matches that of the body's Lagrangian velocity $\Ub^{n+1,k+1}$.
Combining Eqs.~\eqref{eq_constraint_projection} and~\eqref{eq_constraint_force}
yields a succinct update equation for the Eulerian velocity field
\begin{equation}
\label{eq_velocity_correction}
\u^{n+1,k+1} = \widetilde{\u}^{n+1,k+1} + 
\cS_h\left[\X^{n+\half, k}\right] \left(\Ub^{n+1,k+1} - \cJ_h\left[\X^{n+\half,k}\right]\widetilde{\u}^{n+1,k+1}\right),
\end{equation}
which is identical to the update described by Bhalla et al.~\cite{Bhalla13} for neutrally
buoyant (constant density) problems. In fact, we simply reuse an existing implementation~\cite{IBAMR-web-page}
of the DLM or constraint immersed boundary method to carry out our multiphase FSI simulations.
Note that in general, the corrected velocity field will not satisfy the discrete continuity equation,
i.e. $\grad_h \cdot \u^{n+1,k+1} \ne 0$.
One could apply an additional velocity projection and pressure correction step to
ensure that the final velocity is divergence-free~\cite{Bhalla13}, but we have found that
it is not necessary to obtain accurate results. As described previously~\cite{Griffith2009,Nangia2018},
the initial value for pressure at the start of each time step $p^{n+\half,0}$ does not affect the flow dynamics
nor the pressure solution at the end of the time step $p^{n+\half}$;
rather it serves as an initial guess to the iterative solution of the linear system.

Next, we describe a procedure to determine $\Ub^{n+1,k+1}$, which is required to compute
$\fc^{n+1,k+1}$. Since the structure is constrained to have a vanishing deformation rate tensor,
the velocity of each Lagrangian marker can be decomposed as the following rigid body motion
(dropping the time superscripts for now)
\begin{equation}
(\Ub)_{l,m,n} = \Ur + \Wr \times \R_{l,m,n},
\end{equation}
in which $\Ur$ and $\Wr$ represent the linear and angular center of mass velocities, respectively,
and $\R_{l,m,n} = \X_{l,m,n} - \Xcom$ is the radius vector pointing from the center of mass to the
Lagrangian marker position. Two distinct scenarios are considered in the present work:
\begin{enumerate}
	\item \emph{Fully prescribed motion}: \\
	For problems in which the motion of the body is specified as a function of time, we can directly
	set the Lagrangian velocity field at time step $n+1$ as
	\begin{equation}
		\label{eq_prescribed_velocity}
		(\Ub)^{n+1,k+1}_{l,m,n} = \Ur^{n+1} + \Wr^{n+1} \times \R^{n+\half,k}_{l,m,n},
	\end{equation}
	which is then used to update the position of the Lagrangian markers
	\begin{equation}
		\label{eq_prescribed_position}
		\X_{l,m,n}^{n+1,k+1} = \X_{l,m,n}^n + \dt (\Ub)^{n+\half,k+1}_{l,m,n}.
	\end{equation}
	This algorithm can be used to simulate one-way FSI problems such as flows past stationary objects
	or bodies entering or exiting fluid interfaces with constant velocity.
	
	\item \emph{Free-body motion}: \\
	For coupled problems in which the body moves as a result of the fluid-structure interaction,
	we determine the Lagrangian velocity field at time step $n+1$ by \emph{redistributing} the
	linear and angular momentum~\cite{Patankar2000,Shirgaonkar2009,Bhalla13} in the structure domain
	\begin{align}
		\Mb \Ur^{n+1,k+1} &= \sum_{\X_{l,m,n} \in \Vb} \rhos \left(\cJ_h\left[\X^{n+\half,k}\right]\widetilde{\u}^{n+1,k+1}\right)_{l,m,n} \ds_1\ds_2\ds_3, \label{eq_linear_conservation} \\
		\Ib \Wr^{n+1,k+1} &= \sum_{\X_{l,m,n} \in \Vb} \rhos \R^{n+\half,k}_{l,m,n} \times \left(\cJ_h\left[\X^{n+\half,k}\right]\widetilde{\u}^{n+1,k+1}\right)_{l,m,n} \ds_1\ds_2\ds_3. \label{eq_angular_conservation}
	\end{align}
	Here, $\Ib = \sum_{\X_{l,m,n} \in \Vb} \rhos \left(\R_{l,m,n}^{n+\half,k} \cdot \R_{l,m,n}^{n+\half,k} \I - \R_{l,m,n}^{n+\half,k} \otimes \R_{l,m,n}^{n+\half,k} \right)$ is the moment of inertia tensor, in which $\I$ is the $d$-dimensional identity tensor, 
	and $\Mb = \sum_{\X_{l,m,n} \in \Vb} \rhos \ds_1\ds_2\ds_3$ is the mass of the body.
	Note that since we assume a uniform density in the solid region, the contribution from $\rhos$ cancels out
	in the actual implementation of Eqs.~\eqref{eq_linear_conservation} and~\eqref{eq_angular_conservation}.
	Hence, buoyancy effects due to differences in the fluid and solid densities are implicitly accounted for
	by the multiphase fluid solver. Once the rigid body velocity components are determined, the structure's velocity
	and position are updated via Eqs.~\eqref{eq_prescribed_velocity} and~\eqref{eq_prescribed_position}.
\end{enumerate}

We remark that the above formulation assumes that the six rigid degrees of freedom either are \emph{all}
fully prescribed (locked) or \emph{all} undergoing free-body motion (unlocked). This is not a limitation of the
implementation: we are able to mix and match which degrees of freedom are locked and unlocked.
Many of the numerical examples considered in the present work make use of this flexibility. Finally, we make two interesting 
observations in the rigid body projection algorithm:
\begin{enumerate}
\item The fluid-structure coupling is implicit, i.e., we are not iterating 
back-and-forth between a fluid and a rigid body integrator.
\item We do not need to explicitly evaluate the hydrodynamic 
stress on the immersed structure to displace it or solve the fluid equations with internal velocity boundary conditions.
\end{enumerate}
The physical reason behind this implicit coupling can be understood if we consider the hydrodynamic force as an \emph{internal} 
force of the system, which is equal and opposite at the fluid-structure interface. This is the essence of the fast and efficient DLM 
method of Sharma and Patankar~\cite{Sharma2005}.
This makes our method computationally more efficient than certain sharp-interface IB approaches, which can require several \emph{stability-preserving} FSI iterations and complex velocity and pressure reconstruction techniques at the immersed surface to compute hydrodynamic forces and moments~\cite{Calderer2014,Borazjani2008}.       

\section{Prescription of solid density, viscosity and a well-balanced gravitational force}
\label{sec_solid_materials}
In the case of a neutrally buoyant structure within a single phase flow, the density and viscosity within
$\Vb(t)$ is simply taken to be that of the surrounding fluid (i.e. the constant $\rho$ and $\mu$
used in the momentum equation)~\cite{Bhalla13}. However, the choice of the ``virtual" fluid that occupies the
solid region for multiphase flow problems warrants additional discussion.
Specification of $\rhos$ and $\mus$ in this region is required to ensure that the linear system of
equations~\eqref{eq_c_discrete_momentum} and~\eqref{eq_c_discrete_continuity} is well-posed.

For the ``virtual" viscosity, we follow the recommendation of Patel and Natarajan~\cite{Patel2018} and 
set $\mus$ equal to that of the largest (most viscous) of all fluids in the problem. In our experience, this choice leads to
accurate FSI simulations and reasonably fast convergence of the FGMRES solver. In cases where the object is undergoing
\emph{free-body} motion, e.g. a sedimenting sphere, a proper specification of the solid density is vitally important in order 
to capture inertia and buoyancy effects due to the structure's weight. Hence, we must set $\rhos$ based on the physical 
properties of the body we are trying to simulate.

In cases where the immersed body's velocity is \emph{fully prescribed}, we set $\rhos$ equal to that
of the largest (most dense) of all the fluids in the problem. Note that when this object is in contact
with the less dense phase, the gravitational term $ \rho \g$ in the momentum equation will generate
spurious momentum in the solid phase. These spurious velocities will contaminate the flow field throughout the
duration of simulation and lead to inaccurate results. In order to mitigate this erroneous momentum generation,
we compute the gravitational body force using only the \emph{flow density field} $\rho^{\text{flow}}$
(see Eq.~\eqref{eq_ls_flow} and Fig.~\ref{ls_flow_diagram}). Thus for the \emph{fully prescribed} kinematics 
case $\rhos$ enters only in the linear operator but not as a gravitational body force in the solid region. As we showed in our 
previous work, the gravitational force based on $\rho^{\text{flow}}$ is well-balanced by the pressure gradient term~\cite{Nangia2018}. Hence, we 
in-effect recover a well-balanced gravity force for the coupled three-phase flow problem as well; we will show in 
Sec.~\ref{sec_examples} that no parasitic currents are generated at the air-water-structure interface in the
hydrostatic limit.

\section{Software implementation} \label{sec_sfw}
The numerical algorithm described here is implemented
in the IBAMR library~\cite{IBAMR-web-page}, which is an open-source C++
simulation software focused on immersed boundary methods with
adaptive mesh refinement. All of the numerical examples presented here
are publicly available via \url{https://github.com/IBAMR/IBAMR}.
IBAMR relies on SAMRAI \cite{HornungKohn02, samrai-web-page} for Cartesian grid 
management and the AMR framework. Linear and nonlinear solver support in IBAMR is provided by 
the PETSc library~\cite{petsc-efficient, petsc-user-ref, petsc-web-page}.
All of the example cases in the present work made use of distributed-memory
parallelism using the Message Passing Interface (MPI) library.
Between $4$ and $512$ processors were used in all the cases described here.

\section{Wave-structure interaction}
\label{sec_wsi}
In this section, we demonstrate that the present numerical method is capable of modeling complex
wave-structure interaction problems arising in marine and coastal engineering. We begin by describing 
our implementation of a numerical wave tank (NWT). Although NWTs based on VOF methods have been 
detailed in the literature~\cite{Chen14,Higuera2013,Hu2016,Jacobsen2012},
studies based on the level set methodology are sparse~\cite{Kasem2010,Bihs2016,Windt2018}.
Wave generation and wave absorption techniques for NWTs is an active area of research, and there are several strategies recommended 
in the literature (typically in the context of VOF methods)~\cite{Hu2016,Jacobsen2012}.
In this work, we use a combination of Dirichlet wave generation
boundary conditions and a relaxation-based wave damping procedure as our preferred choice. More specifically, 
by imposing inlet velocity boundary conditions based on Stokes wave theory at one end of the domain we are able to 
generate nonlinear water waves, which coherently propagate throughout the computational domain. By smoothly damping the traveling wave over a wavelength long region towards the opposite end, we mitigate 
the wave reflection and wave interference phenomena. 

In Sec.~\ref{sec_wave_theory}, we describe some background theory required to simulate a NWT within the 
present computational methodology. In Sec.~\ref{sec_wave_compute}, we present a number of validation 
cases to demonstrate that the solver is able to accurately produce second-order Stokes waves.
In Sec.~\ref{sec_wave_trap}, we investigate the problem of second-order Stokes wave interaction with 
a submerged trapezoid. The material properties of the liquid (gas) phase are set to be that of water (air):
$\rhol = 1 \times 10^3$, $\mul = 1 \times 10^{-3}$, $\rhog = 1.2$, and $\mug = 1.8 \times 10^{-5}$.
The gravitational acceleration of $g = 9.81$ is directed in the negative $y$-direction for the $2$D simulations
presented in this section.
\subsection{Stokes wave theory and numerics}
\label{sec_wave_theory}
According to second-order Stokes theory~\cite{Dean1991}, the wave elevation $\eta(x,t)$
from a mean water depth $d$ is given by
\begin{equation}
\label{eq_stokes_eta}
\eta(x,t) = \frac{\cH}{2} \cos\left(kx - \omega t\right) + \frac{\pi \cH^2}{8 \lambda}\frac{\cosh\left(kd\right)
		\left[2 + \cosh\left(2 kd\right)\right]}{\sinh^3\left(kd\right)} \cos\left(2kx-2\omega t\right),
\end{equation} 
in which $\cH$ is the peak-to-peak height of the wave, $\cT$ is the time period, $\omega = 2\pi/\cT$ is the angular frequency,
$\lambda$ is the wavelength, and $k = 2 \pi/\lambda$ is the wave number. The horizontal and vertical
components of velocity that generate this wave profile are written as
\begin{align}
&\uw(x,y,t) = \frac{\cH g k}{2 \omega} \frac{\cosh\left[ k(d+y)\right]}{\cosh(kd)} \cos(kx -\omega t) +
	       \frac{3 \cH^2 \omega k}{16}\frac{\cosh\left[2k(d+y)\right]}{\sinh^4(kd)}\cos(2kx-2\omega t), \label{eq_stokes_u}\\
&\vw(x,y,t) = \frac{\cH g k}{2 \omega} \frac{\sinh\left[ k(d+y)\right]}{\cosh(kd)} \sin(kx -\omega t) +
	       \frac{3 \cH^2 \omega k}{16}\frac{\sinh\left[2k(d+y)\right]}{\sinh^4(kd)}\sin(2kx-2\omega t).  \label{eq_stokes_v}
\end{align}
Note that in the above expressions for the theory and numerics presented in this section, 
we are considering a domain with bottom left corner situated at $(0,-d)$, without loss of generality.
Since the water phase is represented by negative signed distance values and the free surface
is initially located at $y = 0$, the elevation of the wave can be computed from $\phi_{i,j}$ via
\begin{equation}
\eta_{i,j} = -\phi_{i,j} + y_{i,j},
\end{equation}
in which $ y_{i,j}$ is the $y$-coordinate of grid cell $\x_{i,j}$. Since $\phi_{i,j}$ represents the signed distance function
to the interface, it is straightforward to show that the computed elevation $\eta_{i,j}$ will 
only be a function of the horizontal grid index $i$, i.e.
$\eta_{i,m} = \eta_{i,n}$ for all $m,n = 0, \ldots, \Ny - 1$.
 
At the inlet (left) boundary, we impose the desired velocities Eqs.~\eqref{eq_stokes_u} and~\eqref{eq_stokes_v} as
boundary conditions acting only in the liquid phase. For the normal velocity component, we compute the face-centered
level set value
based on the analytical elevation value along the computational boundary, $\phi^n_{-\half,j} = -\eta(0,t^n) + y_{-\half,j}$.
The normal velocity boundary condition is then given by
$u^n_{-\half,j} =  \left(1 - \widetilde{H}^n_{-\half,j}\right) \uw\left(0,y_{-\half,j},t^n\right)$,
where the expression for the numerical Heaviside at the boundary reads
\begin{equation}
\label{eq_heaviside_normal_bdry}
\widetilde{H}^n_{-\half,j} = 
\begin{cases} 
       0,  & \phi^n_{-\half,j} < -\ncells \sqrt{\delV},\\
        \frac{1}{2}\left(1 + \frac{1}{\ncells \sqrt{\delV}} \phi^n_{-\half,j} + \frac{1}{\pi} \sin\left(\frac{\pi}{ \ncells  \sqrt{\delV}} \phi^n_{-\half,j}\right)\right) , & |\phi^n_{-\half,j}| \le \ncells  \sqrt{\delV}, \\
        1,  & \textrm{otherwise}.
\end{cases}
\end{equation}
In the above expression, $\sqrt{\delV} = \sqrt{\dx \dy}$ represents a characteristic grid spacing for grids with unequal grid
spacing in each direction, e.g. $\dx \ne \dy$. Similarly for the tangential velocity component, the desired node-centered 
level set values can be computed
as $\phi^n_{-\half,j-\half} = -\eta(0,t^n) + y_{-\half,j-\half}$ with corresponding Heaviside function
$\left(1-\widetilde{H}^n_{-\half,j-\half}\right)$, which are multiplied by $\vw\left(0,y_{-\half,j-\half},t^n\right)$ to obtain desired boundary
condition. We refer readers to~\cite{Nangia2018,Griffith2009} for more details on the imposition of 
normal and tangential velocity boundary conditions in a staggered flow solver. Note that we are simply imposing homogenous 
Neumann conditions for the level set value at \emph{all} domain boundaries.
No-slip boundary conditions are imposed along the bottom 
and right boundary, while homogenous tangential velocity and zero pressure boundary conditions~\footnote{Imposition of 
pressure or normal traction boundary conditions is possible because of the monolithic velocity-pressure solver.} are imposed at the top boundary. 

In order to mitigate the reflection of waves at the right boundary, we place a damping zone at the downstream
end of the computational domain from $x = \xL$ to $x = \xU$.
We follow the approach described by Jacobsen et al.~\cite{Jacobsen2012}, in which
the numerical velocities and level set values are smoothly relaxed at the end of each time step via,
\begin{align}
& u_{i-\half,j} = \alpha_{i-\half,j} u^{\textrm{computed}}_{i-\half,j} 
		    + \left(1 - \alpha_{i-\half,j}\right) u^{\textrm{target}}_{i-\half,j}, \\
& v_{i,j-\half} = \alpha_{i,j-\half} v^{\textrm{computed}}_{i,j-\half} 
		    + \left(1 - \alpha_{i,j-\half}\right) v^{\textrm{target}}_{i,j-\half}, \\
& \phi_{i,j} = \alpha_{i,j} \phi^{\textrm{computed}}_{i,j} 
		    + \left(1 - \alpha_{i,j}\right) \phi^{\textrm{target}}_{i,j}.
\end{align} 
In the above expressions, the superscript ``computed" indicates the staggered grid velocity and cell-centered level set
values computed from the solution methodology described in Sec.~\ref{sec_temporal_scheme}, 
and the superscript ``target" indicates the desired analytical values representing still water of depth $d$. Hence,
$u^{\textrm{target}}_{i-\half,j} = 0$, $v^{\textrm{target}}_{i,j-\half} = 0$, and $\phi^{\textrm{target}}_{i,j} = y_{i,j}$.
The relaxation parameter $\alpha$ is smoothly varied from $1$, at the interface between the non-relaxed portion
of the domain and the damping zone (e.g. $\xL$), to $0$ at the rightmost computational boundary (e.g. $\xU$).
For example at cell centers,
the functional form of alpha reads,
\begin{equation}
 \alpha_{i,j} = 1 - \frac{\exp\left(\bar{x}_{i,j}^{3.5}\right) - 1}{\exp(1) - 1},
\end{equation}
in which $\bar{x}_{i,j} = \left(x_{i,j} - \xL\right)/\left(\xU-\xL\right)$ is the normalized horizontal coordinate varying
from $0$ to $1$ across the length of the damping zone. Analogous expressions are determined for 
$\alpha_{i-\half,j}$ and $\alpha_{i,j-\half}$.
In all of the cases considered in this section, a damping zone of length $10d$ is prescribed.
Next, we present various numerical examples demonstrating the accuracy of the aforementioned
wave generation and damping techniques. 

\subsection{Validation of second-order Stokes waves propagating in a NWT}
\label{sec_wave_compute}
As an initial example we consider a $2$D computational domain of size $\Omega = [0,68d]\times[-d,0.3d]$, which
is occupied by initially quiescent water of depth $d = 0.4$. Air occupies the remainder of the domain from $y = 0$
to $y = 0.3d$. One grid cell of smearing $\ncells = 1$ is used on either side of the air-water interface and surface tension
forces are neglected.
The domain is discretized by a grid of size $\Nx \times \Ny$ and a constant time step size of $\dt = 100/(57 \Nx)$ is used.
The wave parameters are chosen to be $\cH = 0.05d$, $\cT = 9.8995 \sqrt{d/g}$, and $\lambda = 9.232 d$; these are chosen
to satisfy the required dispersion relation for (second-order) Stokes waves~\cite{Dean1991},
\begin{equation}
\label{eq_disp}
	\omega^2 = g k  \tanh(k d).
 \end{equation}
 
\begin{figure}[]
  \centering
    \includegraphics[scale = 0.3]{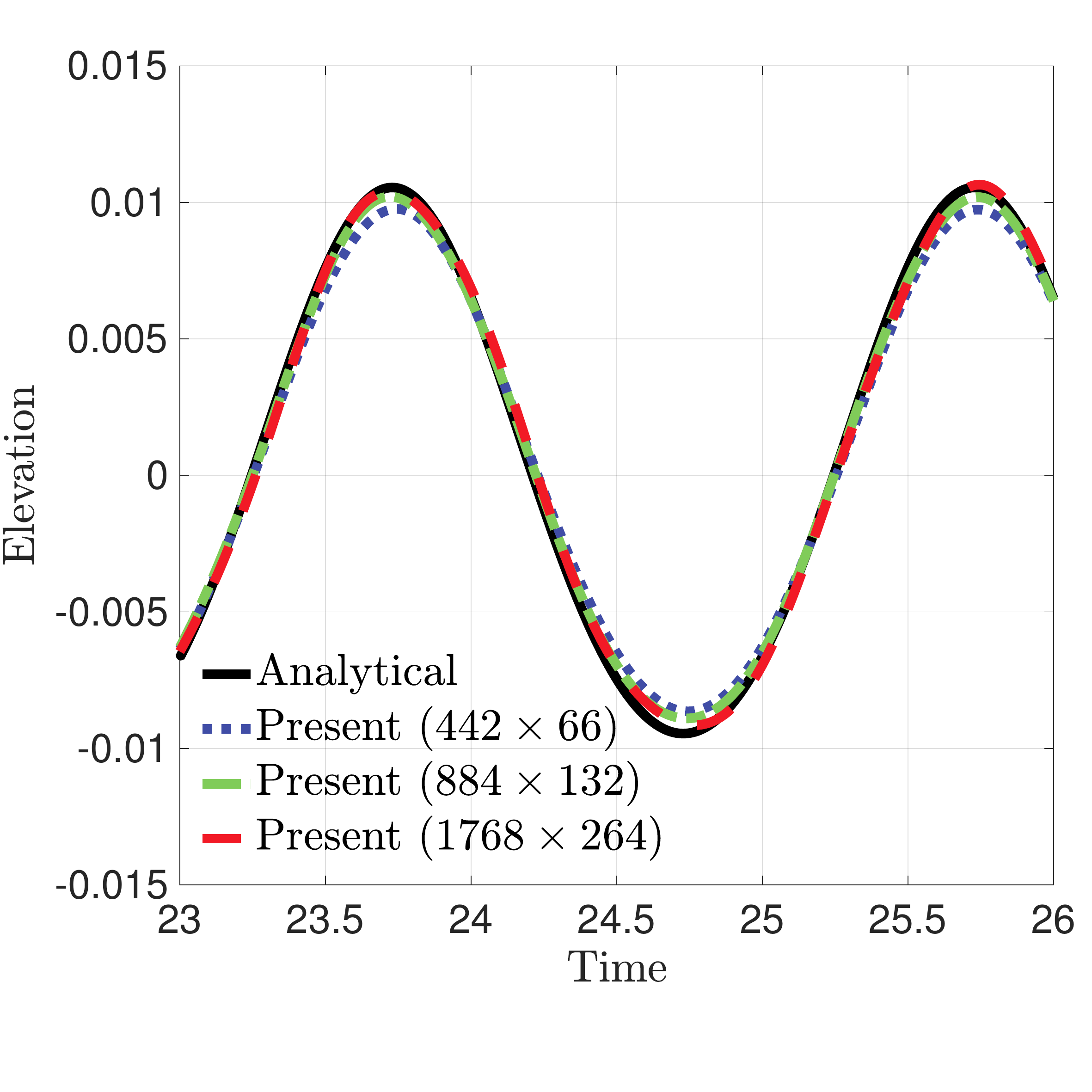}
   \caption{
   Convergence study for the temporal evolution of wave elevation at
   $x = 2.87 \lambda$ for a second-order Stokes wave;
   (---, black) analytical expression given by Eq.~\ref{eq_stokes_eta};
   (\texttt{...}, blue) present simulation for a  $442 \times 66$ grid;
    (\texttt{-}$\cdot$\texttt{-}, green) present simulation for a  $884 \times 132$ grid;
     (\texttt{---}, red) present simulation for a $1768 \times 264$ grid.
 }
  \label{wave_nobody_zoom_s2}
\end{figure}

To quantitatively assess the accuracy of the wave generation boundary conditions, the analytical and simulated 
elevation computed at a probe situated at $x = 2.87 \lambda$, are plotted against time in Fig.~\ref{wave_nobody_zoom_s2}
for three different grid sizes: $442 \times 66$, $884 \times 132$, and $1768 \times 264$. As the resolution increases,
the numerical simulations converge towards the theoretical elevation given by Eq.~\eqref{eq_stokes_eta}.
The errors in maximum elevation attained over the shown time period
decrease as the resolution increases, yielding a convergence rate of $1.23$ between grid sizes $442 \times 66$
and $884 \times 132$ and a convergence rate of $1.17$ between grid sizes $884 \times 132$ and $1768 \times 264$.
There are approximately $N_\cH = 10$ grid cells per wave height and $N_{\lambda} = 240$ grid cells per wavelength
for the finest resolution case considered here, which we hereafter denote as Case A.

\begin{figure}[]
  \centering
    \subfigure[Wave classification phase space]{
    \includegraphics[scale = 0.8]{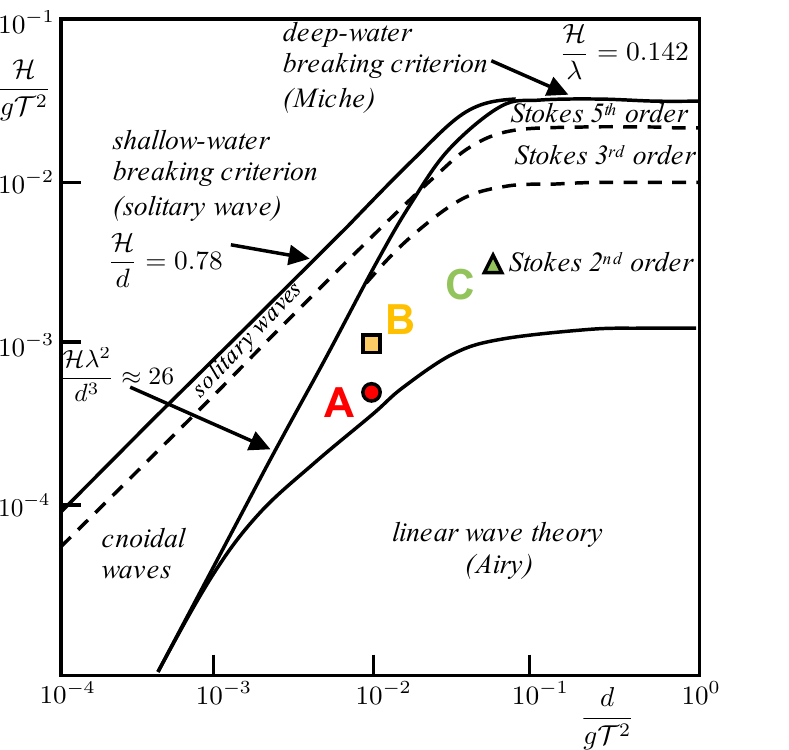}
    \label{wave_phase_diagram}
  }
   \subfigure[Case A]{
    \includegraphics[scale = 0.28]{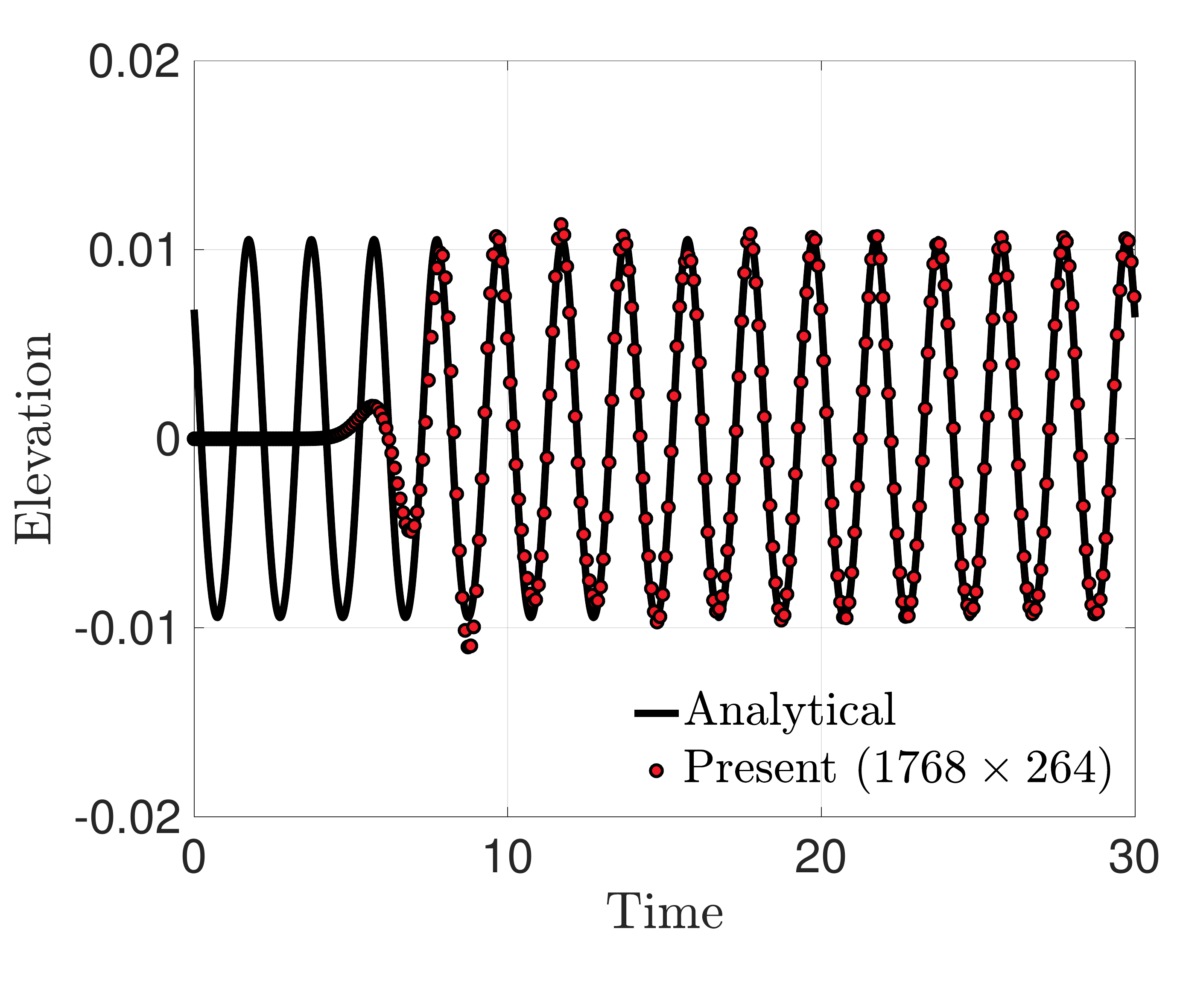}
    \label{wave_caseA}
  }
    \subfigure[Case B]{
    \includegraphics[scale = 0.28]{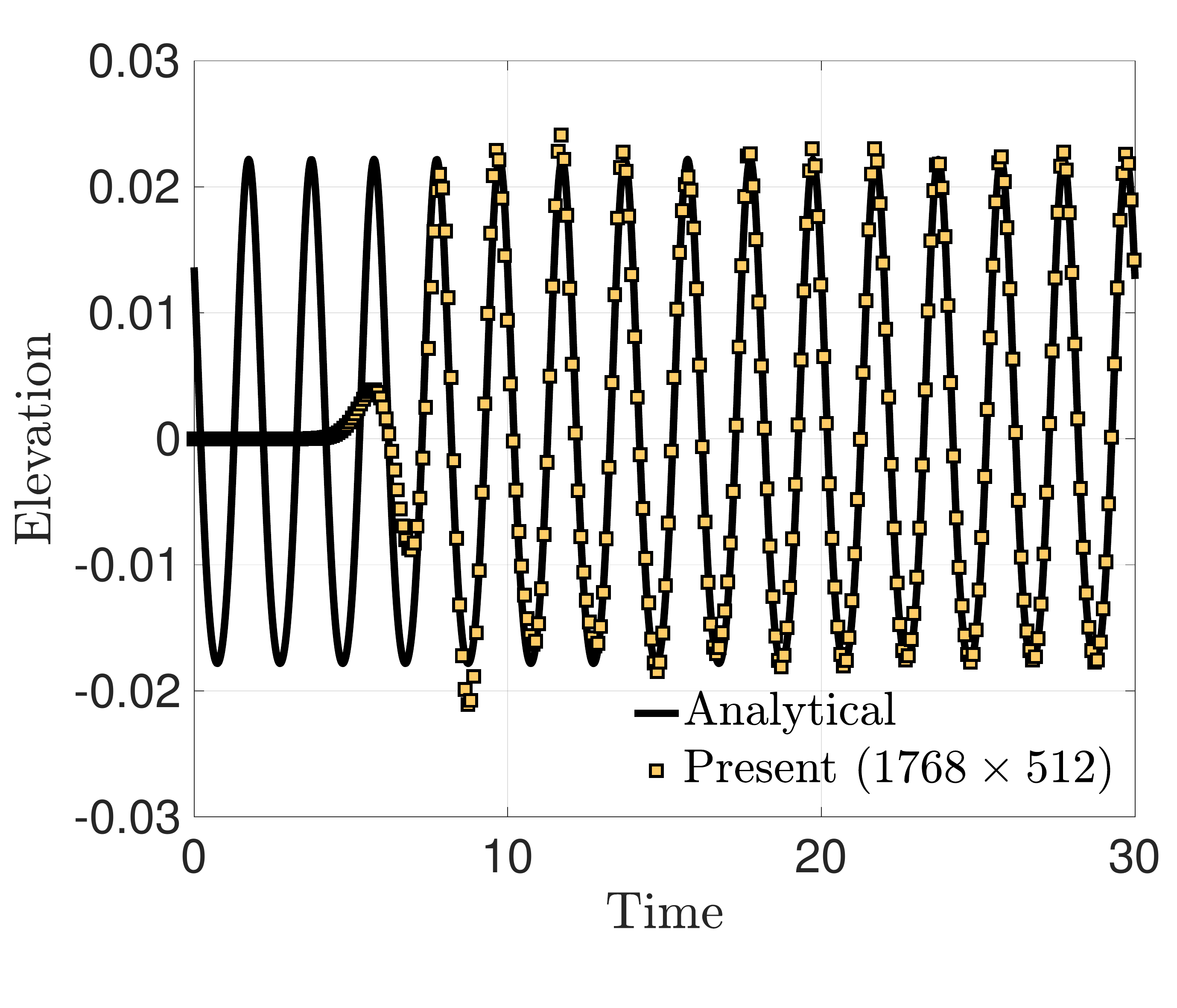}
    \label{wave_caseB}
  }
   \subfigure[Case C]{
    \includegraphics[scale = 0.28]{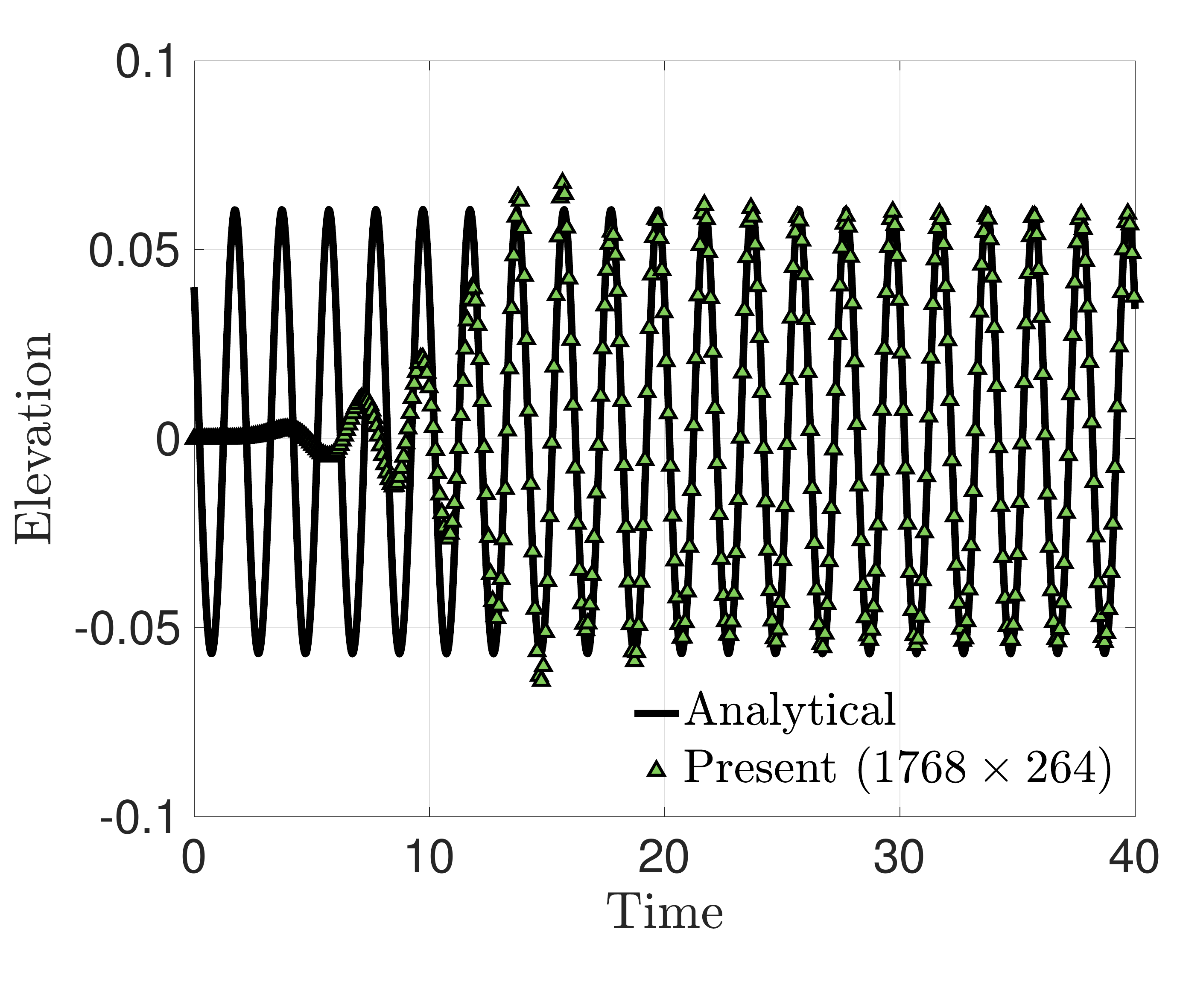}
    \label{wave_caseC}
  }
   \caption{
   \subref{wave_phase_diagram} Locations of ($\bullet$, red) Case A, ($\blacksquare$, yellow) Case B,
   and ($\blacktriangle$, green) Case C on a phase diagram denoting the applicability of wave theories described 
   by Le M{\'e}haut{\'e}~\cite{LeMehaute2013}; figure adapted from Holthuijsen~\cite{Holthuijsen2010}.
   Long-time temporal evolution of wave elevation for second-order Stokes waves with parameters described
   by
   \subref{wave_caseA} Case A,
    \subref{wave_caseB} Case B, and
     \subref{wave_caseC} Case C;
   (---, black) analytical expression given by Eq.~\ref{eq_stokes_eta};
   wave elevation for all cases is measured at $x = 2.87 \lambda$.}
  \label{fig_wave_all_cases}
\end{figure}

For our next example we consider two additional sets of wave parameters, which we denote as Case B and Case C;
see Table~\ref{tab_wave_cases} for a full specification of all three cases. These parameters are chosen such that they
satisfy the dispersion relation Eq.~\eqref{eq_disp} and occupy different locations within the second-order Stokes
regime for the wave classification phase space described by Le M{\'e}haut{\'e}~\cite{LeMehaute2013}
(see Fig.~\ref{wave_phase_diagram}). Figs.~\ref{wave_caseA}--\ref{wave_caseC} show
the long-time temporal evolution of elevation for cases A, B, and C, respectively. In all three cases, the numerical
wave tank produces elevations that are in excellent agreement with Eq.~\ref{eq_stokes_eta}. These examples show
that the present numerical method can be confidently used to simulate second-order Stokes waves across the entire
(second-order Stokes) region of applicability.

\begin{table}
    \centering
    \caption{Parameter specification for the three
    second-order Stokes wave cases considered in Sec.~\ref{sec_wave_compute}.}
    \rowcolors{2}{}{gray!10}
    \begin{tabular}{*6c}
        \toprule
        Parameters & Case A & Case B & Case C\\
        \midrule
        Depth ($d$) & $0.4$ & $0.4$ & $2.35$\\
        Wave height ($\cH$)  & $0.05d$ & $0.1d$ & $0.05d$ \\
        Wave period ($\cT$) & $9.8995 \sqrt{d/g}$ & $9.8995 \sqrt{d/g}$ & $4.0825\sqrt{d/g}$\\
        Wavelength ($\lambda$) & $9.232 d$ & $9.232 d$ & $2.610d$ \\
        Domain size & $68 d \times 1.3 d$ & $68 d \times 1.3 d$ & $68 d \times 1.3 d$\\
        Cells per wavelength ($N_\lambda$) & $240$ & $240$ & $68$ \\
        Cells per wave height ($N_\cH$) & $10$ & $10$ & $10$ \\
        Elevation probe location ($x$) & $2.87 \lambda$ & $2.87 \lambda$ & $2.87 \lambda$ \\
        \bottomrule
    \end{tabular}
    \label{tab_wave_cases}
\end{table}

\subsection{Wave interaction with a submerged trapezoid}
\label{sec_wave_trap}

\begin{figure}[]
  \centering
  \textbf{A: Without submerged body} \\
  \subfigure[$t = 0$]{
    \includegraphics[scale = 0.18]{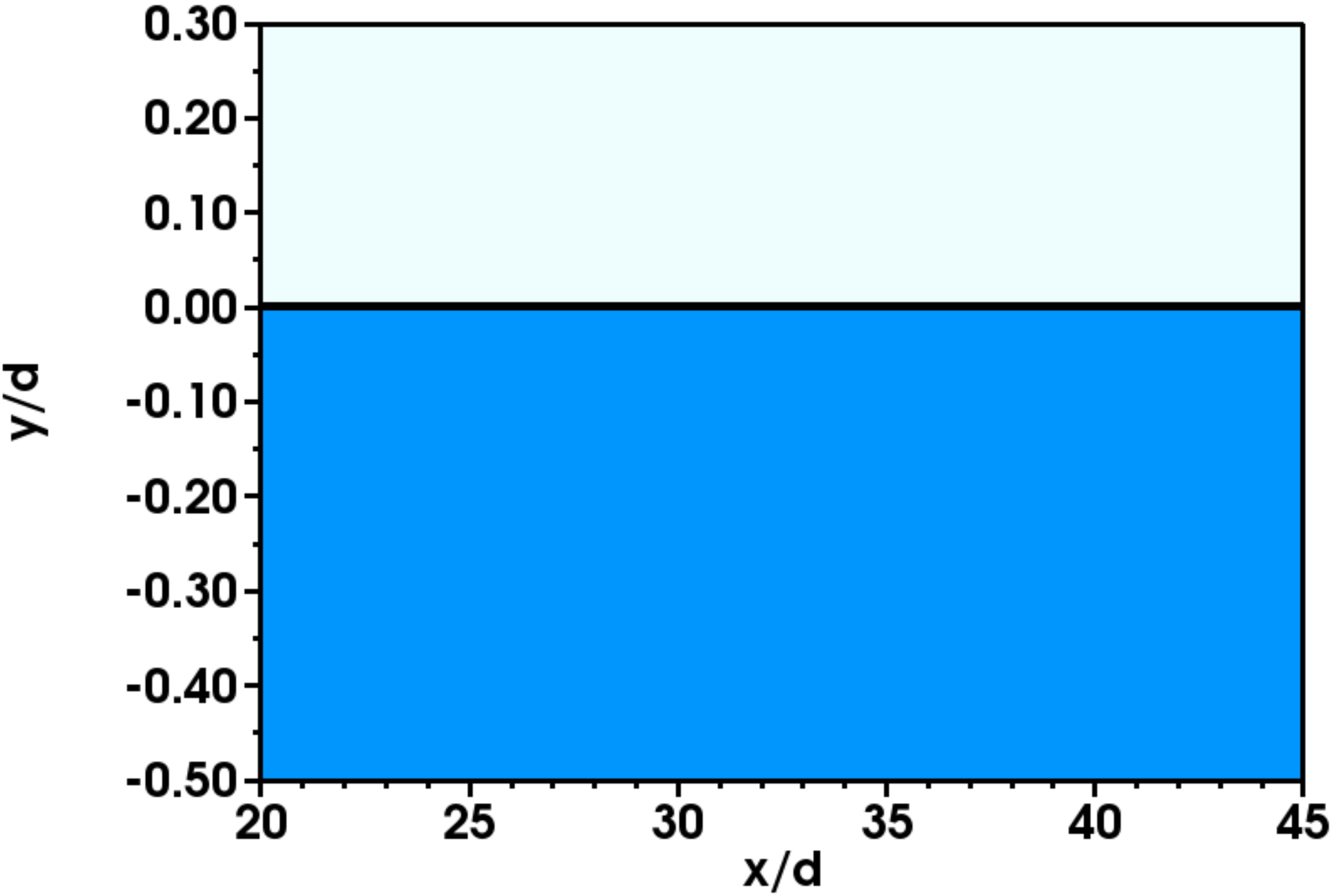}
    \label{Wave_NoBody_t0}
  }
     \subfigure[$t = 22$]{
    \includegraphics[scale = 0.18]{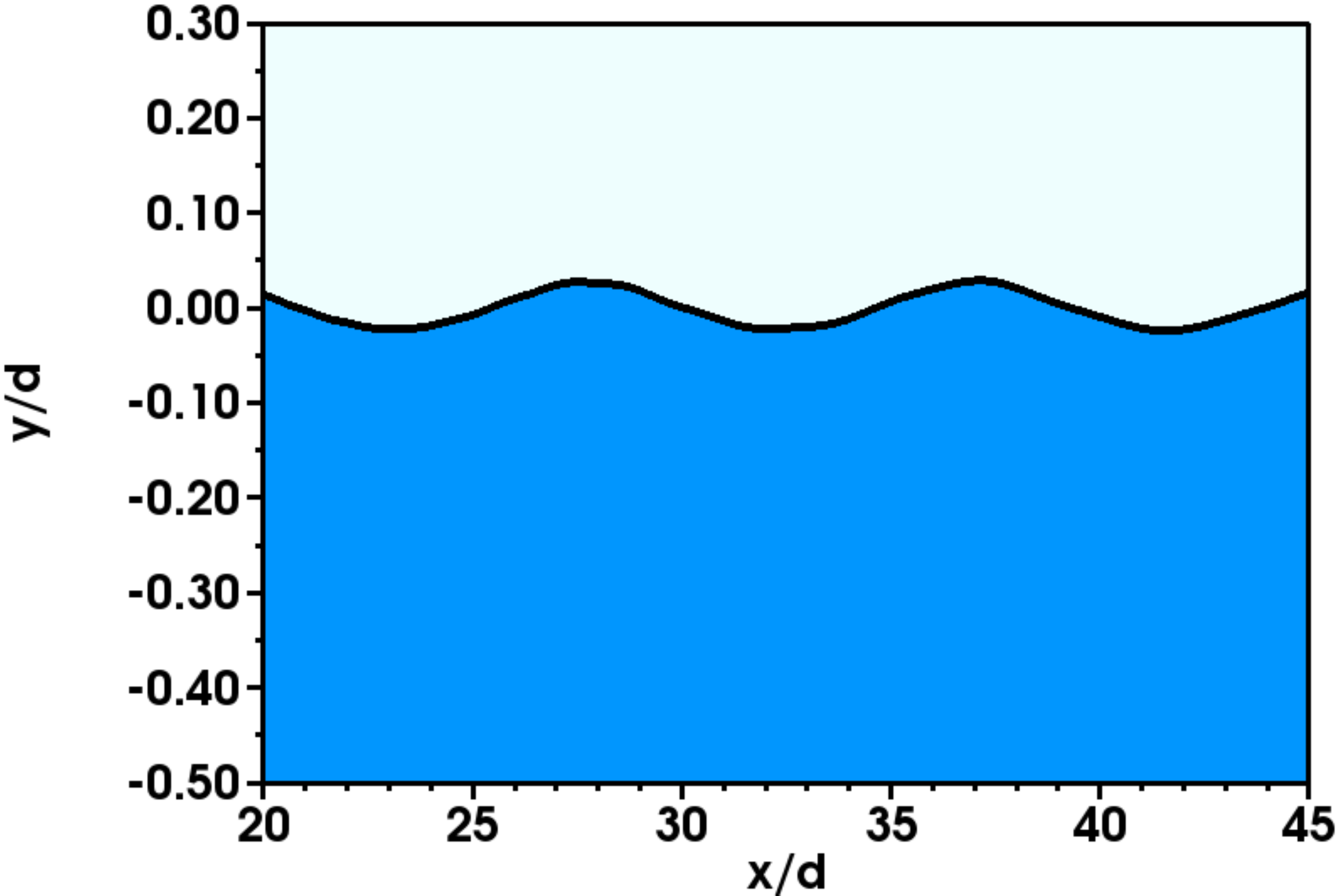}
    \label{Wave_NoBody_t22}
  }
   \subfigure[$t = 22.5$]{
    \includegraphics[scale = 0.18]{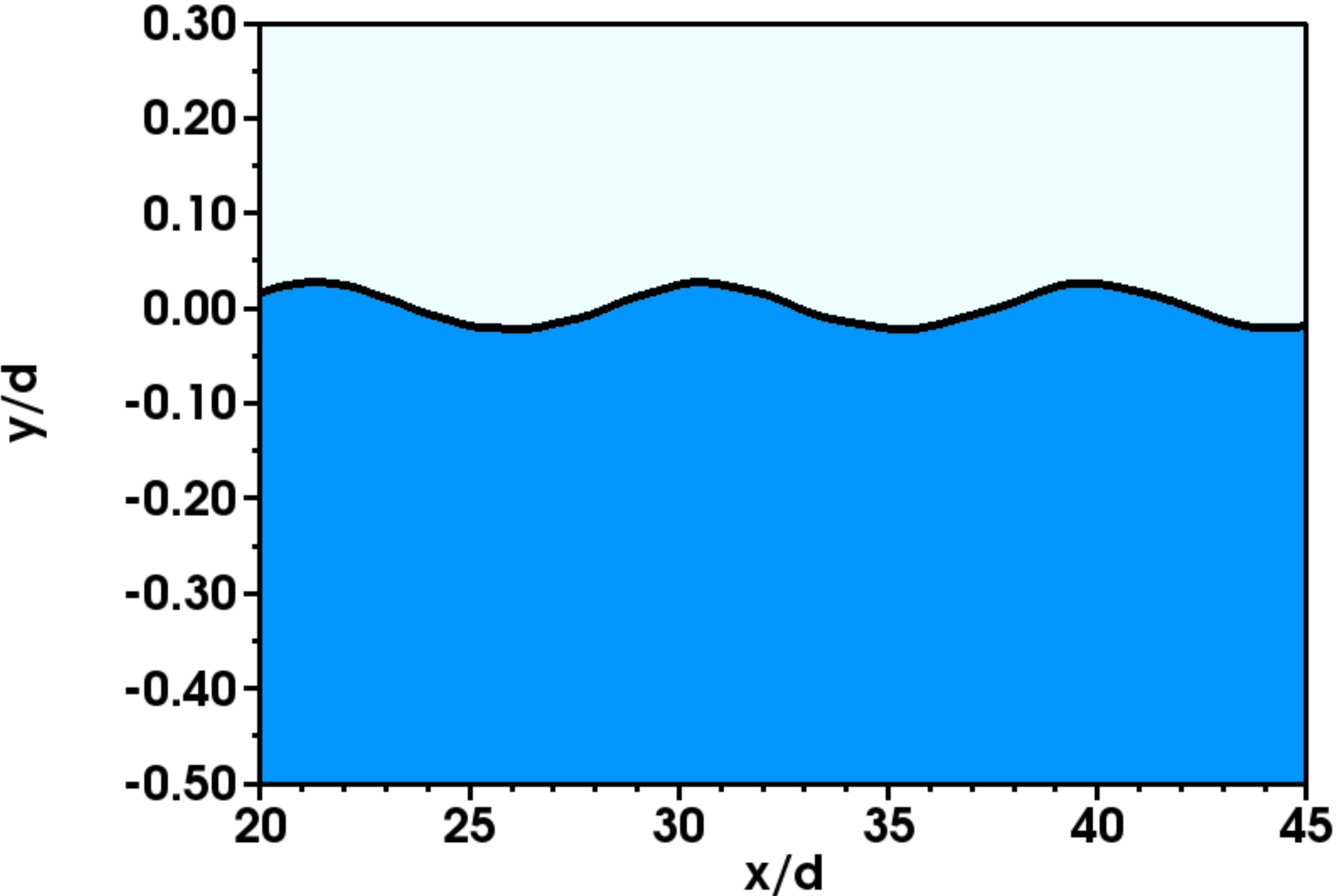}
    \label{Wave_NoBody_t22p5}
  }
     \subfigure[$t = 23$]{
    \includegraphics[scale = 0.18]{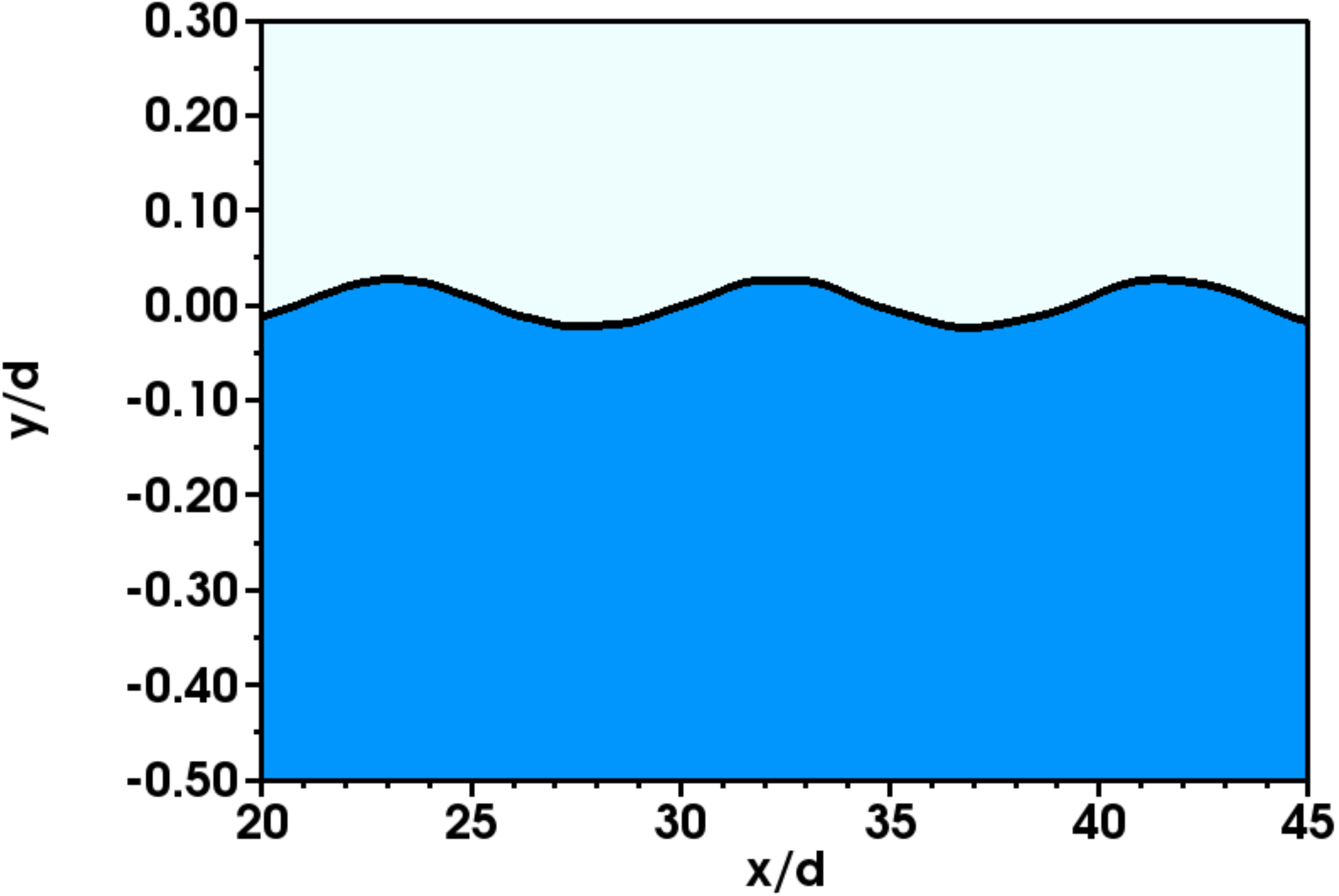}
    \label{Wave_NoBody_t23}
  }
   \subfigure[$t = 23.5$]{
    \includegraphics[scale = 0.18]{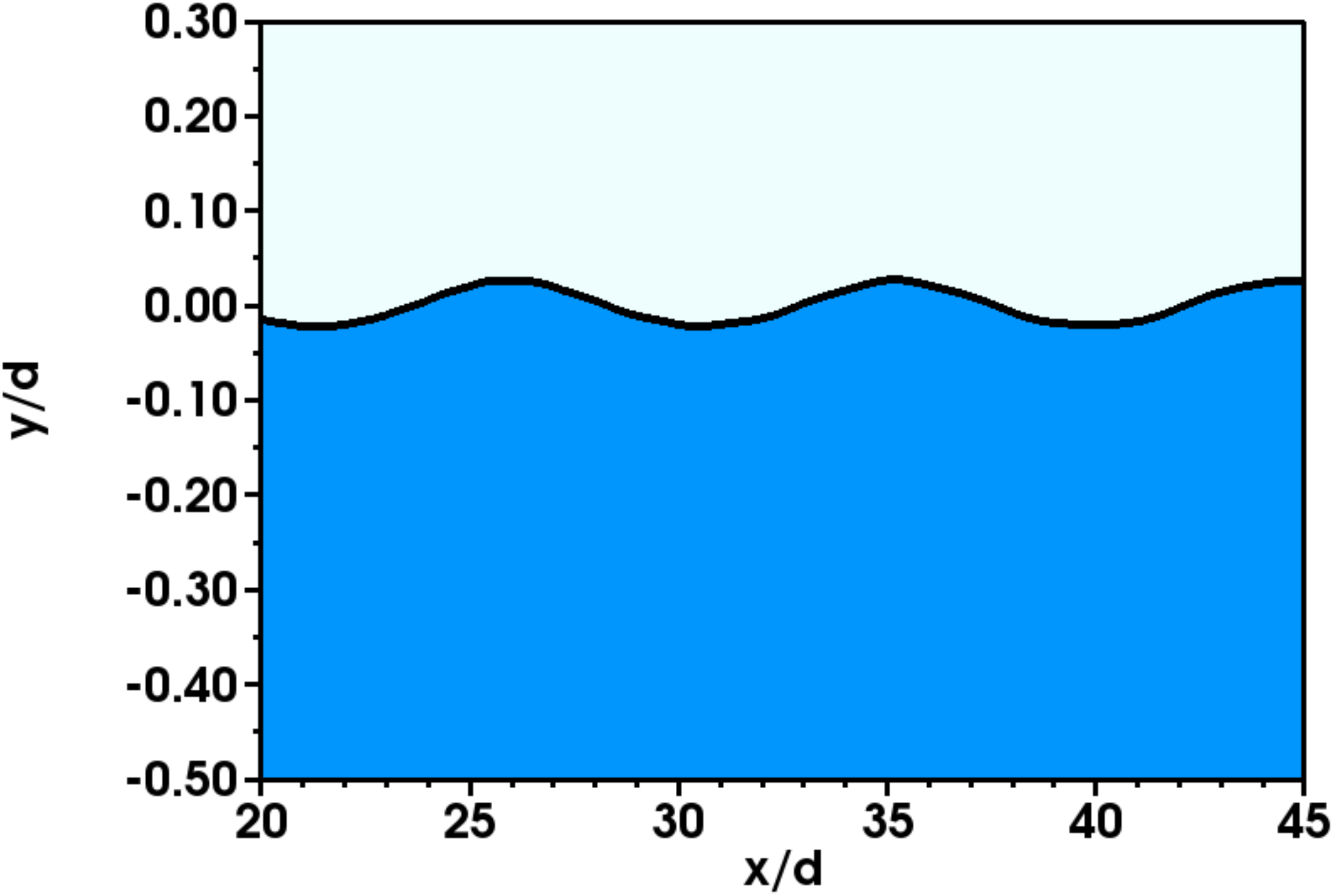}
    \label{Wave_NoBody_t23p5}
  }
     \subfigure[$t = 24$]{
    \includegraphics[scale = 0.18]{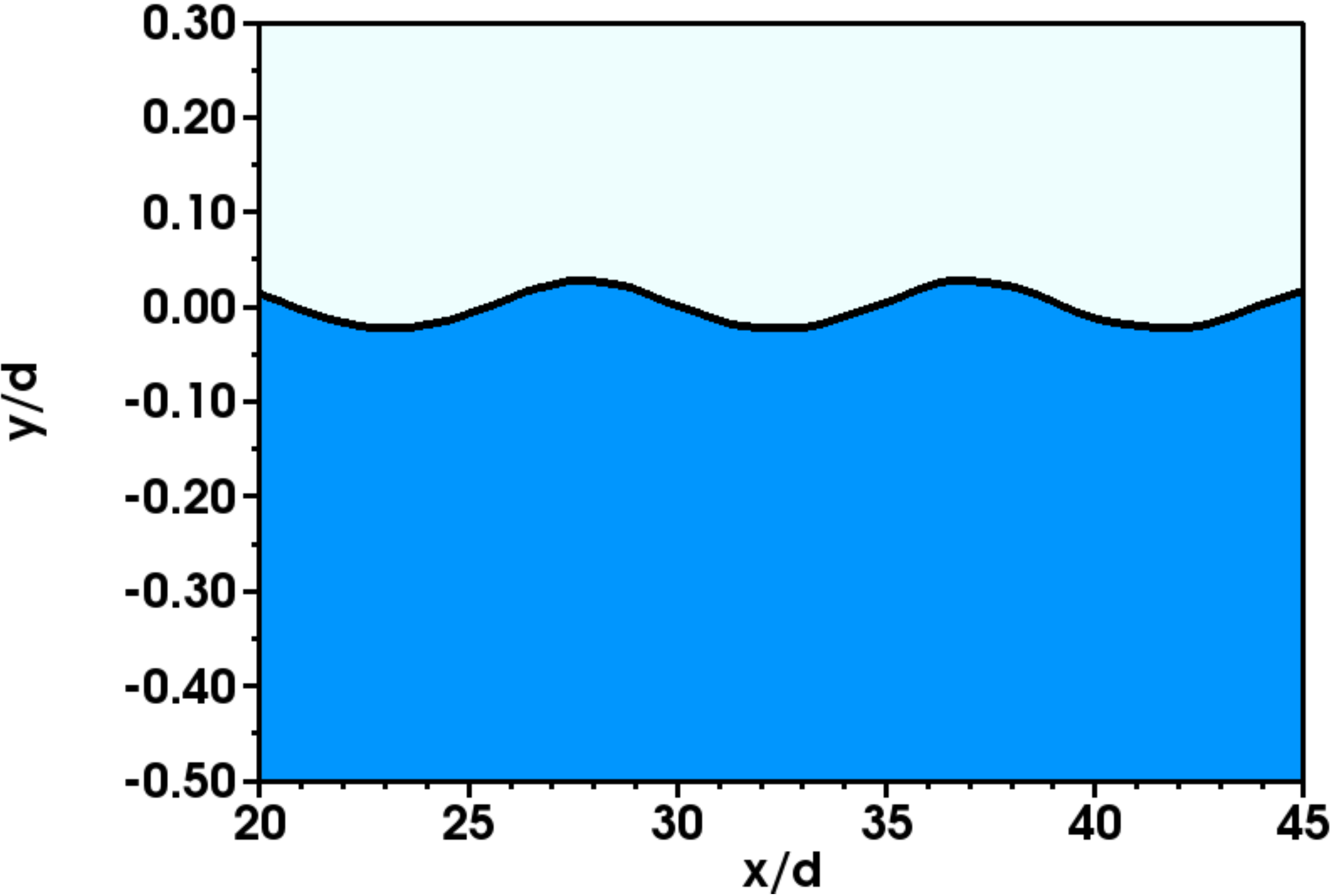}
    \label{Wave_NoBody_t24}
  }
 \textbf{B: With submerged body} \\
  \centering
  \subfigure[$t = 0$]{
    \includegraphics[scale = 0.18]{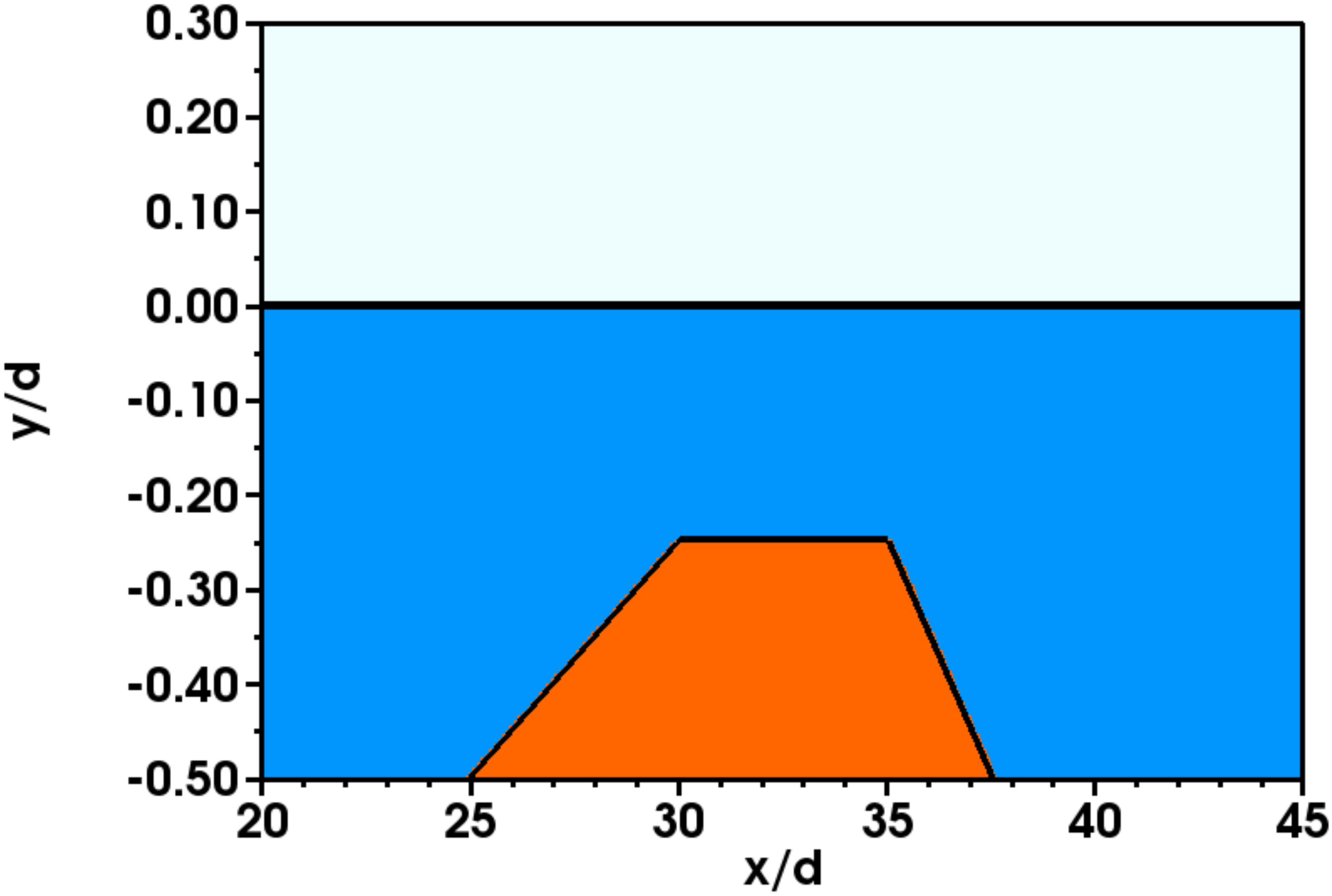}
    \label{Wave_WithBody_t0}
  }
     \subfigure[$t = 22$]{
    \includegraphics[scale = 0.18]{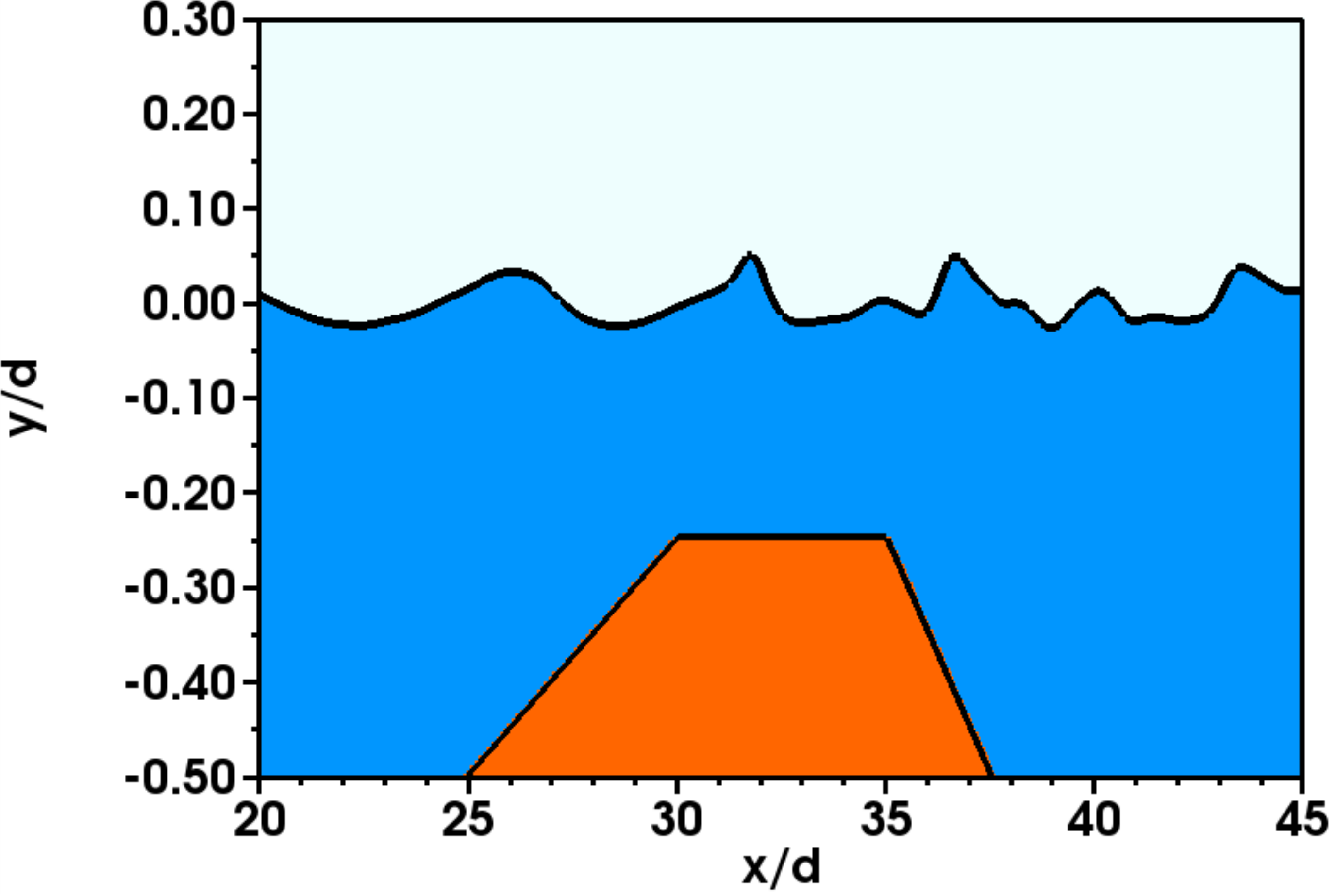}
    \label{Wave_WithBody_t22}
  }
   \subfigure[$t = 22.5$]{
    \includegraphics[scale = 0.18]{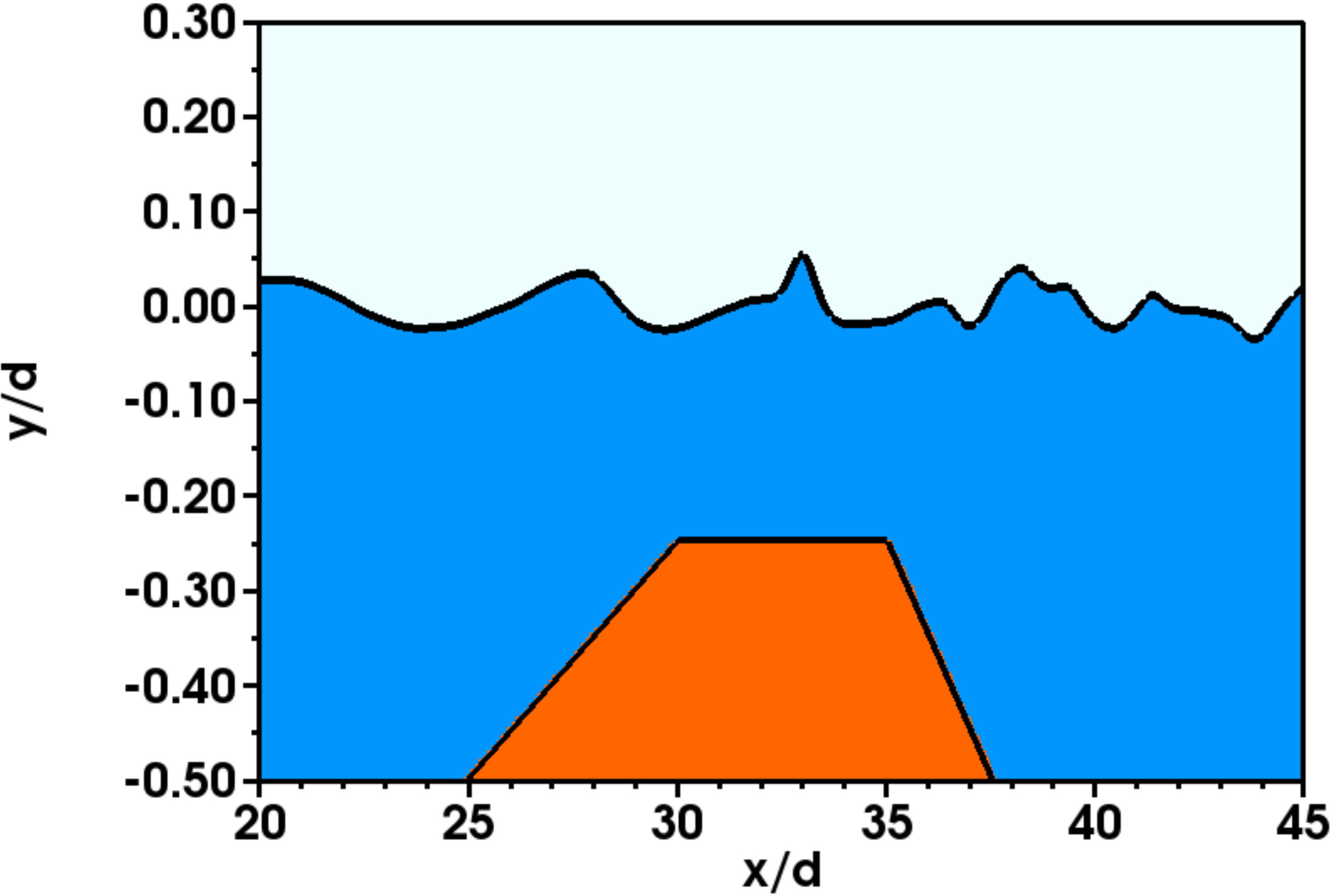}
    \label{Wave_WithBody_t22p5}
  }
     \subfigure[$t = 23$]{
    \includegraphics[scale = 0.18]{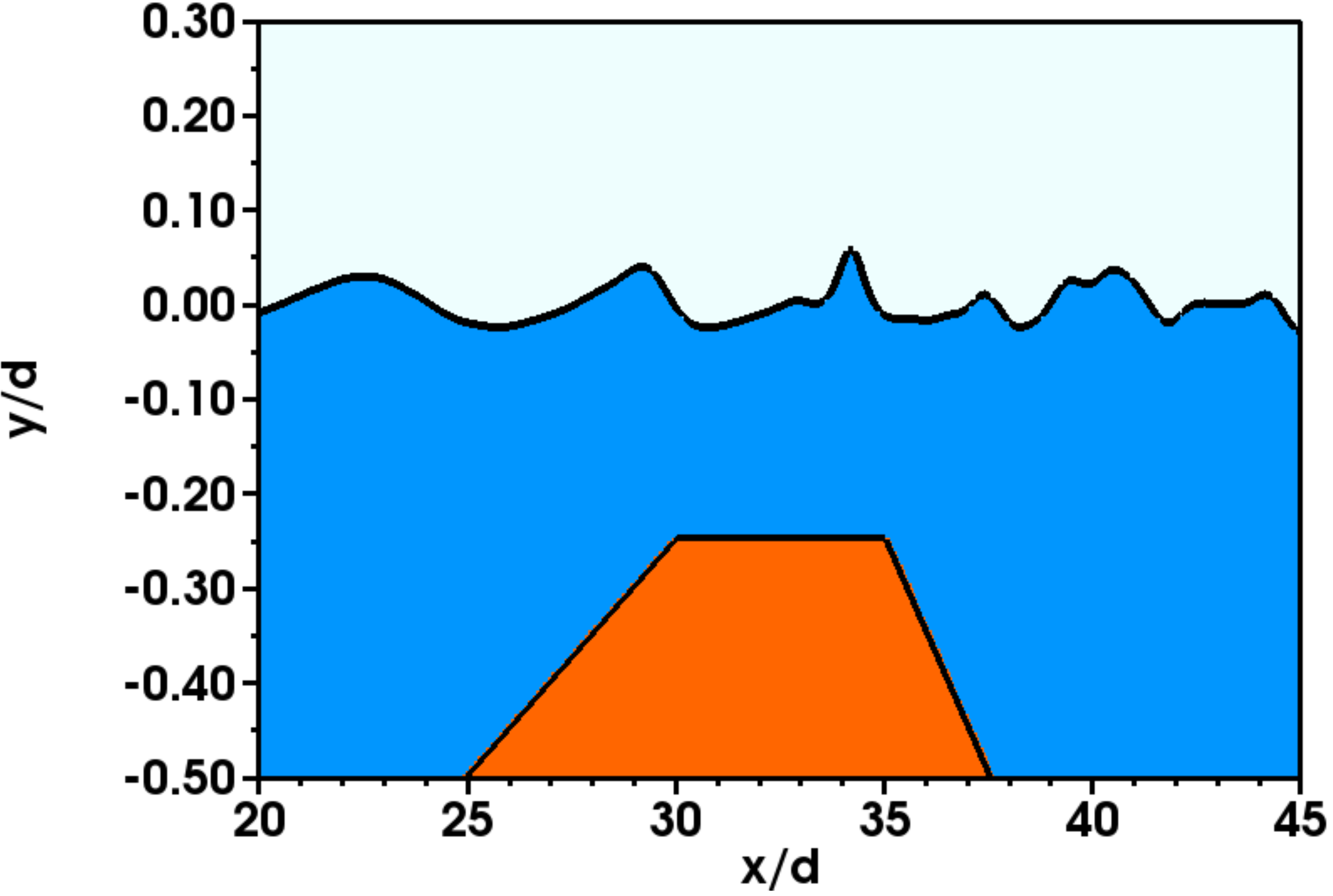}
    \label{Wave_WithBody_t23}
  }
   \subfigure[$t = 23.5$]{
    \includegraphics[scale = 0.18]{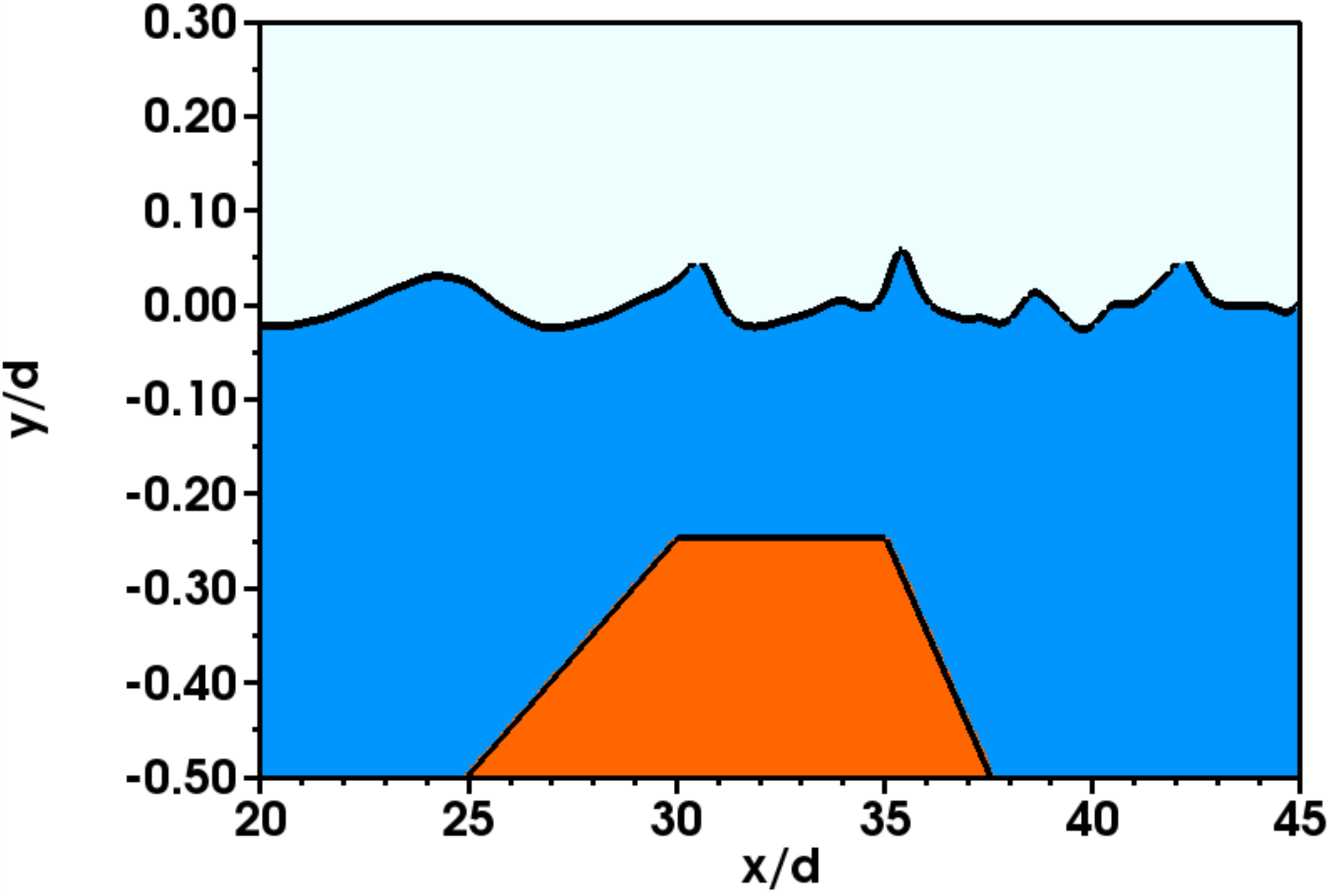}
    \label{Wave_WithBody_t23p5}
  }
     \subfigure[$t = 24$]{
    \includegraphics[scale = 0.18]{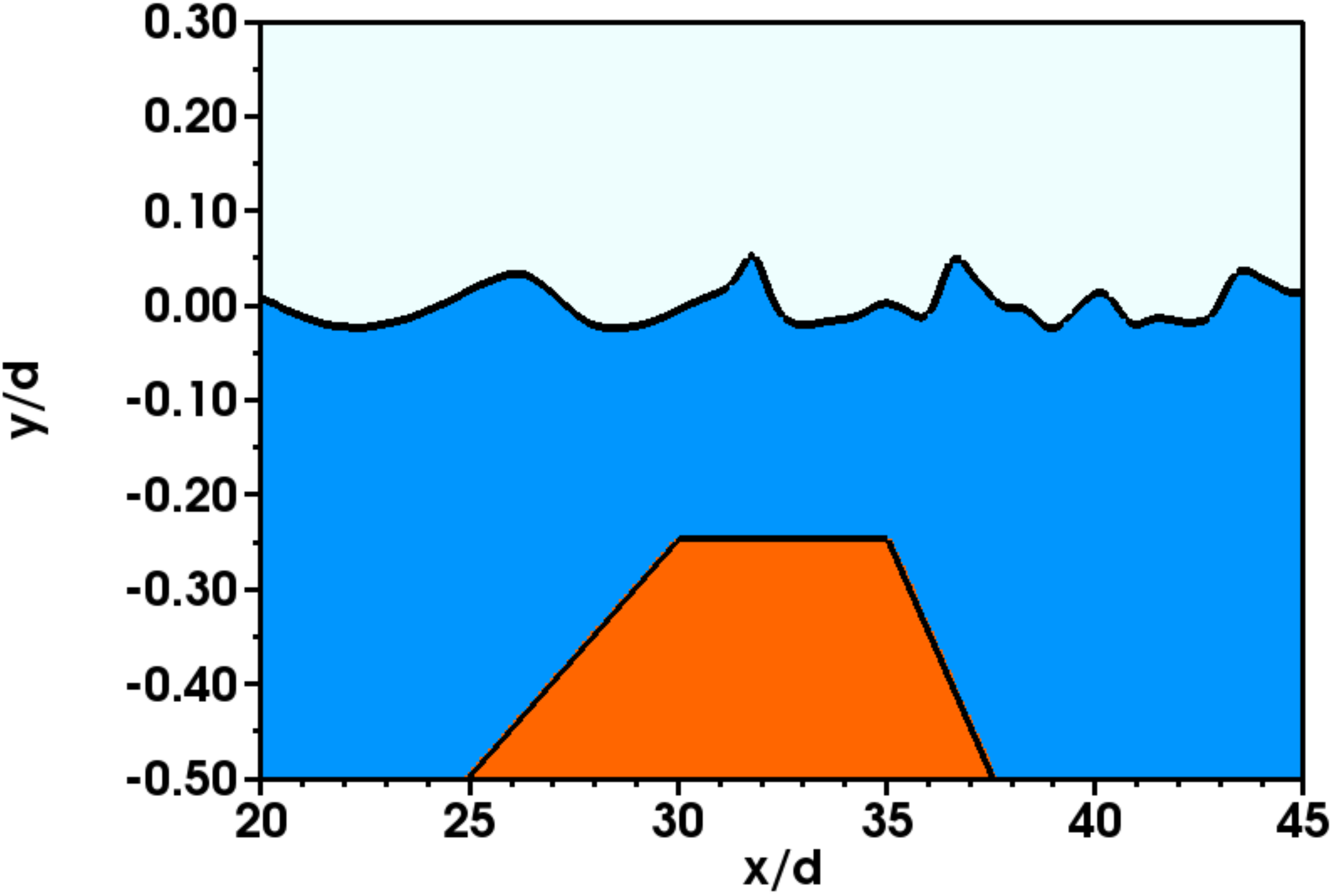}
    \label{Wave_WithBody_t24}
  }
  \caption{\textbf{A: }Temporal evolution of unobstructed second-order Stokes waves at six different 
  		time instances. \textbf{B:} Temporal evolution of second-order Stokes waves interacting with a stationary trapezoidal obstacle at six different 
  time instances.}
  \label{fig_wave_body_viz}
\end{figure}

\begin{figure}[]
  \centering
    \includegraphics[scale = 0.75]{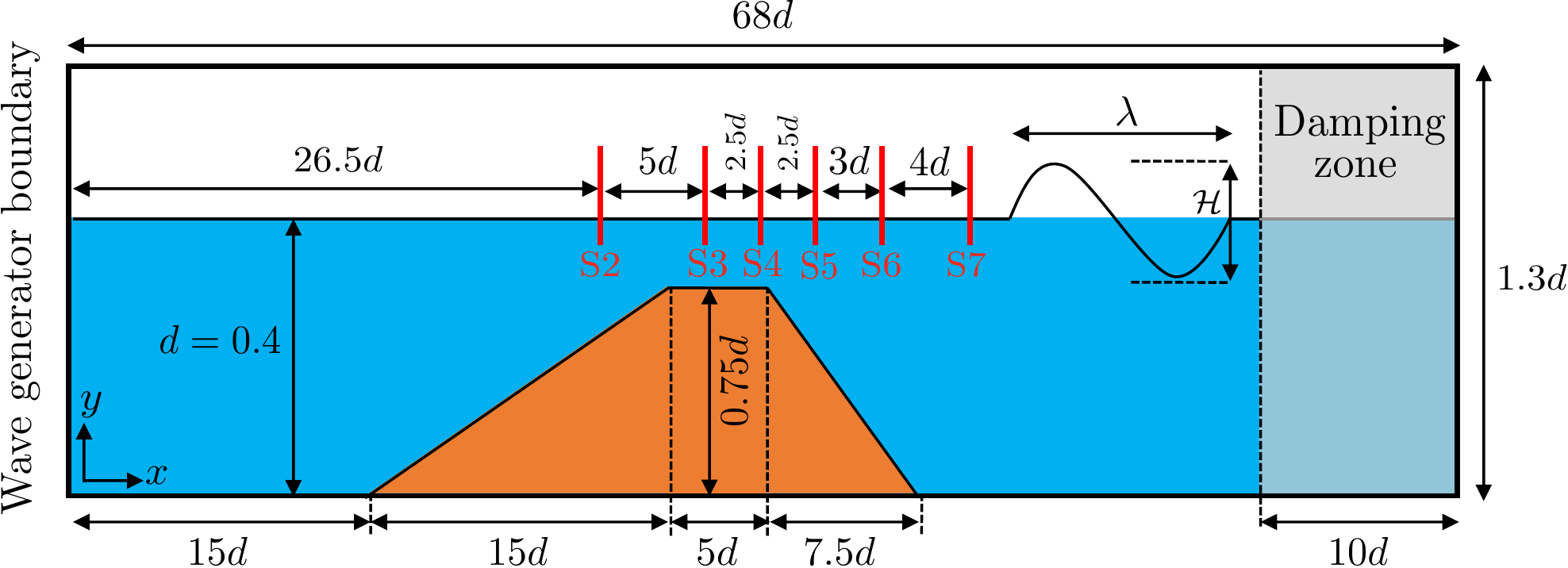}
     \caption{Sketch of the problem set up for an incoming second-order Stokes wave interacting with
     a trapezoidal structure;
     (blue) initially quiescent water of depth $d$;
     (orange) stationary trapezoidal obstacle;
     (grey) region over which velocities and level set values are damped;
     (red) locations of the water elevation measurement stations. Diagram is not to scale.}
  \label{fig_wavetrap_schematic}
\end{figure}

Now we investigate the interaction between second-order Stokes waves and a fully submerged 
trapezoidal shaped structure. The domain size and numerical parameters are identical to those of Case A in the previous section,
and the simulation is carried out on a grid of size $1768 \times 264$. The top and base of the trapezoid have length $5d$
and $27.5d$, respectively, and its height is $0.75d$; see Fig.~\ref{fig_wavetrap_schematic} for a full description of the
problem set up.
All of the trapezoid's translational and rotational degrees of freedom
are locked and it is fully constrained to remain stationary. Since the structure is fully submerged in one fluid, 
the issue of parasitic currents due to gravitational force does not arise in this scenario and the solid density is set equal to 
the water phase density. The primary quantities of interest for this example are the elevation values collected from six stations
placed along the computational domain. This problem has been studied experimentally by
Beji and Battjes~\cite{Beji1993,Beji1994}, and numerically by Kasem and Sasaki~\cite{Kasem2010}.

Visualizations of the evolving unobstructed wavefront from the previous section, and the wave-structure interaction
are shown in Fig.~\ref{fig_wave_body_viz}, top half and bottom half panels, respectively. Both simulations
show temporally cyclic behavior, although irregular amplitude profiles are exhibited by the WSI case. The irregular 
wave amplitude corresponds to the wave shoaling effect (reduction of water depth near the shore) caused by the 
submerged structure. The wave profile results qualitatively agree with those shown in~\cite{Kasem2010}.

\begin{figure}[]
  \centering
    \subfigure[S$2$ station]{
    \includegraphics[scale = 0.25]{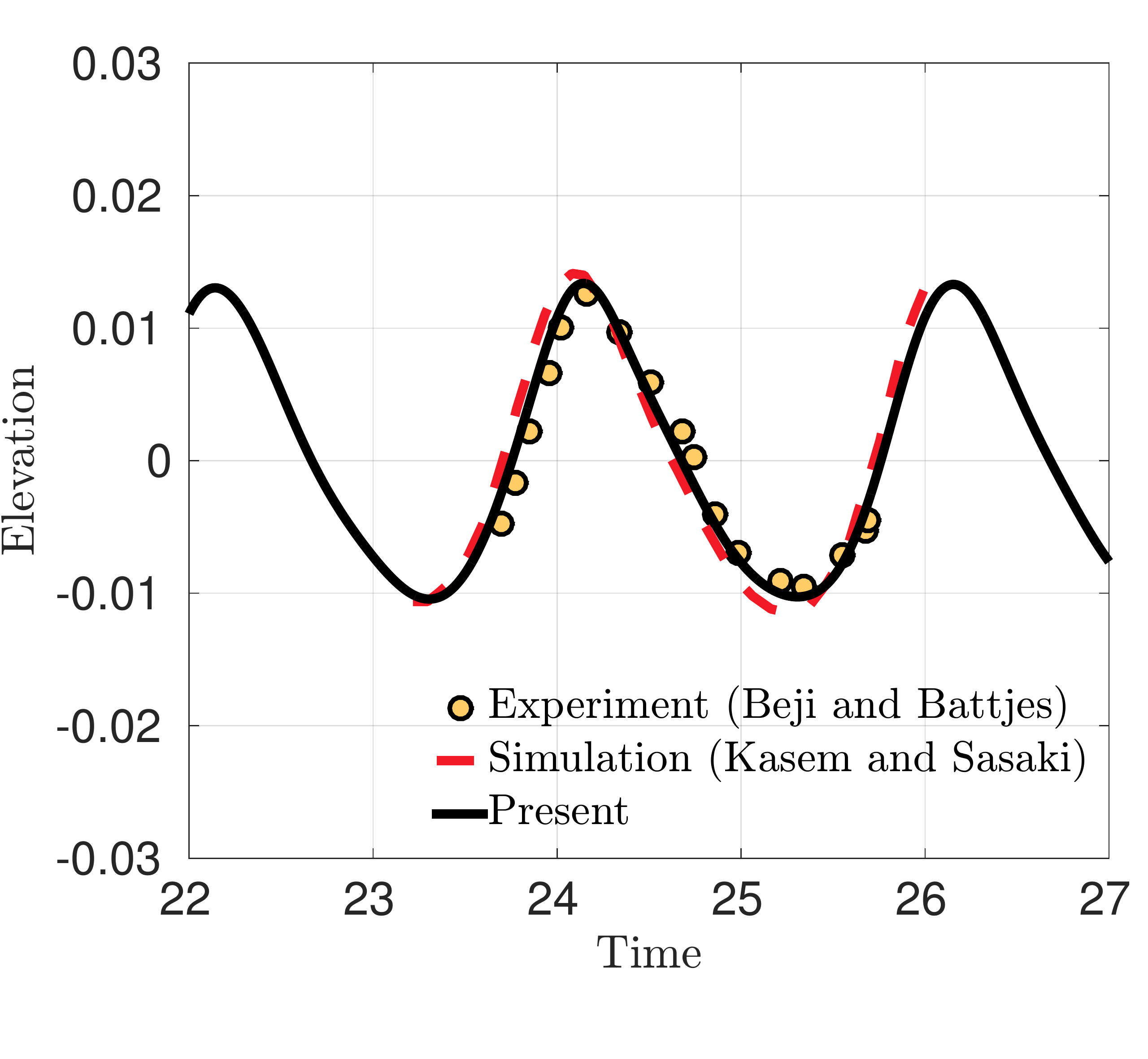}
    \label{s2_wave}
  }
  \subfigure[S$3$ station]{
    \includegraphics[scale = 0.25]{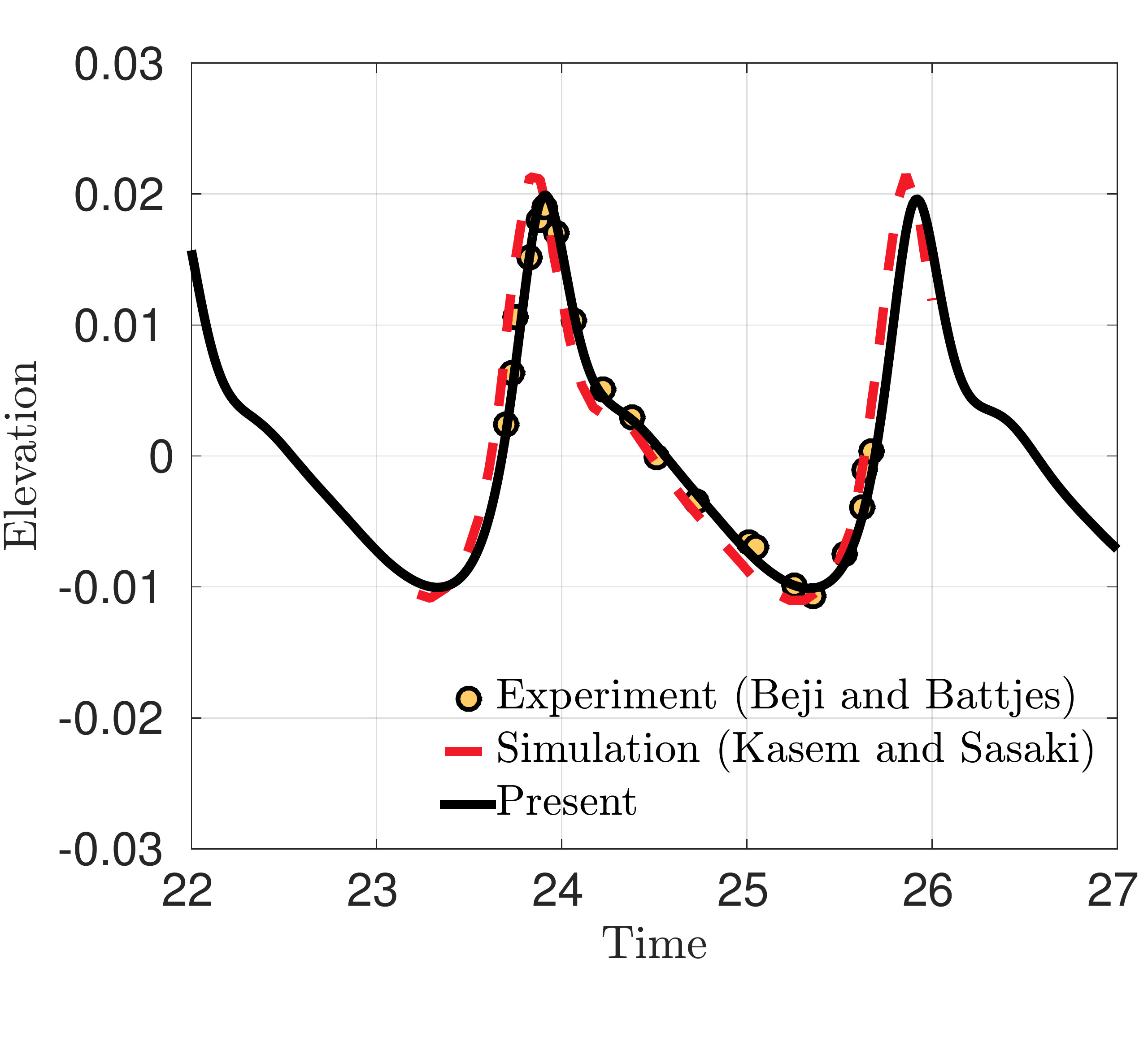}
    \label{s3_wave}
  }
  \subfigure[S$4$ station]{
    \includegraphics[scale = 0.25]{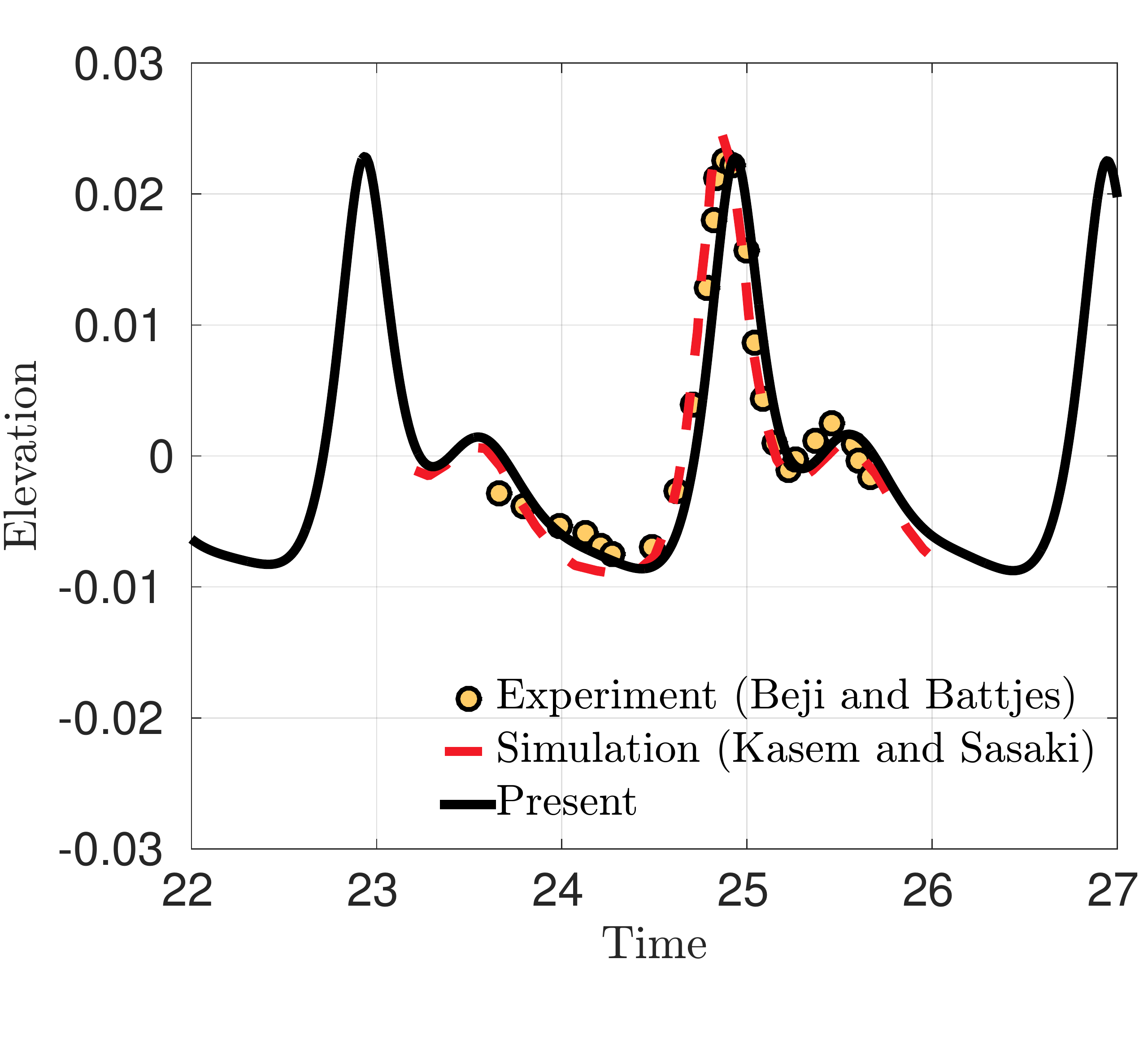}
    \label{s4_wave}
  }
  \subfigure[S$5$ station]{
    \includegraphics[scale = 0.25]{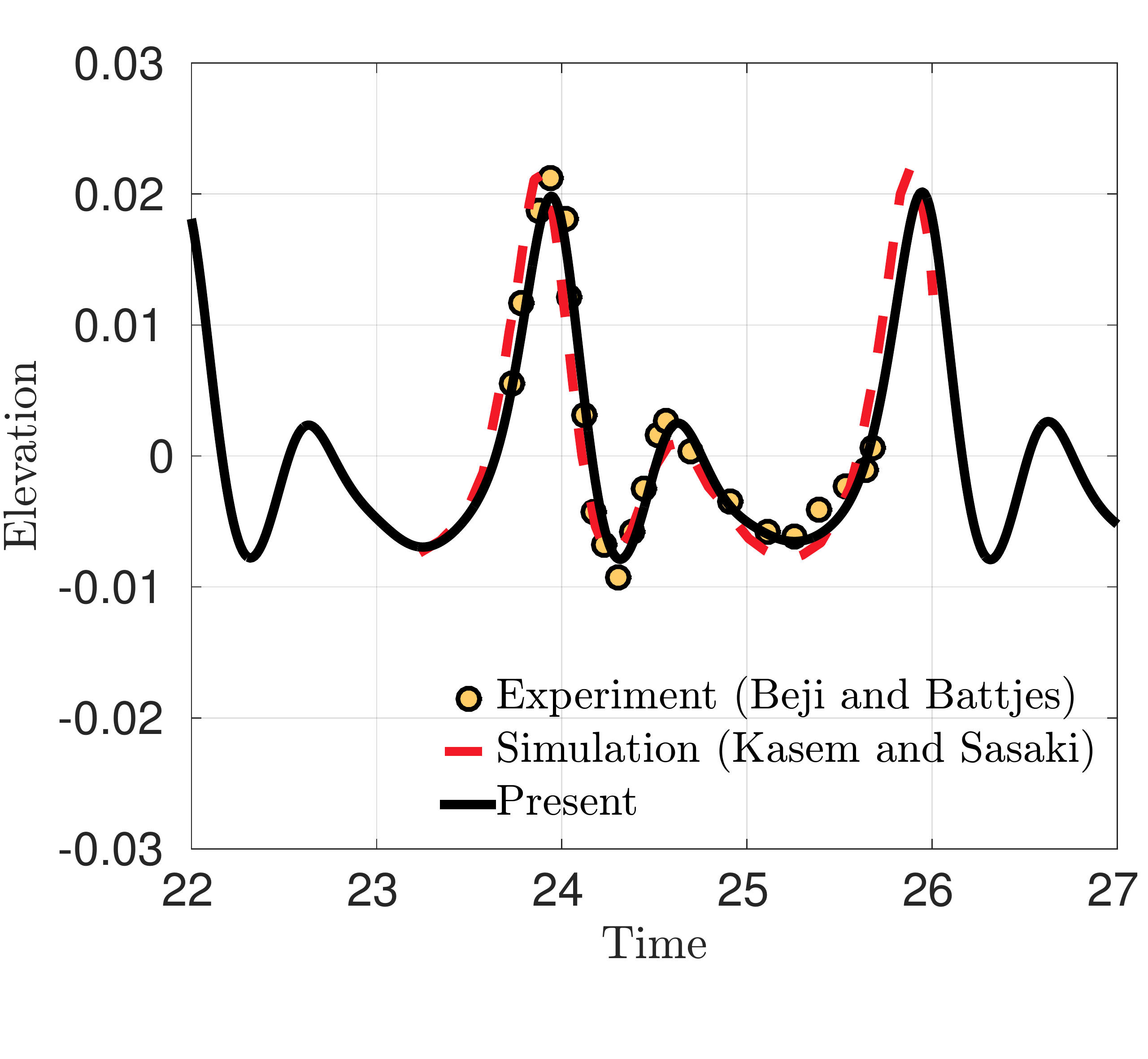}
    \label{s5_wave}
  }
    \subfigure[S$6$ station]{
    \includegraphics[scale = 0.25]{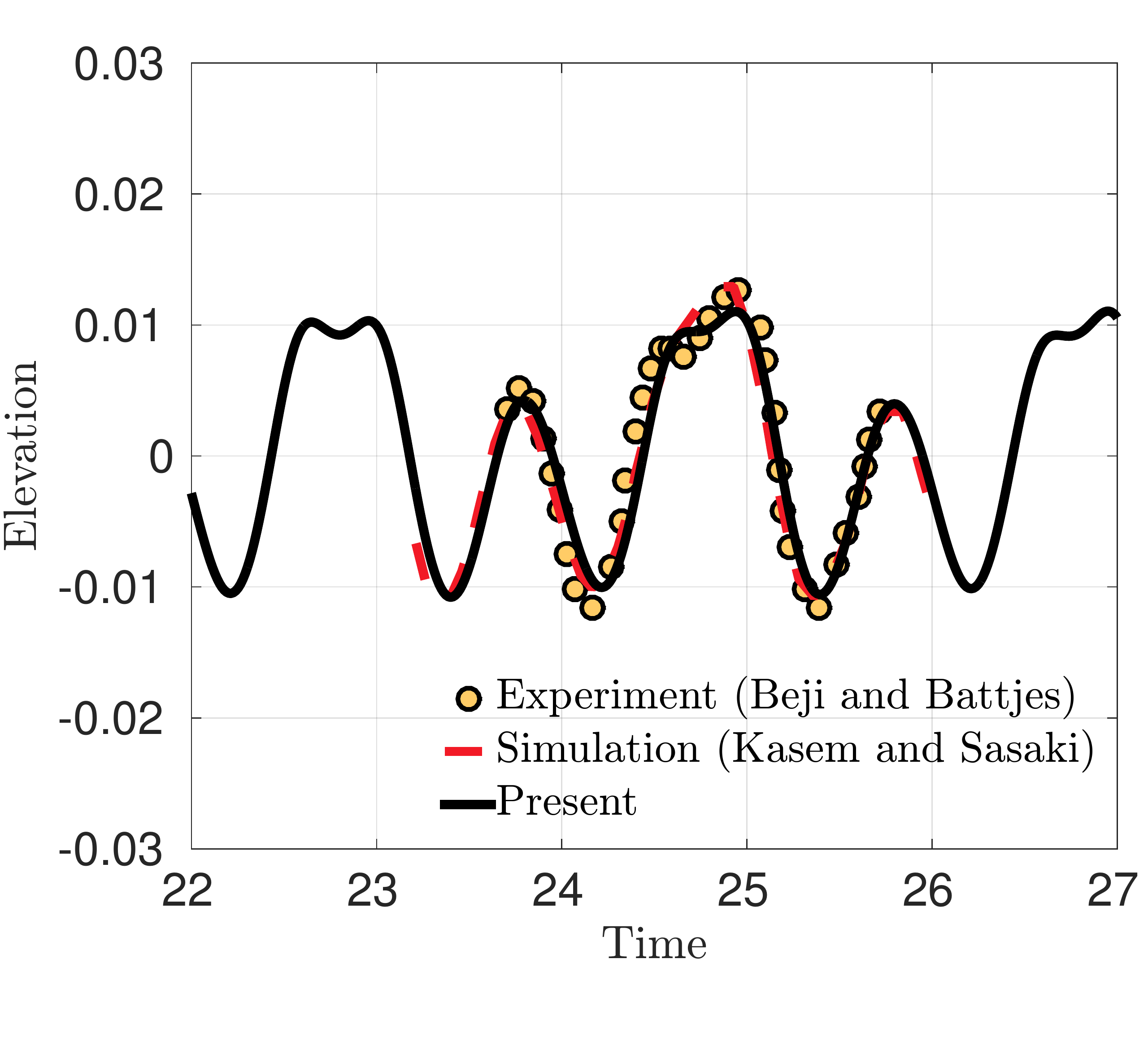}
    \label{s6_wave}
  }
    \subfigure[S$7$ station]{
    \includegraphics[scale = 0.25]{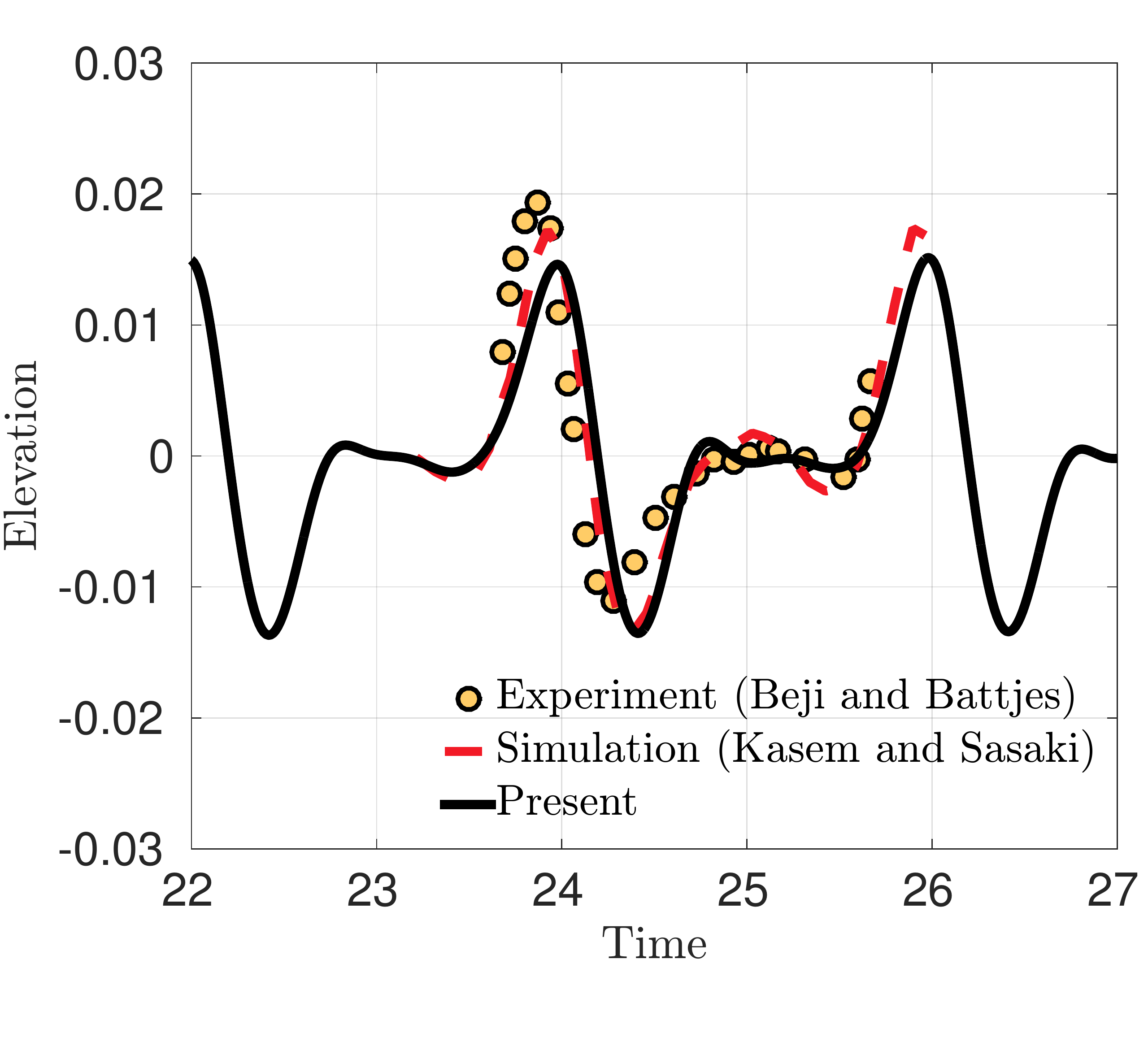}
    \label{s7_wave}
  }
 \caption{Temporal evolution of wave elevation measured at stations
 \subref{s2_wave} S$2$, 
 \subref{s3_wave} S$3$,
 \subref{s4_wave} S$4$,
 \subref{s5_wave} S$5$,
 \subref{s6_wave} S$6$, and
 \subref{s7_wave} S$7$,
 for a 2D second-order Stokes wave interacting with a stationary trapezoidal obstacle
 (see Fig.~\ref{fig_wavetrap_schematic});
 ($\bullet$, yellow) experimental data from Beji and Battjes~\cite{Beji1993,Beji1994};
(\texttt{---}, red)  simulation data from Kasem and Sasaki~\cite{Kasem2010};
(---, black) present simulation data.
 }
  \label{fig_wave_trap_elevation}
\end{figure}

To quantitatively assess the accuracy of the wave-structure interaction, the temporal evolution of 
wave elevation at the six stations are shown in Fig.~\ref{fig_wave_trap_elevation}.
The results are in decent agreement with the experimental results described
in~\cite{Beji1993,Beji1994}. Our results are also in excellent agreement with the simulation results
of Kasem and Sasaki~\cite{Kasem2010}, with minor disagreements 
being explained by slight differences in the level set discretization, advection, and reinitialization approaches.
With the cases described in this section, we have demonstrated that the numerical method described here
can be used to accurately model and solve practical marine engineering problems involving water wave-structure
interaction.

\section{Free-surface piercing and floating structure examples}
\label{sec_examples}
This section investigates several additional 2D and 3D three-phase flow problems to verify the
accuracy of the present numerical method.
The importance of consistent mass and momentum transport for numerical stability~\cite{Nangia2018}
is demonstrated for WSI simulations involving air-water interfaces. 
We also compare our results to benchmark problems drawn from the multiphase flow literature.

In some of the cases considered in this section the net hydrodynamic force
\begin{equation}
\label{eq_hydro_force}
\cF(t) = \oint_{\Sbt} \ndot \left[-p \I + \div \left[\mu \left(\grad \u + \grad \u^T\right) \right]\right] \dS, 
\end{equation}
is a quantity of interest. Here, $\n$ is the outward unit normal to the surface of the immersed body.
An \emph{extrinsic} approach to computing these forces is via the Lagrange multiplier method~\cite{Bhalla13, Nangia17}, which
reads as

\begin{equation}
\label{eq_lmforce}
\cF^{n+1} =   \sum_{\X_{l,m,n} \in \Vb} \rhos \left[\frac{\left(\Ub\right)^{n+1}_{l,m,n} - \left(\Ub\right)^{n}_{l,m,n}}{\dt} -  \frac{\delU^{n+1}_{l,m,n}}{\dt}\right] \ds_1 \ds_2 \ds_3,
\end{equation}
in which the discrete approximations of the quantities on the right-hand side are readily available during each
time step. In a previous work, we have also described an accurate \emph{moving control volume} approach to computing hydrodynamic 
forces and torques on immersed bodies using Eulerian grids and without resolving the irregular surface of the immersed   
structure~\cite{Nangia17}. Both Lagrangian and Eulerian approaches were shown to be equivalent.
\subsection{Cylinder splashing into two fluids}
We first consider a cylinder dropping into a fluid-gas interface of modest density ratio of 1.5. The non-conservative 
flow solver is considered for this problem. A circular cylinder of diameter $D = 2.5 \times 10^{-3}$ 
and density $\rhos = 1.5 \times 10^3$ is placed in a two dimensional computational 
domain of size $\Omega = [0,8D] \times [0, 48D]$ with initial center position $(X_0, Y_0) = (4D, 40D)$. 
The domain is filled halfway from $y = 0$ to $y = 24D$ with a fluid of density $\rhol = 1.25 \times 10^3$; the 
remainder of the tank, from $y = 24D$ to $y = 48D$, is filled with a lighter fluid of density $\rhog = 1 \times 10^3$. 
The viscosity $\mu = 1 \times 10^{-3}$ is held constant throughout all three phases. The domain is 
discretized using a $N \times 6N$ grid and no-slip boundary conditions are imposed 
along $\partial \Omega$. A constant time step size $\dt = 1/(39.0625 N)$ is used. This problem 
has been studied numerically by Ghasemi et al.~\cite{Ghasemi2014}. Surface tension forces 
are neglected and one grid cell of smearing ($\ncells = 1$) is used on either side of the interfaces.

\begin{figure}[]
  \centering
  \subfigure[$T = 0$]{
    \includegraphics[scale = 0.4]{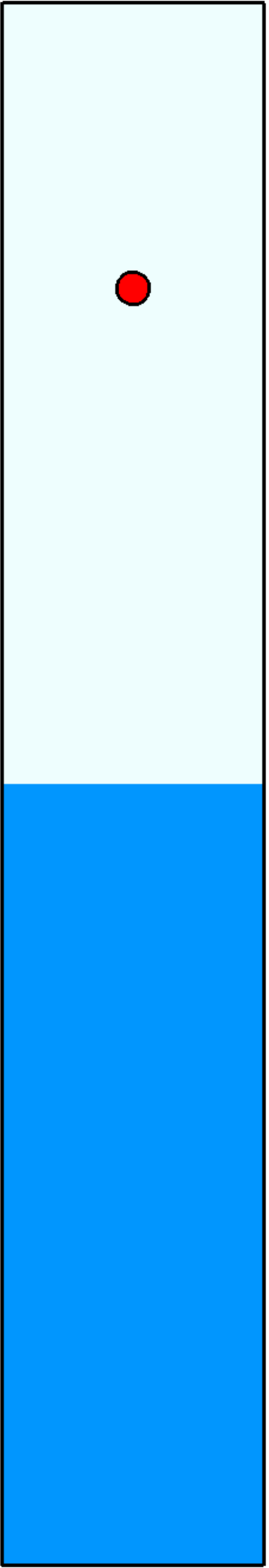}
    \label{2FC_T0}
  }
   \subfigure[$T = 22$]{
    \includegraphics[scale = 0.4]{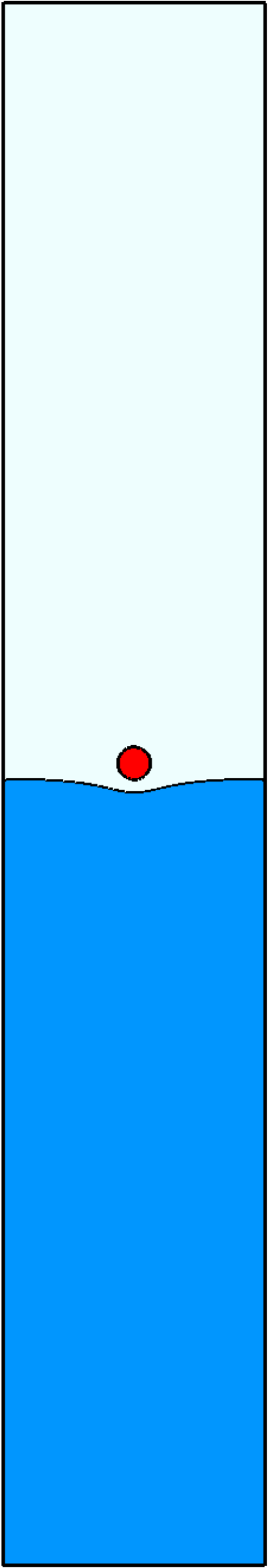}
    \label{2FC_T22}
  }
   \subfigure[$T = 28$]{
    \includegraphics[scale = 0.4]{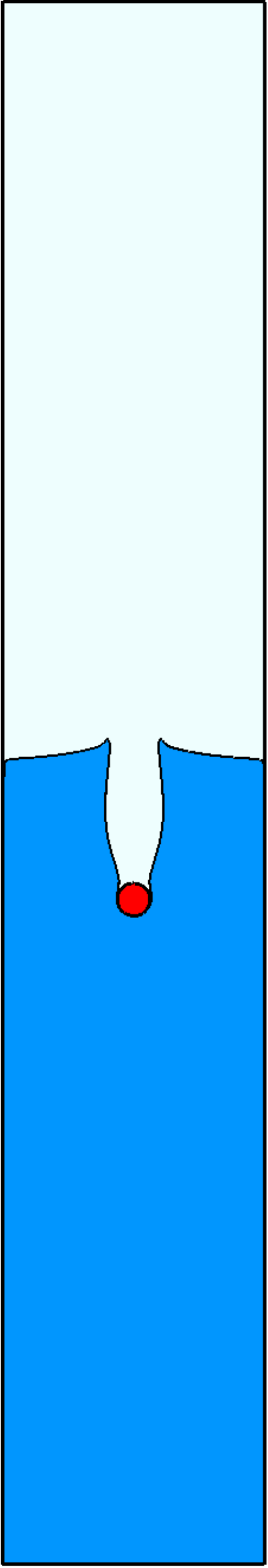}
    \label{2FC_T28}
  }
  \subfigure[$T = 34$]{
    \includegraphics[scale = 0.4]{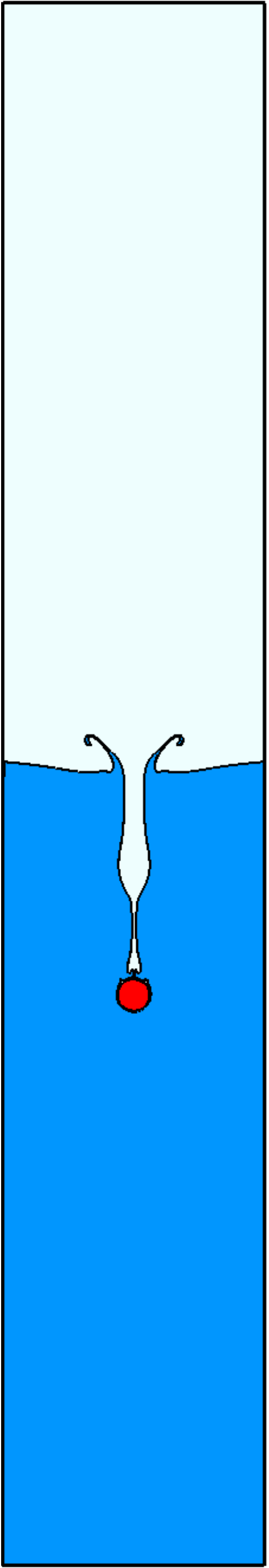}
    \label{2FC_T34}
  }
  \subfigure[$T = 38$]{
    \includegraphics[scale = 0.4]{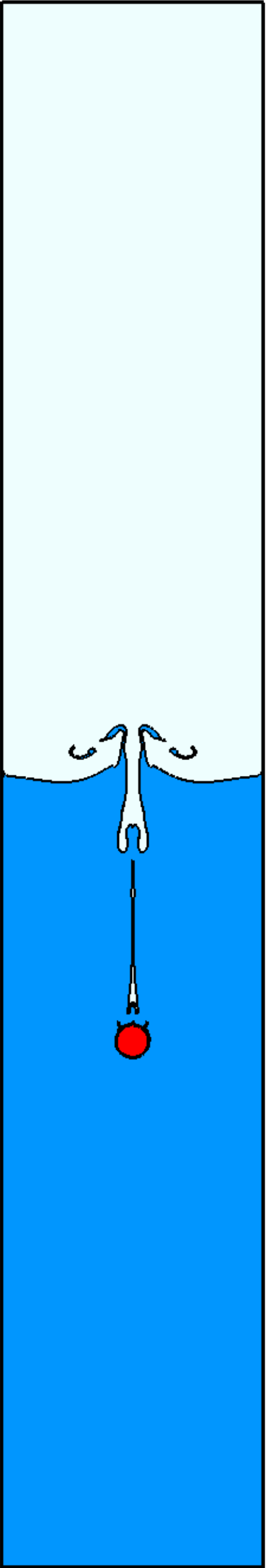}
    \label{2FC_T38}
  }
  \subfigure[$T = 41$]{
    \includegraphics[scale = 0.4]{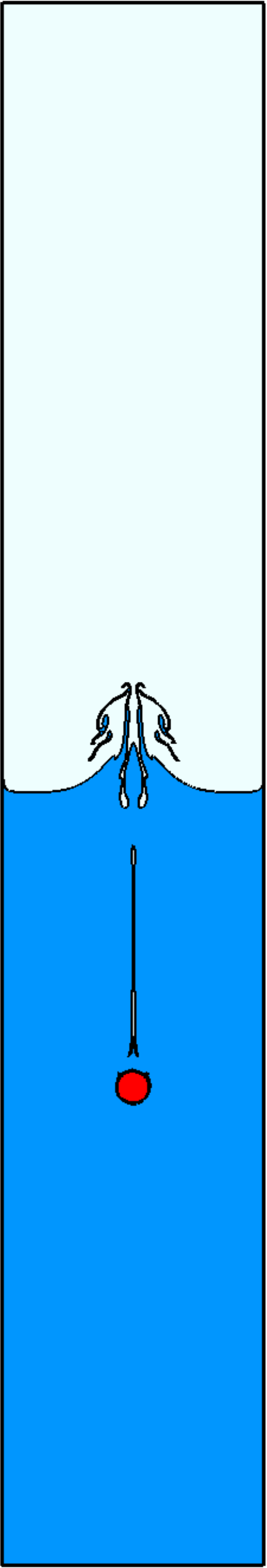}
    \label{2FC_T41}
  }
  \caption{Temporal evolution of a cylinder free-falling into a column containing two fluids at six different time instances with $64$ CPD.
   }
  \label{fig_2FC_viz}
\end{figure}

 \begin{figure}[]
  \centering
    \subfigure[Vertical position]{
    \includegraphics[scale = 0.3]{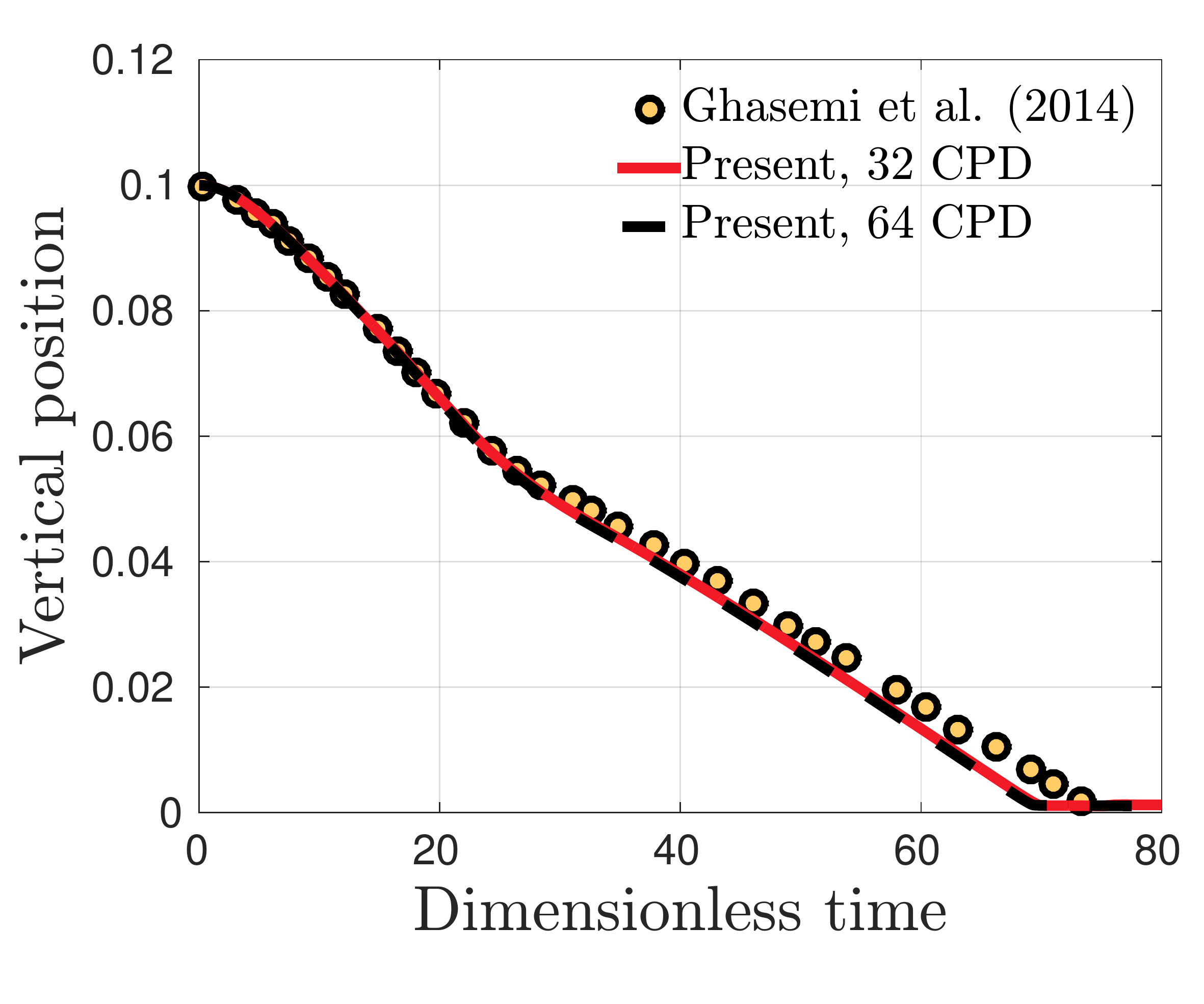}
    \label{2fc_position}
  }
  \subfigure[Vertical velocity]{
    \includegraphics[scale = 0.3]{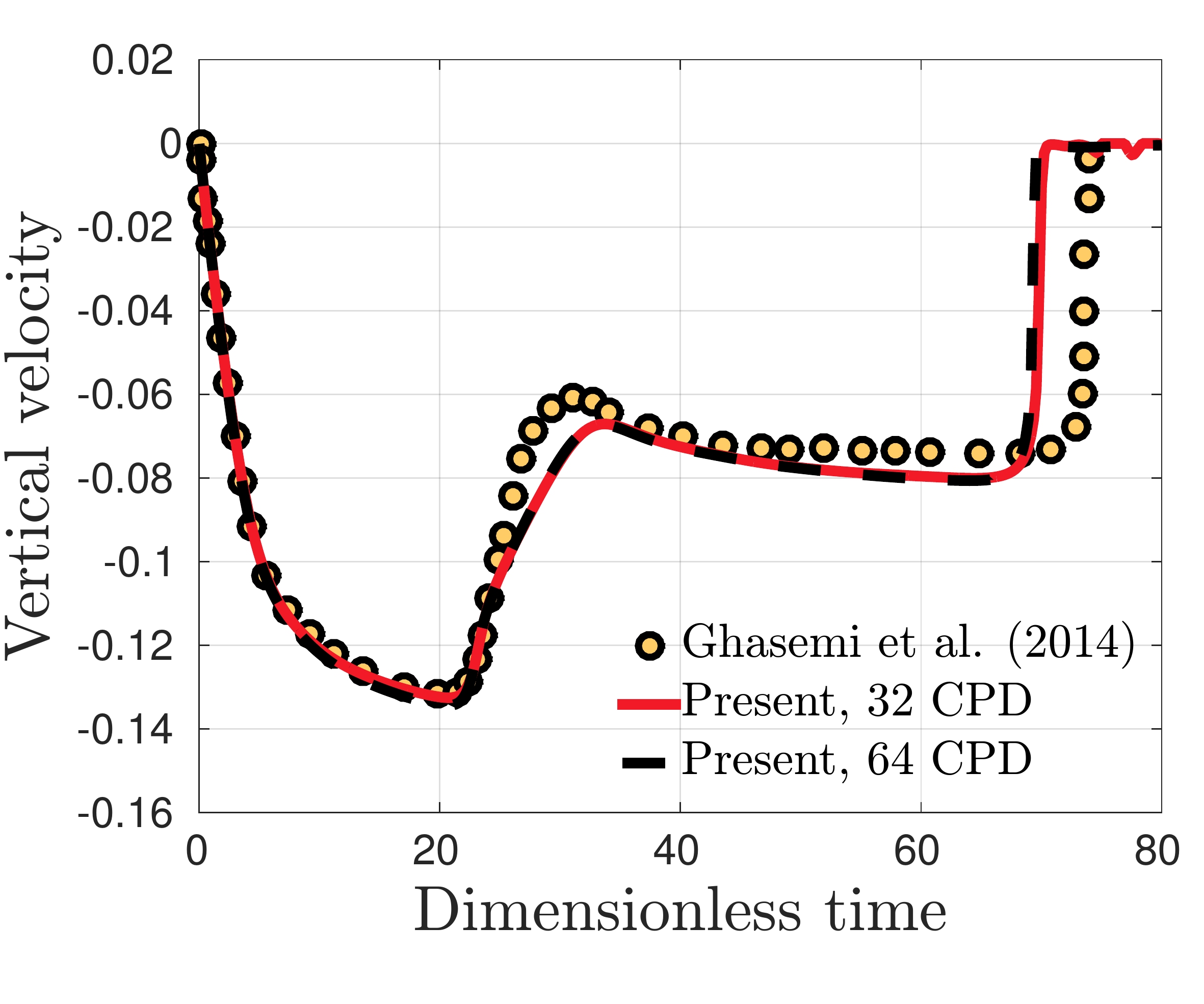}
    \label{2fc_velocity}
  }
   \caption{
 Temporal evolution of \subref{2fc_position} vertical position and \subref{2fc_velocity} vertical velocity for
  a 2D cylinder free-falling into a column containing two fluids.
  ($\bullet$, yellow) simulation data from Ghasemi et al.~\cite{Ghasemi2014};
  (---, red) present simulation data with $32$ grid cells per diameter;
  (\texttt{---}, black) present simulation data for $64$ grid cells per diameter.
}
  \label{fig_2fc_data}
\end{figure}

Fig.~\ref{fig_2FC_viz} shows evolution of the cylinder at various instances in dimensionless time $T = t \sqrt{g/D}$.
At this grid resolution, $N = 512$ or 64 cells per diameter (CPD), a cavity is formed in the wake of the cylinder as it
penetrates the interface. A symmetric jet forms as the cavity collapses, which shoots upwards and breaks up.
Additionally, gas phase entrainment is seen around the cylinder, which is experimentally seen in many three-phase flow
problems.
To demonstrate the quantitative accuracy of the fluid-structure interaction, the vertical position and vertical 
velocity are plotted as a function of $T$ in Fig.~\ref{fig_2fc_data} for both 32 and 64 CPD. 
Both the interface dynamics and the FSI are in decent agreement
with the computational study of Ghasemi et al., with minor disagreements being explained by differences in the interface
tracking approaches (a fully-Eulerian VOF method is used in~\cite{Ghasemi2014} to simulate three phase flows).
Additionally, the rigid body motion in this work is tracked in a Lagrangian reference frame, while a fully-Eulerian FSI approach
is used in~\cite{Ghasemi2014}, which could explain the modest differences in vertical velocities
 after around $T = 65$ in Fig.~\ref{2fc_velocity}; this is when the immersed cylinder
approaches the bottom of the computational domain.
Even though convection is a dominant process here, this numerical test case demonstrates that the non-conservative and 
inconsistent mass and momentum transport scheme can still accurately simulate low density ratio three-phase flows.

\subsection{Heaving cylinder on an air-water interface}

\begin{figure}[]
  \centering
  \subfigure[$T = 5.67$]{
    \includegraphics[scale = 0.28]{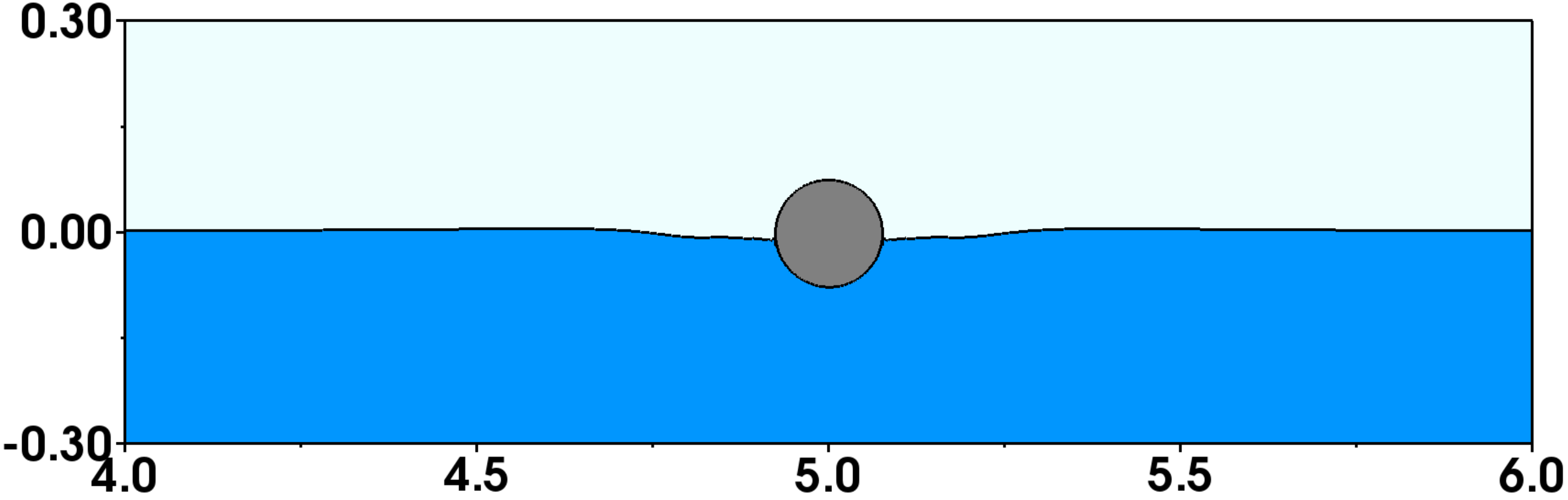}
    \label{ito_densityt05}
  }
     \subfigure[$T = 5.67$]{
    \includegraphics[scale = 0.28]{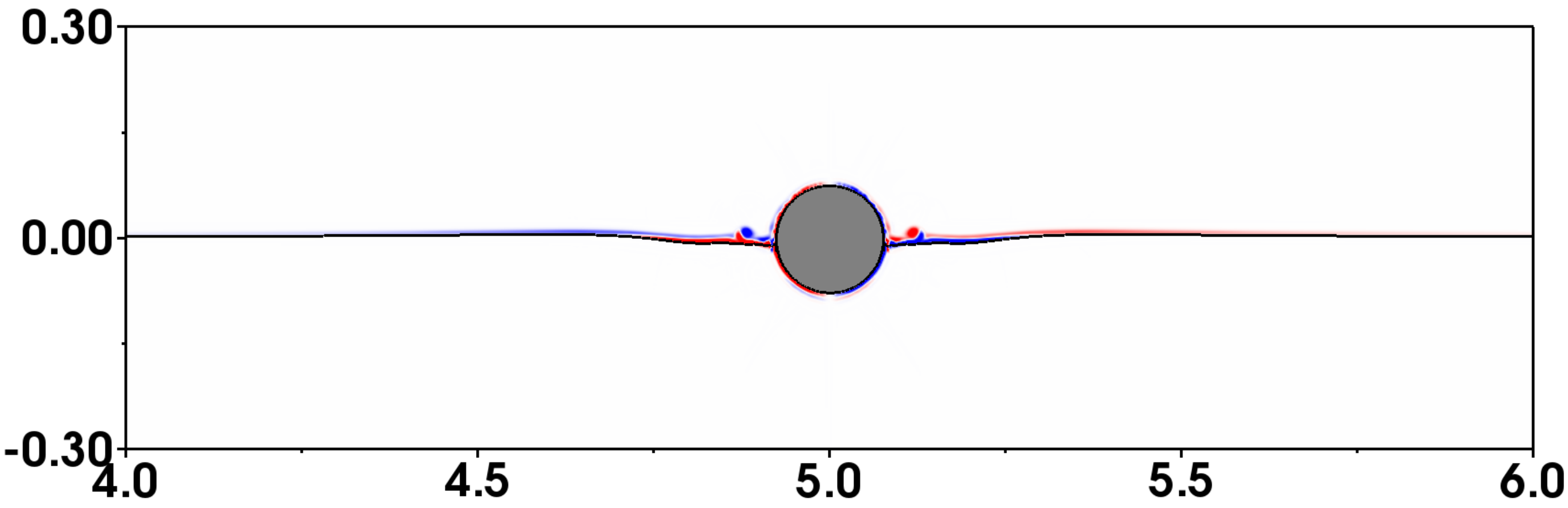}
    \label{ito_omegat05}
  }
   \subfigure[$T = 11.35$]{
    \includegraphics[scale = 0.28]{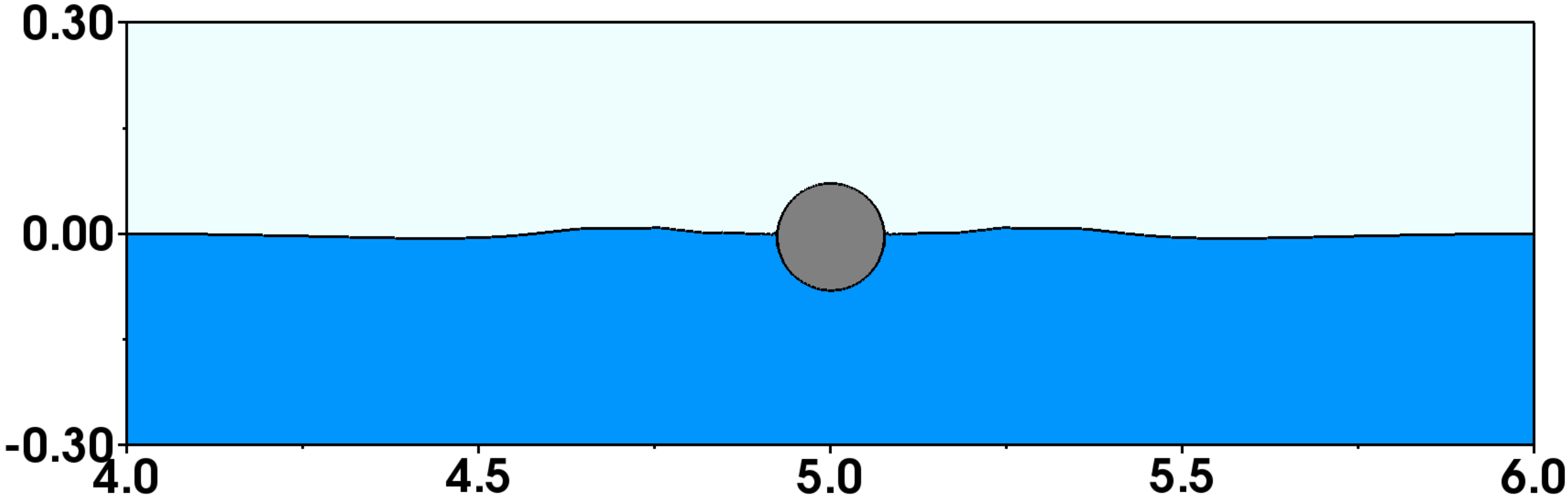}
    \label{ito_densityt10}
  }
     \subfigure[$T = 11.35$]{
    \includegraphics[scale = 0.28]{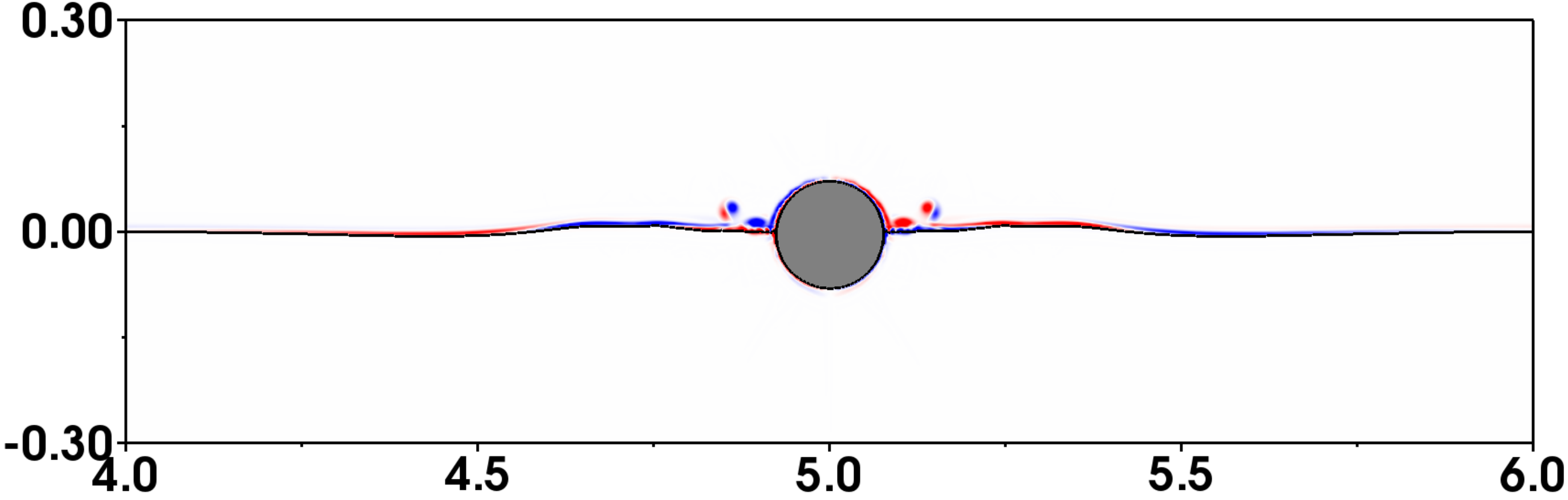}
    \label{ito_omegat10}
  }
   \subfigure[$T = 17.02$]{
    \includegraphics[scale = 0.28]{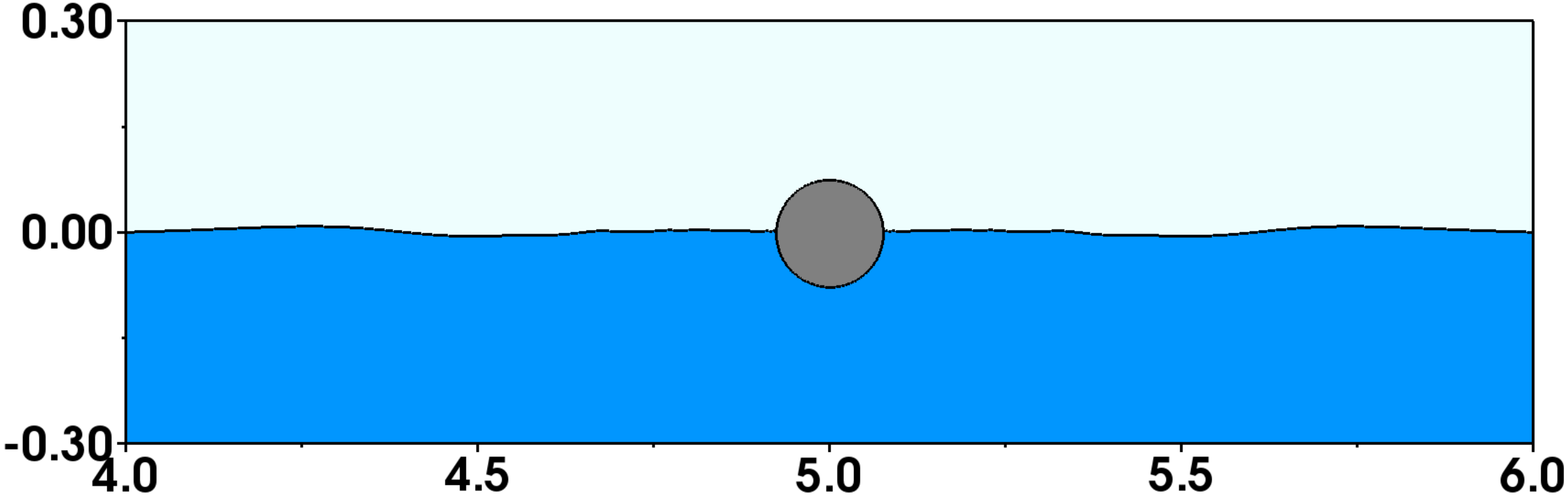}
    \label{ito_densityt15}
  }
   \subfigure[$T = 17.02$]{
    \includegraphics[scale = 0.28]{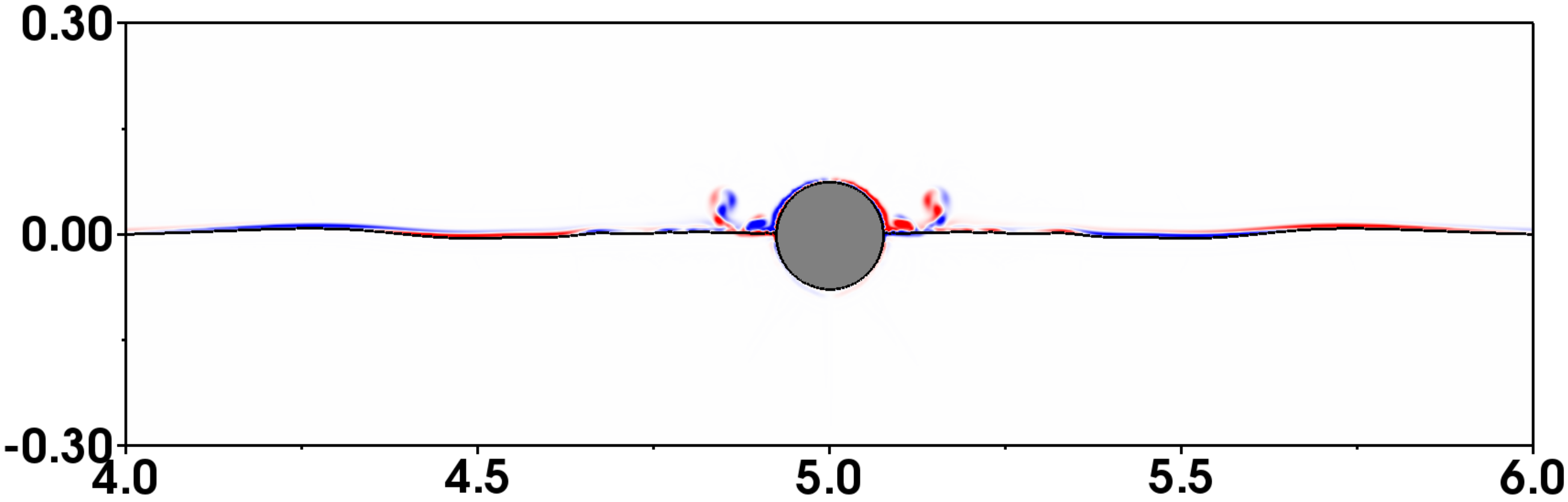}
    \label{ito_omegat15}
  }
  \subfigure[$T = 22.69$]{
    \includegraphics[scale = 0.28]{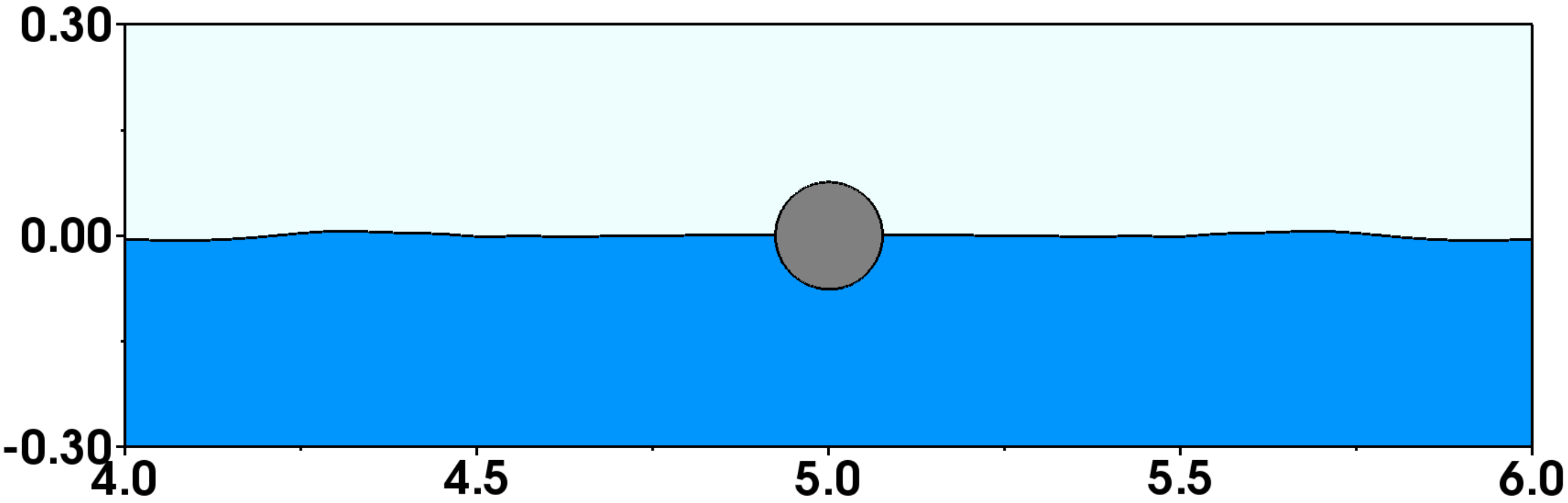}
    \label{ito_densityt20}
  }
  \subfigure[$T = 22.69$]{
    \includegraphics[scale = 0.28]{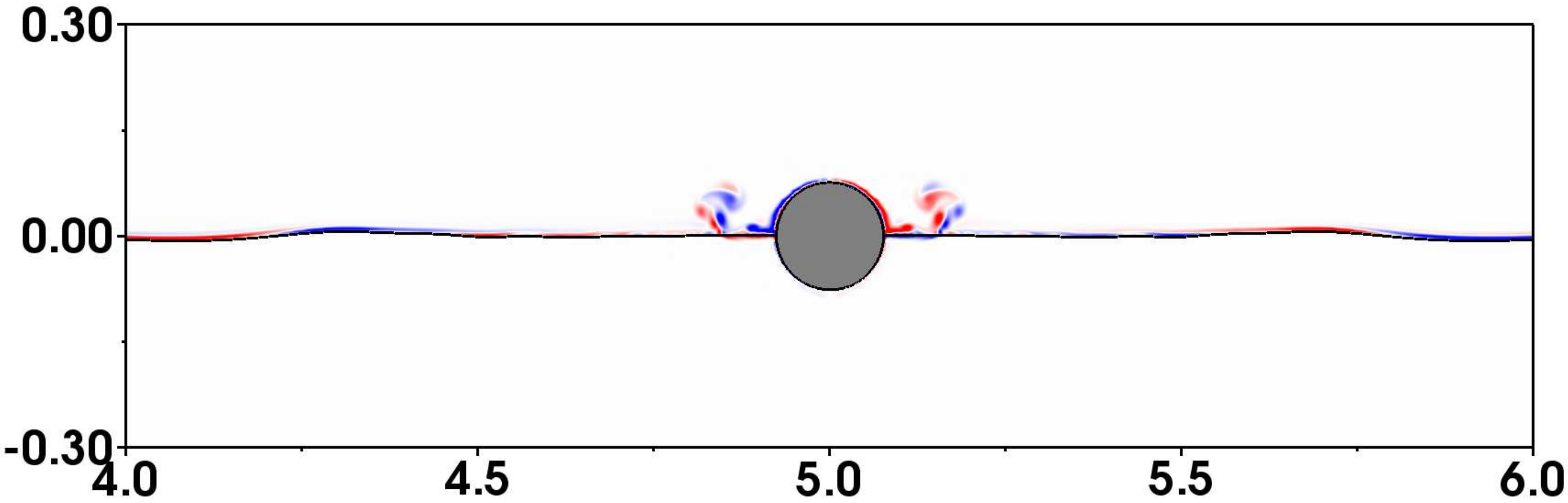}
    \label{ito_omegat20}
  }
\caption{Temporal evolution of a cylinder heaving on an air-water interface at four different time instances with $46$ CPR: (left) density and (right) vorticity generated in the range of $-25$ to $25$.
   }
  \label{fig_ito_viz}
\end{figure}

\begin{figure}[]
  \centering
    \includegraphics[scale = 0.3]{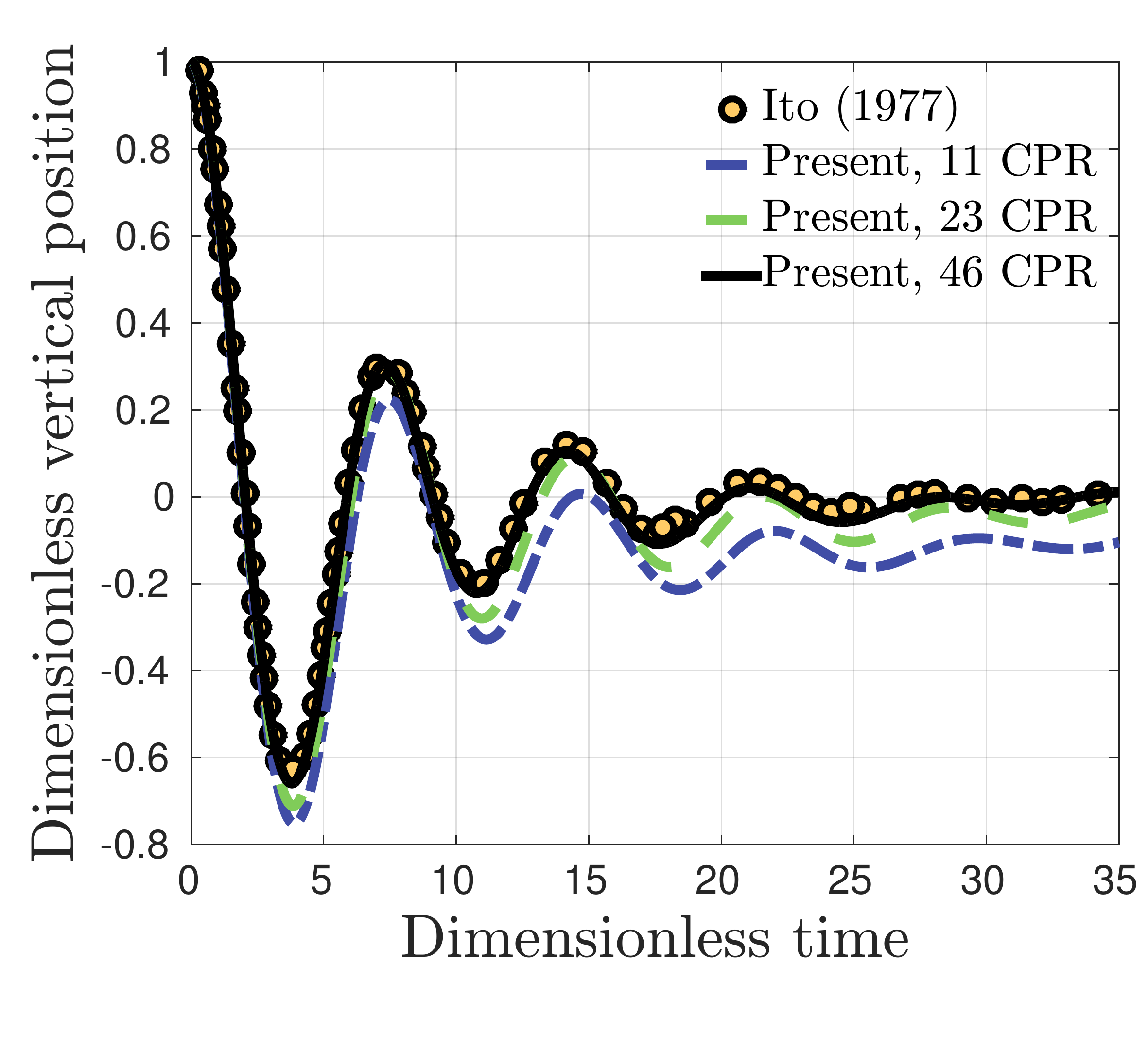}
 \caption{
 Temporal evolution of dimensionless vertical position for a 2D cylinder heaving on an
 air-water interface.
 ($\bullet$, yellow) experimental data from It{\=o}~\cite{Ito1977};
(\texttt{-}$\cdot$\texttt{-}, blue) present simulation data with $11$ grid cells per radius;
(\texttt{---}, green) present simulation data for $23$ grid cells per radius;
  (---, black) present simulation data with $46$ grid cells per radius.
 }
  \label{fig_ito_vertical_position}
\end{figure}

This section investigates the heave decay of a cylinder floating on an air-water interface.
A circular cylinder of radius $R = 0.0762$ is placed within a two-dimensional computational
domain of length $L = 10$ and height $H = 0.2L$. Water occupies the bottom portion of the domain
from $y = -16R$ to $y = 0$, while air occupies the remainder of the tank from $y = 0$ to $y = H - 16R$.
The cylinder is partially submerged in the fluid phase with
initial center position $(X_0, Y_0) = (L/2, R/3)$ and is half as dense as water with $\rhos= 5 \times 10^2$.
Two grid cells of smearing $\ncells = 2$ are used to transition between different material properties on
either side of the interfaces. Only the cylinder's vertical degrees of freedom are unlocked and surface tension forces are neglected .
No-slip boundary conditions are imposed along $\partial \Omega$. The conservative and consistent flow solver is used for this case. 
This problem has been studied both experimentally by It{\=o}~\cite{Ito1977}, and numerically by 
Calderer et al.~\cite{Calderer2014} and Ghasemi et al.~\cite{Ghasemi2014}.

The domain is discretized by grid of size $5N \times N$ and a constant time step size of $\dt = 3/(5N)$ is used.
To assess convergence, we consider three different grid sizes: $N = 300, 600, 1200$, which correspond
to $11$, $23$, and $46$ grid cells per radius (CPR), respectively. Fig.~\ref{fig_ito_viz} shows the evolution
of the cylinder and the air-water interface at various instances in dimensionless time 
$T = t \sqrt{g/R}$ for the 46 CPR simulation.
Modest vorticity is generated as the cylinder bobs up and down, and small ripples can be seen traveling outward
away from the body along the air-water interface.
To quantitatively assess the accuracy and convergence of the fluid-structure interaction, the vertical center of mass
position of the cylinder (nondimensionalized by $Y_0$) is plotted against time in Fig.~\ref{fig_ito_vertical_position}.
As the resolution increases, the numerical simulations converge towards the experimental results of It{\=o}~\cite{Ito1977}.
As expected, the cylinder's heave oscillation eventually damps out as it reaches an equilibrium position on the water.
This case is representative of real-world applications such as wave energy converter devices, and 
demonstrates that the present numerical method can be used to accurately simulate floating objects. We also simulated 
this case with the non-conservative flow solver and obtained similar results (data not shown). Since the flow around the 
heaving cylinder is relatively moderate and decays over time, the non-conservative solver remains stable even for a high 
density ratio of $10^3$. This will not be true for some of the cases shown later. 

\subsection{Heaving sphere on an air-water interface}
As an extension to the case presented in the previous section, 
we now consider the heave decay of a three-dimensional sphere floating on an
air-water interface. A sphere of radius $R = 0.254$ is placed within a three-dimensional computational
domain that is equally long in each direction: $L_x = L_y = L_z = 5$.
Water occupies the bottom portion of the domain
from $z = -12.6R$ to $z = 0$, while air occupies the remainder of the tank from $z = 0$ to $z = L_z - 12.6R$.
The sphere is partially submerged in the fluid phase with
initial center position $(X_0, Y_0, Z_0) = (L_x/2, L_y/2, R/3)$ and is half as dense as water with $\rhos= 5 \times 10^2$.
Two grid cells of smearing $\ncells = 2$ are used to transition between different material properties on
either side of the interfaces.
Only the sphere's vertical degrees of freedom are unlocked and surface tension forces are neglected.
The conservative and consistent flow solver is used for this case. 
No-slip boundary conditions are imposed along $\partial \Omega$. This case has been studied both experimentally by
Beck and Liapis~\cite{Beck1987}, and numerically by Pathak and Raessi~\cite{Pathak16}.

In contrast with the 2D case, these simulations make use of adaptive mesh refinement.
The domain is discretized by $\ell = 4$  grid levels, each with refinement ratio $\nref = 2$.
Two simulations are carried out: one with $\dx_0 = \dy_0 = \dy_0 = 1/10$ yielding a finest grid
spacing of $\dx_\textrm{min} = \dy_\textrm{min} = \dz_\textrm{min} = 1/80$ or $20$ grid cells per radius (CPR),
and one with $\dx_0 = \dy_0 = \dy_0 = 1/20$ yielding a finest grid spacing of 
$\dx_\textrm{min} = \dy_\textrm{min} = \dz_\textrm{min} = 1/160$ or $40$ CPR.
A constant time step size of  $\dt = 0.16 \dx_\textrm{min}$ is used.

\begin{figure}[]
  \centering
  \subfigure[$T = 0.62$]{
    \includegraphics[scale = 0.25]{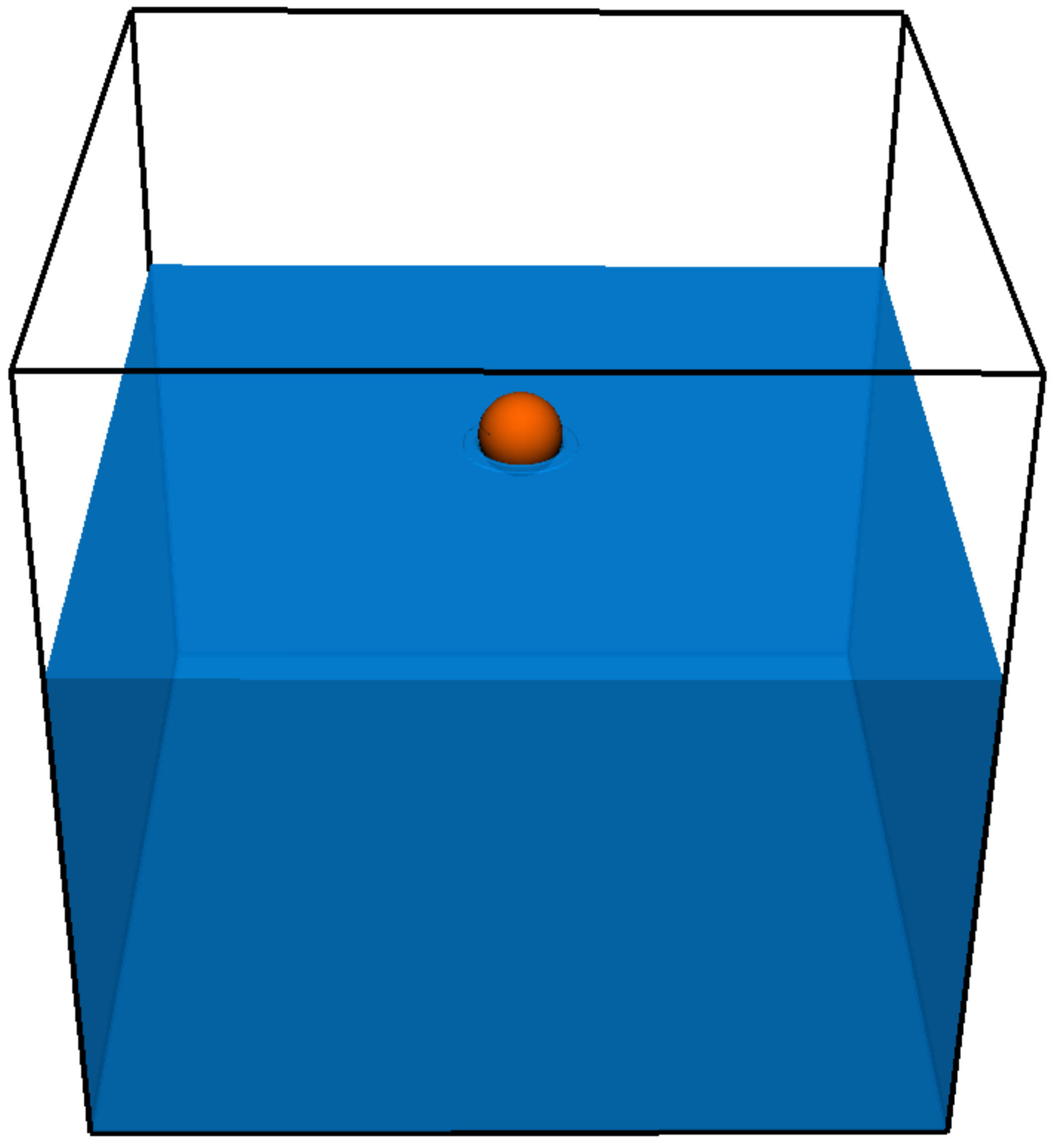}
    \label{heaving_sphere_t0p1}
  }
     \subfigure[$T = 18.64$]{
    \includegraphics[scale = 0.25]{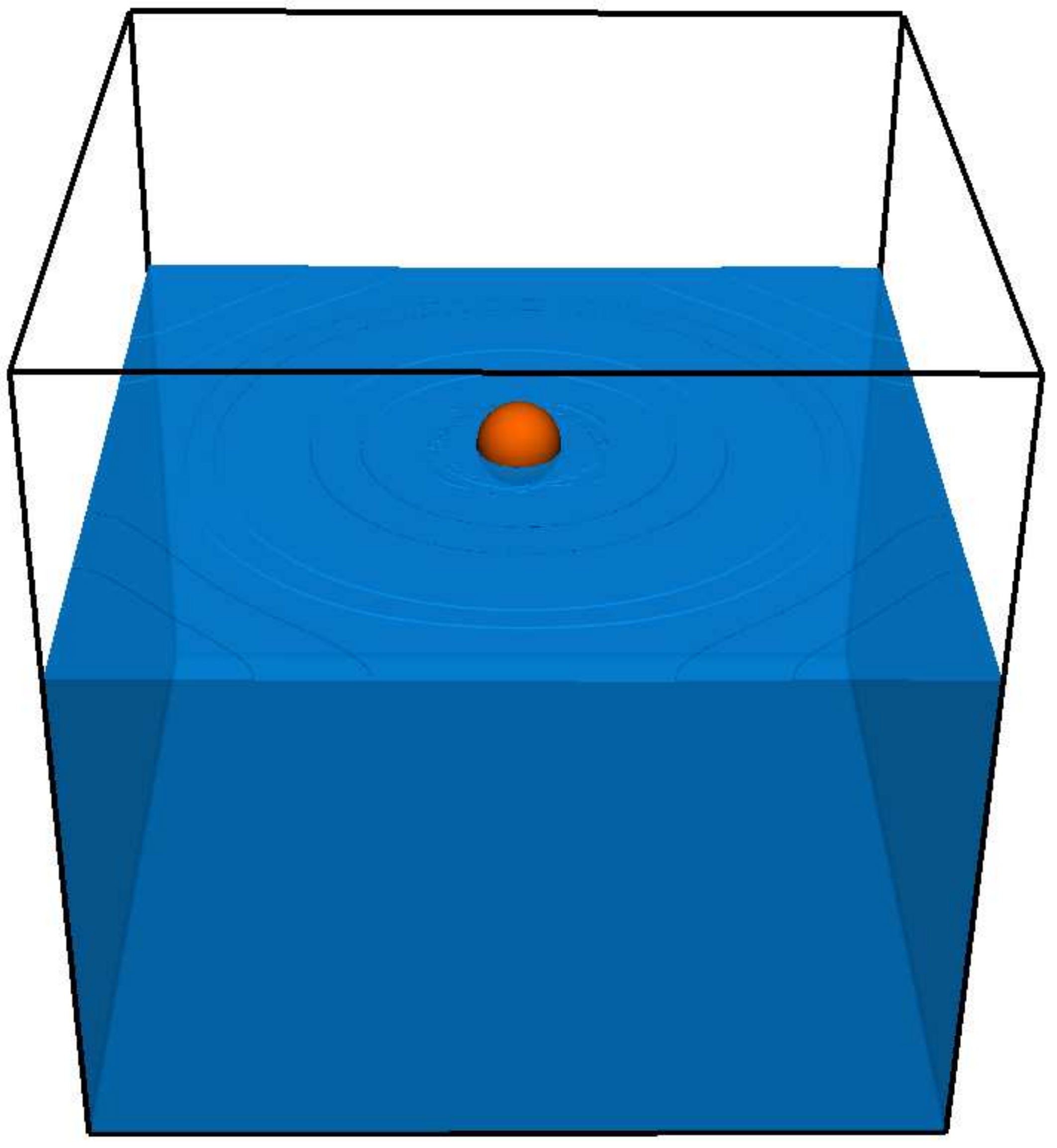}
    \label{heaving_sphere_t3p0}
  }
   \subfigure[Mesh refinement at $T = 0.62$]{
    \includegraphics[scale = 0.25]{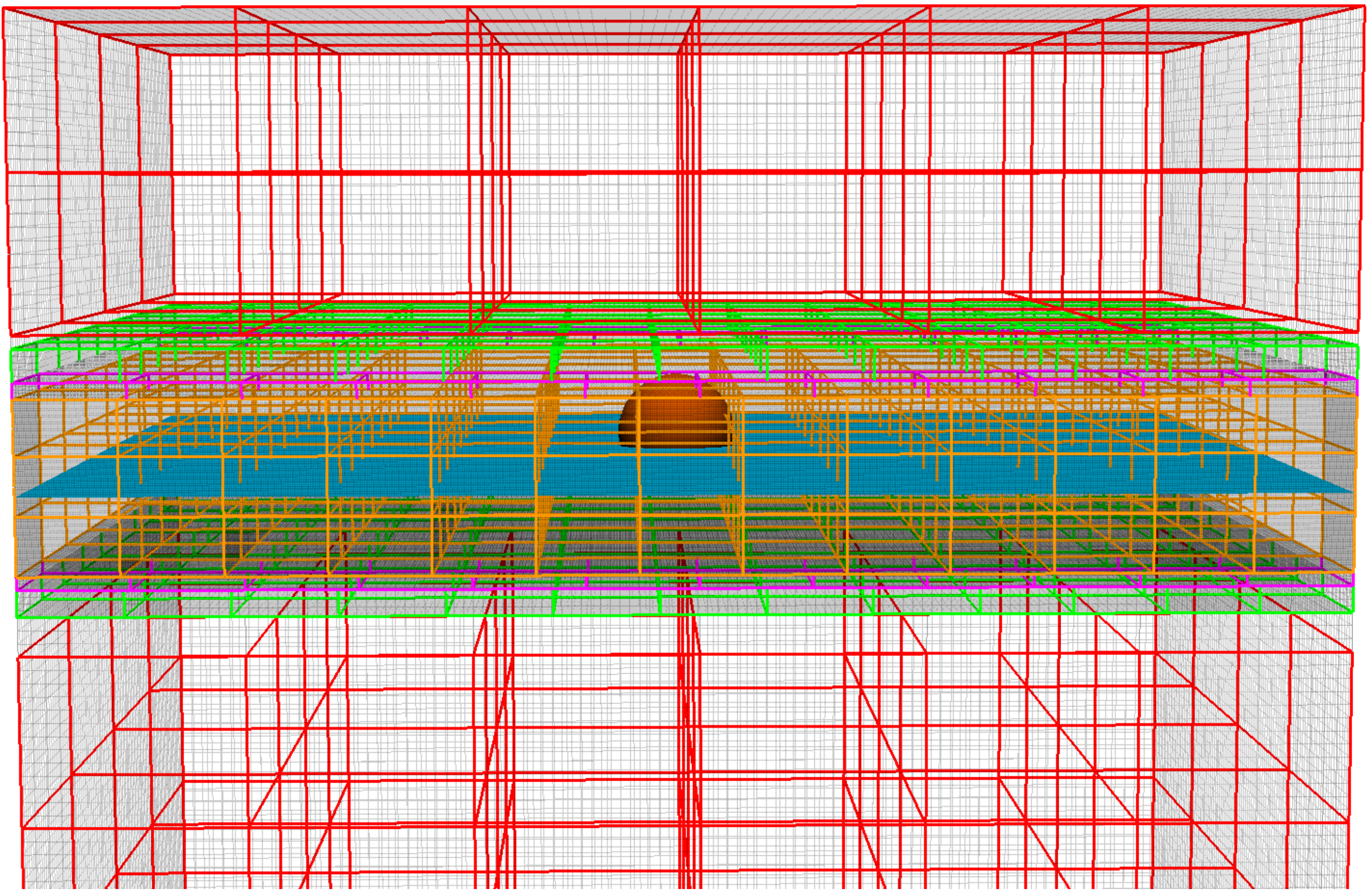}
    \label{heaving_sphere_t0p1_amr}
  }
     \subfigure[Mesh refinement at $T = 18.64$]{
    \includegraphics[scale = 0.25]{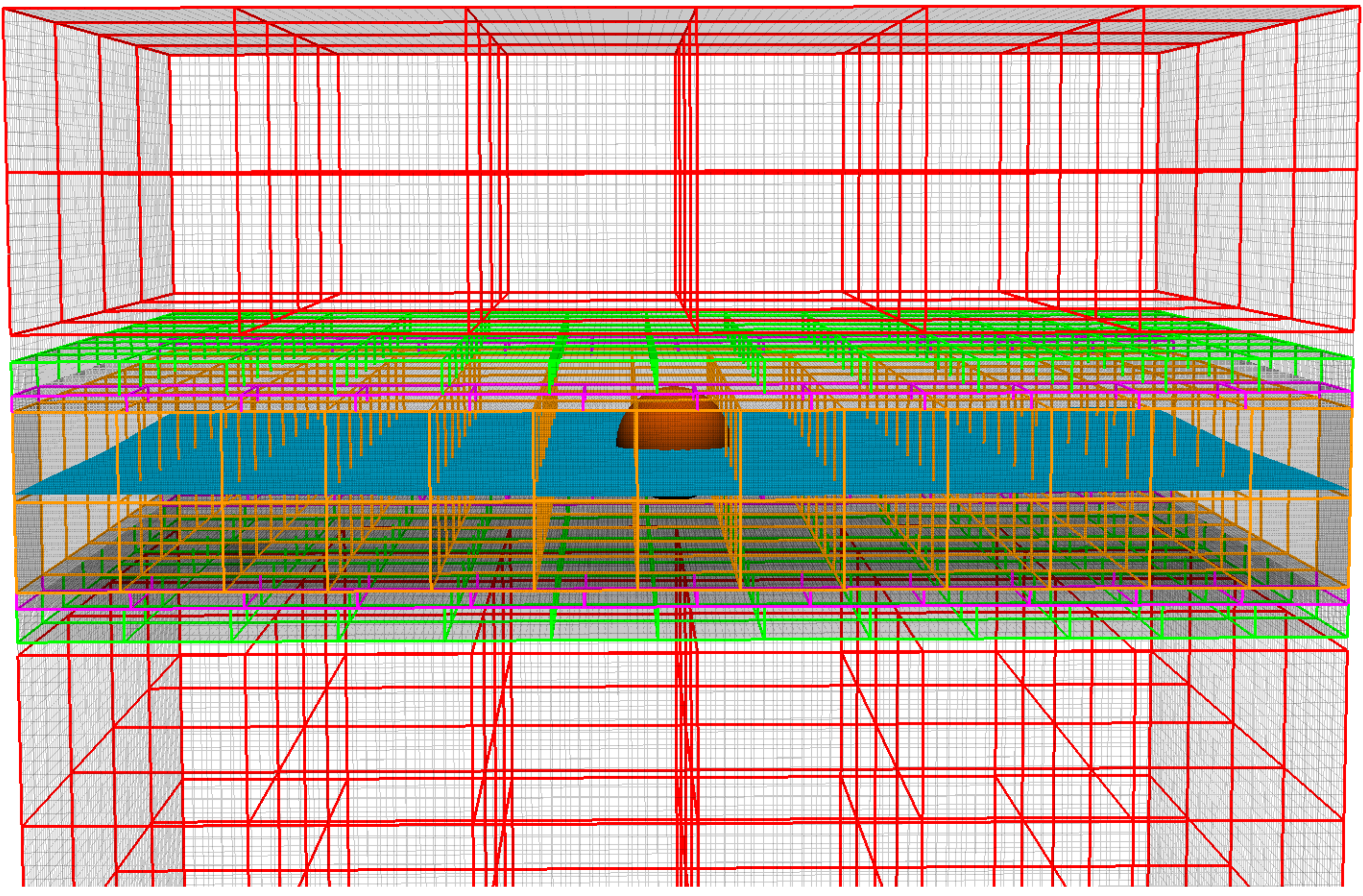}
    \label{heaving_sphere_t3p0_amr}
  } 
\caption{Temporal evolution of a sphere heaving on an air-water interface at two different time instances with $40$ CPR: (top) density and (bottom) locations of the different refined mesh levels from coarsest to finest:
  red, green, pink, orange.
   }
  \label{fig_heaving_sphere_viz}
\end{figure}

\begin{figure}[]
  \centering
    \includegraphics[scale = 0.3]{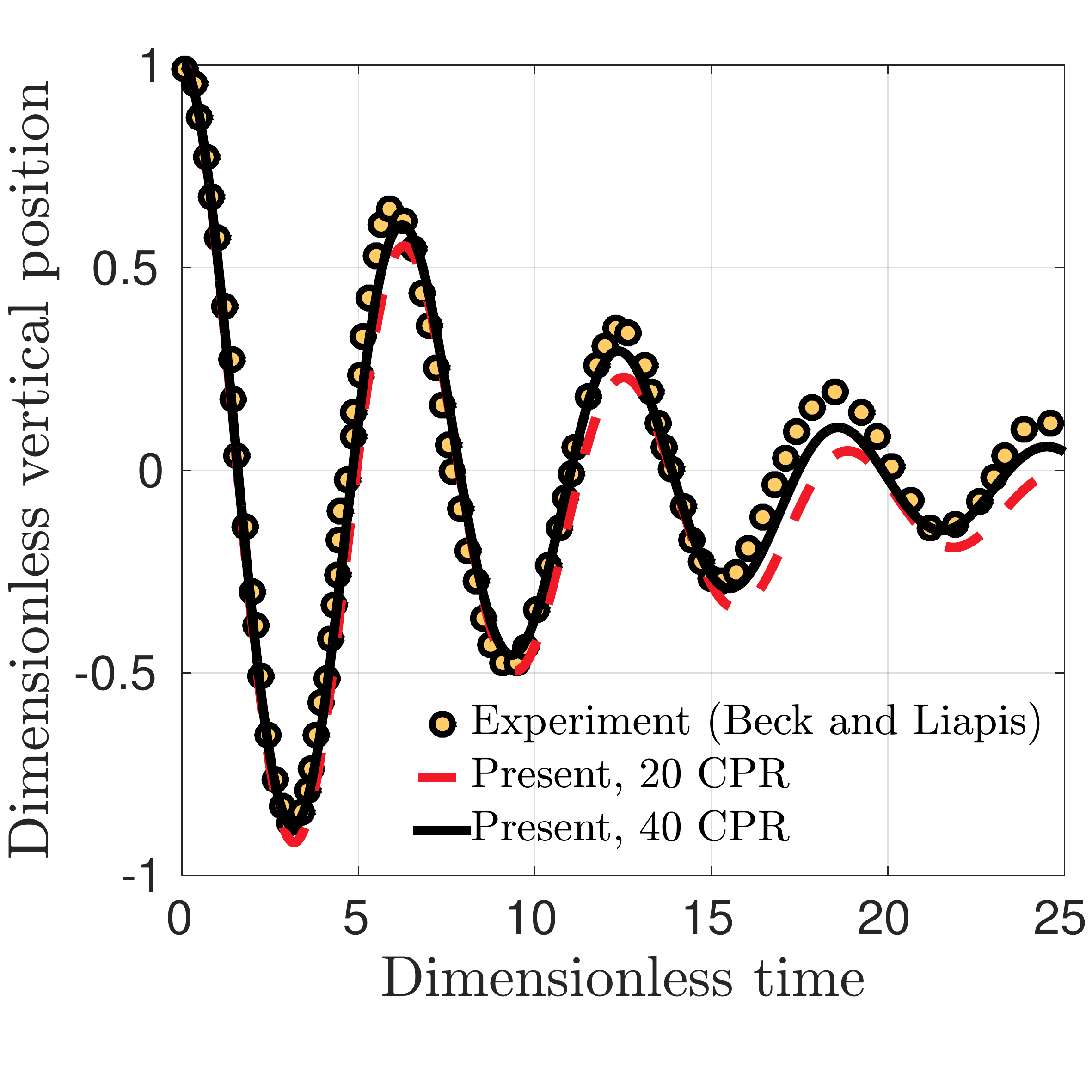}
 \caption{
 Temporal evolution of dimensionless vertical position for a 3D sphere heaving on an
 air-water interface.
 ($\bullet$, yellow) experimental data from Beck and Liapis~\cite{Beck1987};
(\texttt{---}, red) present simulation data with $20$ grid cells per radius;
(---, black) present simulation data for $40$ grid cells per radius.
 }
  \label{fig_heaving_sphere_vertical_position}
\end{figure}

Fig.~\ref{fig_heaving_sphere_viz} shows snapshots of the heaving sphere and the free-surface evolution
at two instances in dimensionless time $T = t \sqrt{g/R}$ for the 40 CPR simulation.
Over time, the sphere generates ripples in the air-water interface that travel radially outward from the body.
The finest mesh level surrounds both the immersed structure and the air-water interface; additional local regions
of refinement are not formed because this particular case does not generate significant vorticity in the air phase.
To quantitatively assess the accuracy and convergence of the wave-structure interaction, the vertical center of mass
position of the sphere (nondimensionalized by $Z_0$) is plotted against time 
in Fig.~\ref{fig_heaving_sphere_vertical_position}.
As the resolution increases, the numerical simulations converge towards the experimental results of Beck and
Liapis~\cite{Beck1987}.
Similar to the previous case, the sphere's oscillation eventually damps out as it reaches an equilibrium position on 
the air-water interface.

\subsection{Static cylinder on an air-water interface} \label{sec_ex_wellbalance}
In the previous cases, the structure underwent free-body motion. Therefore, 
we employed the full gravitational body force $\rho \g$ in the computational domain.
In this section, we demonstrate that using this same treatment of gravitational forcing
for fully constrained motion produces parasitic currents, whereas using a gravitational forcing 
of the form $\rho^{\text{flow}}\g$ eliminates spurious velocity currents in such cases 
(see Sec.~\ref{sec_solid_materials} for discussion).
 
\begin{figure}[]
  \centering
  \subfigure[Initial problem set up]{
    \includegraphics[scale = 0.25]{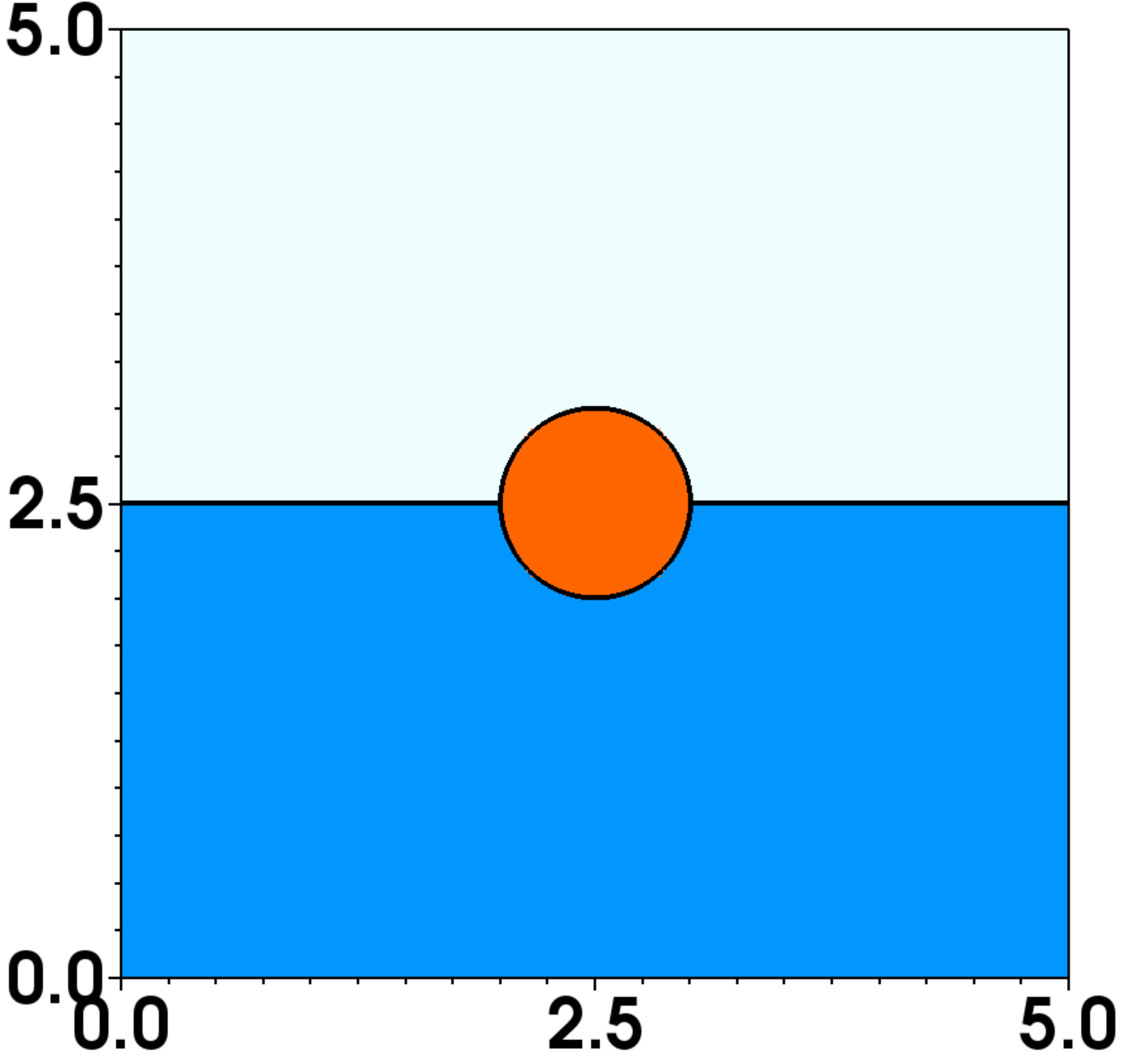}
    \label{Static_Cylinder_Init}
  }
     \subfigure[$L^{\infty}$ norm of velocity vs. time]{
    \includegraphics[scale = 0.25]{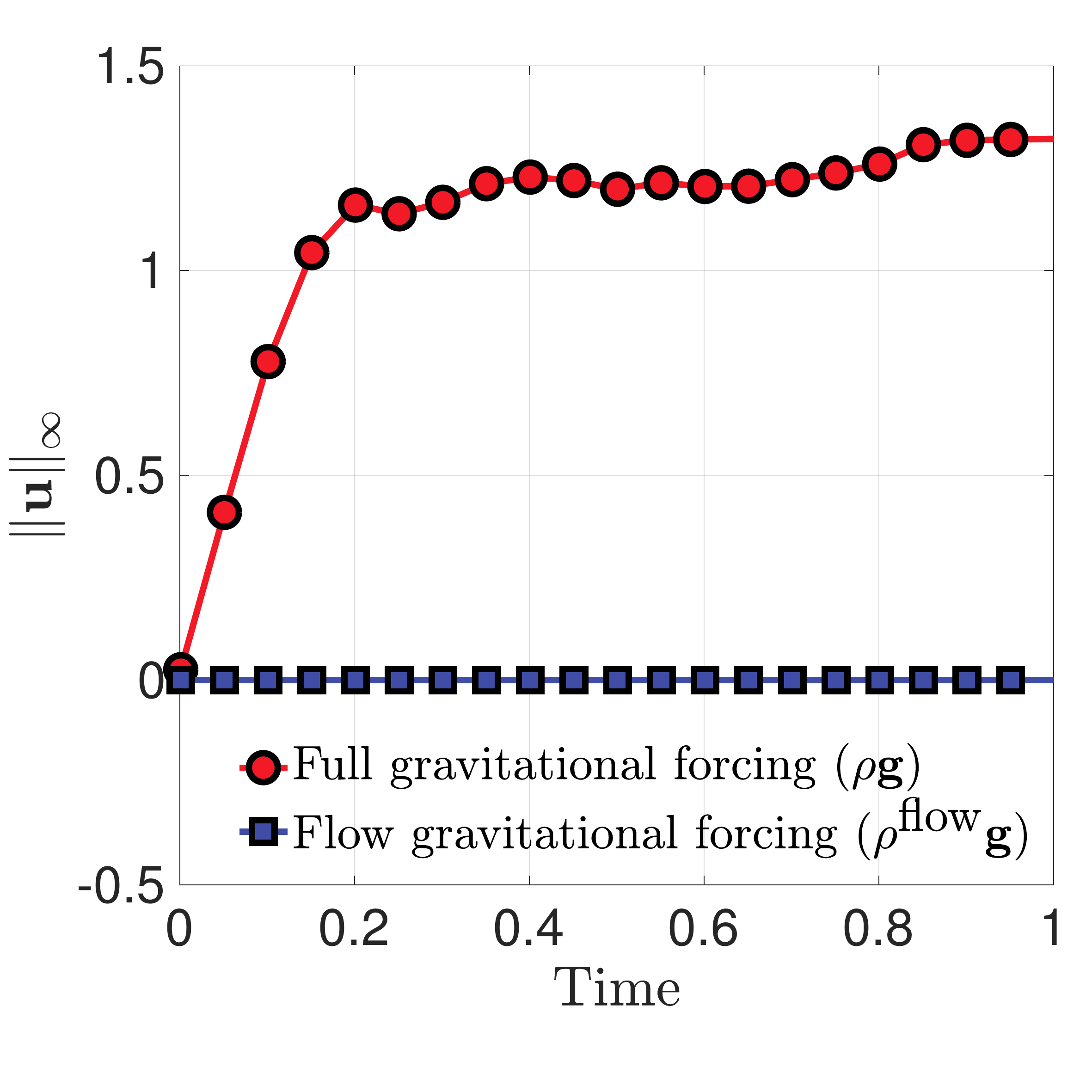}
    \label{Static_Cylinder_Umax}
  }
     \subfigure[$\rho \g$ at $t = 1.0$]{
    \includegraphics[scale = 0.25]{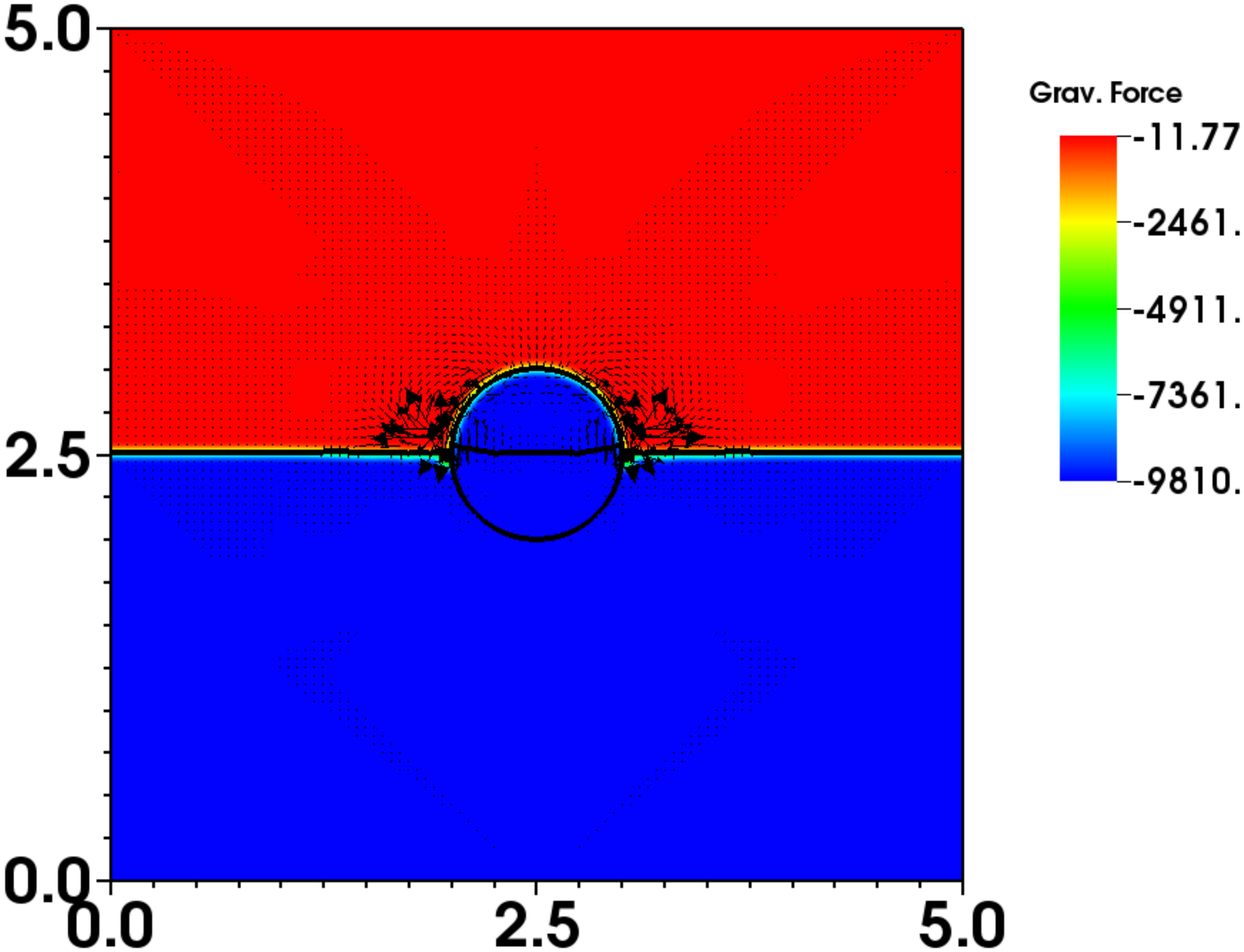}
    \label{Static_Cylinder_FullGrav_t1}
  }
     \subfigure[$\rho^{\text{flow}} \g$ at $t = 1.0$]{
    \includegraphics[scale = 0.25]{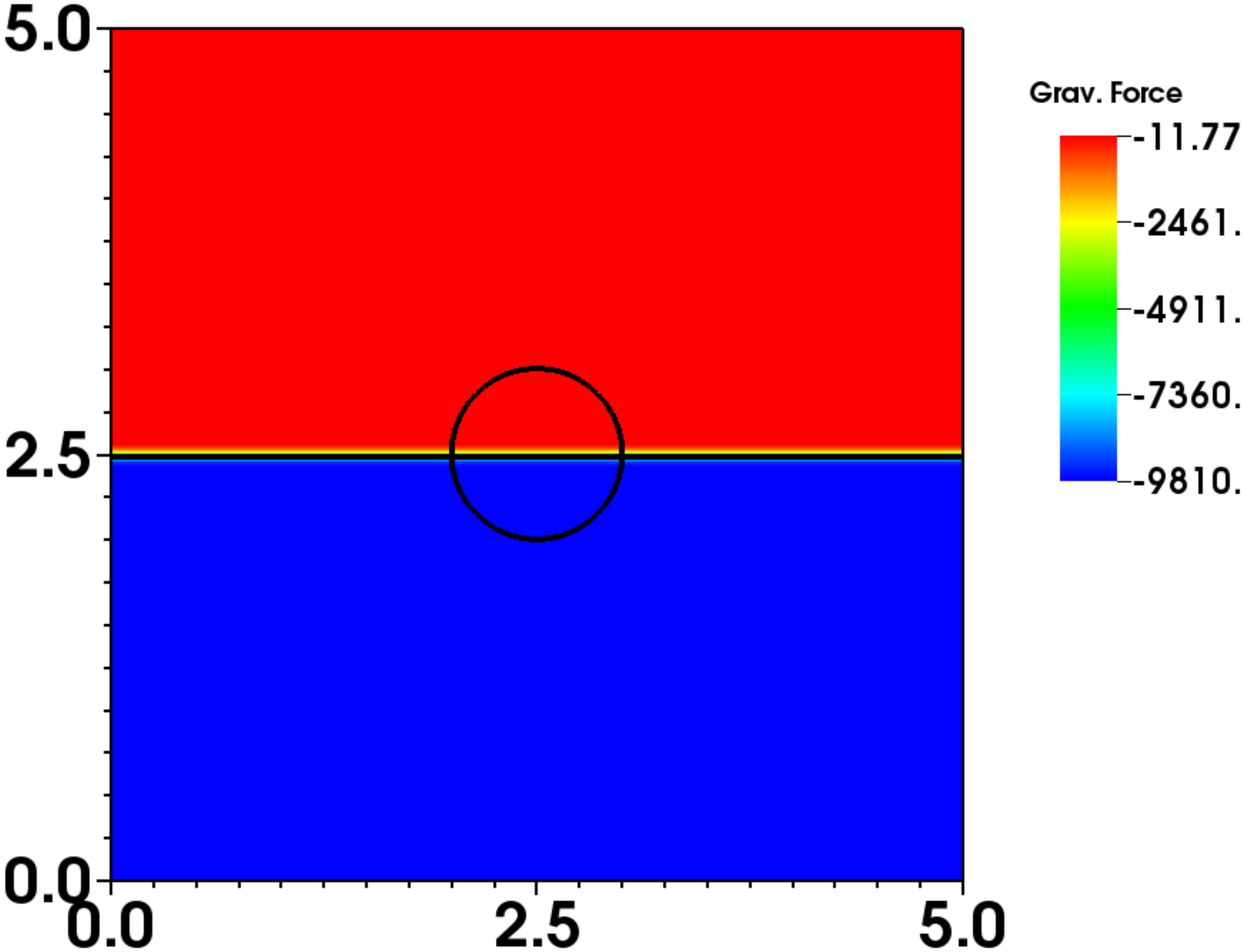}
    \label{Static_Cylinder_FlowGrav_t1}
  }
\caption{\subref{Static_Cylinder_Init} Initial problem set up for a stationary cylinder (orange) placed on a 
quiescent air-water interface.
\subref{Static_Cylinder_Umax} Temporal evolution of $\|\u\|_\infty$ with the ($\bullet$, red) full gravitational
forcing, and the ($\blacksquare$, blue) flow gravitational forcing.
Velocity vectors and gravitational forcing for \subref{Static_Cylinder_FullGrav_t1} $\rho \g$, and
\subref{Static_Cylinder_FlowGrav_t1} $\rho^{\text{flow}} \g$ at $t = 1.0$.
The scale for velocity vectors is identical in both figures}
  \label{fig_static_cylinder}
\end{figure}

To begin, a circular cylinder of diameter $D = 1$ is placed within a 2D computational
domain of size $\Omega = [0,5D]^2$. Water occupies the bottom half of the domain
and air occupies the remainder of the tank. The cylinder is placed at the center of the domain
with initial center of mass $(X_0,Y_0) = (2.5D, 2.5D)$.
In contrast with the previous cases, all of the cylinder's translational and rotational degrees of freedom
are locked and it is fully constrained to remain stationary, i.e. $\Ub = (U_\text{b}, V_\text{b}) = (0,0)$.
As described in Sec.~\ref{sec_solid_materials}, the density and viscosity in the solid region are set
to those of the water phase. Two grid cells of smearing $\ncells = 2$ are used to
transition between different material properties on 
either side of the interfaces, and surface tension forces are neglected.
No-slip boundary conditions are imposed along $\partial \Omega$.
The initial problem set up is shown in Fig.~\ref{Static_Cylinder_Init}.

We consider two forms of the gravitational body force:
\begin{enumerate}
\item the \emph{full} gravitational forcing $\rho \g$, with $\rho$ prescribed using Eq.~\eqref{eq_ls_solid},
\item the \emph{flow} gravitational forcing $\rho^{\text{flow}}\g$, with $\rho$ prescribed using Eq.~\eqref{eq_ls_flow}.
\end{enumerate}
The domain is discretized with a uniform $N \times N$ grid with $N = 200$ and a constant time step size of
$\dt = 1/(5N)$ is used. For this particular case, no flow dynamics should be generated because the cylinder
is held fixed in place and the initially quiescent fluids should maintain hydrostatic equilibrium. Hence, the quantity
of interest is the $L^{\infty}$ norm of velocity $\|\u\|_\infty$, which indicates the largest value (in magnitude) of parasitic 
velocity generated in the domain. Fig.~\ref{Static_Cylinder_Umax} shows the temporal evolution of $\|\u\|_\infty$
as a function of time for both cases. It is seen that the full gravitational forcing leads to nonzero velocity
values that do not dissipate over time. This is because spurious momentum is accumulated in the solid phase
due to its density value $\rhos$ while solving the discretized momentum
equations~\eqref{eq_c_discrete_momentum} and~\eqref{eq_c_discrete_continuity}.
However, the flow gravitational forcing produces velocities on the order of machine precision,
indicating that hydrostatic equilibrium is maintained. Figs~\ref{Static_Cylinder_FullGrav_t1} and~\ref{Static_Cylinder_FlowGrav_t1} show the velocity vectors and gravitational forcing for both cases
at $t = 1.0$. Significant velocity is shown to be generated for the $\rho \g$ case, while
these velocities are absent for the $\rho^{\text{flow}} \g$ case. Based on these test results, for all 
fully prescribed motion cases considered in this paper, we make use of the 
flow gravitational forcing field to ensure accurate and well-balanced results. Finally, we remark that the 
parasitic current generation due to the gravitational body force occurs for both types of flow solvers.

\subsection{Water entry of a circular cylinder}
In this section, we demonstrate the importance of consistent mass and momentum
transport to achieve stability for air-water density ratios. A circular cylinder of radius
$R = 0.055$ is placed within a 2D computational domain of size
$\Omega = [0,40R]\times[0,24R]$. Water occupies the bottom half of the domain
from $y = 0$ to $y = 12R$ and air occupies the remainder of the tank.
The cylinder is placed just above the fluid phase with initial center position
$(X_0, Y_0) = (20R, 14R)$ and has density equal to that of the water phase (i.e. $\rhos = 1 \times 10^3$).
Two grid cells of smearing $\ncells = 2$ are used to transition between different material properties on
either side of the interfaces, and surface tension forces are neglected.
No-slip boundary conditions are imposed along $\partial \Omega$.

Similar to the previous case, all of the cylinder's translational and rotational degrees of freedom
are locked and its motion is fully constrained to be unity in the vertical direction,
i.e. $\Ub = (U_\text{b}, V_\text{b}) = (0,-1)$. As described in Sec.~\ref{sec_solid_materials},
gravitational forces are not evaluated using the density within the solid region for prescribed motion cases.
The primary quantity of interest for this example is the dimensionless vertical hydrodynamic force (slamming coefficient)
given by
\begin{equation}
\label{eq_Cs}
C_\textrm{s} = \frac{\cF \cdot \e_y}{\rhol R V_\text{b}^2}
\end{equation}
as a function of the penetration depth $P_\textrm{d} = V_\text{b}(t - t_\text{impact})/R$, where $t_\text{impact} = R/V_\text{b}$ denotes the
time at which the bottom of the cylinder first touches the air-water interface.
This case has been studied both analytically by von K{\`a}rm{\`a}n~\cite{von1929}, using potential flow theory, and
experimentally by Campbell and Weynberg~\cite{Campbell1980}. This case has also been studied numerically
by a number of authors, including Patel and Natarajan~\cite{Patel2018}, Zhang et al.~\cite{Zhang2010},
and Kleefsman et al.~\cite{Kleefsman2005}.
The domain is discretized by $\ell = 2$ grid levels with refinement ratio $\nref = 2$.
The grid spacing at the coarsest grid level is $\dx_0 = \dy_0 = 1/200$ yielding a finest
grid spacing of $\dx_\textrm{min} = \dy_\textrm{min} = 1/400$ or $22$ grid cells per radius.
A constant time step size of $\dt = 0.02 \dx_\textrm{min}$ is used.

\begin{figure}[]
  \centering
  \subfigure[$t = 0.05$]{
    \includegraphics[scale = 0.25]{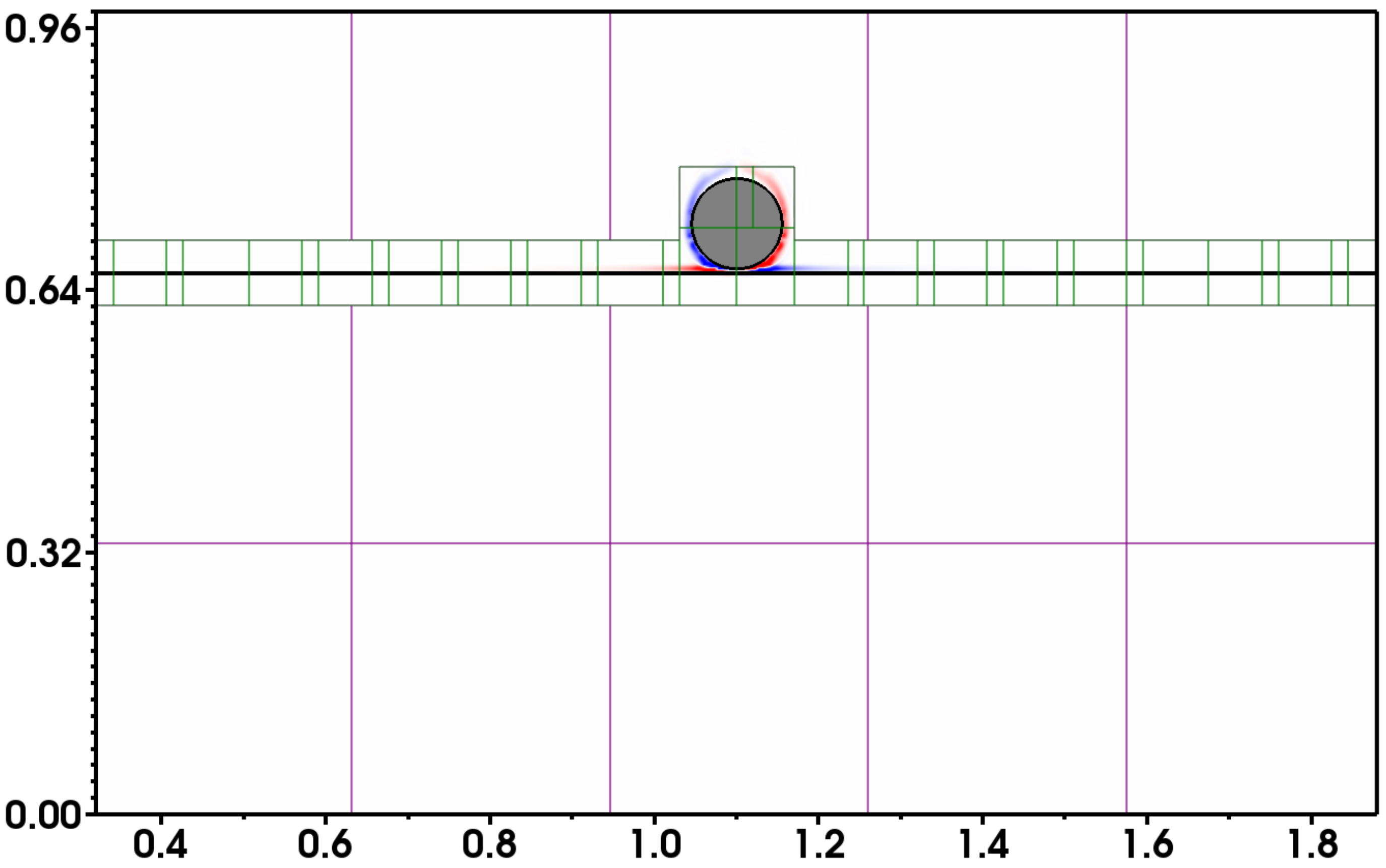}
    \label{Non_Conservative_Water_Entry_t0p05}
  }
     \subfigure[$t = 0.15$]{
    \includegraphics[scale = 0.25]{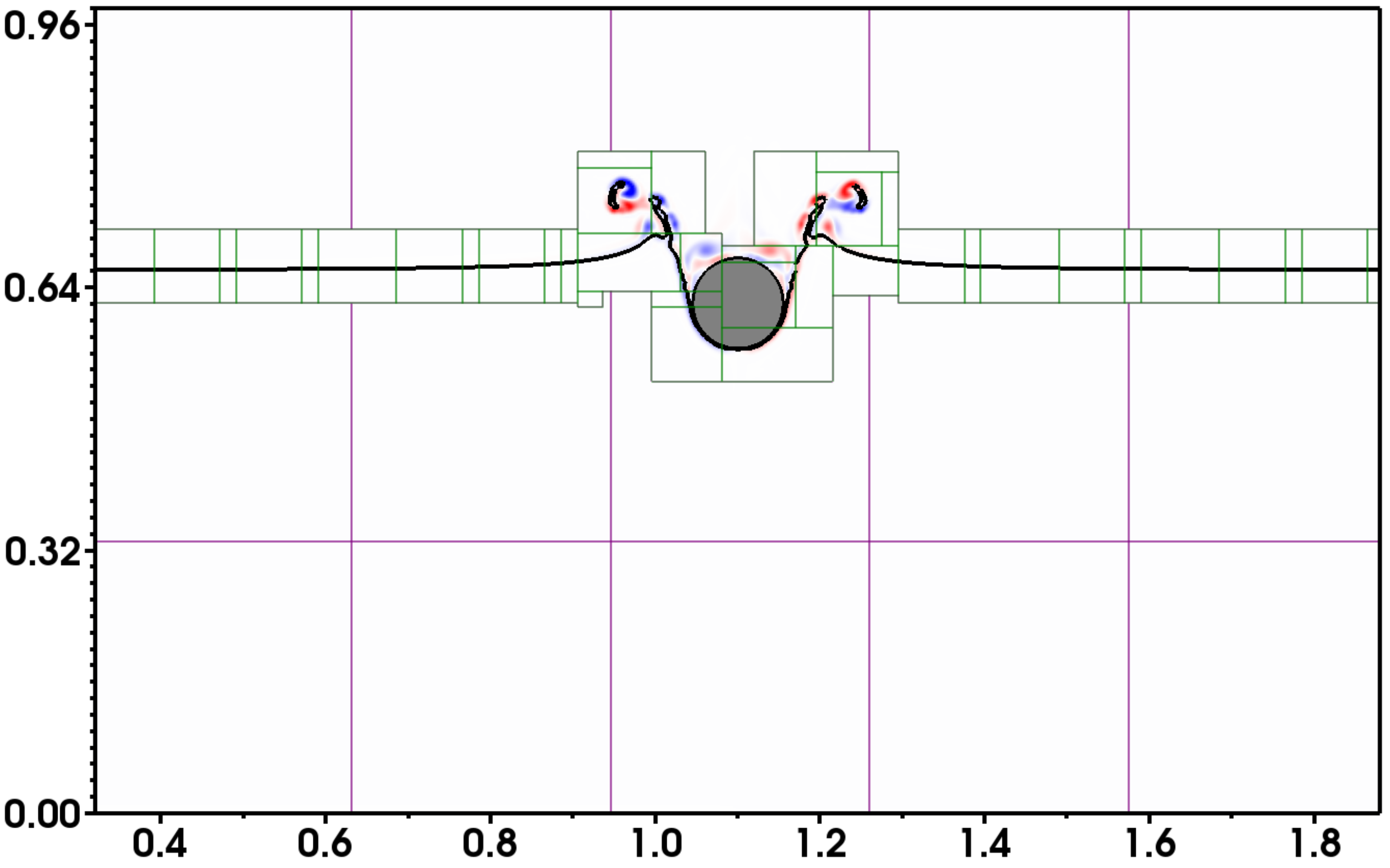}
    \label{Non_Conservative_Water_Entry_t0p15}
  }
     \subfigure[$t = 0.25$]{
    \includegraphics[scale = 0.25]{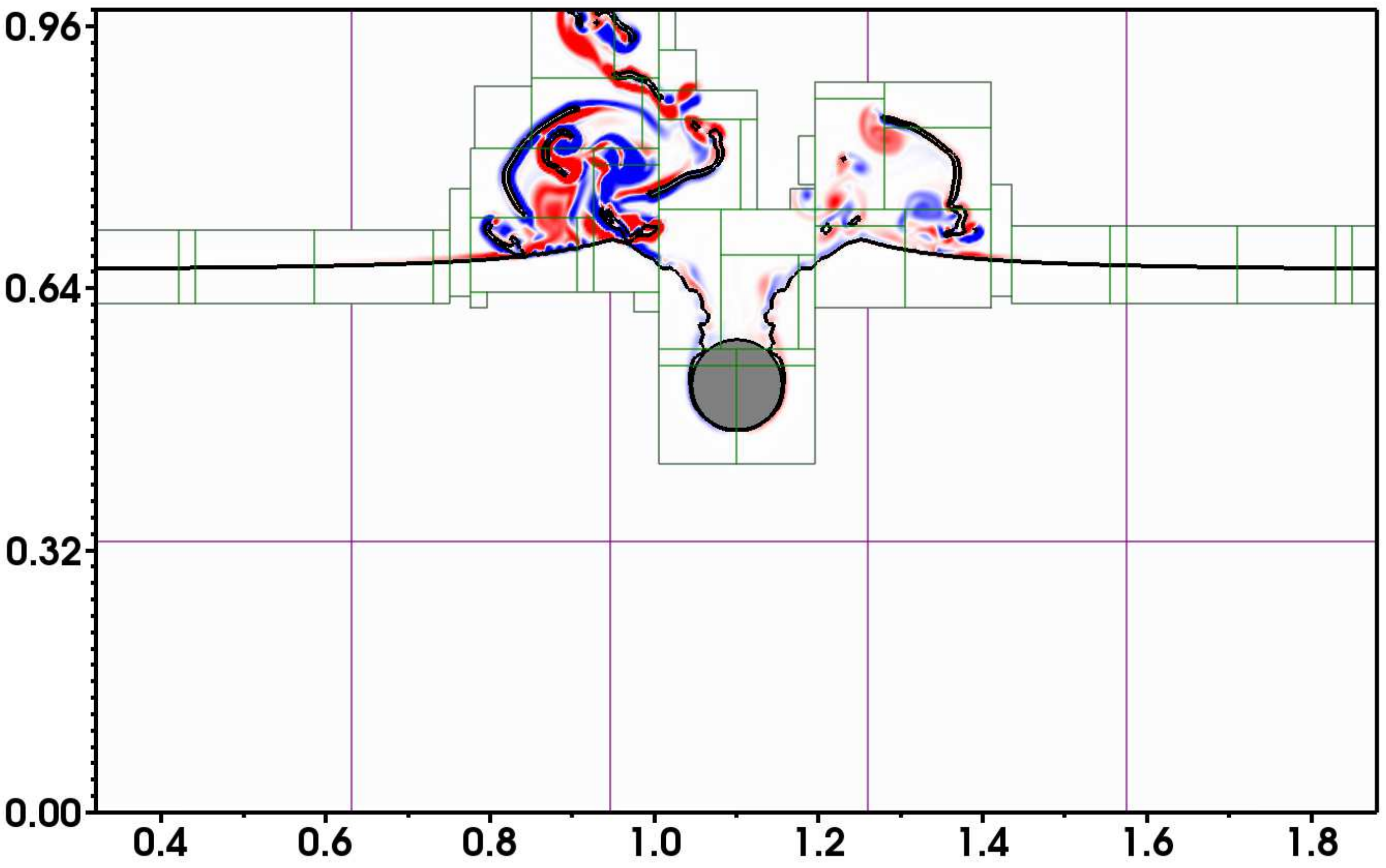}
    \label{Non_Conservative_Water_Entry_t0p25}
  }
\caption{Temporal evolution of a cylinder entering an air-water interface at four different time instances.
Inconsistent transport of mass and momentum is used for these cases. The simulation becomes unstable
shortly after $t = 0.25$.
The plotted vorticity is in the range $-50$ to $50$.
Locations of the different refined mesh levels from coarsest to finest are shown in purple and green.
}
  \label{fig_non_conservative_water_entry}
\end{figure}

\begin{figure}[]
  \centering
  \subfigure[$t = 0.05$]{
    \includegraphics[scale = 0.25]{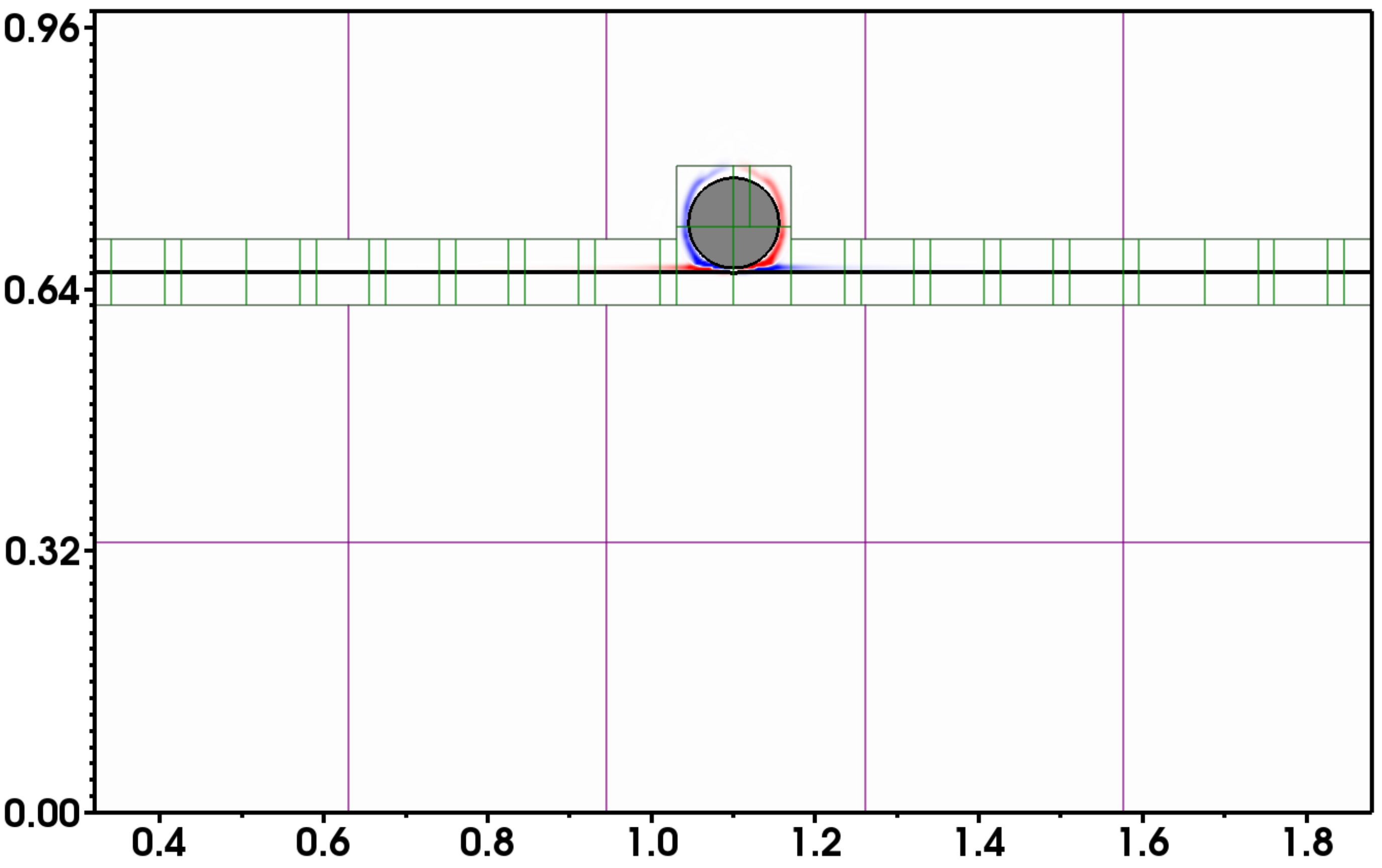}
    \label{Conservative_Water_Entry_t0p05}
  }
     \subfigure[$t = 0.25$]{
    \includegraphics[scale = 0.25]{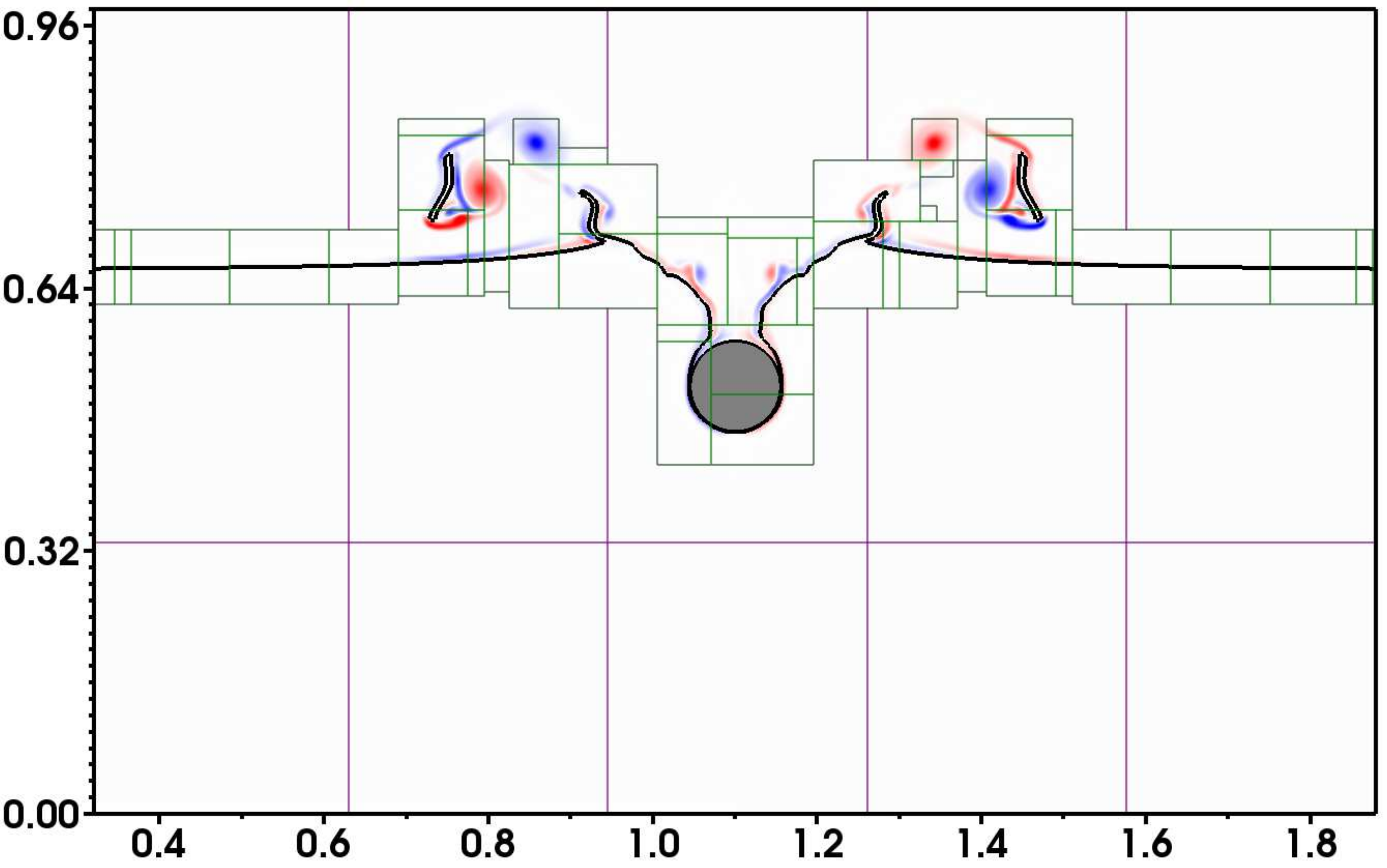}
    \label{Conservative_Water_Entry_t0p25}
  }
     \subfigure[$t = 0.35$]{
    \includegraphics[scale = 0.25]{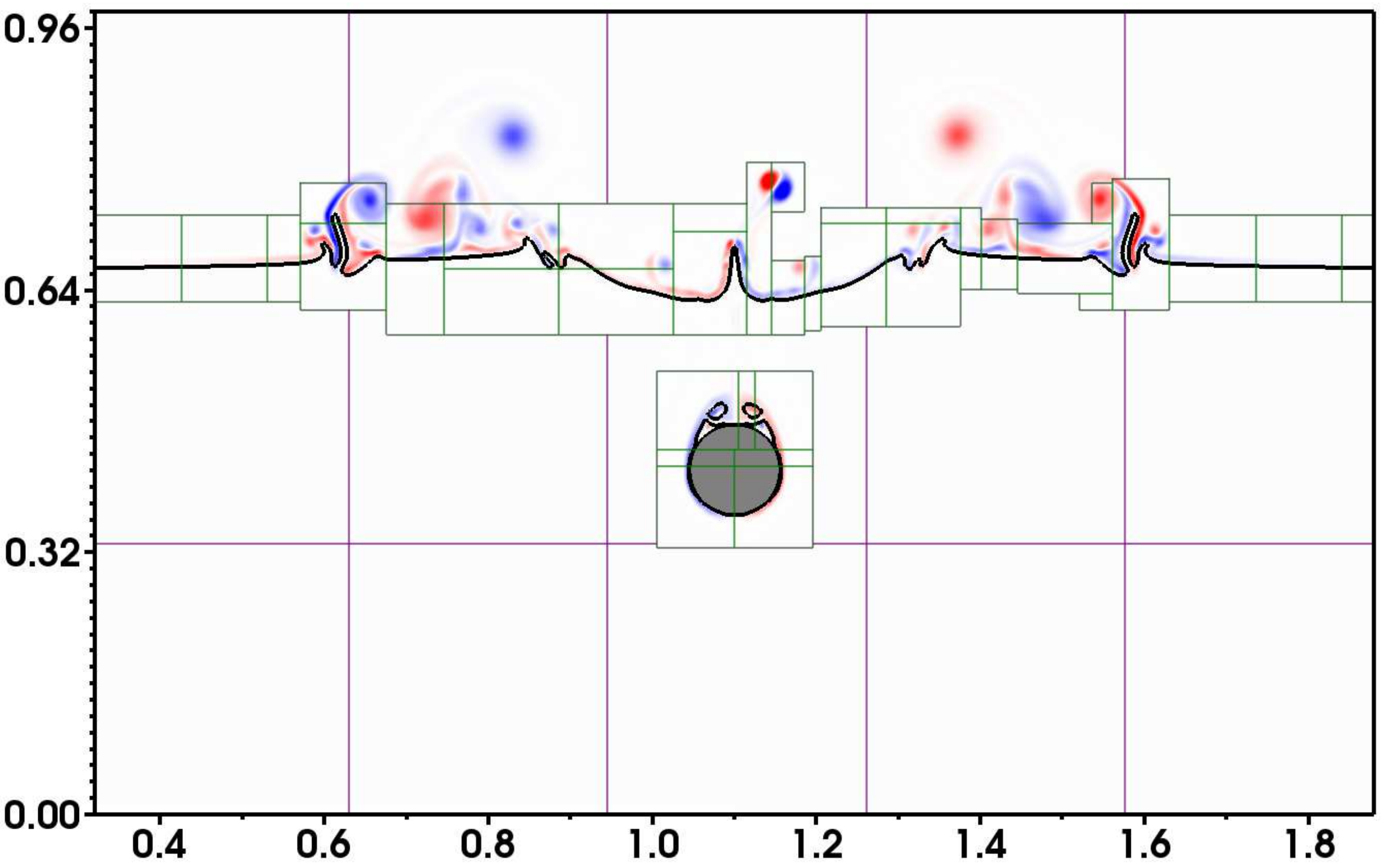}
    \label{Conservative_Water_Entry_t0p35}
  }
     \subfigure[$t = 0.45$]{
    \includegraphics[scale = 0.25]{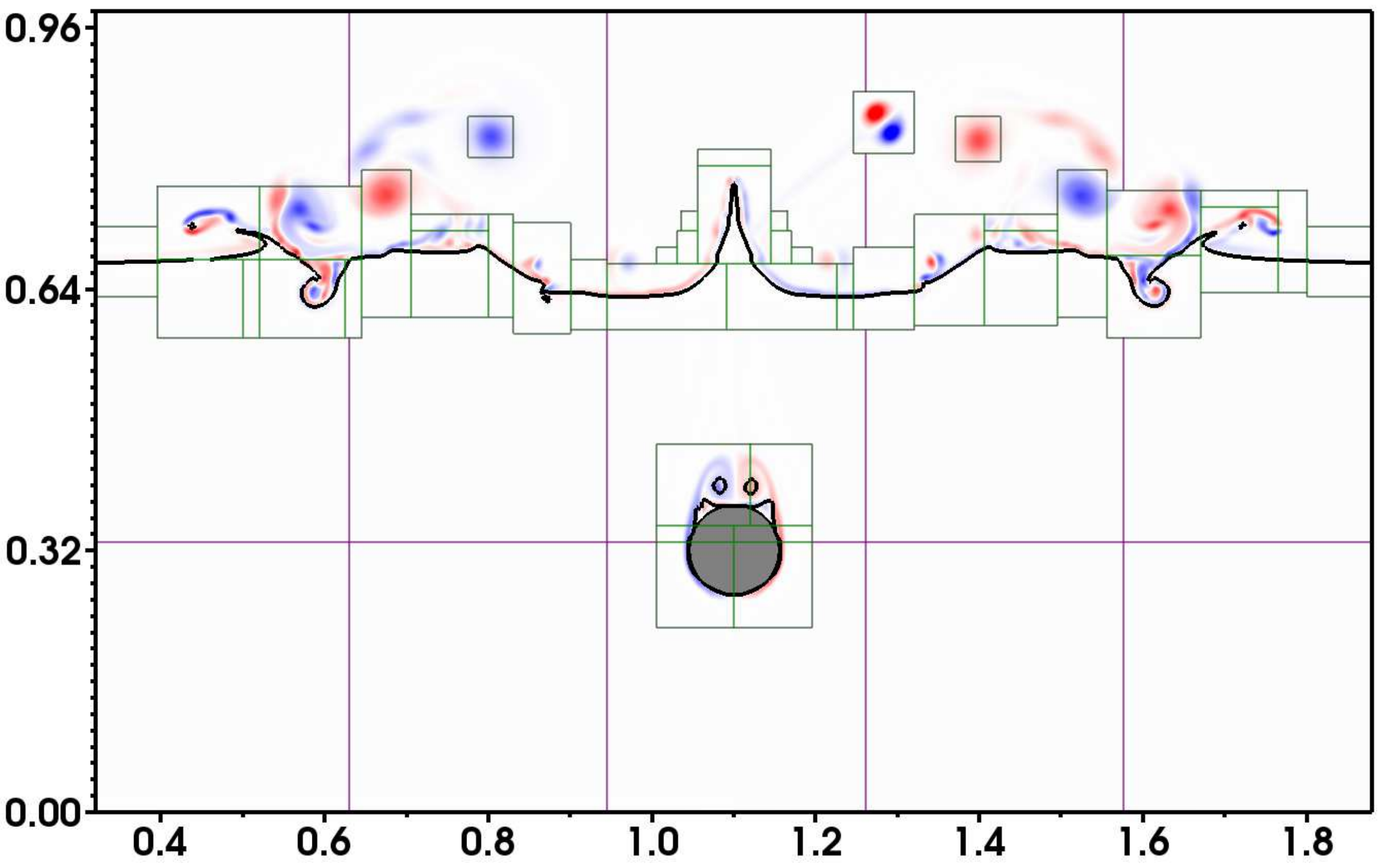}
    \label{Conservative_Water_Entry_t0p45}
  }
\caption{Temporal evolution of a cylinder entering an air-water interface at four different time instances.
Consistent transport of mass and momentum is used for these cases.
The plotted vorticity is in the range $-50$ to $50$.
Locations of the different refined mesh levels from coarsest to finest are shown in purple and green.
}
  \label{fig_conservative_water_entry}
\end{figure}

As a first test, we demonstrate the importance of consistent mass and momentum transport for the stability
of high density ratio multiphase flows. Fig.~\ref{fig_non_conservative_water_entry} shows the evolution
of the cylinder and the air-water interface when simulated with inconsistent mass and momentum transport,
i.e. by using a non-conservative momentum integrator. The simulation
quickly becomes unstable, leading to the generation of unphysical interfaces and high regions of vorticity.
Lowering the time step even further did not resolve these stability problems.
However, the simulation is stable when using consistent mass and momentum
transport (Fig.~\ref{fig_conservative_water_entry}). As the cylinder enters the water, crashing wave-like
structures are generated and air entrainment is seen around the cylinder.

\begin{figure}[]
  \centering
    \includegraphics[scale = 0.3]{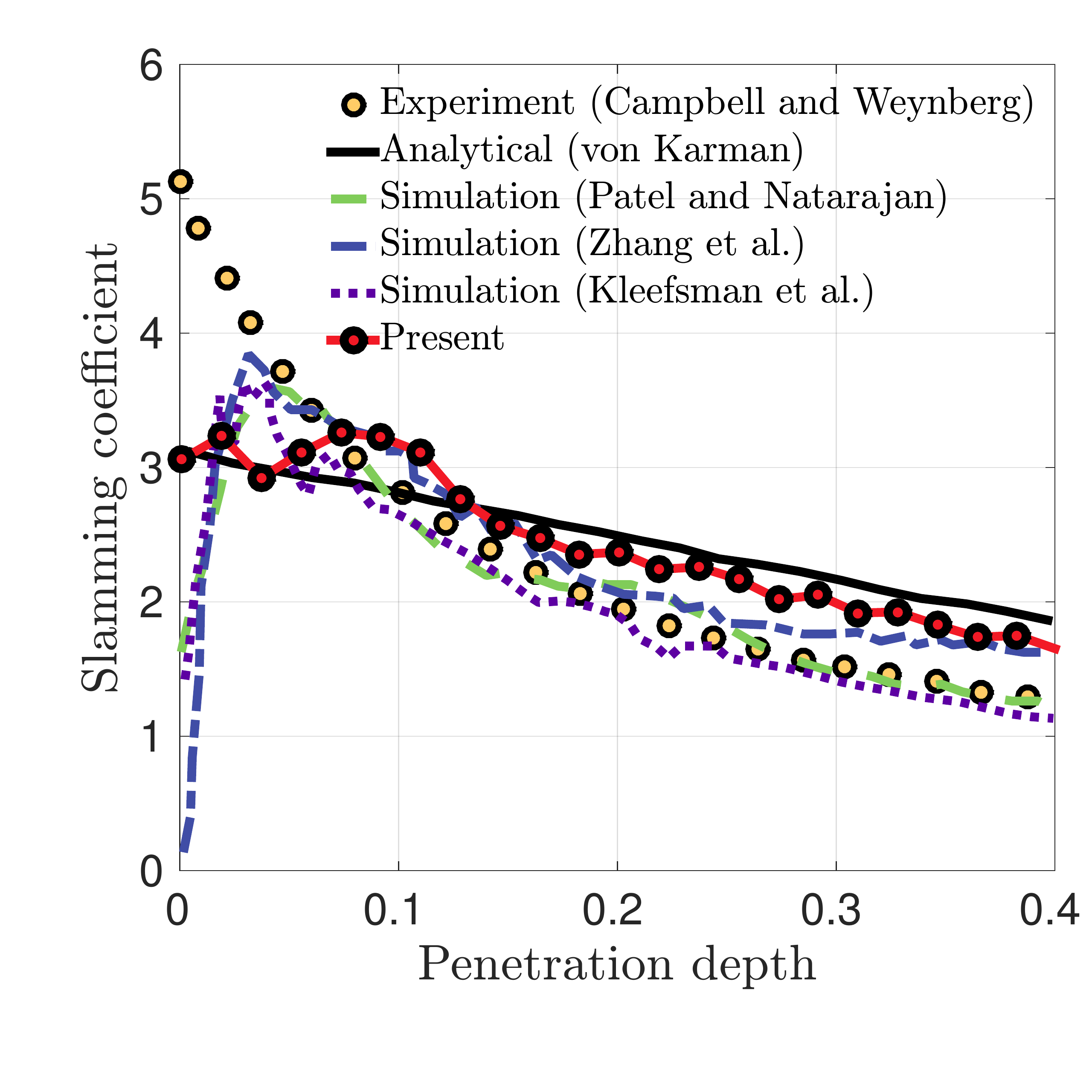}
 \caption{
 Dimensionless slamming coefficient as a function of penetration depth for a 2D cylinder entering an air-water interface
 at constant velocity.
 ($\bullet$, yellow) experimental data from Campbell and Weynberg~\cite{Campbell1980};
 (---, black) analytical formula from von K{\`a}rm{\`a}n~\cite{von1929};
 (\texttt{---}, green) simulation data from Patel and Natarajan~\cite{Patel2018};
 (\texttt{-}$\cdot$\texttt{-}, blue) simulation data from Zhang et al.~\cite{Zhang2010}
  (\texttt{...}, purple) simulation data from Kleefsman et al.~\cite{Kleefsman2005};
(-$\bullet$-, red) present simulation data.
 }
  \label{fig_slamming_coefficient}
\end{figure}

Fig.~\ref{fig_slamming_coefficient} shows the slamming coefficient as a function of penetration depth.
The results are in decent agreement with previous studies,
with minor disagreements being explained by differences in the interface
tracking approaches and/or difference in the fluid-structure coupling techniques.
This test case demonstrates how important consistent mass and momentum transport is
for stable simulation of practical multiphase flows, for which air-water density ratios are ubiquitous.
Moreover, we demonstrate that the present numerical method can be used to accurately simulate
problems involving fully prescribed body motion. 

\subsection{Rolling barge}
In this section, we investigate the roll decay of a rectangular barge floating on an
air-water interface. A rectangle of length $L = 0.3$ and width $W = L/3$ is placed
within a 2D computational domain of size $\Omega = [-25W, 25W] \times [-9W, 16W]$.
Water occupies the bottom portion of the domain up until $y = 0$ with air occupying the
remaining portion. The barge is initially situated with center of mass $(X_0, Y_0) = (0,0)$
at a $15^\circ$ angle of inclination with the horizontal, and has density $\rhos = 1.18\times 10^3$.
Two grid cells of smearing $\ncells = 2$ are used to transition between different material properties on
either side of the interfaces. Only the barge's rotational degrees of freedom are unlocked and surface
tension forces are neglected.
The conservative and consistent flow solver is used for this case.
No-slip boundary conditions are imposed along $\partial \Omega$.
This specific 2D case has been numerically studied by Patel and Natarajan~\cite{Patel2017}.

\begin{figure}[]
  \centering
  \subfigure[$T = 0.0$]{
    \includegraphics[scale = 0.28]{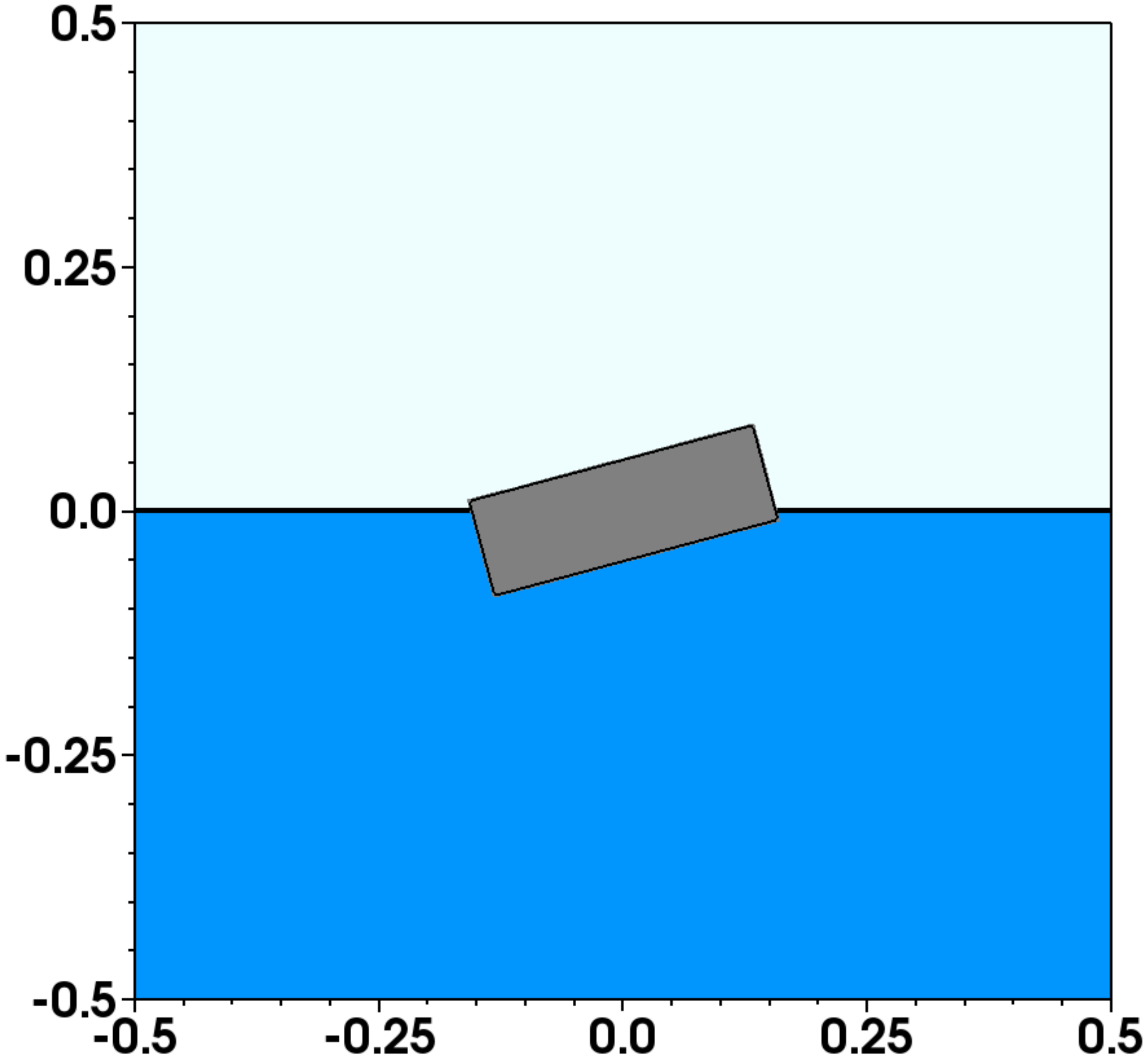}
    \label{barge_t0p0}
  }
     \subfigure[$T = 1.36$]{
    \includegraphics[scale = 0.28]{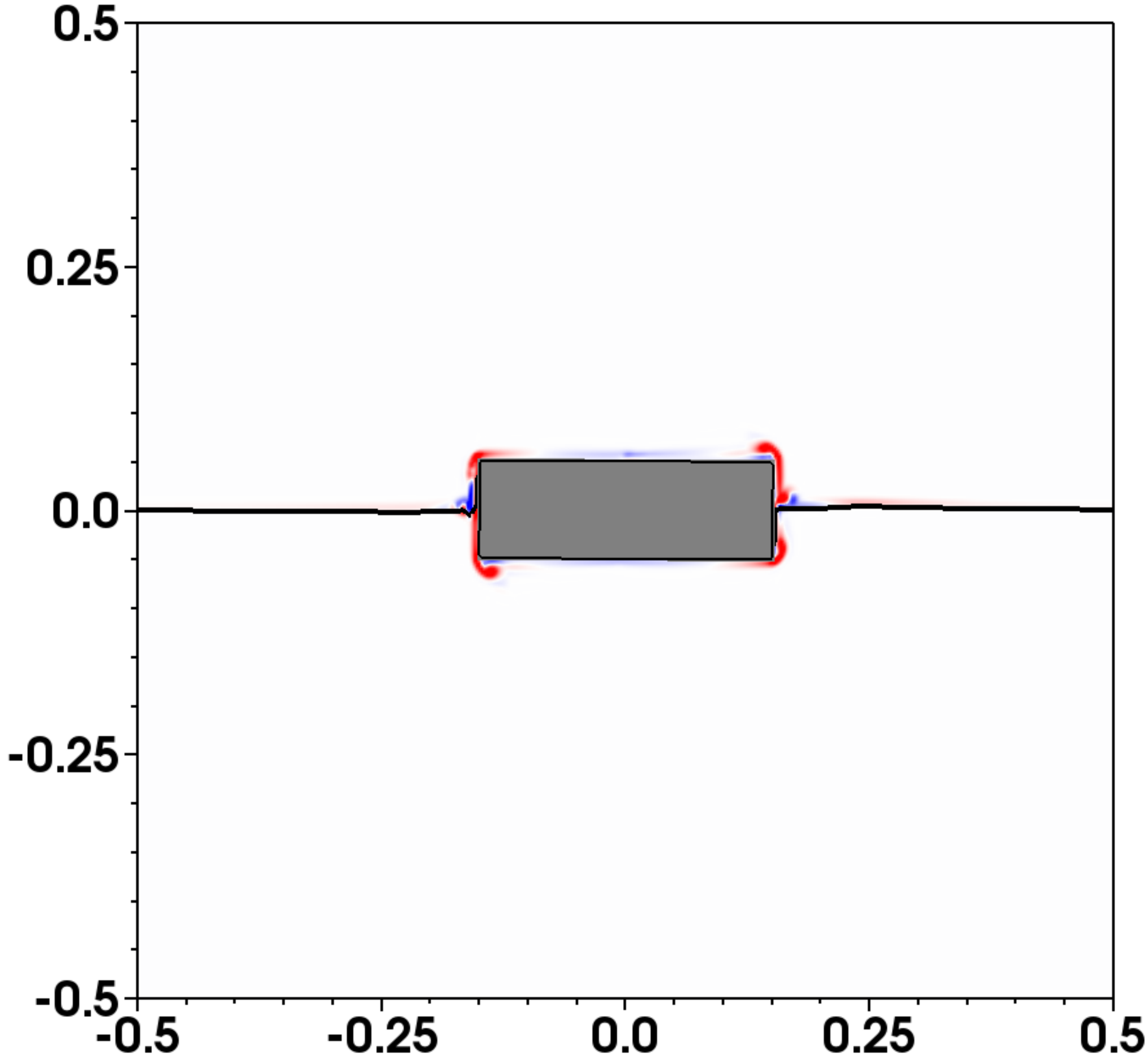}
    \label{barge_t0p25}
  }
     \subfigure[$T = 2.71$]{
    \includegraphics[scale = 0.28]{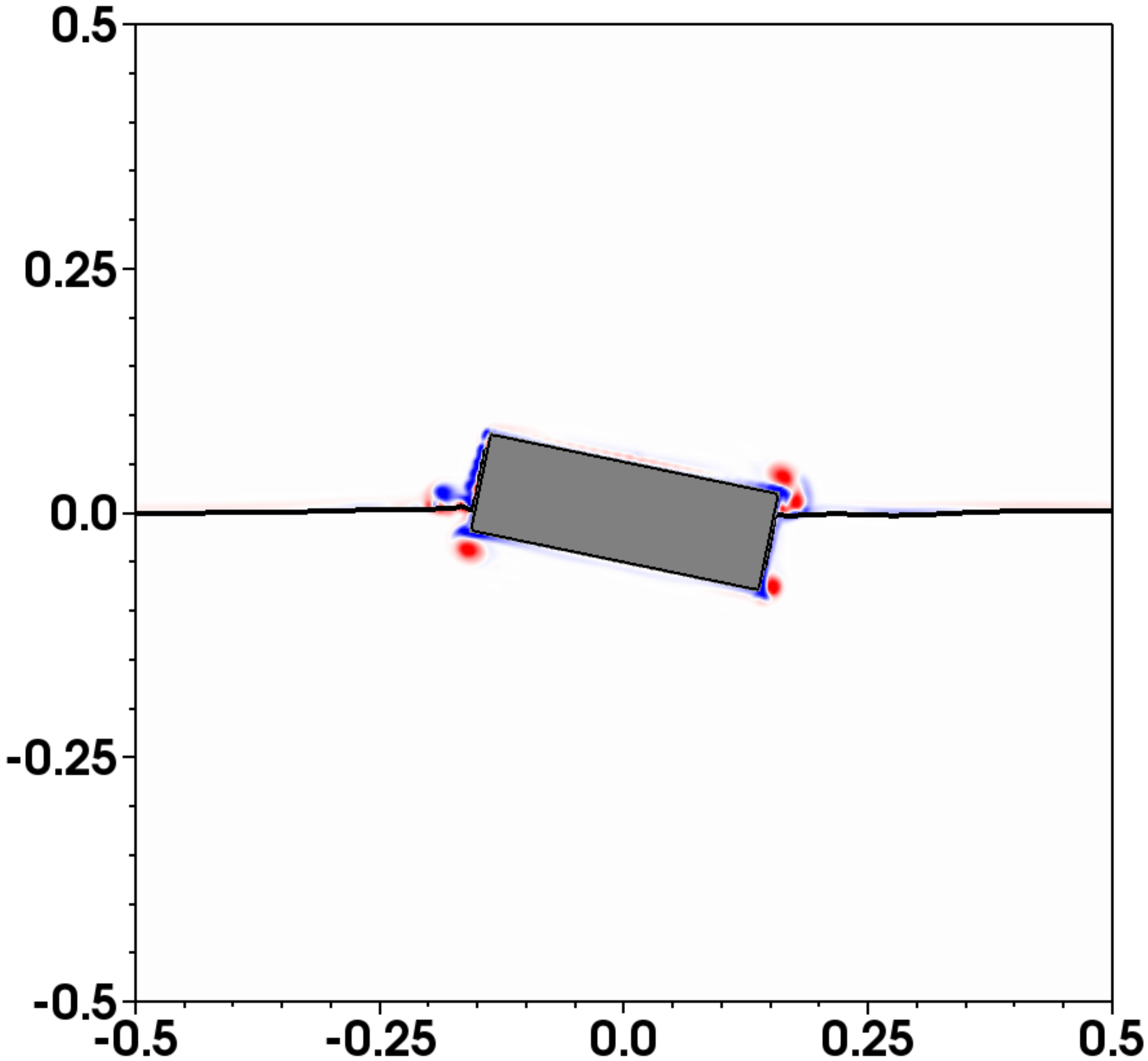}
    \label{barge_t0p5}
  }
     \subfigure[$T = 10.85$]{
    \includegraphics[scale = 0.28]{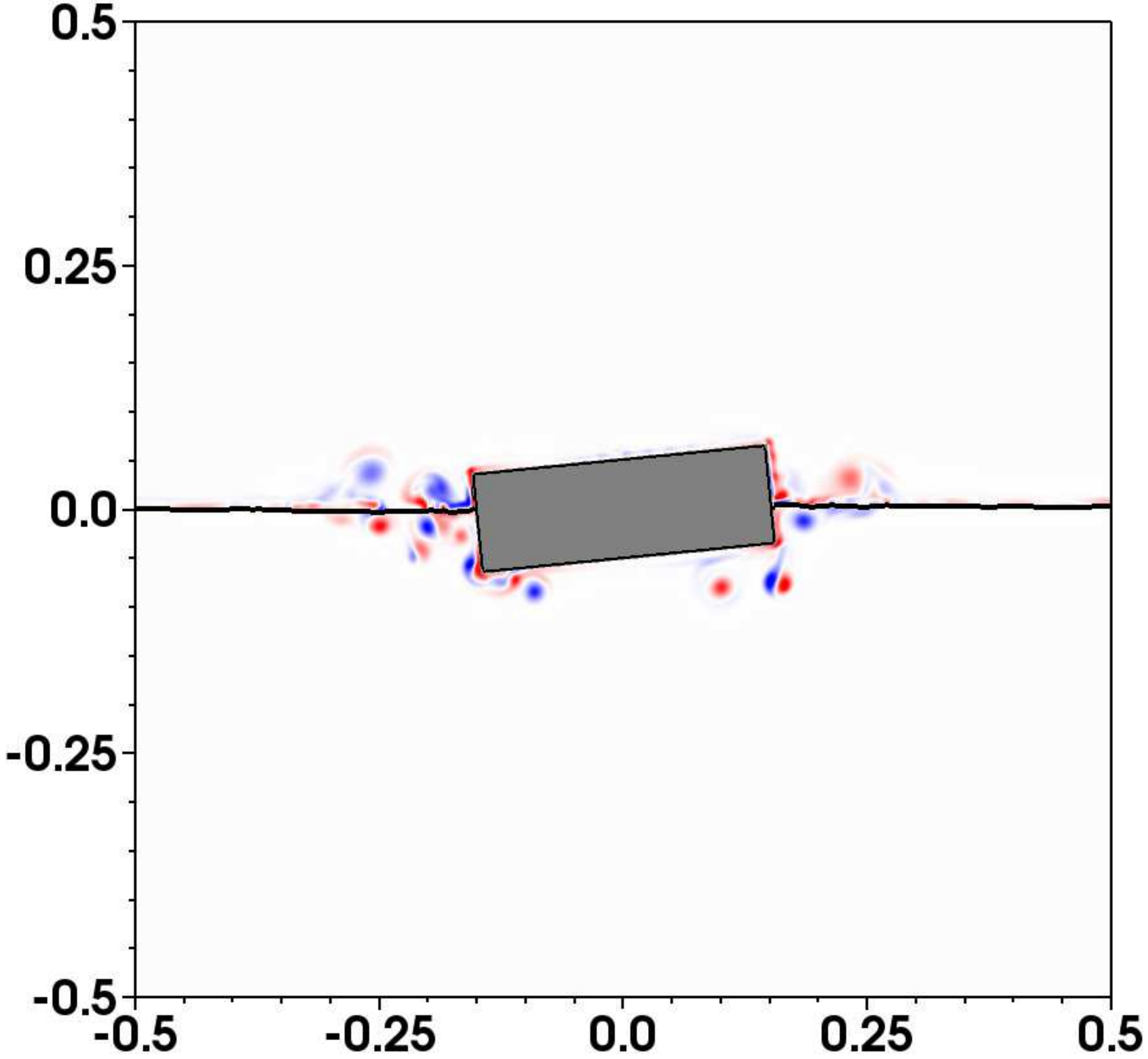}
    \label{barge_t2p0}
  }
  \subfigure[$T = 13.56$]{
    \includegraphics[scale = 0.28]{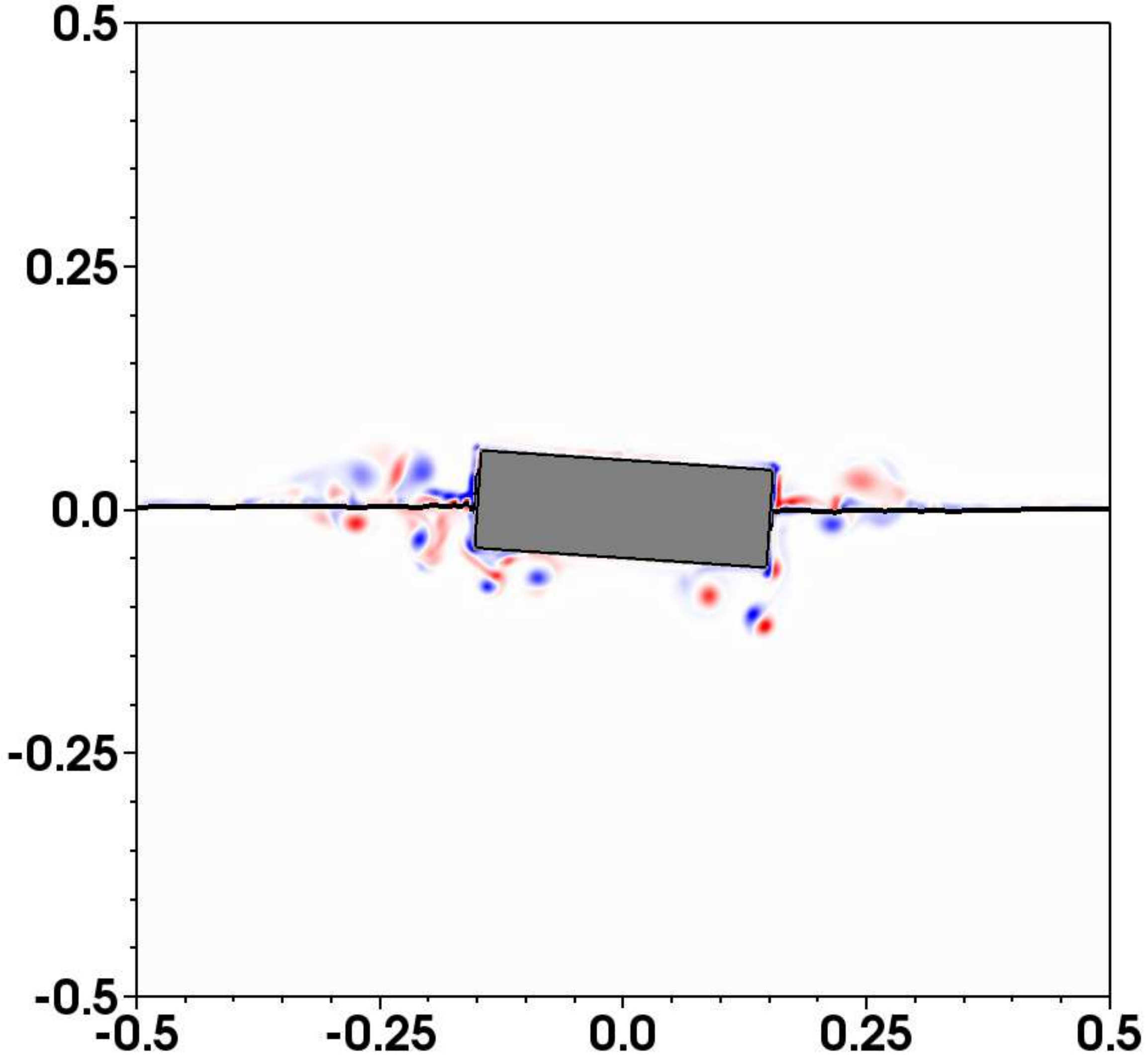}
    \label{barge_t2p5}
  }
   \subfigure[$T = 18.99$]{
    \includegraphics[scale = 0.28]{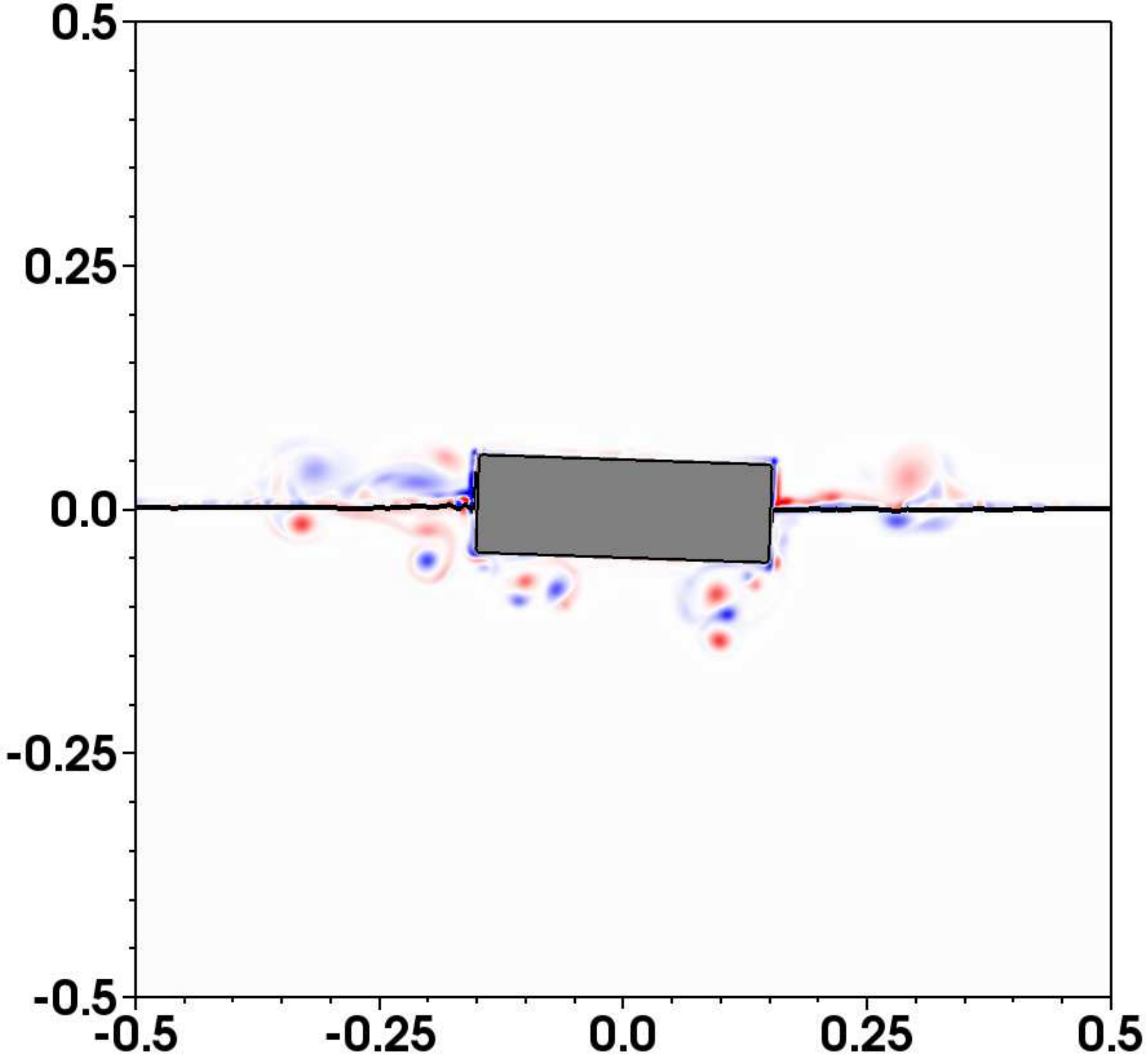}
    \label{barge_t3p5}
  }
    \caption{Temporal evolution of a rectangular barge floating an air-water interface at six different time instances with $40$ CPW. The plotted vorticity is in the range $-50$ to $50$.
   }
  \label{fig_barge_viz}
\end{figure}

\begin{figure}[]
  \centering
    \includegraphics[scale = 0.25]{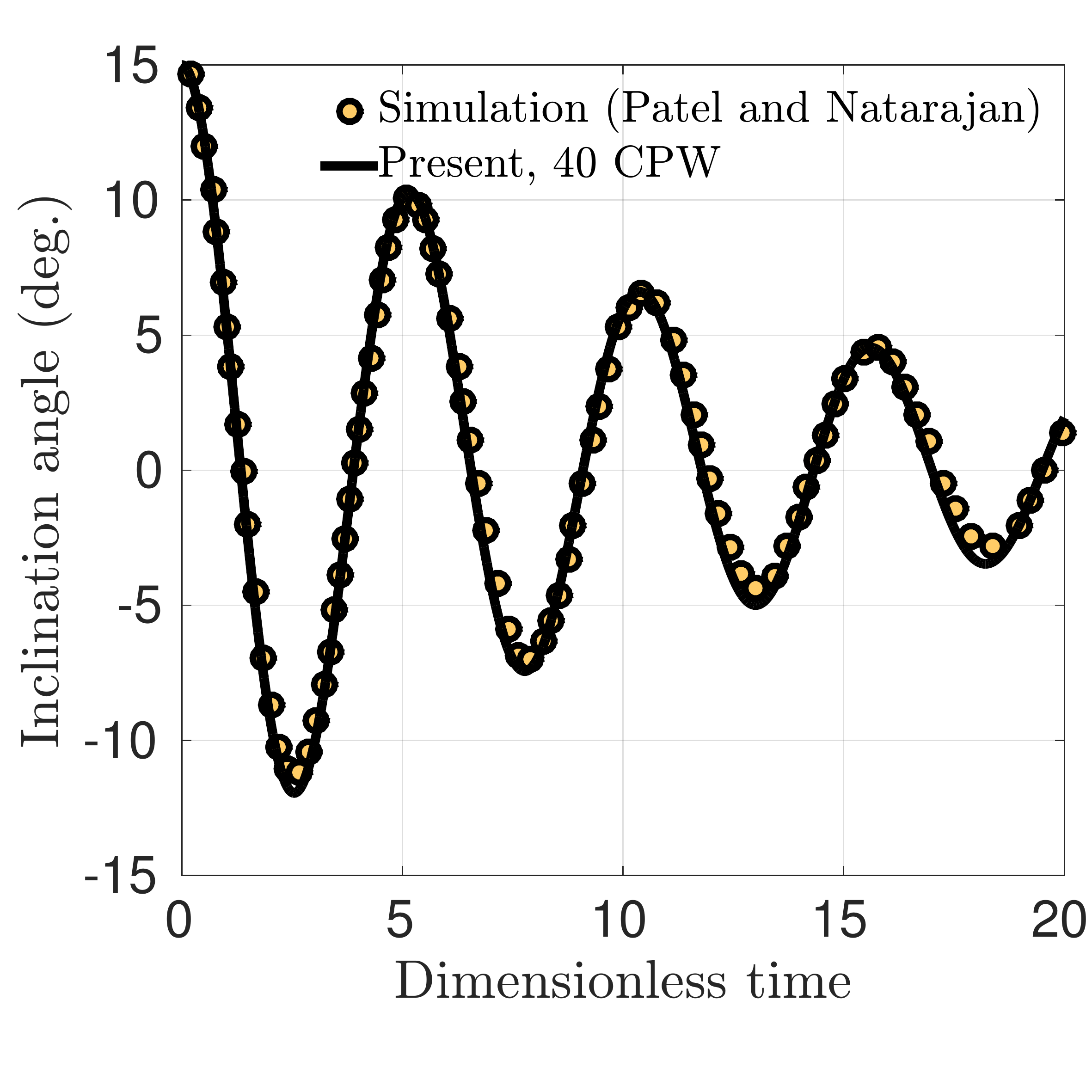}
 \caption{
 Temporal evolution of the inclination angle for a 2D barge floating on an air-water interface.
 ($\bullet$, yellow) simulation data from Patel and Natarajan~\cite{Patel2017};
(---, black) present simulation data for $40$ grid cells per width.
 }
  \label{fig_rolling_barge_inclination_angle}
\end{figure}

The domain is discretized by a grid of size $2N \times N$ with $N = 1000$, yielding $40$ grid cells per
width. A constant time step size of $dt = 1/(4N)$ is used. Fig.~\ref{fig_barge_viz} shows the evolution
of the rotating body and the air-water interface at various instances in dimensionless time $T = t \sqrt{g/L}$.
Vortices are shed from the corners of the barge as it rolls back and forth and small disturbances are
seen along the water. The temporal evolution of the angle of inclination (Fig.~\ref{fig_rolling_barge_inclination_angle})
are in excellent agreement with the numerical results of Patel and Natarajan~\cite{Patel2017}. Over time, the
roll angle damps out due to the viscous forces of the water phase. This case demonstrates that the present
numerical method can be used to accurately simulate floating objects undergoing rigid, rotational motion.

\subsection{Wedge free-falling into water}
In this section, we investigate the problem of a wedge-shaped object impacting
a pool of water. A 2D triangular body with top length $L = 1.2$ is placed within
a computational domain of size $\Omega = [0, 10L] \times [0, 2.5L]$. The wedge
is oriented with one of its vertices pointing downwards, making a $25^\circ$ deadrise
angle with the horizontal. Water occupies the bottom third of the domain, while air occupies
the remainder of the tank. The bottom point of the wedge is placed with initial position
$(X_0, Y_0) = (5L, 23L/12)$ and the wedge has density $\rhos = 466.6$.
Two grid cells of smearing $\ncells = 2$ are used to transition between different material properties on
either side of the interfaces, and surface tension forces are neglected.
No-slip boundary conditions are imposed along $\partial \Omega$.
Only the wedge's vertical degrees of freedom are unlocked.

\begin{figure}[]
  \centering
   \subfigure[Inconsistent, $t = 0.45$]{
    \includegraphics[scale = 0.25]{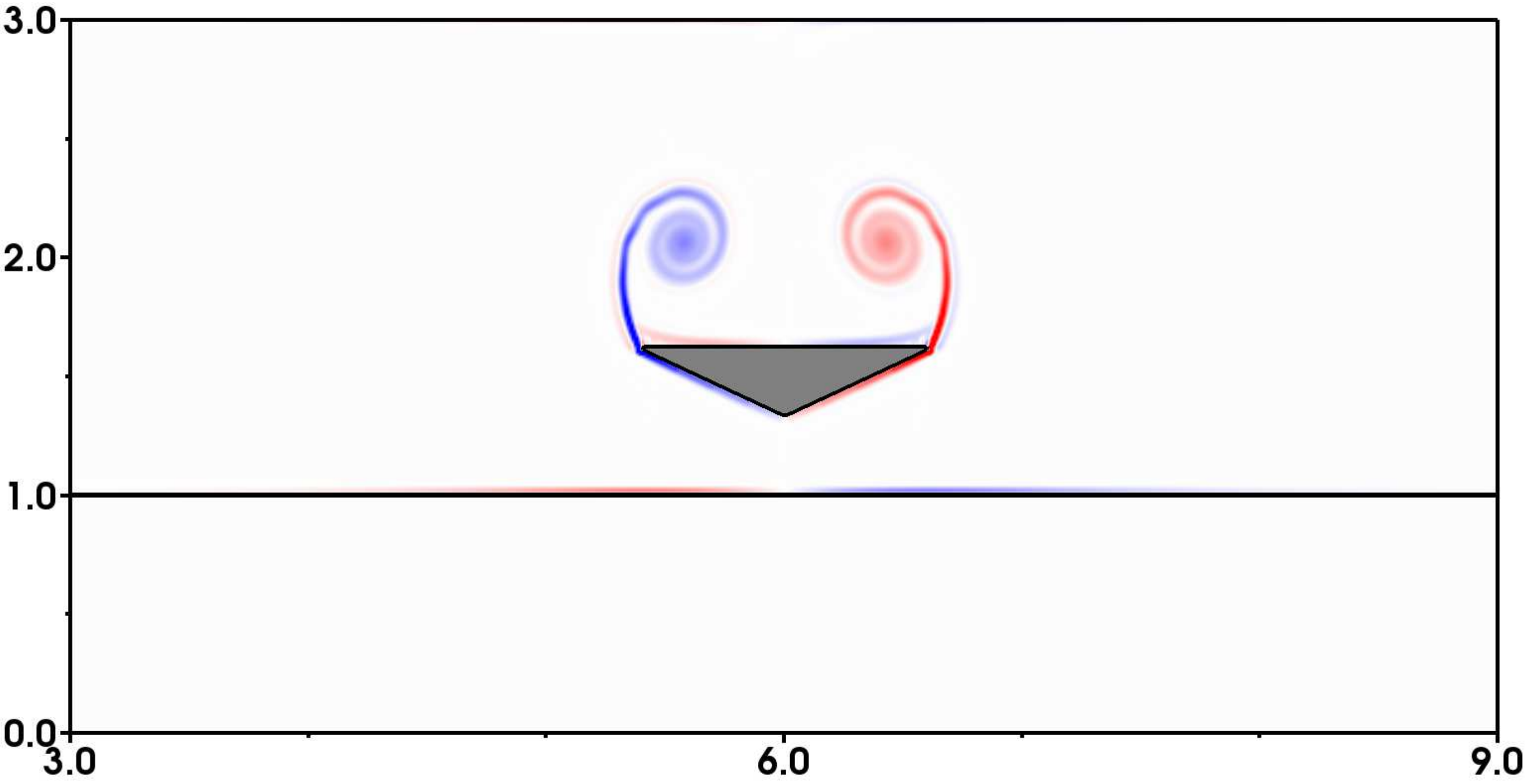}
    \label{nc_wedge2d_t0p45}
  }
   \subfigure[Inconsistent, $t = 0.5625$]{
    \includegraphics[scale = 0.25]{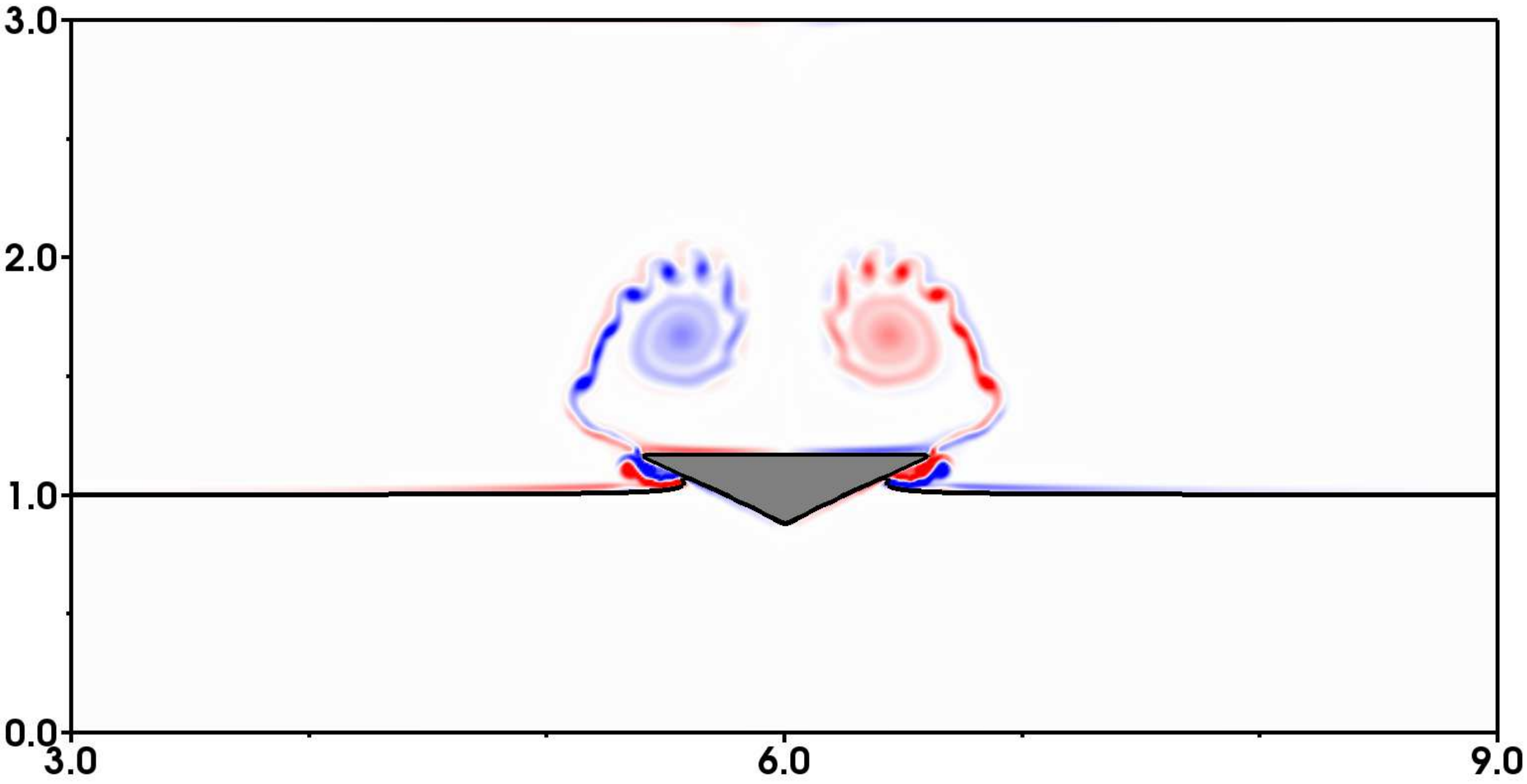}
    \label{nc_wedge2d_t0p5625}
  }
  \subfigure[Inconsistent, $t = 0.75$]{
    \includegraphics[scale = 0.25]{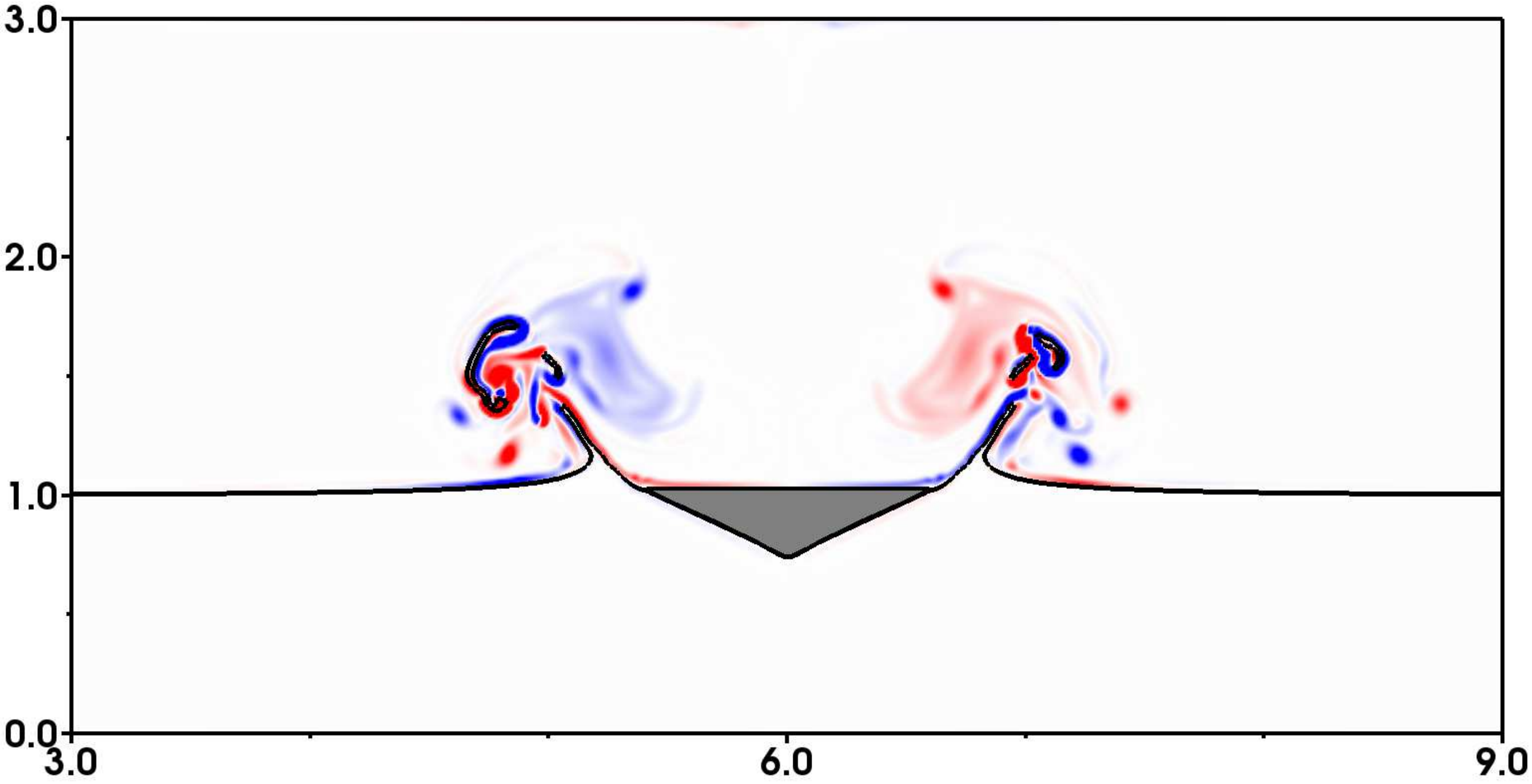}
    \label{nc_wedge2d_t0p75}
  }
    \subfigure[Inconsistent, $t = 0.785$]{
    \includegraphics[scale = 0.25]{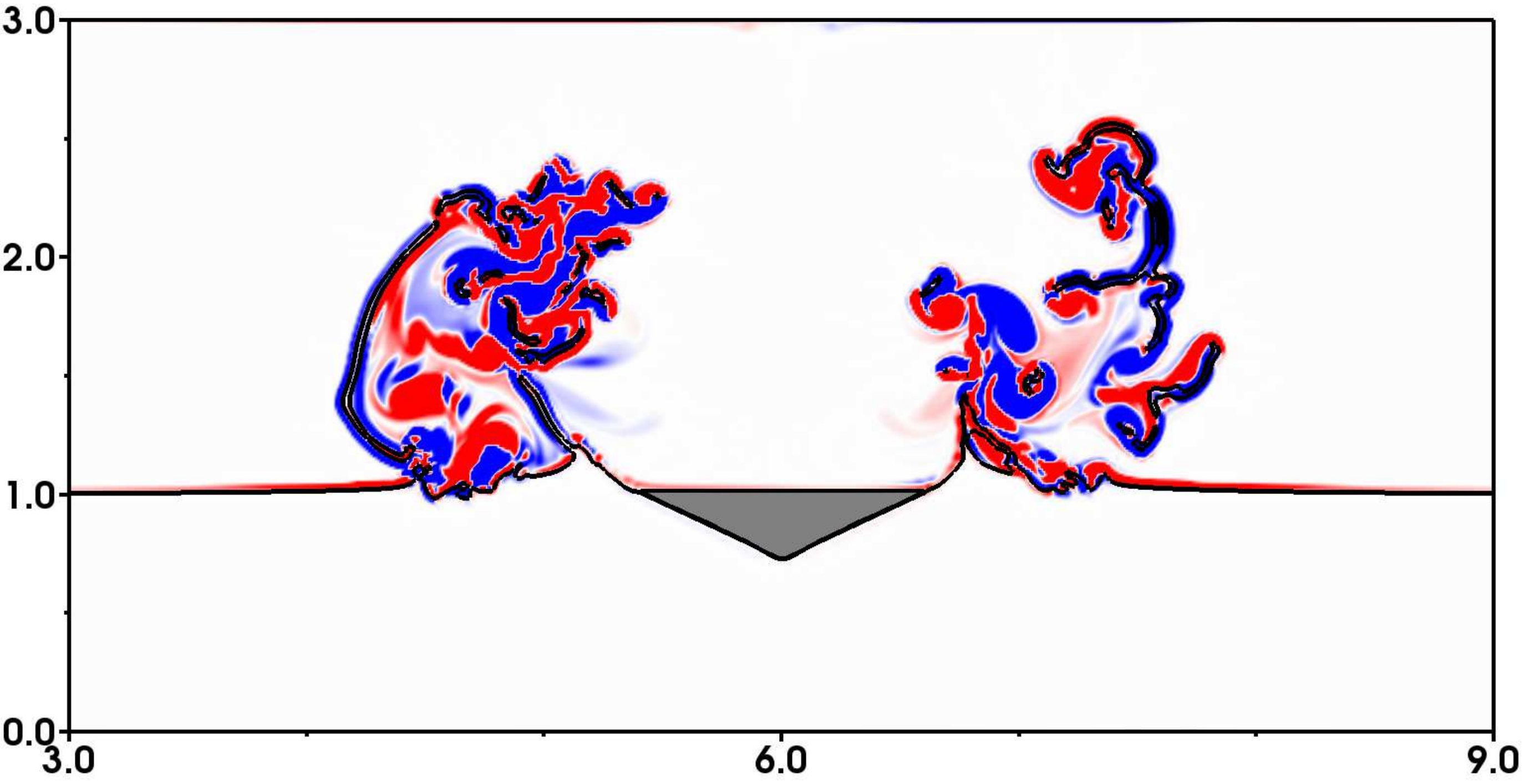}
    \label{nc_wedge2d_t0p785}
  }
  \caption{Temporal evolution of a 2D wedge free-falling into an air-water interface at four different time instances.
Inconsistent transport of mass and momentum is used for these cases. The simulation becomes unstable
shortly after $t = 0.785$. The plotted vorticity is in the range $-300$ to $300$.
}
  \label{fig_nc_wedge2d_viz}
\end{figure}

\begin{figure}[]
  \centering
\subfigure[Consistent, $t = 0.45$]{
    \includegraphics[scale = 0.25]{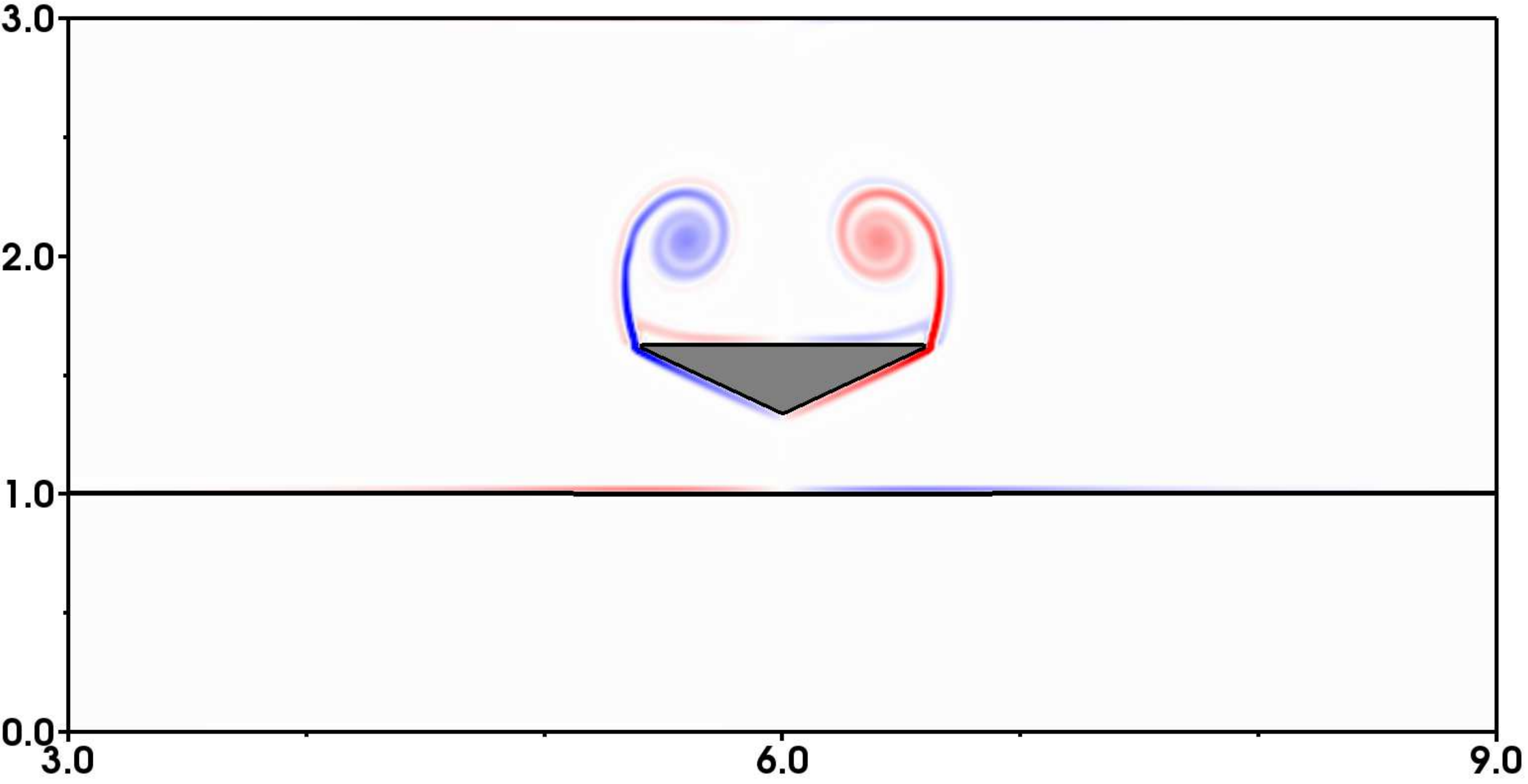}
    \label{c_wedge2d_t0p45}
  }
    \subfigure[Consistent, $t = 0.5625$]{
    \includegraphics[scale = 0.25]{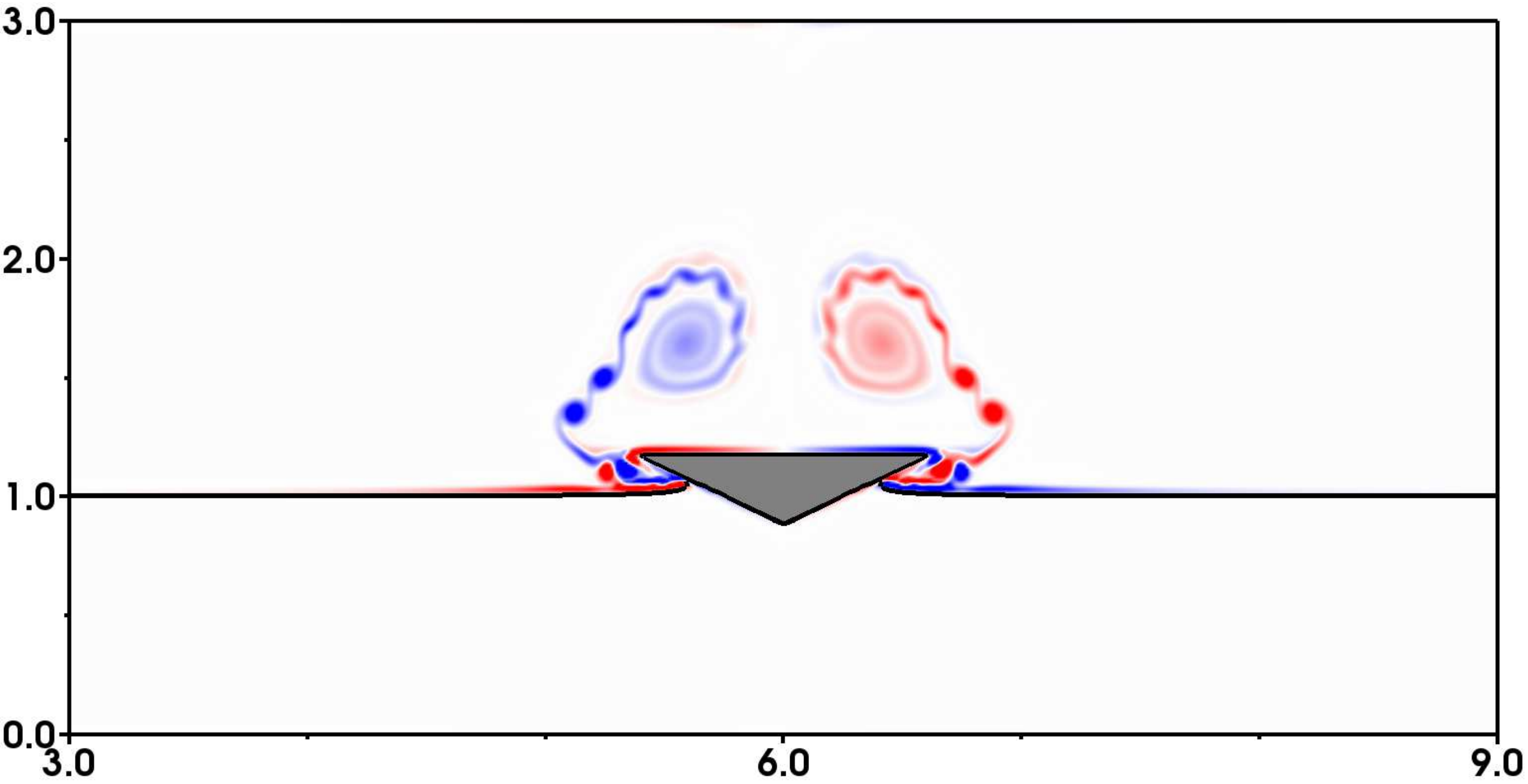}
    \label{c_wedge2d_t0p5625}
  }
  \subfigure[Consistent, $t = 0.875$]{
    \includegraphics[scale = 0.25]{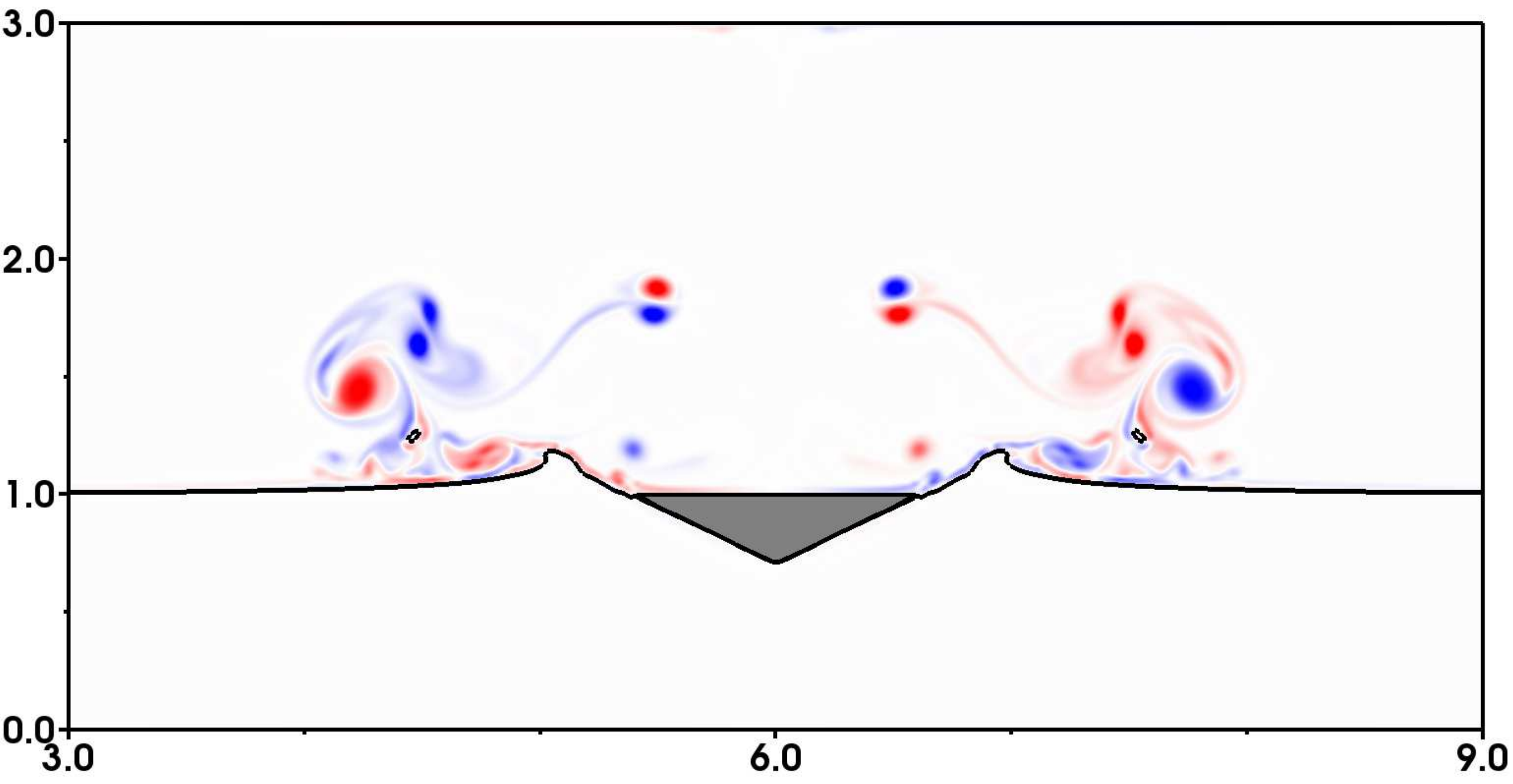}
    \label{c_wedge2d_t0p875}
  }
  \subfigure[Consistent, $t = 1.25$]{
    \includegraphics[scale = 0.25]{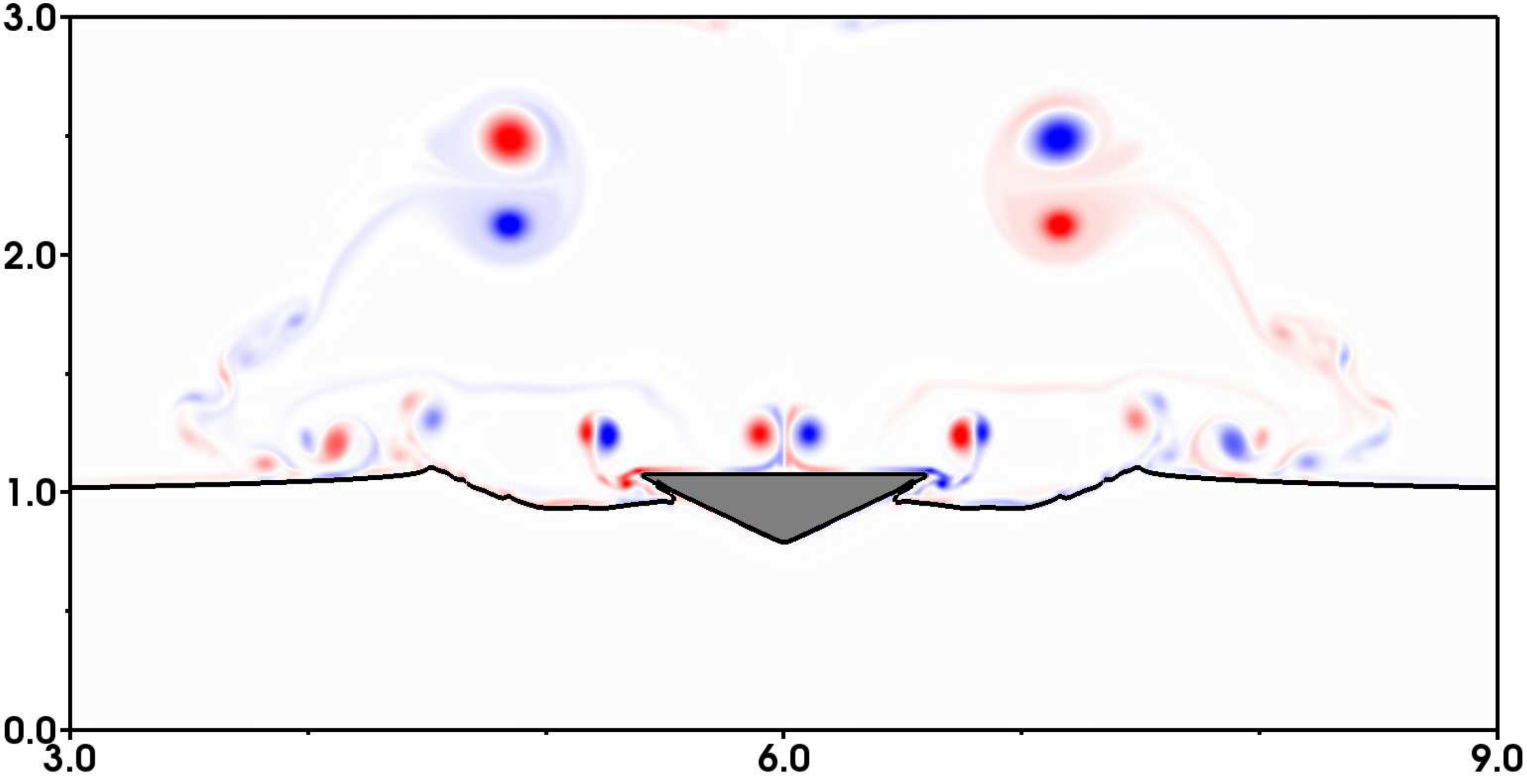}
    \label{c_wedge2d_t1p25}
  }
  \caption{Temporal evolution of a 2D wedge free-falling into an air-water interface at four different time instances.
Consistent transport of mass and momentum is used for these cases.
The plotted vorticity is in the range $-300$ to $300$.
}
  \label{fig_c_wedge2d_viz}
\end{figure}

For this 2D case, the domain is discretized by a $4N \times N$ grid with $N = 300$, and a constant
time step size of $\dt = 3/(160 N)$ is used. We again demonstrate the importance of consistent mass
and momentum transport by comparing inconsistent and consistent formulations. Fig.~\ref{fig_nc_wedge2d_viz}
shows the evolution of the wedge and the air-water interface when simulated using the non-conservative
momentum integrator. The simulation becomes unstable as the wedge
impacts the water, leading to the generation of unphysical interfaces and high vorticity regions.
Lowering the time step even further did not resolve these stability problems.
However, the simulation remains stable when using consistent mass and momentum transport, achieved by the
conservative momentum integrator. Fig.~\ref{fig_c_wedge2d_viz} shows physical interface
deformation and reasonable vorticity generation from the vertices of the wedge.

We now turn our attention to the free-fall and water impact of a 3D wedge.
The three-dimensional wedge geometry has a square top of size $L \times L$ with $L = 1.2$
and is placed within a computational domain of size $\Omega = [0, 10L]  \times [0, 5L/3] \times [0, 5L/2]$.
The wedge is placed such that there is a gap of size $5L/12$ between the vertical sides of the wedge and
the lateral walls of the computational domain. The rest of the simulation parameters are analogous to those of
the previous 2D case. Consistent mass and momentum transport is used.

In contrast with the two-dimensional case, we make heavy use of adaptive mesh refinement to selectively
resolve regions of interest. The domain is discretized by $\ell = 2$ grid levels with refinement ratio $\nref = 2$.
The grid spacing at the coarsest level is $\dx_0 = \dy_0 = \dz_0 = 1/25$, yielding a finest grid spacing
of $\dx_\textrm{min} = \dy_\textrm{min} = \dz_\textrm{min} = 1/50$.
A constant time step size of $\dt = \dx_\textrm{min}/160$ is used.
This 3D case has been studied both experimentally by Yettou et al.~\cite{Yettou2006}, and numerically
by Calderer et al.~\cite{Calderer2014} and Pathak and Raessi~\cite{Pathak16}.

\begin{figure}[]
\centering
\subfigure[Mesh refinement at $t = 0.0$]{
    \includegraphics[scale = 0.3]{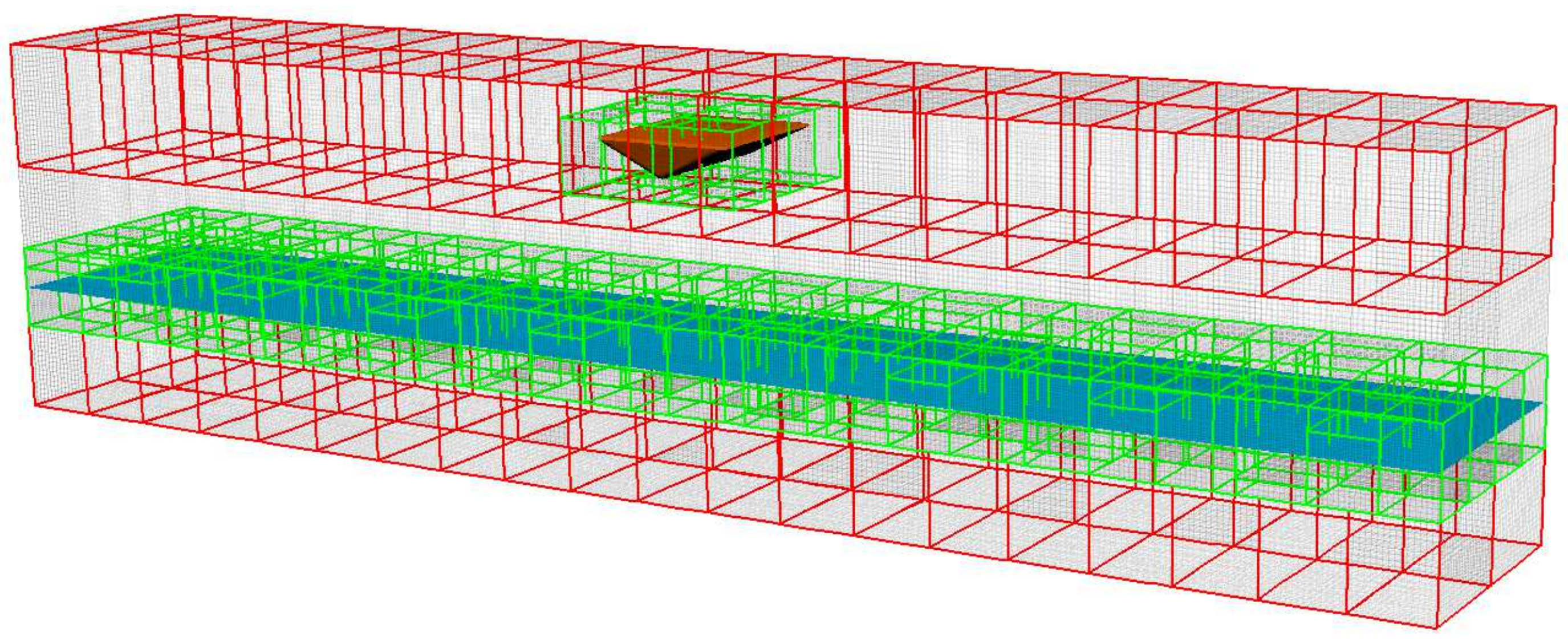}
    \label{wedge3d_t0_amr}
  }
      \subfigure[Mesh refinement at $t = 1.5$]{
    \includegraphics[scale = 0.3]{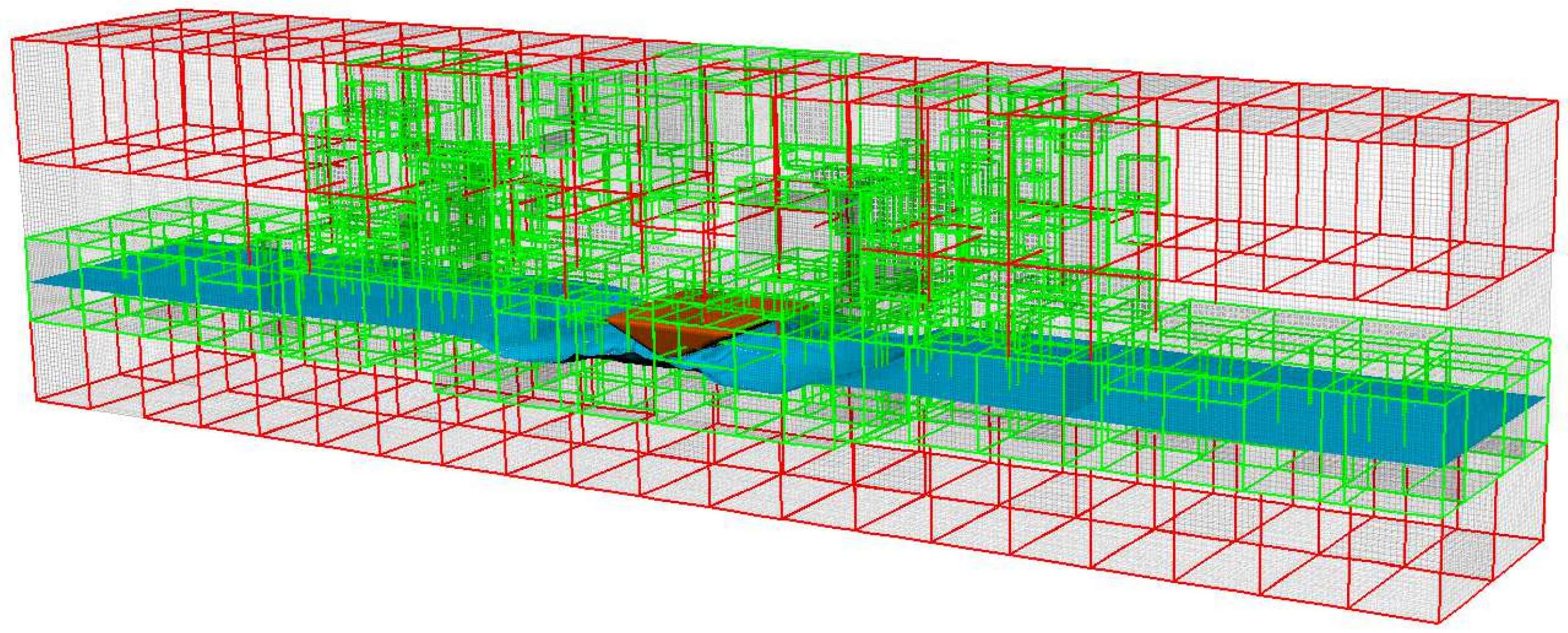}
    \label{wedge3d_t15_amr}
  }
  \caption{Locations of the different refined mesh levels from coarsest to finest for the 3D free-falling wedge simulation. The coarse mesh is outlined by red boxes while the fine mesh is outlined by green boxes.
   }
  \label{fig_wedge3d_amr_viz}
\end{figure}

\begin{figure}[]
\centering
\subfigure[$t = 0.0$]{
    \includegraphics[scale = 0.45]{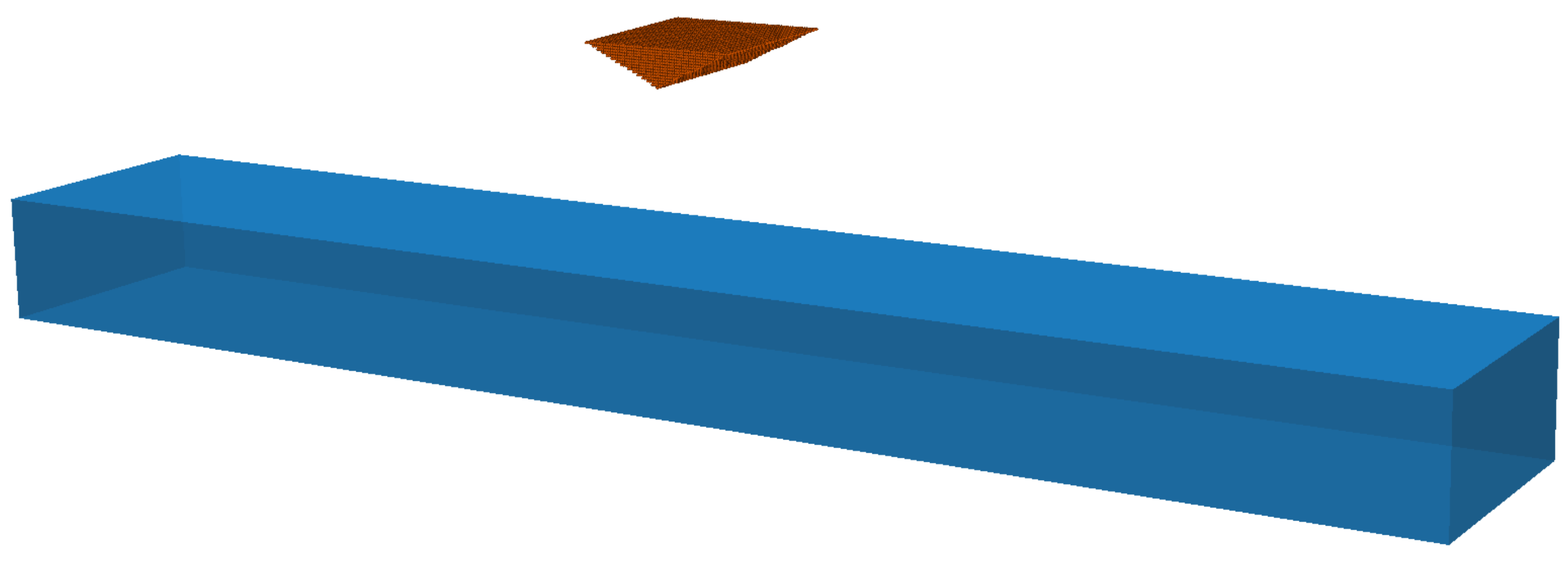}
    \label{wedge3d_t0}
  }
\subfigure[$t = 0.5625$]{
    \includegraphics[scale = 0.45]{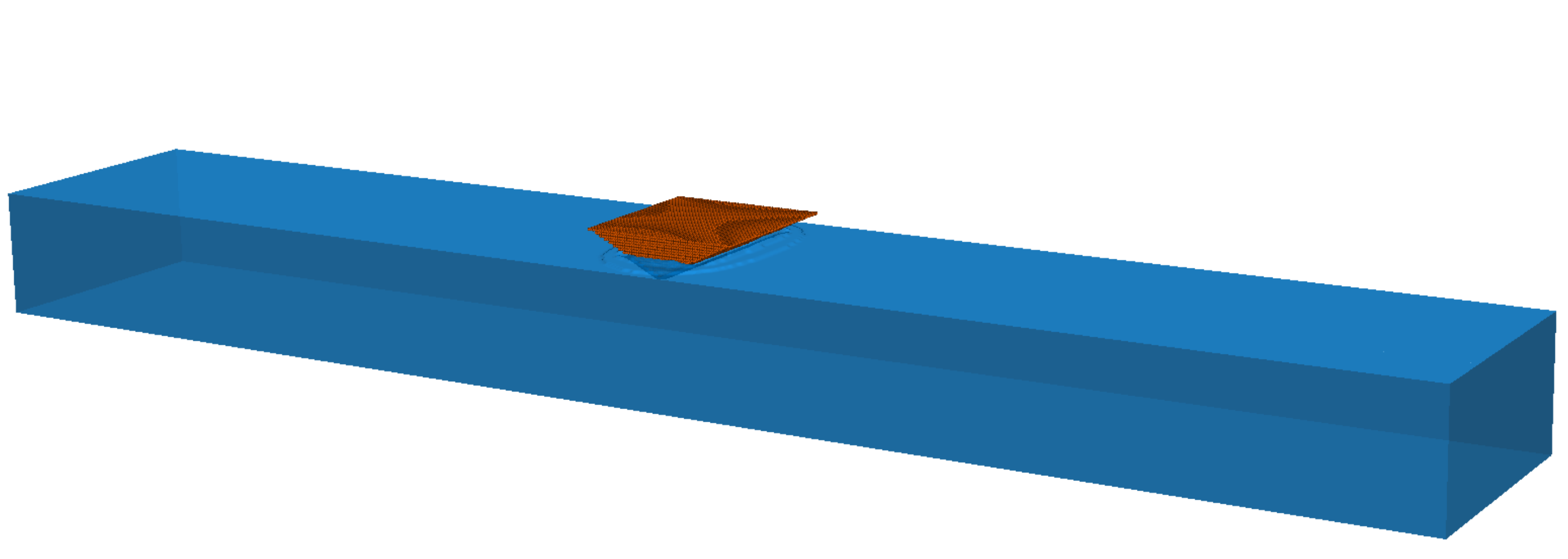}
    \label{wedge3d_t5625}
  }
  \subfigure[$t = 1.0$]{
    \includegraphics[scale = 0.45]{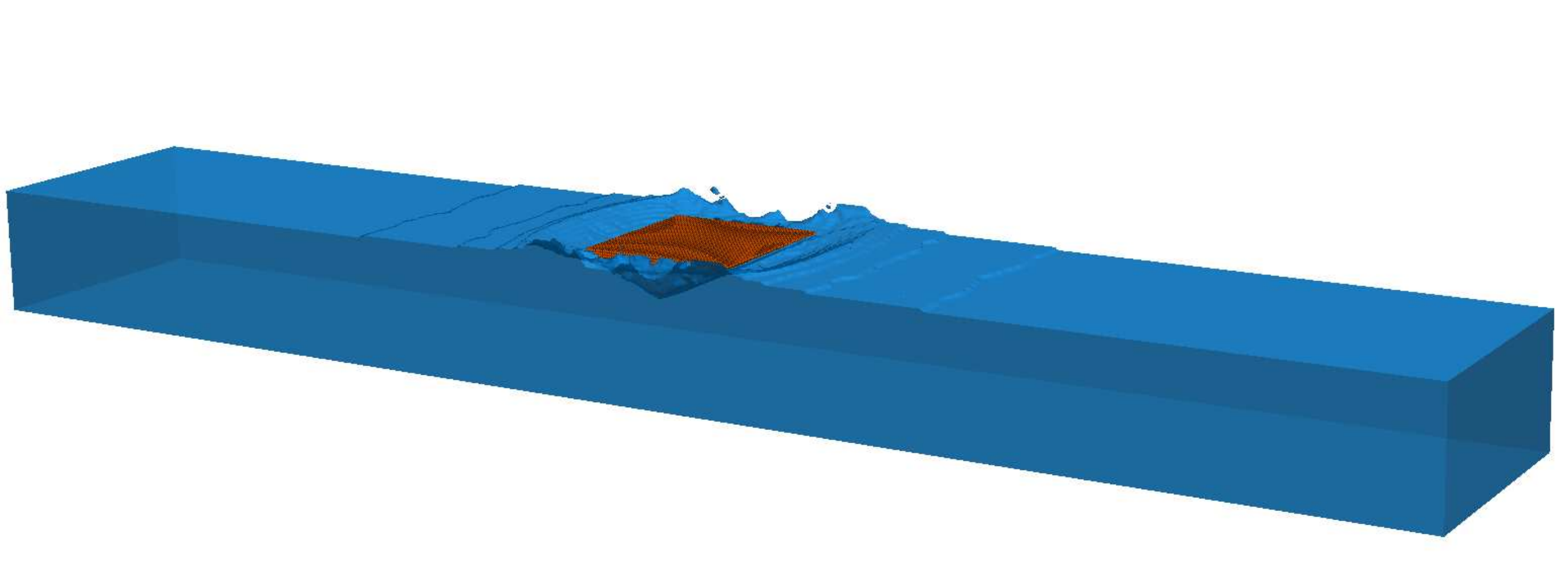}
    \label{wedge3d_t1}
  }
    \subfigure[$t = 1.5$]{
    \includegraphics[scale = 0.45]{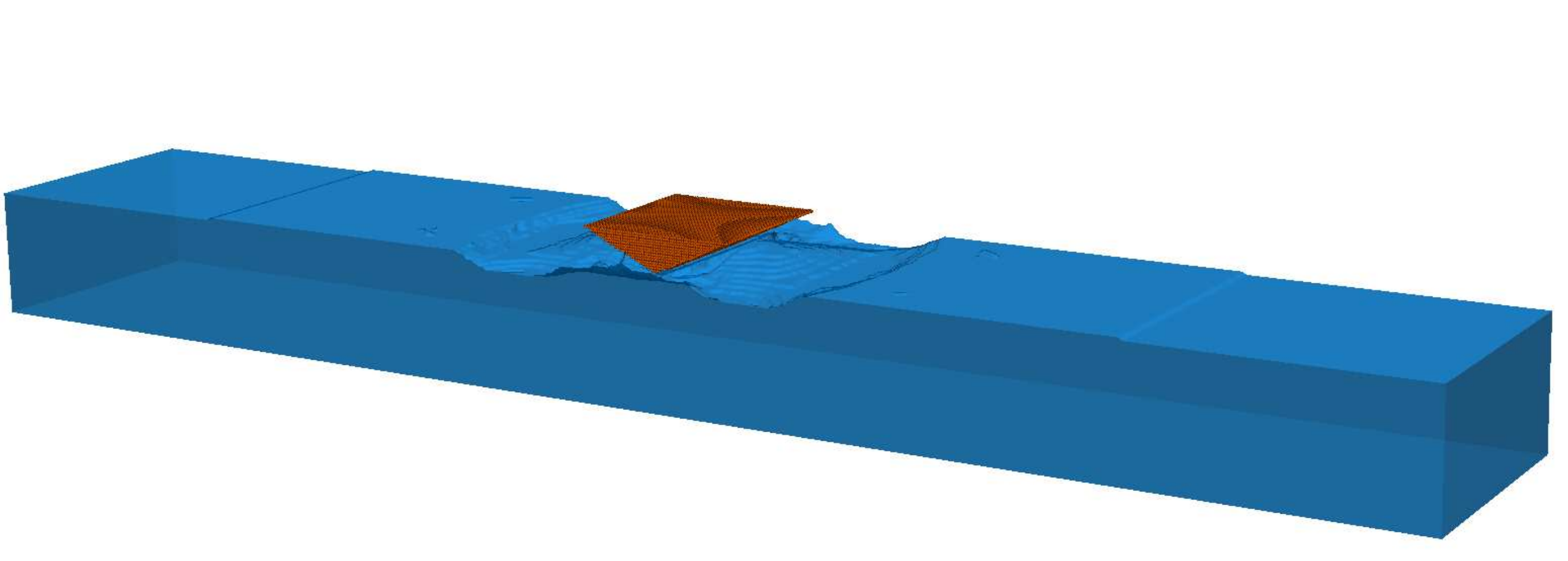}
    \label{wedge3d_t15}
  }
  \caption{Temporal evolution of a 3D wedge free-falling into an air-water interface at four different time instances.
}
  \label{fig_wedge3d_viz}
\end{figure}

\begin{figure}[]
  \centering
    \subfigure[Vertical position]{
    \includegraphics[scale = 0.25]{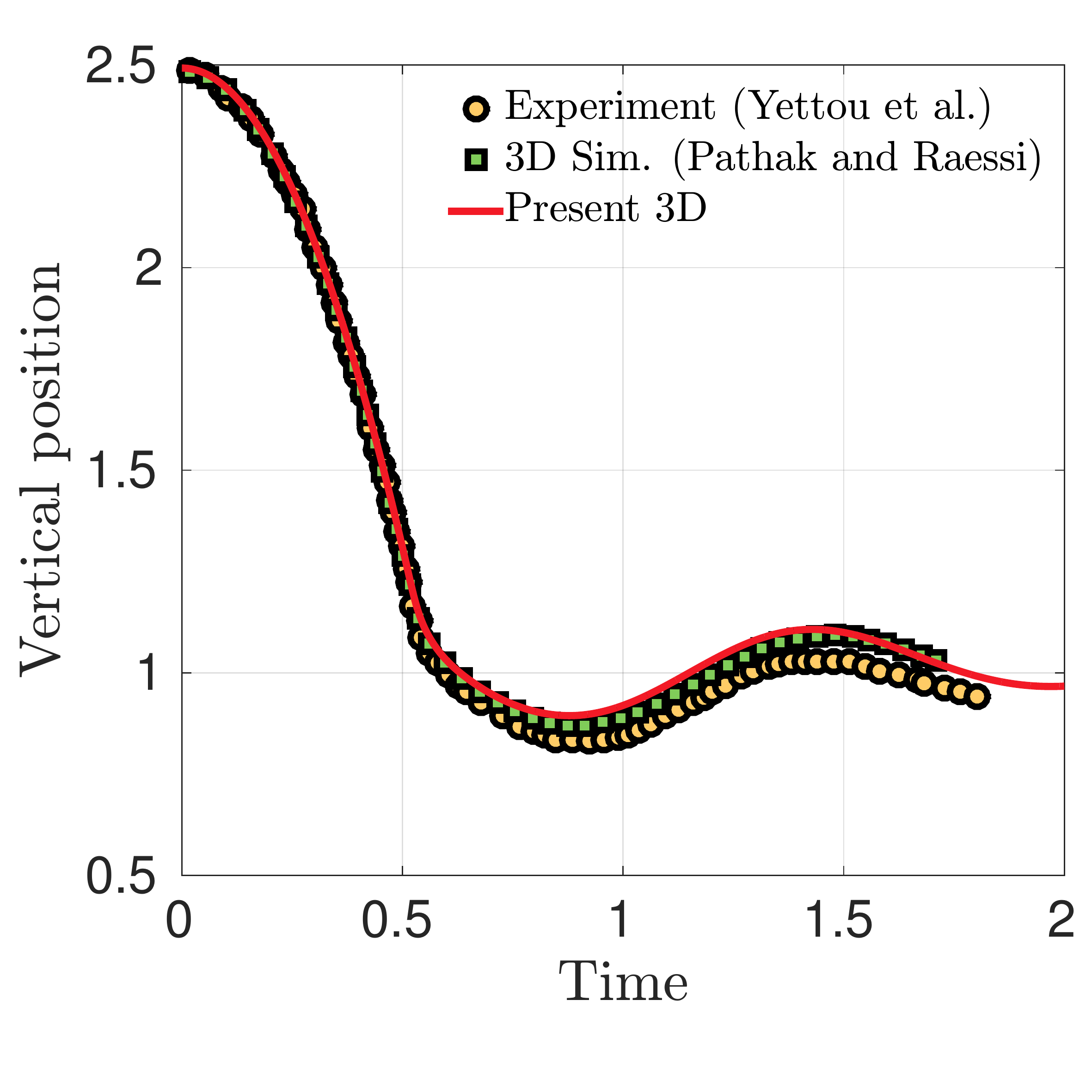}
    \label{wedge_position}
  }
  \subfigure[Vertical velocity]{
    \includegraphics[scale = 0.25]{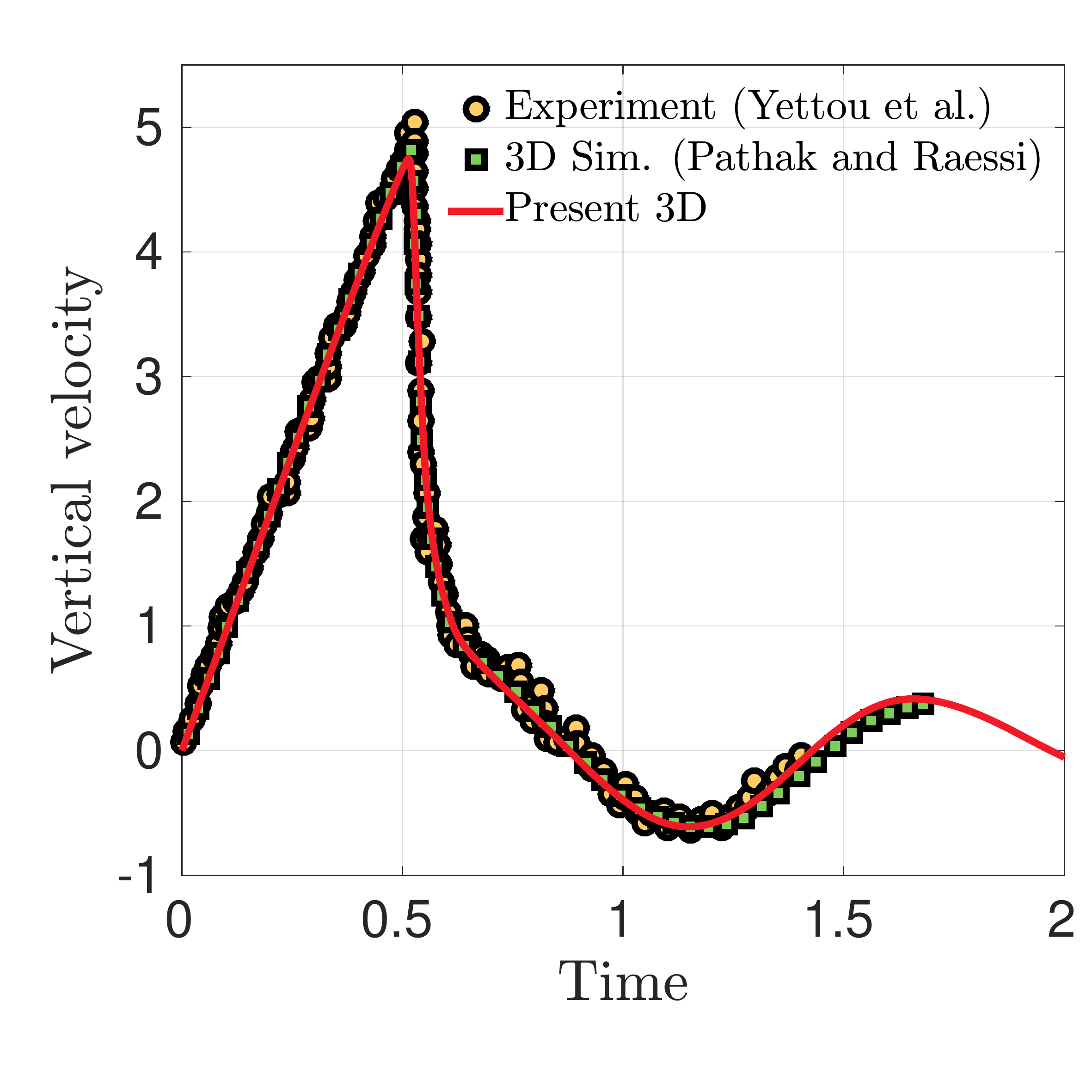}
    \label{wedge_velocity}
  }
   \caption{
 Temporal evolution of \subref{wedge_position} vertical position and \subref{wedge_velocity} vertical velocity for
  a 3D wedge free-falling into an air-water interface.
  ($\bullet$, yellow) experimental data from Yettou et al.~\cite{Yettou2006};
  ($\blacksquare$, green) simulation data from Pathak and Raessi~\cite{Pathak16};
  (---, red) present simulation data.
  }
  \label{fig_wedge}
\end{figure}

Fig~\ref{fig_wedge3d_amr_viz} shows the locations of the different mesh levels at two snapshots in time.
At the initial time, (Fig.~\ref{wedge3d_t0_amr}) the region surrounding the air-solid and air-water
interfaces are covered by the finer mesh. As the wedge impacts the water (Fig.~\ref{wedge3d_t15_amr}) additional
finer mesh regions are generated to resolve the vorticity generated in the air phase. Fig.~\ref{fig_wedge3d_viz}
shows the evolution of the wedge at various instances in time. Upon impact, the wedge generates complex splashing
and ripple phenomena along the water as it bobs up and down. To quantitatively assess the accuracy of the wave-structure
interaction, the vertical center of mass position and velocity of the wedge are plotted against time in Fig.~\ref{fig_wedge}.
The results are in excellent agreement with the 3D simulation carried out by Pathak and Raessi~\cite{Pathak16} and
experimental data compiled by Yettou et al.~\cite{Yettou2006}. Additionally, we note that while
Calderer et al.~\cite{Calderer2014} required $\ncells = 8$ smeared transition cells to obtain a stable simulation,
we only use $\ncells = 2$. In fact in our formulation $\ncells$ do not affect solver stability 
at all --- only solver convergence properties are affected by this parameter. 
We obtain sharp flow structures and interfacial dynamics using $\ncells = 1$ or $2$; larger smearing leads to diffuse interfaces and  
vortex dynamics around them. We attribute our simulation robustness to consistent mass and momentum
transport~\cite{Nangia2018}, which is not being used in~\cite{Calderer2014}~\footnote{We also tried $\ncells = 8$ with the non-conservative solver, 
but were unable to circumvent numerical instabilities.}.
Finally, we remark that this case demonstrates the robustness and computational efficiency of our FSI coupling approach. We did not need to consider any special numerical treatment to ensure stability for this particular example --- the solution methodology used for this case is identical to the one used for \emph{all} the cases considered in the present work. This is in contrast to the approach described by Calderer et al., in which a special Aitken acceleration technique was employed to reduce the FSI iteration count to 4-5 in order to achieve a strong FSI coupling and stable solutions 
for this problem~\cite{Calderer2014}.
This case is representative of many real world WSI
applications involving the interaction between heavy, buoyant objects interacting with air-water interfaces. The present
numerical method can be robustly applied to these types of problems.

\subsection{3D water column impacting a stationary obstacle}
\begin{figure}[]
  \centering
  \subfigure[Initial problem set up]{
    \includegraphics[scale = 0.4]{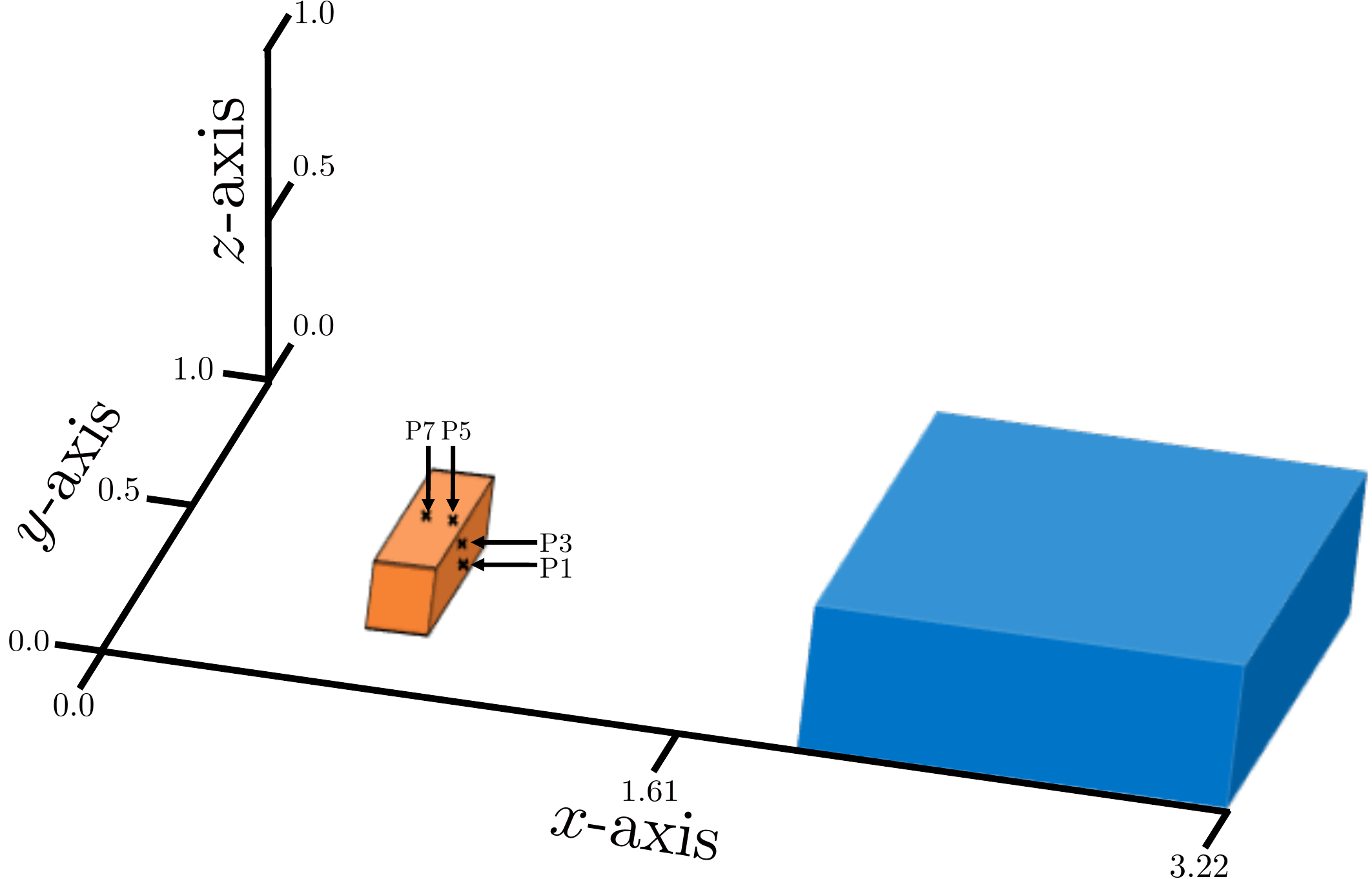}
    \label{dam_break_3d_view}
  }
   \subfigure[Top view]{
    \includegraphics[scale = 0.4]{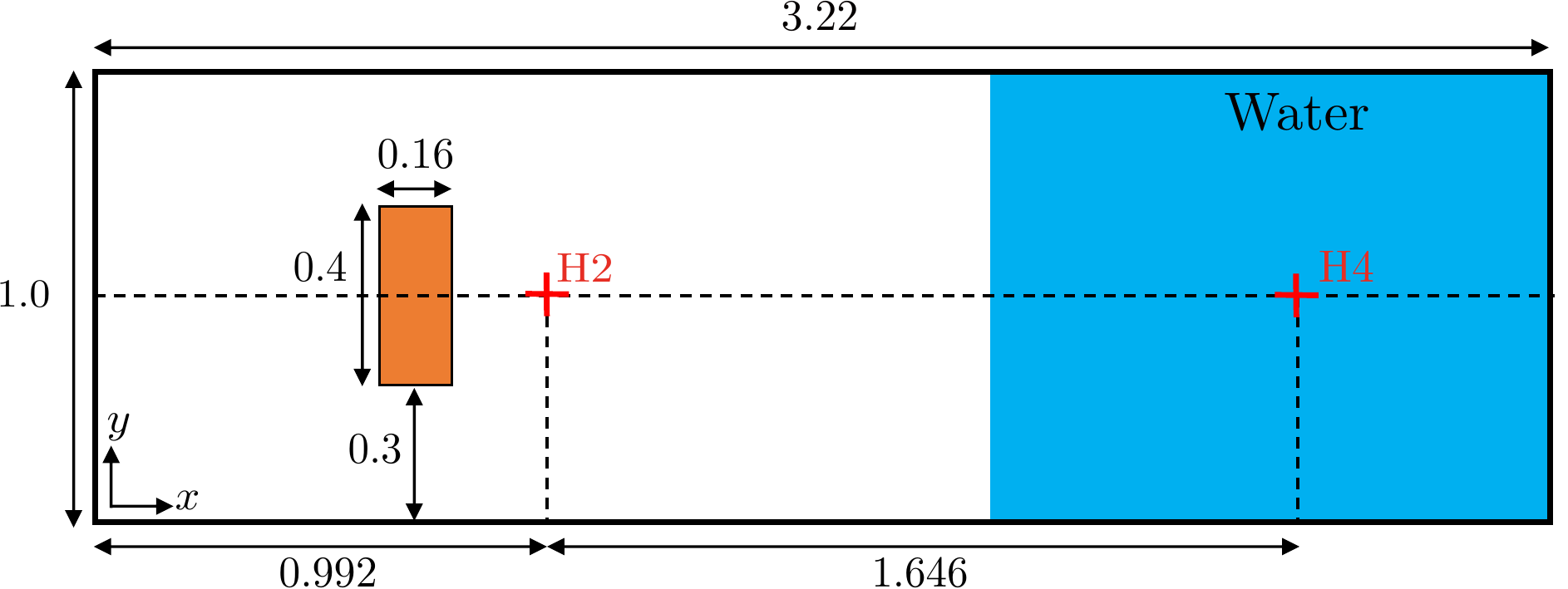}
    \label{dam_break_top_view}
  }
  \subfigure[Side view]{
    \includegraphics[scale = 0.4]{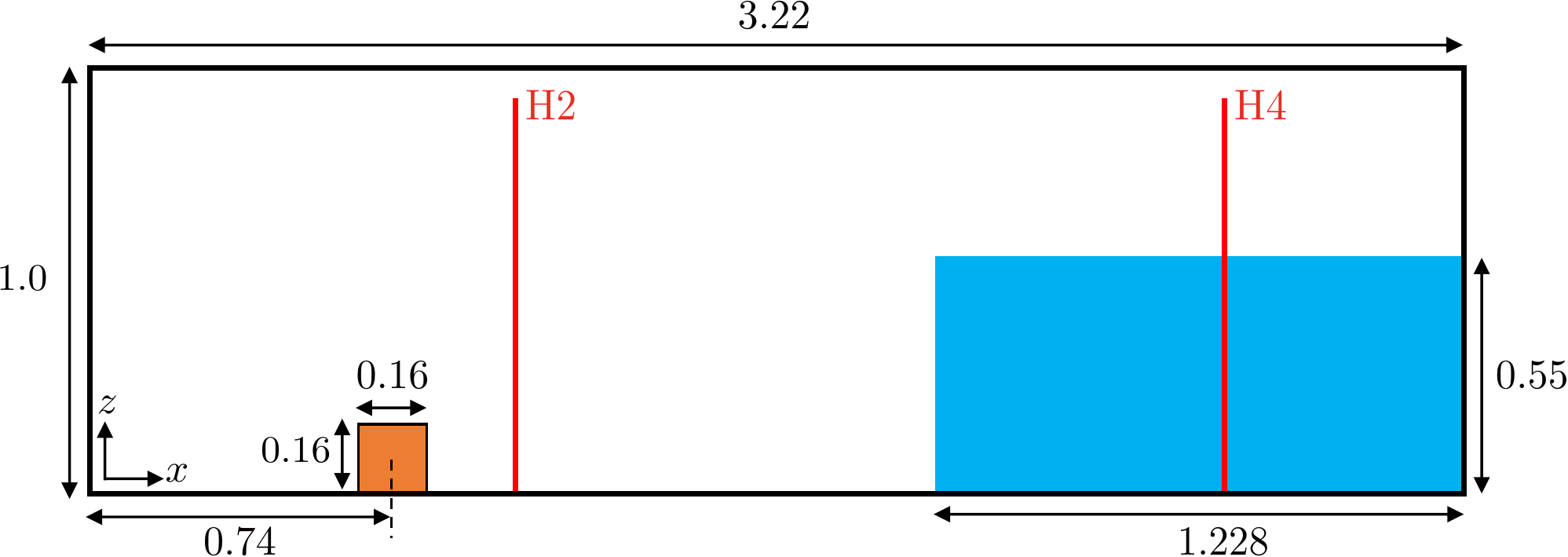}
    \label{dam_break_side_view}
  }
  \caption{Sketches of the initial problem set up for a water column impacting a stationary rectangular object;
  \subref{dam_break_3d_view} 3D view specifying the size of the domain and the locations of the pressure
  sensors P$1$, P$3$, P$5$, and P$7$ ($\times$, black) on the surface of the body;
  \subref{dam_break_top_view} top view and \subref{dam_break_side_view} side view indicating the dimensions and
  locations of the water column (blue) and the obstacle (orange), along with the locations of the water height probes
  H$2$ and H$4$.}
  \label{fig_dam_break_schematic}
\end{figure}

In this section, we investigate the problem of a water column impacting
a stationary box. A rectangular obstacle is placed in a 3D computational
domain of size $\Omega = [0, 3.22] \times [0, 1] \times [0, 1]$, which initially contains
a rectangular water column at its far end; see Fig.~\ref{fig_dam_break_schematic} for a full description
of the initial problem set up. The structure is held stationary and therefore
gravitational forces are not evaluated using the density within the solid region (see Sec.~\ref{sec_solid_materials}).
Two grid cells of smearing $\ncells = 2$ are used to transition between different material properties on
either side of the interfaces, and surface tension forces are included with coefficient equal to that of air-water:
$\sigma = 0.0728$.
No-slip boundary conditions are imposed along $\partial \Omega$.
The domain is discretized by a $161 \times 50 \times 50$ grid with constant time step size $\dt = 1\times 10^{-4}$.
This problem has been studied experimentally at the Maritime Research Institute Netherlands (MARIN),
and numerically by Kleefman et al.~\cite{Kleefsman2005} and Pathak and Raessi~\cite{Pathak16}.

\begin{figure}[]
\centering
\subfigure[$t = 0.00$]{
    \includegraphics[scale = 0.28]{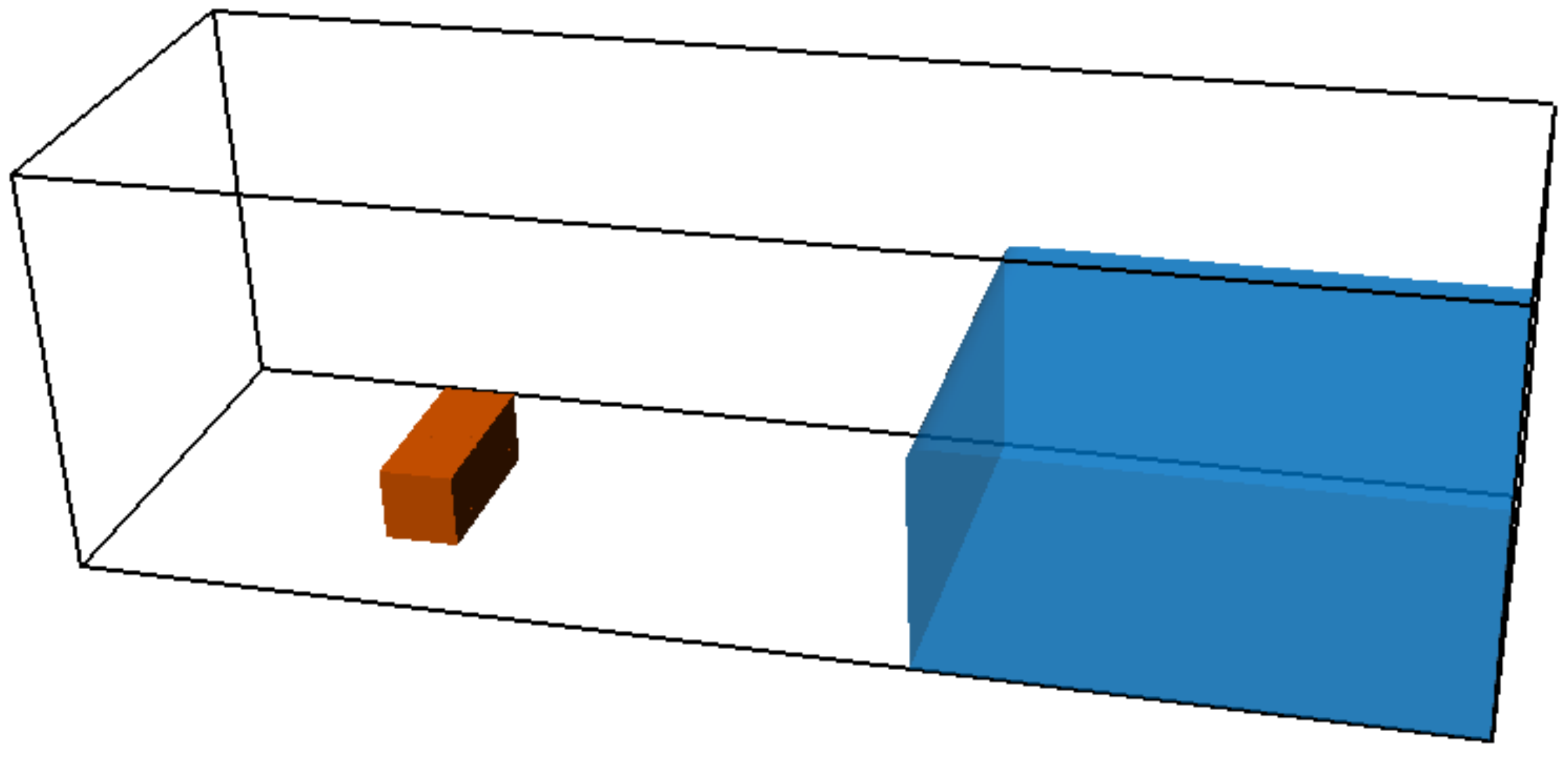}
    \label{DamBeak3D_t0p0}
  }
\subfigure[$t = 0.36$]{
    \includegraphics[scale = 0.28]{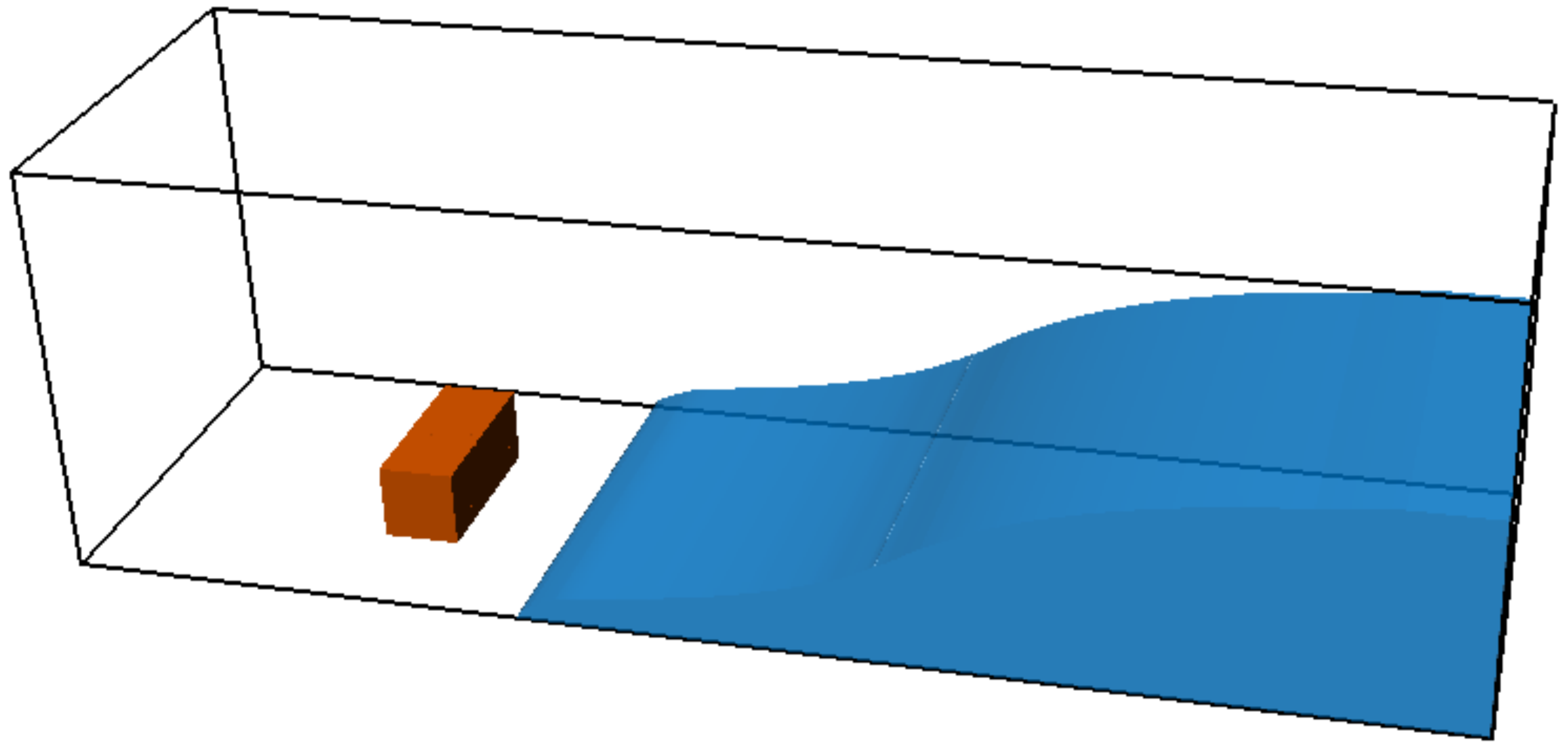}
    \label{DamBeak3D_t0p36}
  }
  \subfigure[$t = 0.68$]{
    \includegraphics[scale = 0.28]{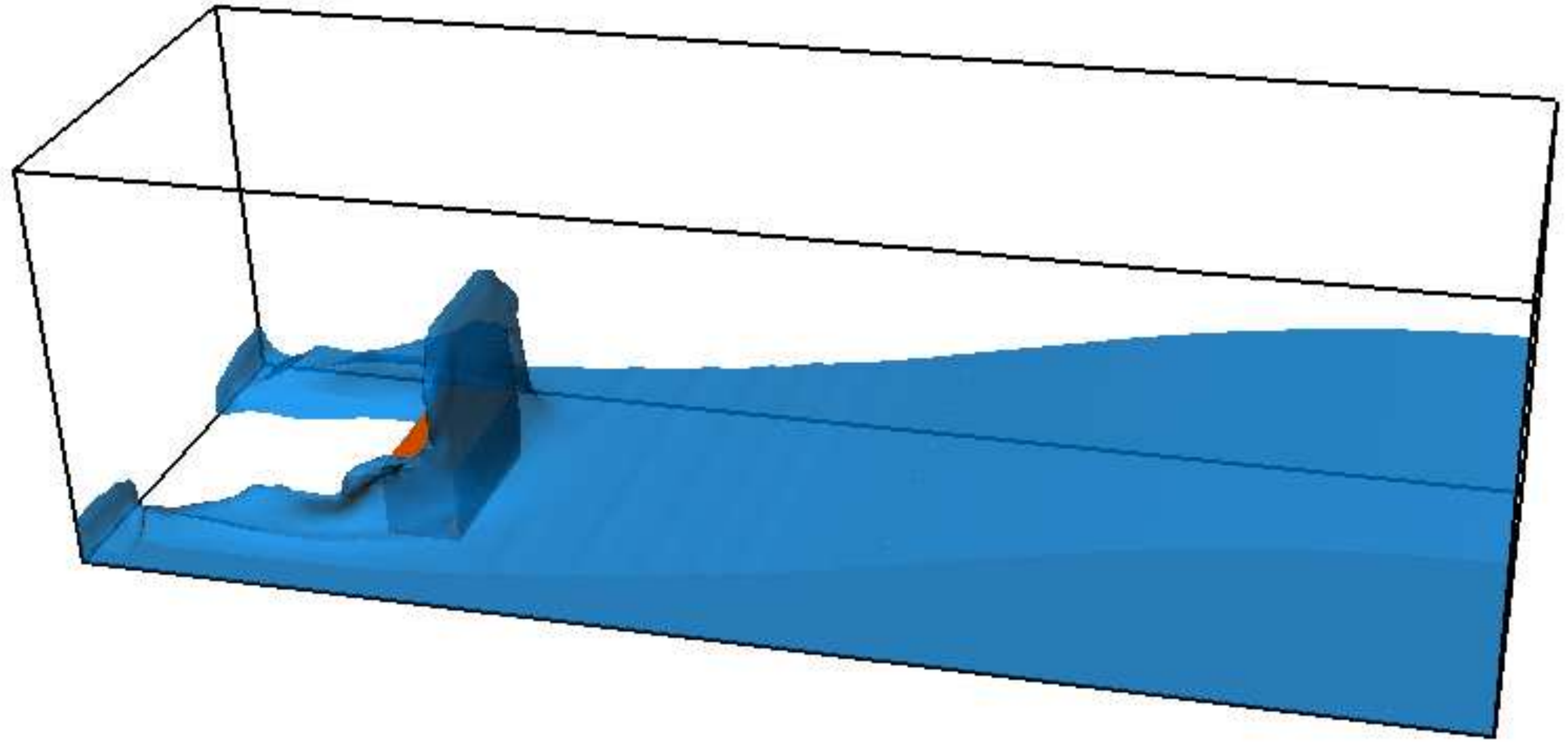}
    \label{DamBeak3D_t0p68}
  }
    \subfigure[$t = 0.96$]{
    \includegraphics[scale = 0.28]{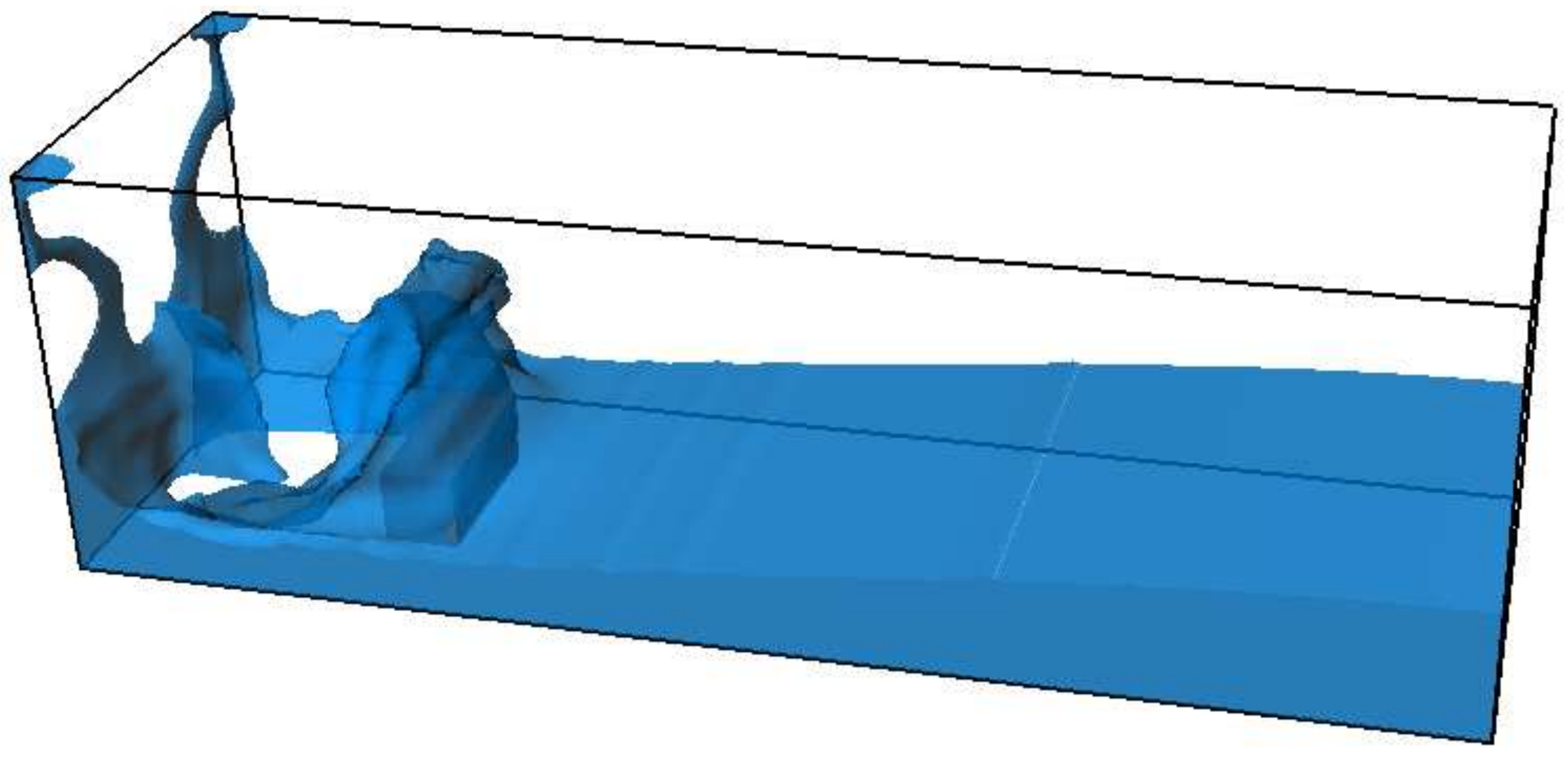}
    \label{DamBeak3D_t0p96}
  }
  \subfigure[$t = 1.28$]{
    \includegraphics[scale = 0.28]{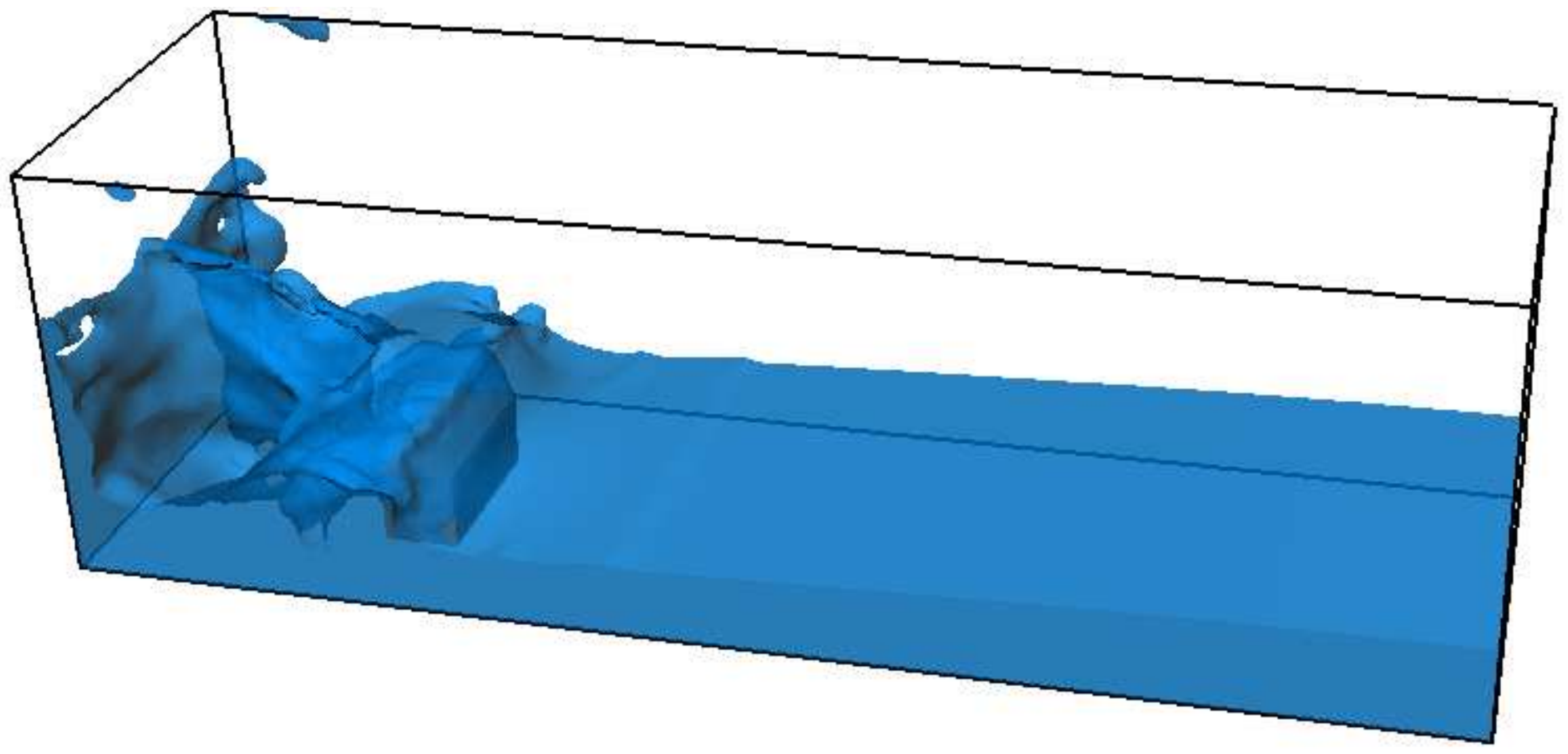}
    \label{DamBeak3D_t1p28}
  }
   \subfigure[$t = 1.43$]{
    \includegraphics[scale = 0.28]{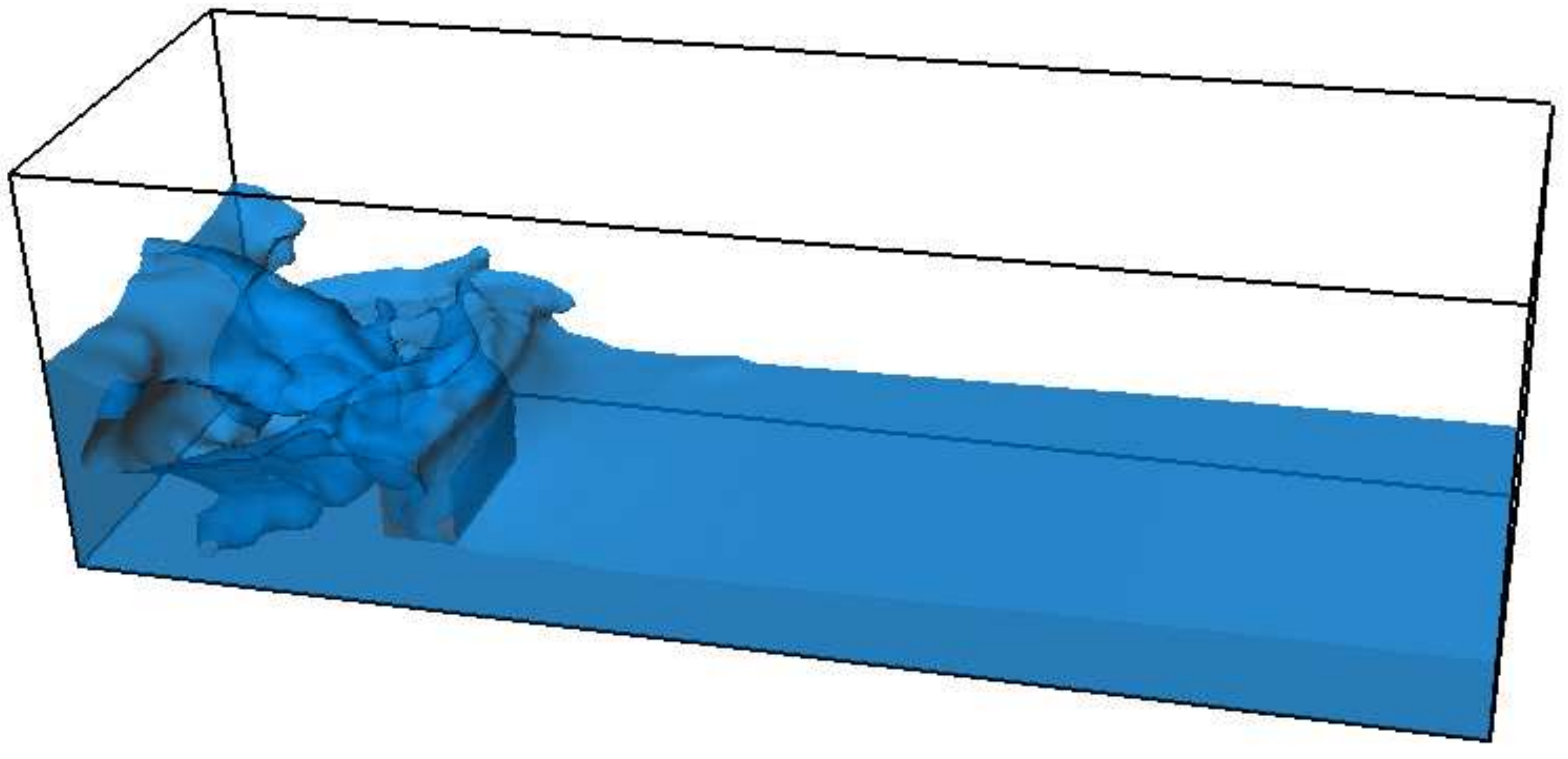}
    \label{DamBeak3D_t1p43}
  }
   \subfigure[$t = 1.72$]{
    \includegraphics[scale = 0.28]{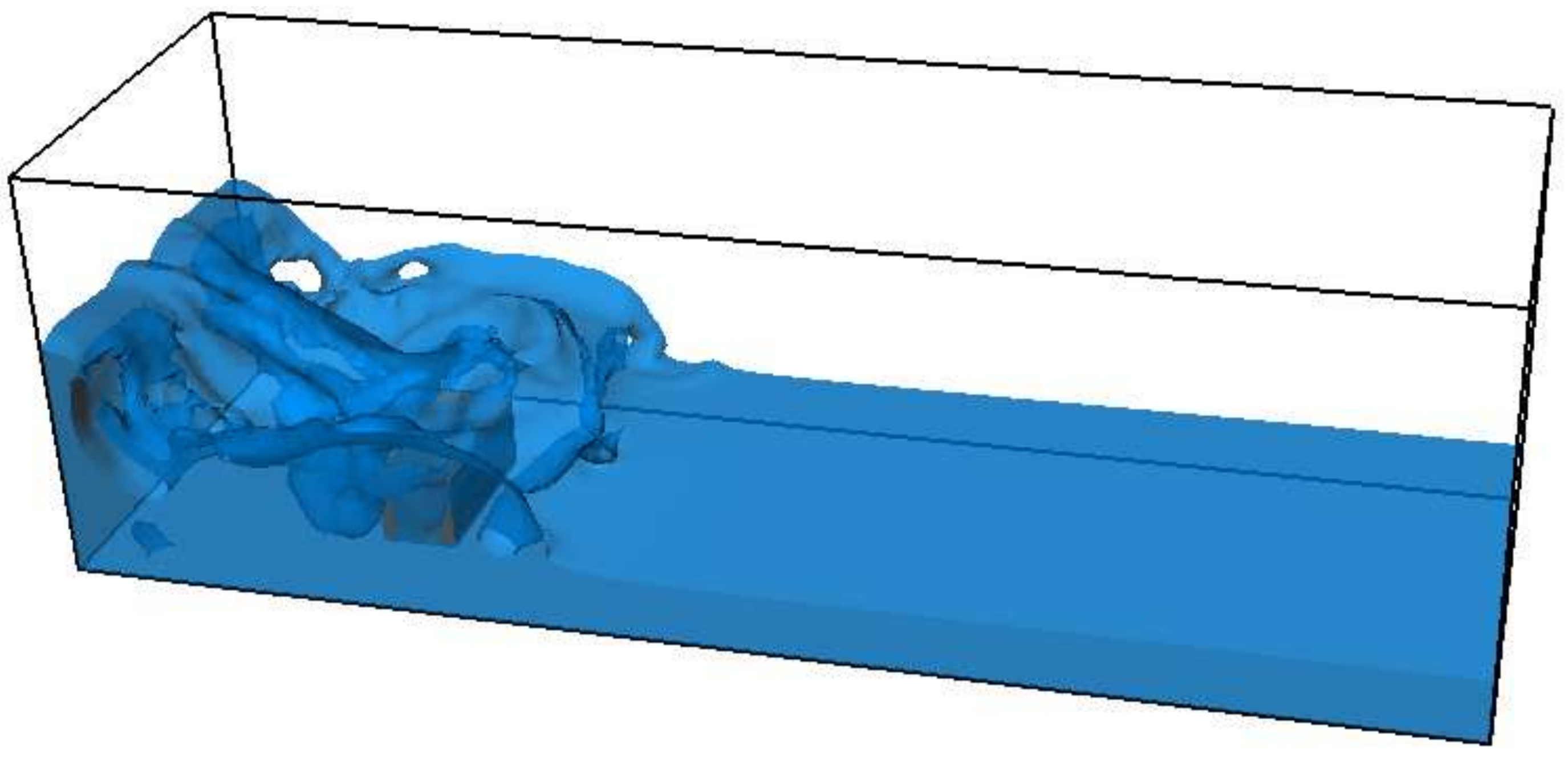}
    \label{DamBeak3D_t1p72}
  }
   \subfigure[$t = 2.98$]{
    \includegraphics[scale = 0.28]{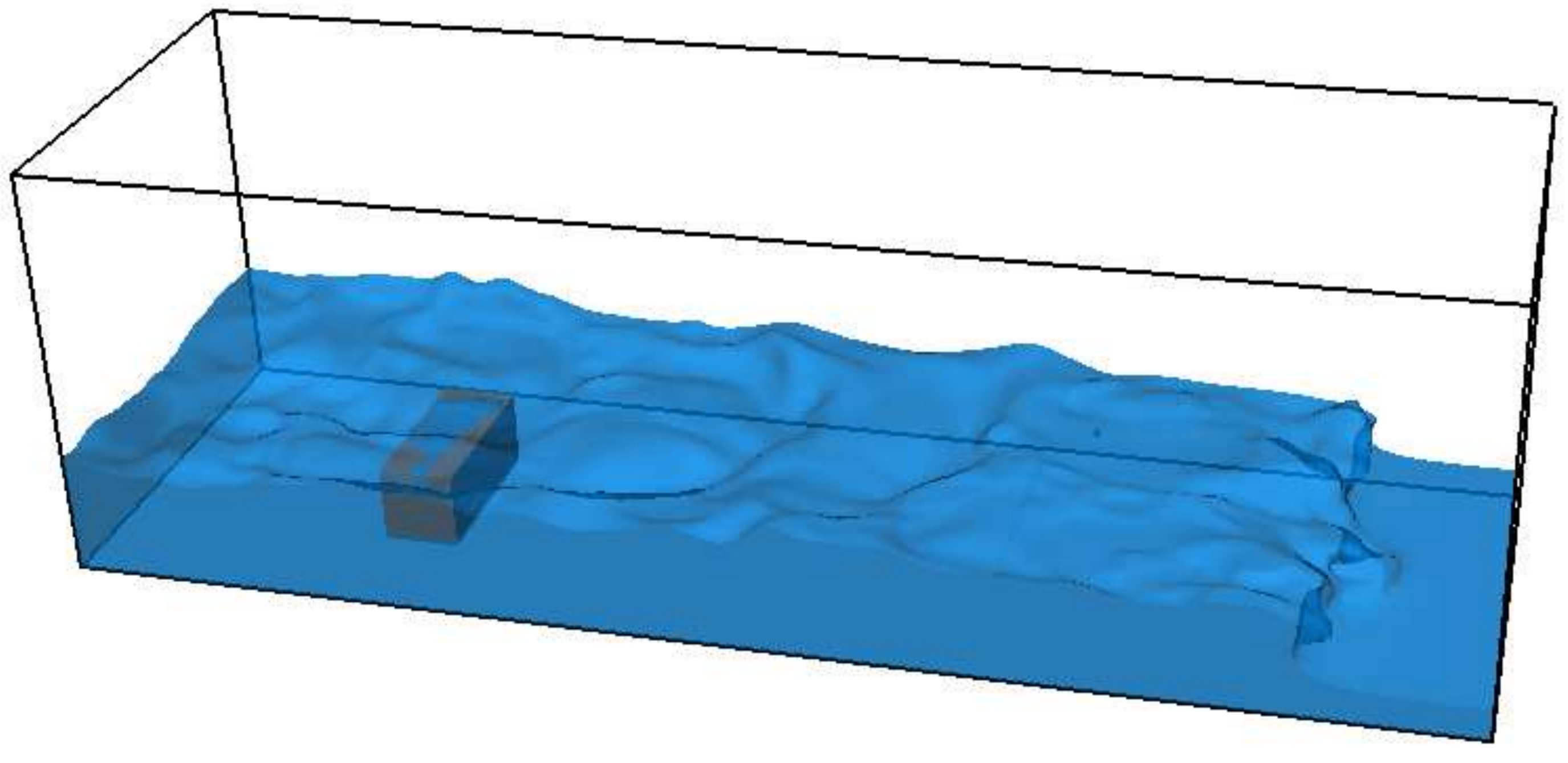}
    \label{DamBeak3D_t2p98}
  }
  
  \caption{Temporal evolution of a water column impacting a rectangular object at
  eight different time instances.
}
  \label{fig_dambreak3d_viz}
\end{figure}

Fig.~\ref{fig_dambreak3d_viz} shows the evolution of the water column
at various instances in time. Under the effects of gravity, the column spreads across the
domain's lower boundary, eventually crashing into the rectangular structure. The structure obstructs
the fluid flow, causing the water to divert upwards and over the body. Small ripples are also generated in
the bulk flow. The water eventually hits the left side of the computational domain, causing a flow reversal
and secondary crash over the body.

\begin{figure}[]
  \centering
    \subfigure[P$1$ probe]{
    \includegraphics[scale = 0.25]{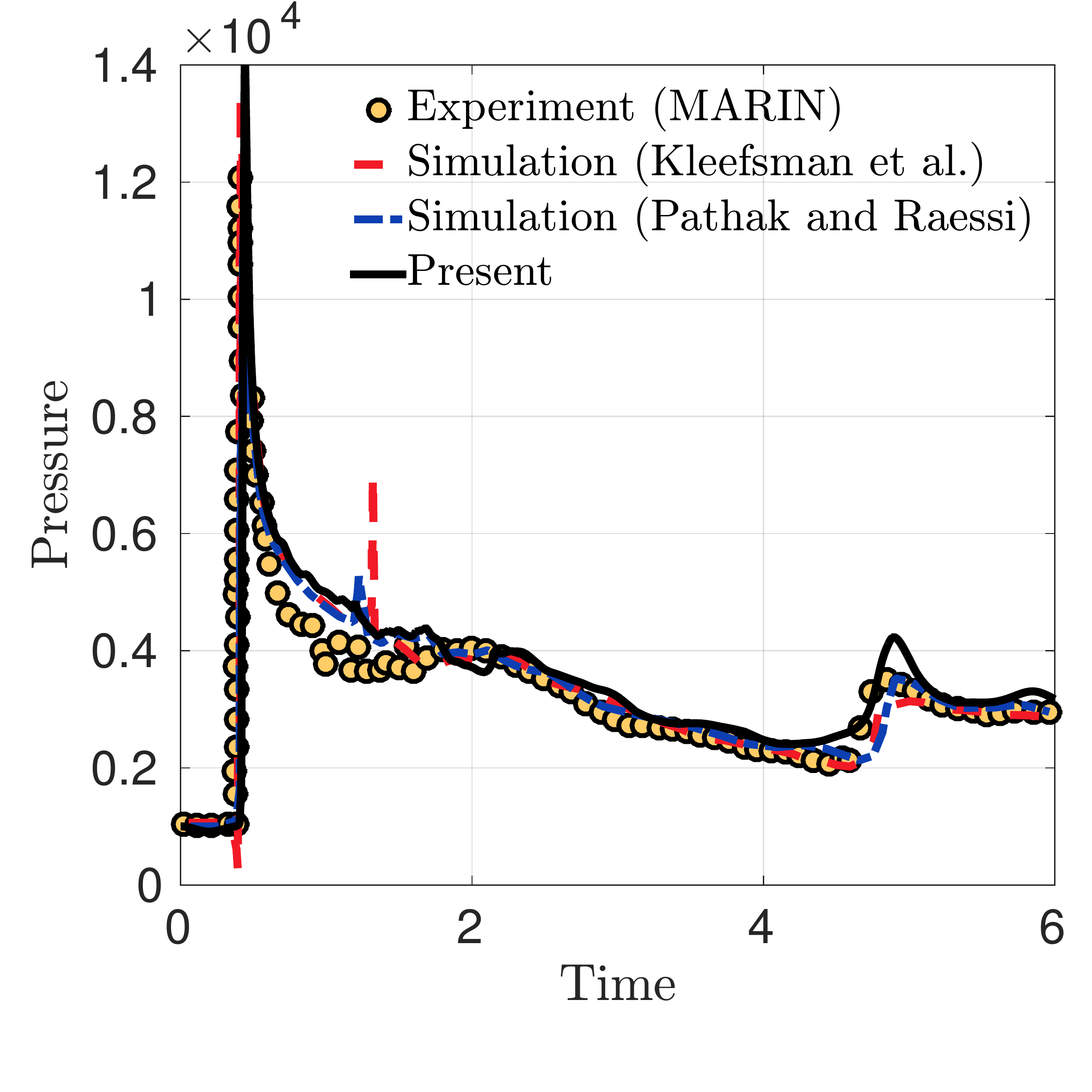}
    \label{p1_sensor}
  }
  \subfigure[P$3$ probe]{
    \includegraphics[scale = 0.25]{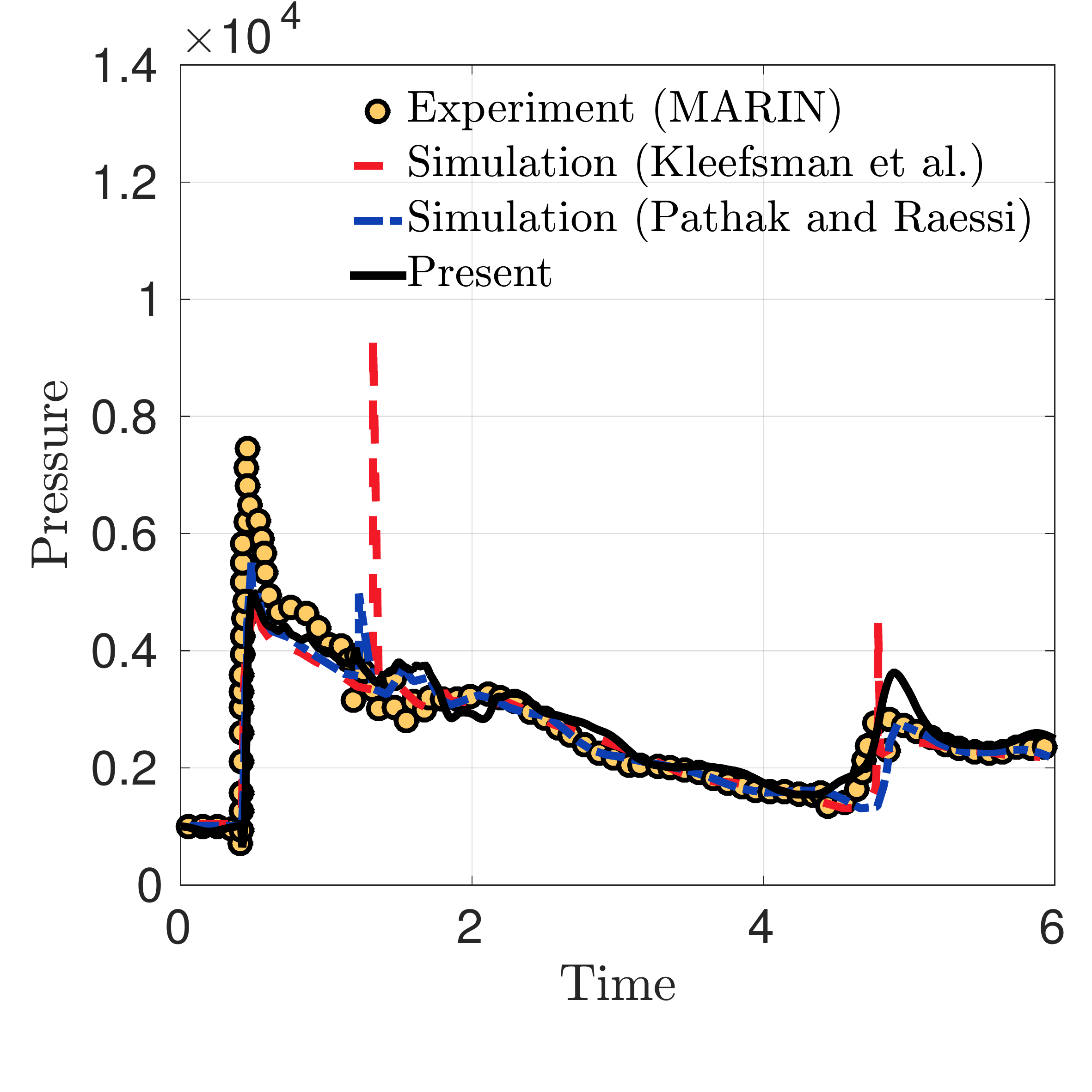}
    \label{p3_sensor}
  }
  \subfigure[P$5$ probe]{
    \includegraphics[scale = 0.25]{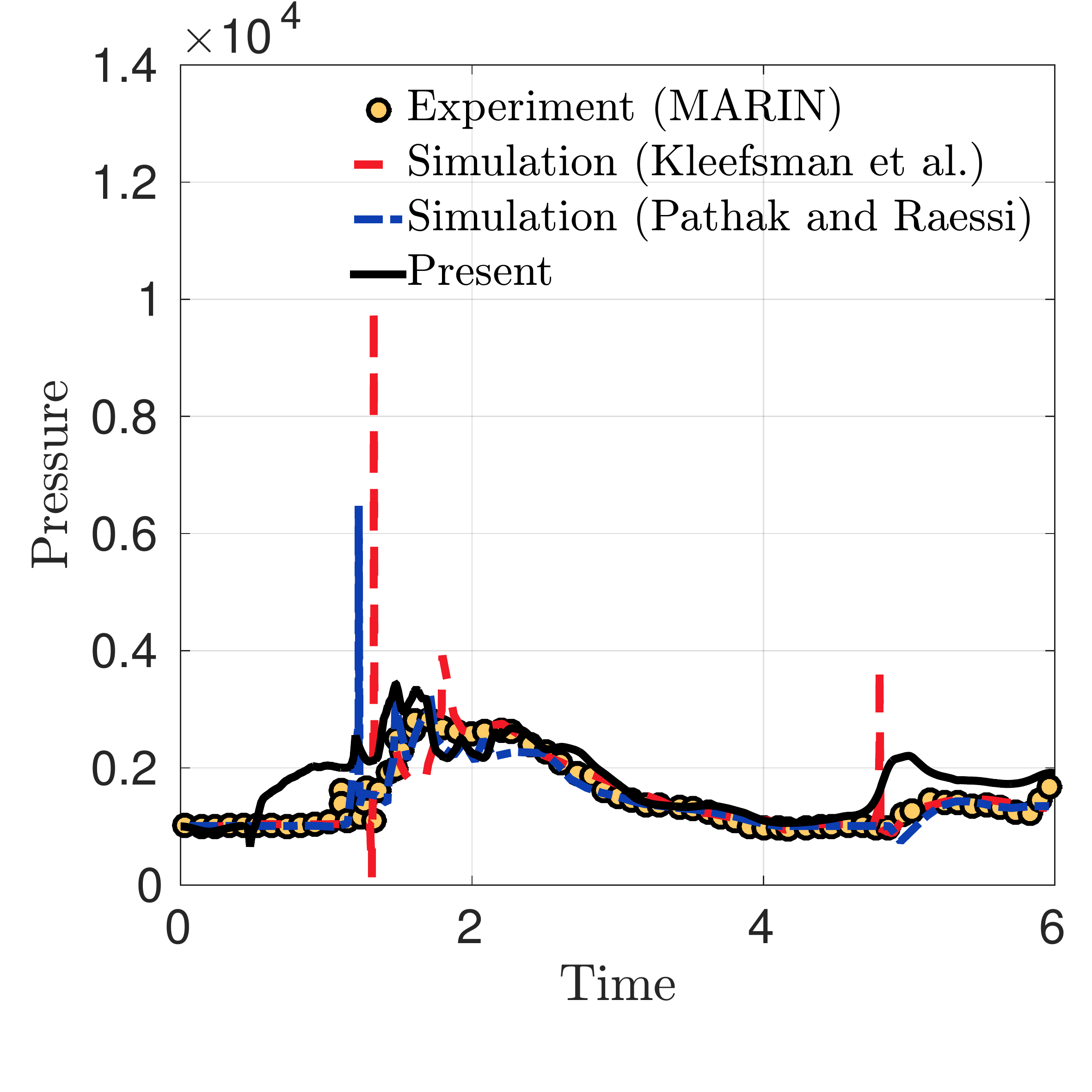}
    \label{p5_sensor}
  }
  \subfigure[P$7$ probe]{
    \includegraphics[scale = 0.25]{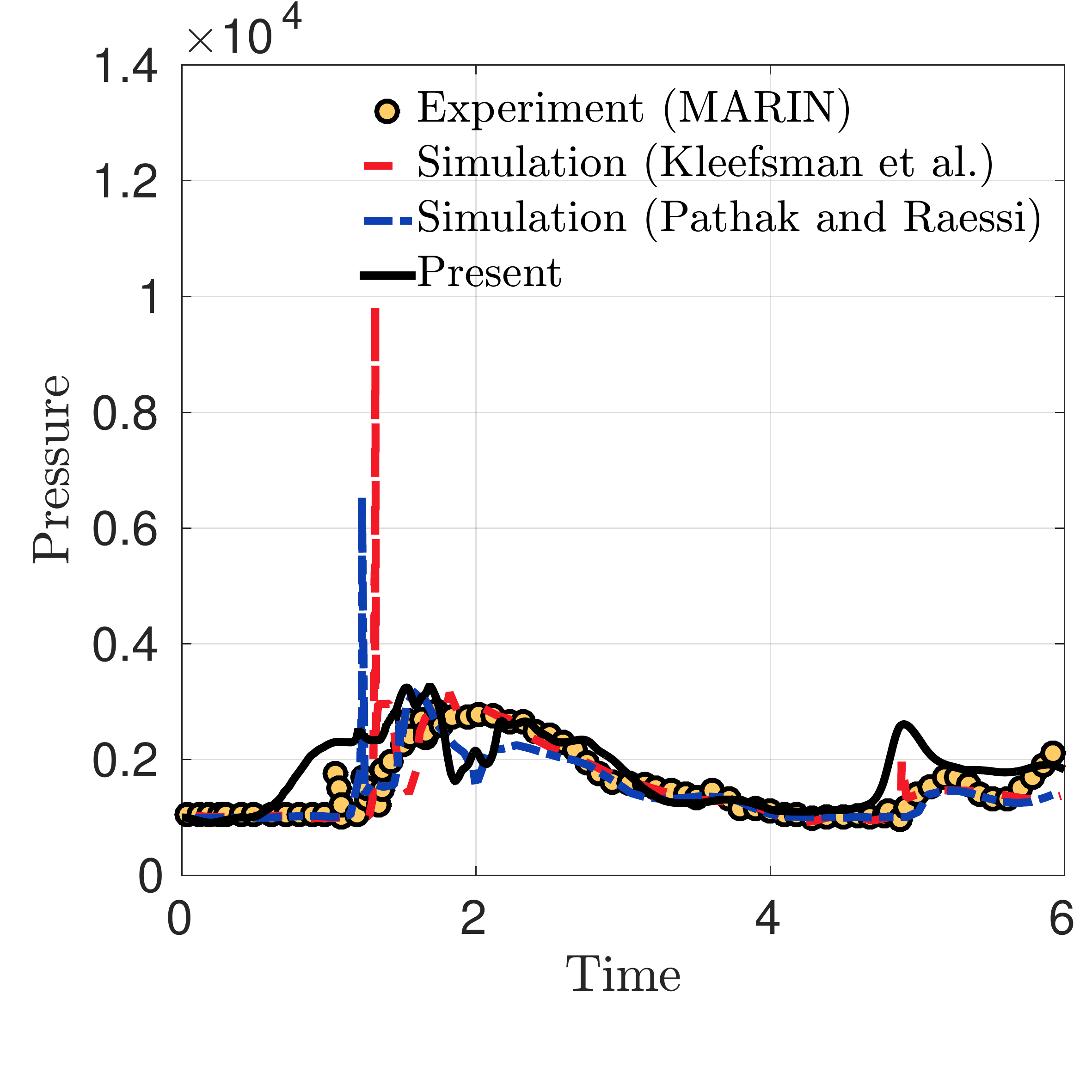}
    \label{p7_sensor}
  }
 \caption{Temporal evolution of pressure measured at probes
 \subref{p1_sensor} P$1$, 
 \subref{p3_sensor} P$3$,
 \subref{p5_sensor} P$5$, and
 \subref{p7_sensor} P$7$,
 for a 3D water column impacting a stationary obstacle
 (see Fig.~\ref{fig_dam_break_schematic} and Table~\ref{tab_probes});
 ($\bullet$, yellow) experimental data from MARIN;
(\texttt{---}, red)  simulation data from Kleefsman et al.~\cite{Kleefsman2005};
 (\texttt{-}$\cdot$\texttt{-}, blue) simulation data from Pathak and Raessi~\cite{Pathak16};
(---, black) present simulation data.
 }
  \label{fig_dam_break_pressure}
\end{figure}

\begin{figure}[]
  \centering
    \subfigure[H$2$ probe]{
    \includegraphics[scale = 0.25]{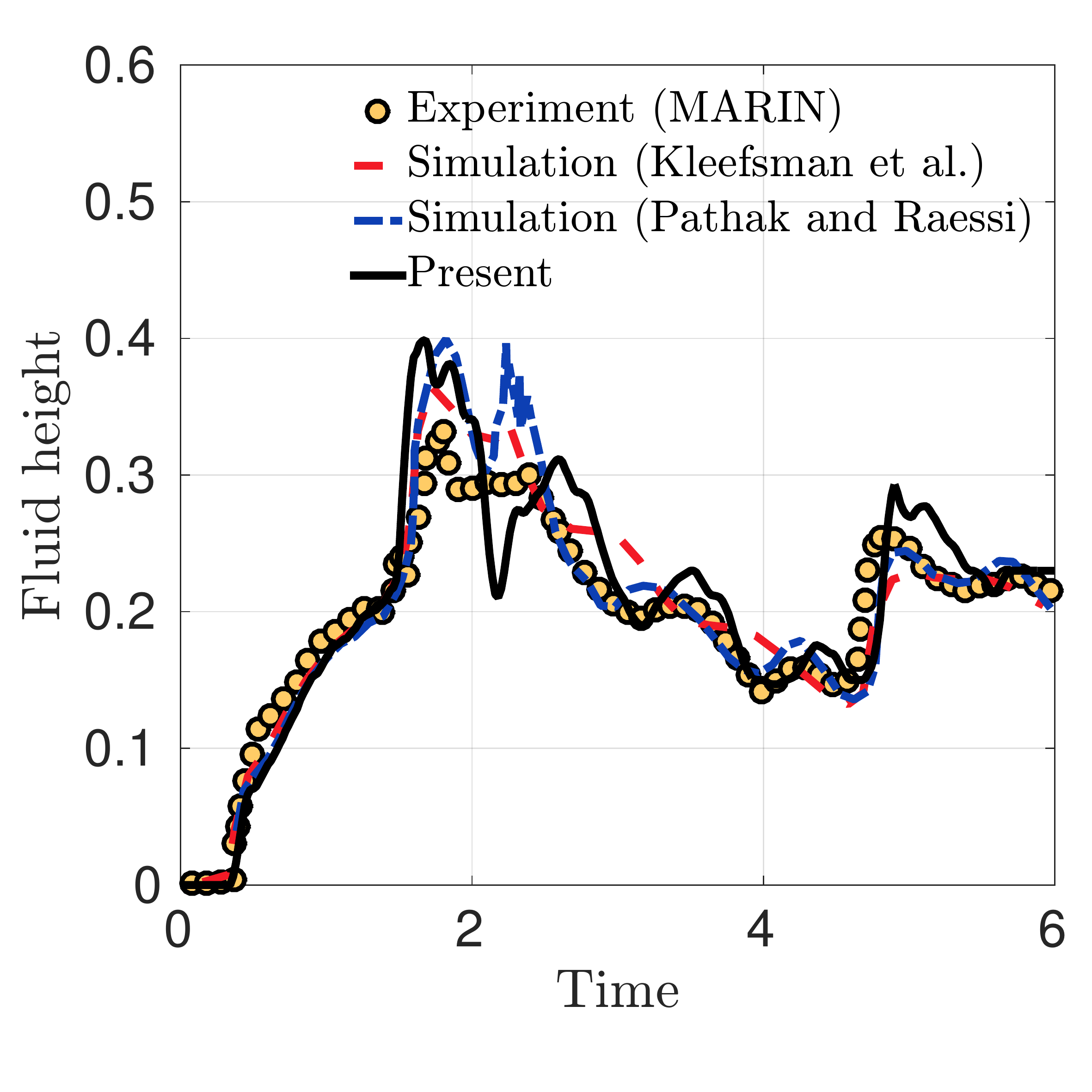}
    \label{h2_sensor}
  }
  \subfigure[H$4$ probe]{
    \includegraphics[scale = 0.25]{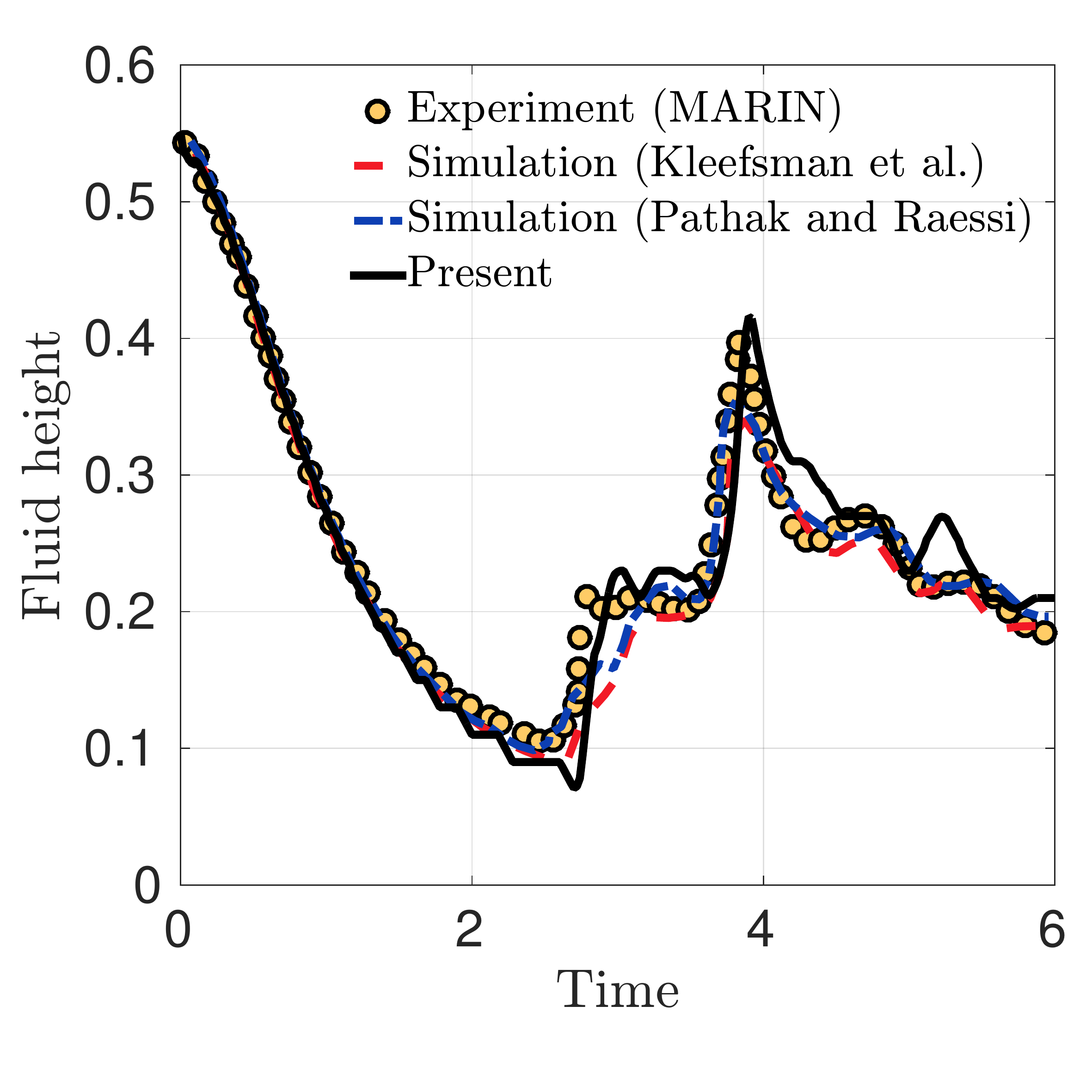}
    \label{h4_sensor}
  }
 \caption{Temporal evolution of water height measured at probes
 \subref{h2_sensor} H$2$, and
 \subref{h4_sensor} H$4$
 for a 3D water column impacting a stationary obstacle
 (see Fig.~\ref{fig_dam_break_schematic} and Table~\ref{tab_probes});
 ($\bullet$, yellow) experimental data from MARIN;
(\texttt{---}, red)  simulation data from Kleefsman et al.~\cite{Kleefsman2005};
 (\texttt{-}$\cdot$\texttt{-}, blue) simulation data from Pathak and Raessi~\cite{Pathak16};
(---, black) present simulation data.
 }
  \label{fig_dam_break_height}
\end{figure}

The primary quantities of interest for this example are the pressure values collected from 
four probes placed on the surface of the obstacle and the water height collected from two
probes placed along the domain (Fig.~\ref{fig_dam_break_schematic}).
The coordinates of the pressure and height probes are shown in Table~\ref{tab_probes}.
Note that the pressure probes are located at a specific spatial location in the computational
domain, whereas the water height probes are lines extending upward from a given 2D coordinate in the $xy$-plane.
The temporal evolution of pressure and water height at these probes are shown in Figs.~\ref{fig_dam_break_pressure}
and~\ref{fig_dam_break_height}. The results are in decent agreement with the experimental data and the simulations
carried out in~\cite{Pathak16,Kleefsman2005}, with minor disagreements being explained by differences in the interface
tracking approaches and/or variations in post-processing pressure and water height data from the 
simulations~\footnote{We relied on the VisIt~\cite{VisIt} software to extract the probe data from the parallel HDF5 files.}.
With this particular case, we have demonstrated that complex surface tension and gravity driven
splashing dynamics are accurately simulated by the present numerical method.

\begin{table}
    \centering
    \caption{Locations of the pressure and water height probes 
    for the 3D water column impacting a stationary obstacle.
    Pressure probe locations are given by $3$D coordinates while
    water height probe locations are given by
    $2$D coordinates in the $xy$-plane}
    \rowcolors{2}{}{gray!10}
    \begin{tabular}{*6c}
        \toprule
        Probe & Measurement & Location\\
        \midrule
        P$1$ & Pressure & $(0.82, 0.475, 0.02)$\\
        P$3$ & Pressure & $(0.82, 0.475, 0.1)$\\
        P$5$ & Pressure & $(0.84, 0.525, 0.16)$\\
        P$7$ & Pressure & $(0.92, 0.525, 0.16)$\\
        H$2$ & Water height & $(0.992, 0.5)$\\
        H$4$ & Water height & $(2.638, 0.5)$\\
        \bottomrule
    \end{tabular}
    \label{tab_probes}
\end{table}

\section{Conclusions}
In this study, we coupled the robust multiphase flow solver of Nangia et al.~\cite{Nangia2018} with
the DLM-based immersed boundary method of Bhalla et al.~\cite{Bhalla13} to enable fast simulations of
high density ratio wave-structure interaction problems. We demonstrated that our method is applicable to a wide
range of WSI problems involving air-water interfaces, and can adequately resolve complex wave and splashing dynamics.
We were able to achieve substantially reduced computational costs by making use of adaptive mesh refinement to
capture important flow features. Various types of rigid body dynamics are modeled, including prescribed, free-translational
and free-rotational motion.

For fully-constrained motion, we showed that the ``virtual" density within the body domain can produce
spurious velocities inside the structure; this erroneous momentum eventually contaminates the flow
field and can yield inaccurate results.
To mitigate these parasitic currents, we described a well-balanced formulation of the gravitational body force based
on the density of the ``flowing" phases.
Additionally, we demonstrated the importance of consistent mass and momentum transport to eliminate numerical instabilities
for high density ratio flows, which have long plagued the multiphase flow community.

We also presented a level set method based numerical wave tank implementation. Second-order Stokes waves were
generated by using inlet velocity boundary conditions. Wave reflection and wave interference effects were mitigated 
by the use of a wave damping zone. Although not shown in this paper, we also generated waves using a relaxation 
procedure instead of inlet velocity boundary conditions, but found the results to be very similar. Generation of more 
complex waves including fifth order Stokes~\cite{Fenton1985}, cnoidal~\cite{Fenton1999}, focused~\cite{Bredmose2010}, and random waves based upon sea and 
ocean spectra~\cite{Hasselmann1973} are already underway. Moreover, all of the code development is open-source.  

Because we presented results for relatively simple geometries, constructive solid geometry concepts to compute the signed 
distance function to the surface of the immersed body sufficed. Our code also has the ability to compute the
signed distance functions from CAD and STL files directly. Moreover, the implementation allows for a finite element representation 
of the immersed body instead of unconnected Lagrangian markers. These extensions allow for complex geometries 
(such as WECs) to be represented on Cartesian grids.  Further, the solution methodology can be augmented with a 
RANS or LES turbulence model~\cite{Spalart1992,Smagorinsky1963,Deardorff1970}, which would enable simulations
of many important industrial and engineering applications such as high inertia vehicles, wave-energy converter devices, and windmills.

\section*{Acknowledgements}
A.P.S.B. acknowledges helpful discussions with Ashish Pathak for some of the example 
cases presented in this work.  N.N. and N.A.P.~acknowledge computational resources
provided by Northwestern University's Quest high performance computing
service. A.P.S.B. acknowledges the College of Engineering's Fermi high performance 
computing service at the San Diego State University. 
N.N.~acknowledges research support from the National Science Foundation 
Graduate Research Fellowship Program (NSF award DGE-1324585).
N.A.P.~acknowledges support from the National Science Foundation's
SI2 program (NSF awards OAC 1450327 and OAC 1450374).
A.P.S.B.~acknowledges research support provided by the San Diego State University.
This work also used the Extreme Science and Engineering Discovery Environment (XSEDE) Bridges, at the
Pittsburgh Supercomputing Center, and Comet at the San Diego Supercomputer Center
through allocation TG-ASC170023, which is supported by National 
Science Foundation grant number ACI-1548562~\cite{XSEDE2014}.



\section*{Bibliography}
\begin{flushleft}
 \bibliography{VCINS_DF}
\end{flushleft}

\end{document}